\documentclass[aps,prd,showpacs,superscriptaddress,twocolumn,nofootinbib]{revtex4} 

\usepackage{graphicx}
\usepackage{natbib}
\usepackage{slashed}
\usepackage{hyperref}

\usepackage[tight,FIGTOPCAP]{subfigure}

\def\ttbar {$t{\bar{t}}$}
\def\ppbar {$p{\bar{p}}$}

\def\Wbbbar {$Wb{\bar{b}}$}
\def\Wccbar {$Wc{\bar{c}}$}
\def\Wbb {$Wb{\bar{b}}$}
\def\Wcc {$Wc{\bar{c}}$}
\def\Wc {$Wc$}

\def\sigt {$\sigma_t$}
\def\sigs {$\sigma_s$}

\def\Vtb {$V_{tb}$}

\def \WHlvbb {$ WH \rightarrow \ell \nu b{\bar{b}} $}

\def\gevcc {GeV/$c^2$}
\def\gevc {GeV/$c$}
\def\fb {fb$^{-1}$}
\def\lumi {\mathcal{L}_\mathrm{int}}
\def\ETn {E_{\rm T}}
\def\ET {$E_{\rm T}$}
\def\pt {$p_{\rm T}$}

\def\EtMiss {\slashed{E}_{\rm T}}  
\def\EtMissSig {\slashed{E}_{\rm T,sig}}
\def\EtMissVec {\vec{\slashed{E}}_{\rm T}}  
\def\met {$\not\!\! E_{\rm T} $}

\def\LambdaQCD {$\Lambda_{\mathrm{QCD}}$}

\def\Checkmark {X}

\begin{document}
\title{Observation of Single Top Quark Production and Measurement of {\boldmath $|V_{tb}|$} with CDF}

\affiliation{Institute of Physics, Academia Sinica, Taipei, Taiwan 11529, Republic of China} 
\affiliation{Argonne National Laboratory, Argonne, Illinois 60439} 
\affiliation{University of Athens, 157 71 Athens, Greece} 
\affiliation{Institut de Fisica d'Altes Energies, Universitat Autonoma de Barcelona, E-08193, Bellaterra (Barcelona), Spain} 
\affiliation{Baylor University, Waco, Texas  76798} 
\affiliation{Istituto Nazionale di Fisica Nucleare Bologna, $^{dd}$University of Bologna, I-40127 Bologna, Italy} 
\affiliation{Brandeis University, Waltham, Massachusetts 02254} 
\affiliation{University of California, Davis, Davis, California  95616} 
\affiliation{University of California, Los Angeles, Los Angeles, California  90024} 
\affiliation{University of California, San Diego, La Jolla, California  92093} 
\affiliation{University of California, Santa Barbara, Santa Barbara, California 93106} 
\affiliation{Instituto de Fisica de Cantabria, CSIC-University of Cantabria, 39005 Santander, Spain} 
\affiliation{Carnegie Mellon University, Pittsburgh, PA  15213} 
\affiliation{Enrico Fermi Institute, University of Chicago, Chicago, Illinois 60637}
\affiliation{Comenius University, 842 48 Bratislava, Slovakia; Institute of Experimental Physics, 040 01 Kosice, Slovakia} 
\affiliation{Joint Institute for Nuclear Research, RU-141980 Dubna, Russia} 
\affiliation{Duke University, Durham, North Carolina  27708} 
\affiliation{Fermi National Accelerator Laboratory, Batavia, Illinois 60510} 
\affiliation{University of Florida, Gainesville, Florida  32611} 
\affiliation{Laboratori Nazionali di Frascati, Istituto Nazionale di Fisica Nucleare, I-00044 Frascati, Italy} 
\affiliation{University of Geneva, CH-1211 Geneva 4, Switzerland} 
\affiliation{Glasgow University, Glasgow G12 8QQ, United Kingdom} 
\affiliation{Harvard University, Cambridge, Massachusetts 02138} 
\affiliation{Division of High Energy Physics, Department of Physics, University of Helsinki and Helsinki Institute of Physics, FIN-00014, Helsinki, Finland} 
\affiliation{University of Illinois, Urbana, Illinois 61801} 
\affiliation{The Johns Hopkins University, Baltimore, Maryland 21218} 
\affiliation{Institut f\"{u}r Experimentelle Kernphysik, Karlsruhe Institute of Technology, D-76131 Karlsruhe, Germany} 
\affiliation{Center for High Energy Physics: Kyungpook National University, Daegu 702-701, Korea; Seoul National University, Seoul 151-742, Korea; Sungkyunkwan University, Suwon 440-746, Korea; Korea Institute of Science and Technology Information, Daejeon 305-806, Korea; Chonnam National University, Gwangju 500-757, Korea; Chonbuk National University, Jeonju 561-756, Korea} 
\affiliation{Ernest Orlando Lawrence Berkeley National Laboratory, Berkeley, California 94720} 
\affiliation{University of Liverpool, Liverpool L69 7ZE, United Kingdom} 
\affiliation{University College London, London WC1E 6BT, United Kingdom} 
\affiliation{Centro de Investigaciones Energeticas Medioambientales y Tecnologicas, E-28040 Madrid, Spain} 
\affiliation{Massachusetts Institute of Technology, Cambridge, Massachusetts  02139} 
\affiliation{Institute of Particle Physics: McGill University, Montr\'{e}al, Qu\'{e}bec, Canada H3A~2T8; Simon Fraser University, Burnaby, British Columbia, Canada V5A~1S6; University of Toronto, Toronto, Ontario, Canada M5S~1A7; and TRIUMF, Vancouver, British Columbia, Canada V6T~2A3} 
\affiliation{University of Michigan, Ann Arbor, Michigan 48109} 
\affiliation{Michigan State University, East Lansing, Michigan  48824}
\affiliation{Institution for Theoretical and Experimental Physics, ITEP, Moscow 117259, Russia} 
\affiliation{University of New Mexico, Albuquerque, New Mexico 87131} 
\affiliation{Northwestern University, Evanston, Illinois  60208} 
\affiliation{The Ohio State University, Columbus, Ohio  43210} 
\affiliation{Okayama University, Okayama 700-8530, Japan} 
\affiliation{Osaka City University, Osaka 588, Japan} 
\affiliation{University of Oxford, Oxford OX1 3RH, United Kingdom} 
\affiliation{Istituto Nazionale di Fisica Nucleare, Sezione di Padova-Trento, $^{ee}$University of Padova, I-35131 Padova, Italy} 
\affiliation{LPNHE, Universite Pierre et Marie Curie/IN2P3-CNRS, UMR7585, Paris, F-75252 France} 
\affiliation{University of Pennsylvania, Philadelphia, Pennsylvania 19104}
\affiliation{Istituto Nazionale di Fisica Nucleare Pisa, $^{ff}$University of Pisa, $^{gg}$University of Siena and $^{hh}$Scuola Normale Superiore, I-56127 Pisa, Italy} 
\affiliation{University of Pittsburgh, Pittsburgh, Pennsylvania 15260} 
\affiliation{Purdue University, West Lafayette, Indiana 47907} 
\affiliation{University of Rochester, Rochester, New York 14627} 
\affiliation{The Rockefeller University, New York, New York 10021} 
\affiliation{Istituto Nazionale di Fisica Nucleare, Sezione di Roma 1, $^{ii}$Sapienza Universit\`{a} di Roma, I-00185 Roma, Italy} 

\affiliation{Rutgers University, Piscataway, New Jersey 08855} 
\affiliation{Texas A\&M University, College Station, Texas 77843} 
\affiliation{Istituto Nazionale di Fisica Nucleare Trieste/Udine, I-34100 Trieste, $^{jj}$University of Trieste/Udine, I-33100 Udine, Italy} 
\affiliation{University of Tsukuba, Tsukuba, Ibaraki 305, Japan} 
\affiliation{Tufts University, Medford, Massachusetts 02155} 
\affiliation{Waseda University, Tokyo 169, Japan} 
\affiliation{Wayne State University, Detroit, Michigan  48201} 
\affiliation{University of Wisconsin, Madison, Wisconsin 53706} 
\affiliation{Yale University, New Haven, Connecticut 06520} 
\author{T.~Aaltonen}
\affiliation{Division of High Energy Physics, Department of Physics, University of Helsinki and Helsinki Institute of Physics, FIN-00014, Helsinki, Finland}
\author{J.~Adelman}
\affiliation{Enrico Fermi Institute, University of Chicago, Chicago, Illinois 60637}
\author{B.~\'{A}lvarez~Gonz\'{a}lez$^w$}
\affiliation{Instituto de Fisica de Cantabria, CSIC-University of Cantabria, 39005 Santander, Spain}
\author{S.~Amerio$^{ee}$}
\affiliation{Istituto Nazionale di Fisica Nucleare, Sezione di Padova-Trento, $^{ee}$University of Padova, I-35131 Padova, Italy} 

\author{D.~Amidei}
\affiliation{University of Michigan, Ann Arbor, Michigan 48109}
\author{A.~Anastassov}
\affiliation{Northwestern University, Evanston, Illinois  60208}
\author{A.~Annovi}
\affiliation{Laboratori Nazionali di Frascati, Istituto Nazionale di Fisica Nucleare, I-00044 Frascati, Italy}
\author{J.~Antos}
\affiliation{Comenius University, 842 48 Bratislava, Slovakia; Institute of Experimental Physics, 040 01 Kosice, Slovakia}
\author{G.~Apollinari}
\affiliation{Fermi National Accelerator Laboratory, Batavia, Illinois 60510}
\author{J.~Appel}
\affiliation{Fermi National Accelerator Laboratory, Batavia, Illinois 60510}
% ...

\collaboration{CDF Collaboration\footnote{With visitors from $^a$University of Massachusetts Amherst, Amherst, Massachusetts 01003,
$^b$Universiteit Antwerpen, B-2610 Antwerp, Belgium, 
$^c$University of Bristol, Bristol BS8 1TL, United Kingdom,
$^d$Chinese Academy of Sciences, Beijing 100864, China, 
$^e$Istituto Nazionale di Fisica Nucleare, Sezione di Cagliari, 09042 Monserrato (Cagliari), Italy,
$^f$University of California Irvine, Irvine, CA  92697, 
$^g$University of California Santa Cruz, Santa Cruz, CA  95064, 
$^h$Cornell University, Ithaca, NY  14853, 
$^i$University of Cyprus, Nicosia CY-1678, Cyprus, 
$^j$University College Dublin, Dublin 4, Ireland,
$^k$University of Edinburgh, Edinburgh EH9 3JZ, United Kingdom, 
$^l$University of Fukui, Fukui City, Fukui Prefecture, Japan 910-0017,
$^m$Kinki University, Higashi-Osaka City, Japan 577-8502,
$^n$Universidad Iberoamericana, Mexico D.F., Mexico,
$^o$University of Iowa, Iowa City, IA  52242,
$^p$Kansas State University, Manhattan, KS 66506,
$^q$Queen Mary, University of London, London, E1 4NS, England,
$^r$University of Manchester, Manchester M13 9PL, England,
$^s$Muons, Inc., Batavia, IL 60510, 
$^t$Nagasaki Institute of Applied Science, Nagasaki, Japan, 
$^u$University of Notre Dame, Notre Dame, IN 46556,
$^v$Obninsk State University, Obninsk, Russia,
$^w$University de Oviedo, E-33007 Oviedo, Spain, 
$^x$Texas Tech University, Lubbock, TX  79609, 
$^y$IFIC(CSIC-Universitat de Valencia), 56071 Valencia, Spain,
$^z$Universidad Tecnica Federico Santa Maria, 110v Valparaiso, Chile,
$^{aa}$University of Virginia, Charlottesville, VA  22906,
$^{bb}$Bergische Universit\"at Wuppertal, 42097 Wuppertal, Germany,
$^{cc}$Yarmouk University, Irbid 211-63, Jordan,
$^{kk}$On leave from J.~Stefan Institute, Ljubljana, Slovenia
}}
\noaffiliation

\date{April 5, 2010}

\begin{abstract}
We report the observation of electroweak single top quark production
in 3.2~\fb\ of \ppbar\ collision data collected by the Collider
Detector at Fermilab at $\sqrt{s}=1.96$~TeV.  Candidate events in the
$W$+jets topology with a leptonically decaying $W$ boson
are classified as signal-like by four parallel
analyses based on likelihood functions, matrix elements, neural
networks, and boosted decision trees.  These results are combined using
a super discriminant analysis based on genetically evolved neural
networks in order to improve the sensitivity.  This combined result
is further combined
with that of a search for a single top quark signal
in an orthogonal sample of events with missing transverse 
energy plus jets and no charged lepton.  We observe a signal
consistent with the standard model prediction but inconsistent with
the background-only model by 5.0 standard deviations, with a median
expected sensitivity in excess of 5.9 standard deviations.  We measure
a production cross section of $2.3^{+0.6}_{-0.5}(\mathrm{stat+sys})$~pb, extract
the CKM matrix element value \mbox{$|V_{tb}|=0.91^{+0.11}_{-0.11}
(\mathrm{stat+sys})\pm 0.07(\mathrm{theory})$}, and set a lower limit
$|V_{tb}|>0.71$ at the 95\% confidence level, assuming $m_t=175$~GeV/$c^2$.

\end{abstract}


\noaffiliation

\maketitle

\tableofcontents
\clearpage

\section{\label{sec:Introduction} Introduction}

The top quark is the most massive known elementary particle.  Its mass, $m_t$, 
is \mbox{$173.3 \pm 1.1$~\gevcc~\cite{topmass10}}, about forty times
larger than that of the bottom quark, the second-most massive standard
model (SM) fermion.  The top quark's large mass, at the scale of
electroweak symmetry breaking, hints that it may play a role in the
mechanism of mass generation.
The presence of the top quark
was established in 1995 by the CDF and D0 collaborations with
approximately 60~pb$^{-1}$ of \ppbar\ data collected per collaboration at
$\sqrt{s}=1.8$~TeV~\cite{Abe:1995hr,Abachi:1995iq} in Run~I
at the Fermilab Tevatron.  The production mechanism
used in the observation of the top quark was \ttbar\ pair production via
the strong interaction.  

Since then, larger data samples have enabled detailed study of the
top quark.  The \ttbar\ production cross section~\cite{Abulencia:2006in}, 
the top quark's mass~\cite{topmass10},
the top quark decay branching fraction to $Wb$~\cite{Acosta:2005hr},
and the polarization of $W$ bosons in top quark
decay~\cite{Abulencia:2005xf} have been measured precisely.
  Nonetheless, many properties of the
top quark have not yet been tested as precisely.  In particular, the
Cabibbo-Kobayashi-Maskawa (CKM) matrix element \Vtb\ remains poorly
constrained by direct measurements~\cite{Amsler:2008zz}.  The strength of
the coupling, $|V_{tb}|$,
governs the decay rate of the top quark and its decay width into
$Wb$; other decays are expected to have much smaller branching
fractions.  Using measurements of the other CKM matrix elements,
and assuming a three-generation SM with a $3\times 3$ 
unitary CKM matrix, $|V_{tb}|$ is expected to be very close to unity.

\begin{figure}[t]
\includegraphics[width=\columnwidth]{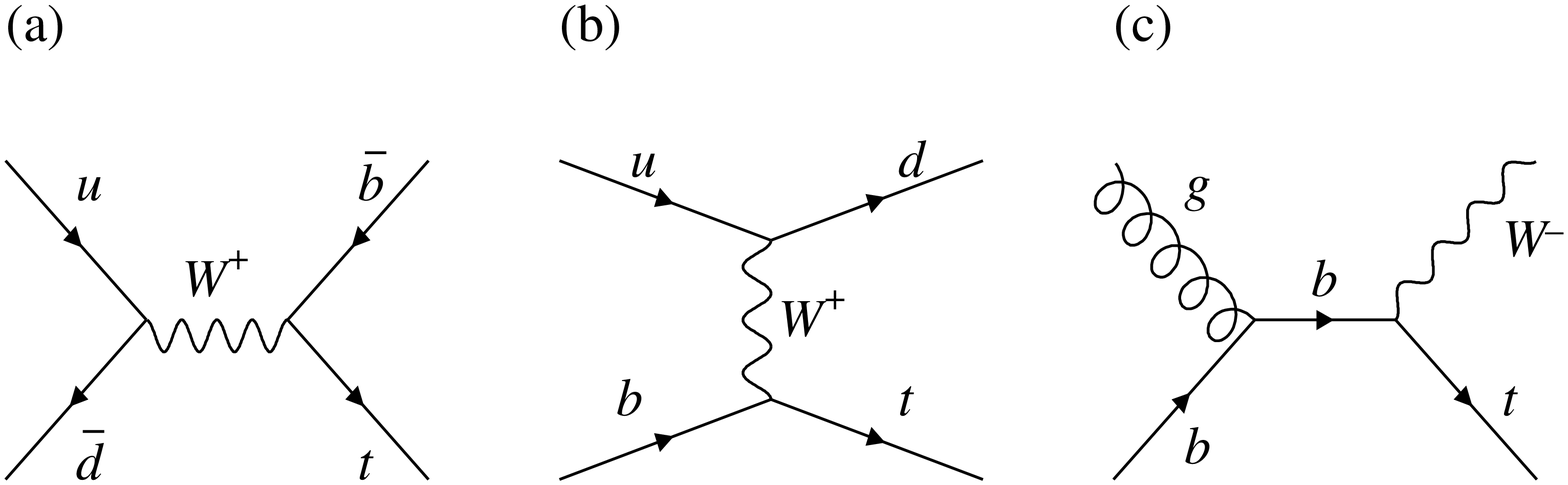}
\caption{Representative Feynman diagrams of single top quark
production.  Figures (a) and (b) are
$s$- and $t$-channel processes, respectively, while figure (c) is
associated $Wt$ production, which contributes a small amount to the expected
cross section at the Tevatron.}
\label{fig:sttfeyn}
\end{figure}

Top quarks are also expected to be produced singly in \ppbar\ collisions
via weak, charged-current interactions.  The dominant processes at the
Tevatron are the $s$-channel process, shown in Fig.~\ref{fig:sttfeyn}(a), and the 
$t$-channel process~\cite{Willenbrock:1986cr}, 
shown in Fig.~\ref{fig:sttfeyn}(b).  The
next-to-leading-order (NLO) cross sections for these two processes are
\sigs$=0.88\pm 0.11$~pb and \sigt$=1.98\pm 0.25$~pb,
respectively~\cite{Harris:2002md,Sullivan:2004ie}.  This cross section is the sum
of the single $t$ and the single ${\bar{t}}$ predictions.  Throughout this paper,
charge conjugate states are implied; all cross sections and yields are shown summed
over charge conjugate states.
A calculation has been performed resumming
soft gluon corrections and calculating finite-order expansions through 
next-to-next-to-next-to-leading order (NNNLO)~\cite{Kidonakis:2006bu}, yielding
\sigs$=0.98\pm 0.04$~pb and \sigt$=2.16\pm 0.12$~pb, also assuming $m_t=175$~GeV/$c^2$.  
Newer calculations are also available~\cite{Kidonakis:2007wg,Kidonakis:2009mx,Kidonakis:2010tc}.
  A third process, the
associated production of a $W$ boson and a top quark, shown in Fig.~\ref{fig:sttfeyn}(c),
 has a very small expected cross section at the Tevatron.

Measuring the two cross sections $\sigma_s$ and $\sigma_t$
 provides a direct determination of $|V_{tb}|$, allowing 
an overconstrained test of the unitarity of the CKM matrix, as well as an indirect
determination of the top quark's lifetime.  We assume that the top quark decays
to $Wb$ 100\% of the time in order to measure the production cross sections.
This assumption does not constrain $|V_{tb}|$ to be near unity, but instead it is
the same as assuming $|V_{tb}|^2 \gg |V_{ts}|^2 + |V_{td}|^2$.
Many extensions to the SM predict measurable deviations of
\sigs\ or \sigt\ from their SM values.  One of the simplest of these
is the hypothesis that a fourth generation of fermions exists beyond
the three established ones.  Aside from the constraint that its
neutrino must be heavier than $M_Z/2$~\cite{Z-Pole} and that the
quarks must escape current experimental limits,
 the existence of a fourth
generation of fermions remains possible.  If these additional
sequential fermions exist, then a $4\times 4$ version of the CKM
matrix would be unitary, and the $3\times 3$ submatrix may not necessarily be
unitary.  The presence of a fourth generation would in general
reduce $|V_{tb}|$,
thereby reducing single top quark production cross sections \sigs\ and \sigt.
Precision electroweak constraints provide some information on possible values of
$|V_{tb}|$ in this extended scenario~\cite{Alwall:2006bx}, but a direct
measurement provides a test with no additional model dependence.

Other new physics scenarios predict larger values of \sigs\ and \sigt\
than those expected in the SM.  A flavor-changing $Ztc$ coupling, for
example, would manifest itself in the production of
$p{\bar{p}}\rightarrow t{\bar{c}}$ events, which may show up in either
the measured value of \sigs\ or \sigt\, depending on the relative
acceptances of the measurement channels.  An additional charged gauge
boson $W^\prime$ may also enhance the production cross sections.  A review
of new physics models affecting the single top quark production cross
section and polarization properties is given in~\cite{Tait:2000sh}.

Even in the absence of new physics, assuming the SM constraints on $|V_{tb}|$, a measurement of
the $t$-channel single top production cross section provides a test of the
$b$ parton distribution function of the proton.

Single top quark production is one of the
background processes in the search for the Higgs boson $H$ in the \WHlvbb\ channel, since they share the same
final state, and a direct measurement of single top quark production
may improve the sensitivity of the Higgs boson search.  
Furthermore, the backgrounds to the
single top quark search are backgrounds to the Higgs boson search.  Careful understanding of these
backgrounds lays the groundwork for future Higgs boson searches.
Since the single top quark processes
have larger cross sections than the Higgs boson signal in the 
\WHlvbb\ mode~\cite{Assamagan:2004mu}, and since the
single top signal is more distinct from the backgrounds than the Higgs boson signal is,
we must pass the milestone of observing single top quark production along the way to testing for
Higgs boson production.

Measuring the single top quark cross section is well motivated but it is also
extremely challenging at the Tevatron.  The total production cross section
is expected to be about one-half of that of \ttbar\ production~\cite{Kidonakis:2003qe}, and with only
one top quark in the final state instead of two, the signal is far less distinct from
the dominant background processes than \ttbar\ production is.  The rate at which a $W$ boson is produced along
with jets, at least one of which must have a displaced vertex which
passes our requirements for $B$ hadron identification 
(we say in this paper that such jets are $b$-tagged), is approximately
twelve times the signal rate.   The {\it a priori} uncertainties on the background processes
are about a factor of three larger than the expected signal rate.  In order to expect to observe
single top quark production, the background rates must be small and well constrained, and
the expected signal must be much larger than the uncertainty on the background.
A much more pure sample of signal events therefore must be 
separated from the background processes in order to make
observation possible.

Single top quark production is characterized
by a number of kinematic properties.  The top quark mass is known, and precise predictions
of the distributions of observable quantities for the top quark and the recoil products are also available.
Top quarks produced singly via the weak interaction are expected to be nearly
100\% polarized~\cite{Mahlon:1996pn,Stelzer:1998ni}.  The background $W$+jets and \ttbar\ processes
have characteristics which differ from those of single top quark production.
Kinematic properties, coupled with the $b$-tagging requirement, provide the keys to
purification of the signal.  Because signal events differ from background events
in several ways, such as in the distribution of the invariant mass of
the final state objects assigned to be the decay products of the top quark and the
rapidity of the recoiling jets, and because the task of observing single top quark production
requires the maximum separation, we apply multivariate
techniques.   The techniques described in this paper together achieve a 
signal-to-background ratio of more than 5:1 in a subset of events with a 
significant signal expectation.  This high purity is needed in order to
overcome the uncertainty in the background prediction.   

The effect of the background
uncertainty is reduced by fitting for both the signal and the background rates together
to the observed data distributions, a technique which is analogous to fitting the background
in the sidebands of a mass peak, but which is applied in this case to multivariate discriminant
distributions.  Uncertainties are incurred in this procedure -- the shapes of the background
distributions are imperfectly known from simulations.  We check in detail the modeling of the distributions of
the inputs and the outputs of the multivariate techniques, using events passing our selection
requirements, and also separately
using events in control samples depleted in signal.  We also check the modeling of the correlations between
pairs of these variables.  In general we find excellent
agreement, with some imperfections.  We assess uncertainties on the shapes of the
discriminant outputs both from {\it a priori} uncertain parameters in the modeling, as well
as from discrepancies observed in the modeling of the data by the Monte Carlo simulations.
These shape uncertainties are included
in the signal rate extraction and in the calculation of the significance.

Both the CDF and the D0 Collaborations have searched for single top quark
production in $p{\bar{p}}$ collision data taken at $\sqrt{s}=1.96$~TeV in Run~II
at the Fermilab Tevatron.
The D0 Collaboration reported evidence for the production of 
single top quarks in 0.9~fb$^{-1}$ of
data~\cite{Abazov:2006gd,Abazov:2008kt}, and observation of the process in
2.3~fb$^{-1}$~\cite{Abazov:2009ii}.  More recently, D0 has conducted a measurement of
the single top production cross section in the $\tau$+jets final state
using 4.8~fb$^{-1}$ of data~\cite{Abazov:2009nu}.
  The CDF Collaboration reported 
evidence in 2.2~fb$^{-1}$ of 
data~\cite{Aaltonen:2008sy} and observation in 3.2~fb$^{-1}$ of 
data~\cite{Aaltonen:2009jj}.  This paper
describes in detail the four $W$+jets analyses of~\cite{Aaltonen:2009jj};
the analyses are based on
multivariate likelihood functions (LF),
artificial neural networks (NN),
matrix elements (ME),
and boosted decision trees (BDT).  These analyses select events 
with a high-$p_{\rm T}$ charged lepton, large missing transverse energy $\EtMiss$,
and two or more jets, at least one of which is $b$-tagged.
Each analysis separately measures the single top quark production cross section and calculates
the significance of the observed excess.  We report here a single set of results and therefore
must combine the information from each of the four analyses.
Because there is 100\% overlap in the
data and Monte Carlo events selected by the analyses, a natural combination technique is 
to use the individual analyses' discriminant outputs as inputs to a super discriminant 
function evaluated for each event.
 The distributions of this
super discriminant are then interpreted in the same way as those of each of the four component
analyses.

A separate analysis is conducted on events without an identified charged lepton, in a data sample
which corresponds to 2.1~fb$^{-1}$ of data. 
Missing transverse energy plus jets, one of which is $b$-tagged, 
is the signature used for this fifth analysis (MJ), which is described
in detail in~\cite{MET_jets}.  There is no overlap of events selected by the MJ analysis
and the $W$+jets analyses.  The results of this analysis are combined with the results of
the  super discriminant analysis to yield the final results:
the measured total cross section $\sigma_s+\sigma_t$, $|V_{tb}|$, 
the separate cross sections $\sigma_s$ and $\sigma_t$, and the statistical significance of the
excess.  With the combination of all analyses, we observe single top quark production
with a significance of 5.0 standard deviations.

The analyses described in this paper were blind to the selected data when they were optimized
for their expected sensitivities. 
Furthermore, since the publication of the 2.2~fb$^{-1}$ $W$+jets results~\cite{Aaltonen:2008sy},
the event selection requirements, the multivariate discriminants for the analyses shared with that
result, and the systematic uncertainties remain unchanged; new data were added without further
optimization or retraining.  When the 2.2~fb$^{-1}$ results were validated, they were done so in a blind
fashion. The distributions of all relevant variables were first checked for accurate modeling by our
simulations and data-based background estimations in 
control samples of data that do not overlap with the selected signal sample.  Then the 
distributions of the discriminant input variables, and also other variables,
were checked in the sample of events passing the selection requirements.  After that,
the modeling of the low signal-to-background portions of the final
output histograms was checked.  Only after all of these validation steps were completed
were the data in the most sensitive regions revealed.
Two new analyses, BDT and MJ, have been added for this paper, and they were validated in a similar way.

This paper is organized as follows: Section~\ref{sec:Detector} describes the CDF~II detector,
Section~\ref{sec:Selection} describes the event selection, 
Section~\ref{sec:SignalModel} describes the simulation of 
 signal events and the acceptance of the signal,
Section~\ref{sec:Background} describes the background rate and kinematic shape modeling,
Section~\ref{sec:btagger} describes a neural-network flavor separator which helps separate
  $b$ jets from others,
Section~\ref{sec:Multivariate} describes the four $W$+jets multivariate analysis techniques,
Section~\ref{sec:Systematics} describes the systematic uncertainties we assess, 
Section~\ref{sec:Interpretation} describes the statistical techniques for extraction of
the signal cross section and the significance,
Section~\ref{sec:Combination} describes the super discriminant, 
Section~\ref{sec:1D} presents our results for the cross section, $|V_{tb}|$, and the significance,
Section~\ref{sec:2D} describes an extraction of $\sigma_s$ and $\sigma_t$ in a joint fit, and
Section~\ref{sec:Summary} summarizes our results.

\section{\label{sec:Detector} The CDF II Detector}

\begin{figure*}[t]
\includegraphics[width=0.7\textwidth]{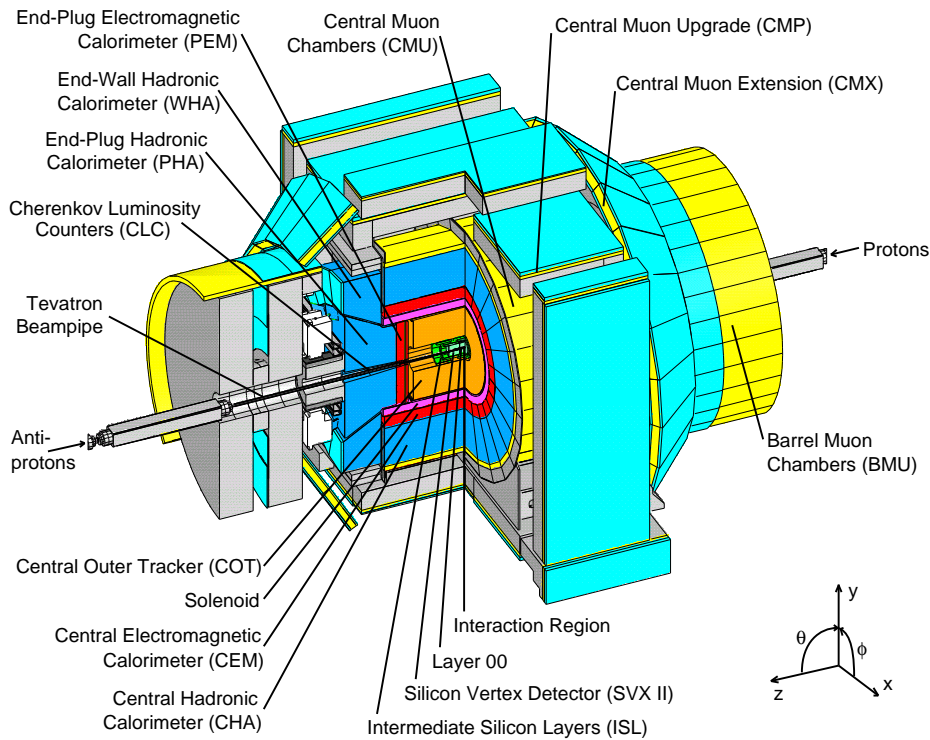}
\caption{Cutaway isometric view of the CDF~II detector.}
\label{fig:cdfisolabel}
\end{figure*}

The CDF~II detector~\cite{Abulencia:2005ix,Acosta:2004yw,Acosta:2004hw} is a general-purpose particle detector with
azimuthal and forward-backward symmetry.  Positions and angles are
expressed in a cylindrical coordinate system, with the $z$~axis directed along the proton beam.
The azimuthal
angle $\phi$ around the beam axis is defined with respect to a horizontal ray running outwards
from the center of the Tevatron, and radii are measured with respect to the beam
axis.   The polar angle $\theta$ is defined
with respect to the proton beam direction, and the pseudorapidity $\eta$ is
defined to be $\eta=-\ln\left[\tan(\theta/2)\right]$.
 The transverse energy (as measured by the calorimetry) and momentum 
(as measured by the tracking systems) of a
particle are defined as $E_{\rm T}=E\sin\theta$ and $p_{\rm T}=p\sin\theta$, respectively.
Figure~\ref{fig:cdfisolabel} shows a cutaway isometric view of the CDF~II detector.

 A silicon tracking system and an
open-cell drift chamber are used to measure the momenta of charged
particles. The CDF~II silicon tracking system consists of three
subdetectors: a layer of single-sided silicon microstrip detectors,
located immediately outside the beam pipe (layer~00)~\cite{Hill:2004qb}, 
a five-layer, double-sided silicon microstrip
detector (SVX~II) covering the region between 2.5 to 11 cm from the
beam axis~\cite{Sill:2000zz}, and intermediate silicon layers
(ISL)~\cite{Affolder:2000tj} located at radii between 19~cm and 29~cm
which provide linking between track segments in the drift chamber and
the SVX~II. The typical intrinsic hit resolution of the silicon detector
is 11~$\mu$m. The impact parameter resolution is $\sigma$($d_0$)~$\approx$~40~$\mu$m,
of which approximately 35~$\mu$m is due to the transverse size of the Tevatron interaction region.
The entire system reconstructs tracks in three dimensions
with the precision needed to identify displaced vertices
associated with $b$ and $c$ hadron decays.

The central outer tracker
(COT)~\cite{Affolder:2003ep}, the main tracking detector of CDF~II, is
an open-cell drift chamber, 3.1~m in length.  It is segmented into eight concentric
superlayers.  The drift medium is a mixture of argon and ethane. Sense wires are
arranged in eight alternating axial and $\pm$ 2$^{\circ}$ stereo
superlayers with twelve layers of wires in each. The active volume covers the
radial range from 40~cm to 137~cm.  The tracking efficiency of the COT
is nearly 100\% in the range $|\eta|\leq$ 1, and with the addition of
silicon coverage, the tracks can be detected within the range $|\eta|<1.8$.

The tracking systems are located within a
superconducting solenoid, which has a diameter of 3.0~m, and which generates a 1.4~T magnetic field
parallel to the beam axis. The magnetic field is used to measure the
charged particle momentum transverse to the beamline. The momentum resolution
is $\sigma$(\pt)/\pt~$\approx$~0.1\%$\cdot$\pt\ for tracks within $|\eta|\leq$1.0 
and degrades with increasing $|\eta|$.

Front electromagnetic lead-scintillator sampling calorimeters~\cite{Balka:1987ty,Albrow:2001jw}
and rear hadronic iron-scintillator sampling calorimeters~\cite{Bertolucci:1987zn}
surround the solenoid and measure the energy flow of interacting particles.  They
are segmented into projective towers, each one covering a small range
in pseudorapidity and azimuth.  The full array has an angular coverage
of $|\eta|<3.6$.   The central region $|\eta|<1.1$ is covered by
the central electromagnetic calorimeter (CEM) and the central and end-wall hadronic
calorimeters (CHA and WHA). The forward region $1.1<|\eta|<3.6$ is covered by the
end-plug electromagnetic calorimeter (PEM) and the end-plug hadronic
calorimeter (PHA).   Energy deposits in the electromagnetic calorimeters
are used for electron identification and energy measurement.
The energy resolution for an electron with transverse energy \ET~(measured in GeV) is given by
$\sigma$(\ET)/\ET~$\approx$~13.5\%/$\sqrt{\ETn}~\oplus$~1.5\% and
$\sigma$(\ET)/\ET~$\approx$~16.0\%/$\sqrt{\ETn}~\oplus$~1\% for electrons identified
in the CEM and PEM respectively. Jets are identified and measured through the energy
they deposit in the electromagnetic and hadronic calorimeter towers. The calorimeters
provide jet energy measurements with resolution of approximately 
$\sigma$(\ET)~$\approx$~0.1$\cdot$\ET~+~1.0~GeV \cite{CDF:inclJet}.
The CEM and PEM calorimeters have two dimensional readout strip detectors located
at shower maximum~\cite{Balka:1987ty,Apollinari:1998bg}.  These detectors provide
higher resolution position measurements of electromagnetic showers than are available
from the calorimeter tower segmentation alone, and also provide local
energy measurements.  The shower maximum detectors
contribute to the identification of electrons and photons, and help separate
them from $\pi^0$ decays.

Beyond the calorimeters resides the muon system, which provides muon
detection in the range $|\eta|<1.5$. For the analyses presented in
this article, muons are detected in four separate subdetectors. Muons
with $p_{\rm T}>1.4$~GeV$/c$ penetrating the five absorption lengths
of the calorimeter are detected in the four layers of planar multi-wire
drift chambers of the central muon detector
(CMU)~\cite{Ascoli:1987av}. Behind an additional 60~cm of steel, a
second set of four layers of drift chambers, the central muon upgrade
(CMP)~\cite{Blair:1996kx,Abulencia:2005ix}, detects muons with $p_{\rm
T}>2.2$~GeV$/c$.  The CMU and CMP cover the same part of the central
region $|\eta|<0.6$. The central muon extension
(CMX)~\cite{Blair:1996kx,Abulencia:2005ix} extends the pseudorapidity coverage of the
muon system from 0.6 to 1.0 and thus completes the coverage over the
full fiducial region of the COT. Muons with $1.0<|\eta|<1.5$
are detected by the barrel muon chambers (BMU)~\cite{Artikov:2004ew}.

The Tevatron collider luminosity is determined with multi-cell gas Cherenkov detectors~\cite{CLC}
located in the region $3.7<|\eta|<4.7$ which measure the average number
of inelastic $p{\bar{p}}$ collisions per bunch crossing.  The total uncertainty
on the luminosity is $\pm 6.0$\%, of which 4.4\% comes from the
acceptance and the operation of the luminosity monitor and 4.0\% comes from the
uncertainty of the inelastic $p{\bar{p}}$ cross section~\cite{inelppbarxs}.

\section{\label{sec:Selection} Selection of Candidate Events}

\begin{figure}[t]
\begin{center}
\includegraphics[width=0.95\columnwidth]{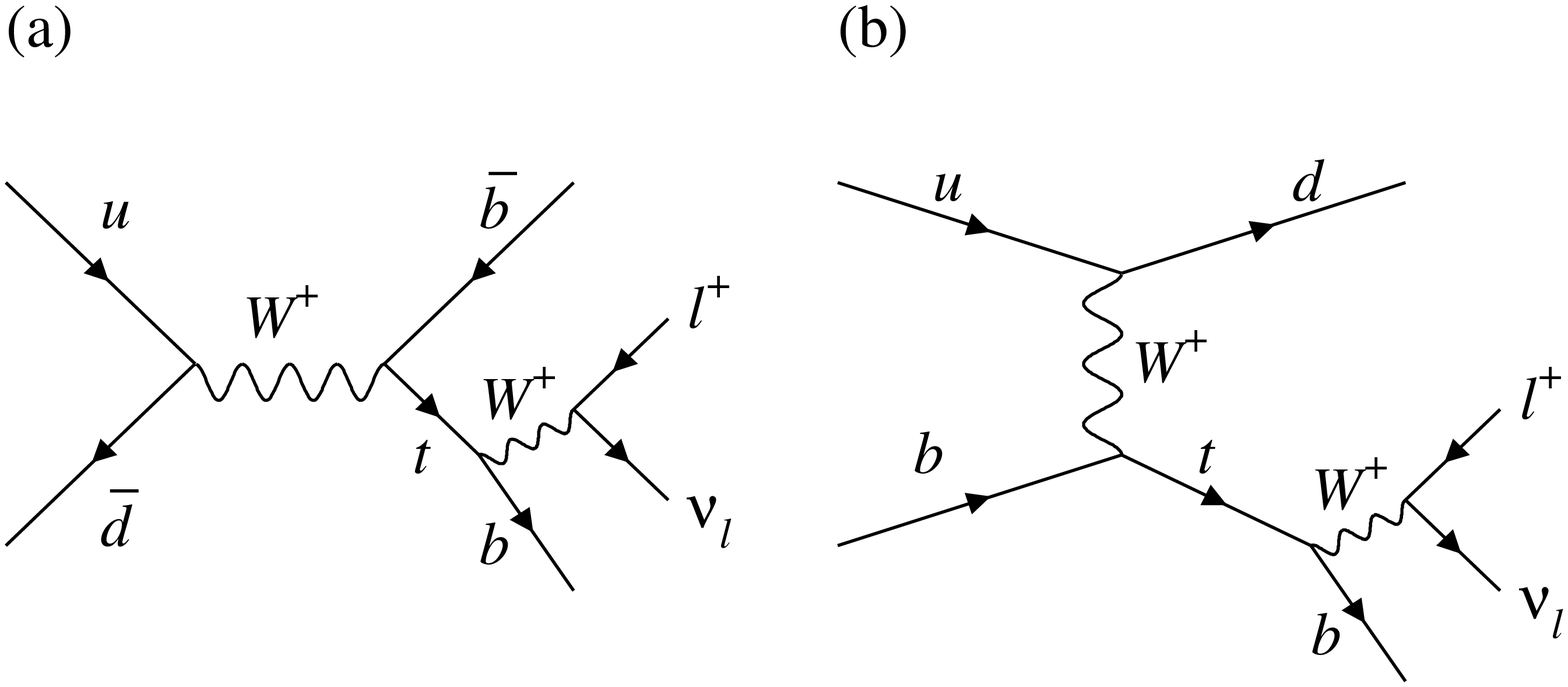}
\caption{Feynman diagrams showing the final states of the dominant
$s$-channel (a) and $t$-channel (b) processes, with leptonic $W$ boson decays.
Both final states contain a charged lepton, a neutrino,
and two jets, at least one of which originates from a $b$ quark.}
\label{fig:singletop}
\end{center}
\end{figure}

Single top quark events (see Fig.~\ref{fig:singletop}) have jets,
a charged lepton, and a neutrino in the final state.  The top
quark decays into a $W$ boson and a $b$ quark before hadronizing.  The quarks
recoiling from the top quark, and the $b$ quark from
top quark decay, hadronize to form jets, motivating our event selection
which requires two or three energetic jets (the third can come from a
radiated gluon), at least one of which is $b$-tagged, and the
decay products of a $W$ boson. In order to reduce background from
multi-jet production via the strong interaction, we focus our event selection
on the decays of the $W$ boson to $e\nu_{e}$ or $\mu\nu_{\mu}$ in
these analyses.  Such events have one charged lepton (an electron or a
muon), missing transverse energy resulting from the undetected
neutrino, and at least two jets.  These events constitute the $W$+jets
sample.  We also include the acceptance for signal and background events in which
$W\rightarrow\tau\nu_{\tau}$, and
the MJ analysis also is sensitive to $W$ boson decays to $\tau$ leptons.

Since the \ppbar\ collision rate at the Tevatron exceeds
the rate at which events can be written to tape by five orders of magnitude, CDF has
an elaborate trigger system with three levels.  The first
level uses special-purpose hardware~\cite{Thomson:2002xp} to reduce
the event rate from the effective beam-crossing frequency of 1.7~MHz to
approximately 15~kHz, the maximum rate at which the detector can be read out.  The second
level consists of a mixture of dedicated hardware and fast software
algorithms and takes advantage of the full information read out of the detector~\cite{Downing:2006xb}.
At this level the trigger rate is reduced further to less than 800~Hz.  At the third
level, a computer farm running fast versions of the offline event reconstruction algorithms
refines the trigger selections based on quantities that are nearly the same as those
used in offline analyses~\cite{GomezCeballos:2004jk}.  In particular, detector calibrations are applied before
the trigger requirements are imposed.  The third level trigger selects events 
for permanent storage at a rate of up to 200~Hz.

Many different trigger criteria are evaluated at each level, and events passing
specific criteria at one level are considered by a subset of trigger algorithms at the
next level.  A cascading set of trigger requirements is known as a trigger path.
This analysis uses the trigger paths which select events with high-\pt\ electron or
muon candidates.  The acceptance of these triggers for tau leptons is included in our rate estimates
but the triggers are not optimized for identifying tau leptons.  
An additional trigger path, which requires 
significant $\EtMiss$ plus at least two high-\pt\ jets, is also
used to add $W$+jets candidate events with non-triggered leptons, which include charged leptons
outside the fiducial volumes of the electron and muon detectors, as well as tau leptons.

The third-level central electron trigger requires a COT track with
\pt$>9$~\gevc\ matched to an energy cluster in the CEM with
\ET$>18$~GeV.  The shower profile of this cluster as measured by the shower-maximum detector is required to
be consistent with those measured using test-beam electrons.
 Electron candidates with $|\eta|>1.1$ are required to deposit more
than 20~GeV in a cluster in the PEM, and the ratio of hadronic energy to electromagnetic
energy $E_{\rm PHA} / E_{\rm PEM}$ for this cluster is required to be less than 0.075.  The
third-level muon trigger requires a COT track with \pt$>$18~\gevc\ matched to a track 
segment in the muon chambers.  The $\EtMiss$+jets trigger path requires $\EtMiss>35$~GeV and two jets with
\ET$>10$~GeV.   

After offline reconstruction, we impose further requirements
on the electron candidates in order to improve the purity of the sample.
A reconstructed track
with \pt$>9$~\gevc\ must match to a cluster in the CEM with
\ET$>$ 20~GeV. Furthermore, we require $E_{\rm HAD} / E_{\rm EM} < 0.055 + 0.00045 \times E$/GeV 
and the ratio of the energy of the cluster
to the momentum of the track $E/p$ has to be smaller than $2.0\,c$ for
track momenta $\le 50$ \gevc.  For electron candidates with tracks with
$p>50$~\gevc, no requirement on $E/p$ is made as the misidentification rate is small.
Candidate objects which fail these requirements
are more likely to be hadrons or jets than those that pass.

 Electron candidates in the forward
direction (PHX) are defined by a cluster in the PEM with $E_{\rm T}>$ 20~GeV
and $E_{\rm HAD} / E_{\rm EM} < 0.05$.  The cluster position and the
primary vertex position are combined to form a search trajectory in
the silicon tracker and seed the pattern recognition of the tracking
algorithm.  

\begin{figure*}
\begin{center}
\subfigure[]{
\includegraphics[width=0.95\columnwidth]{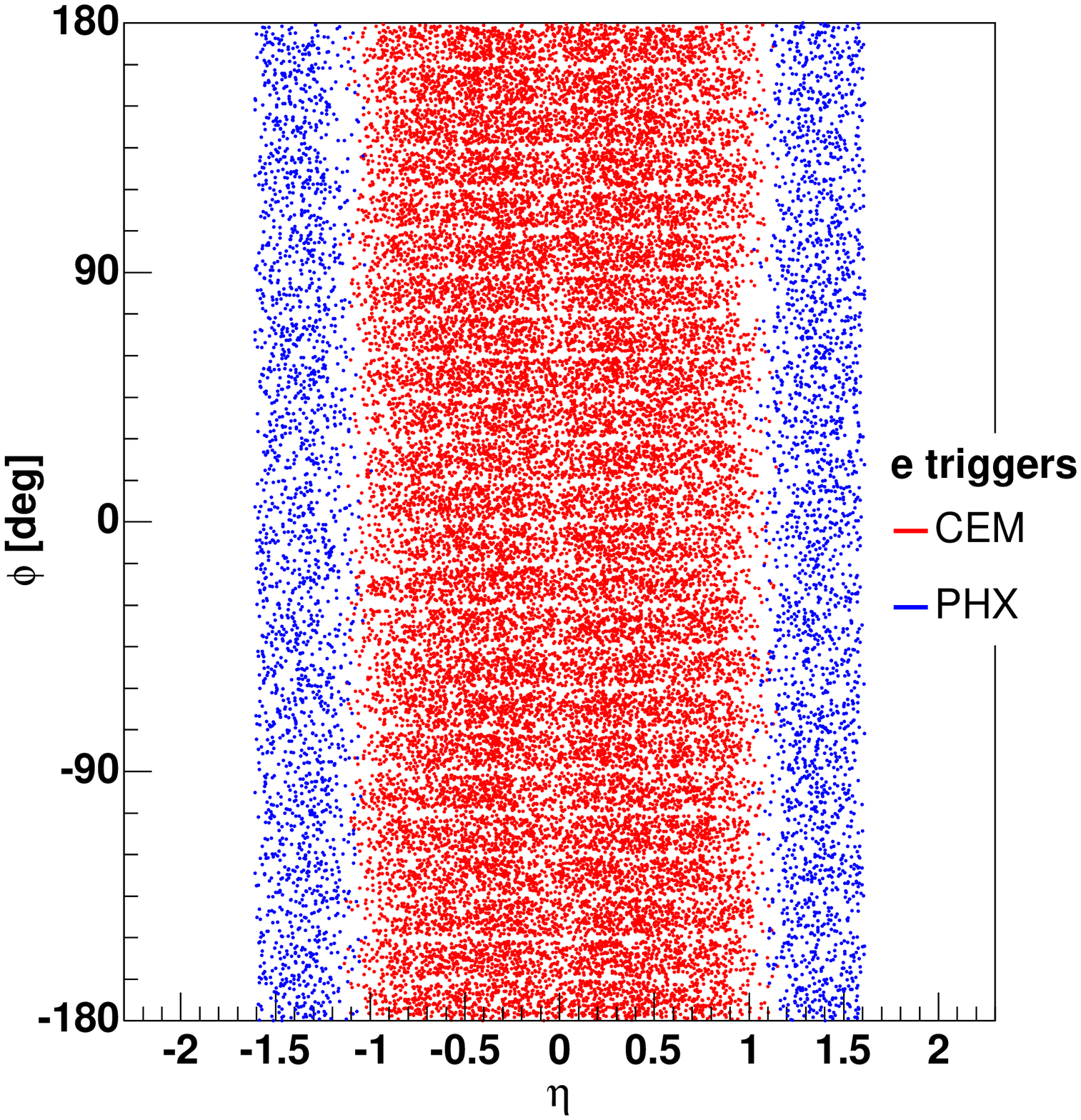}
\label{fig:eletypes}}
\subfigure[]{
\includegraphics[width=0.95\columnwidth]{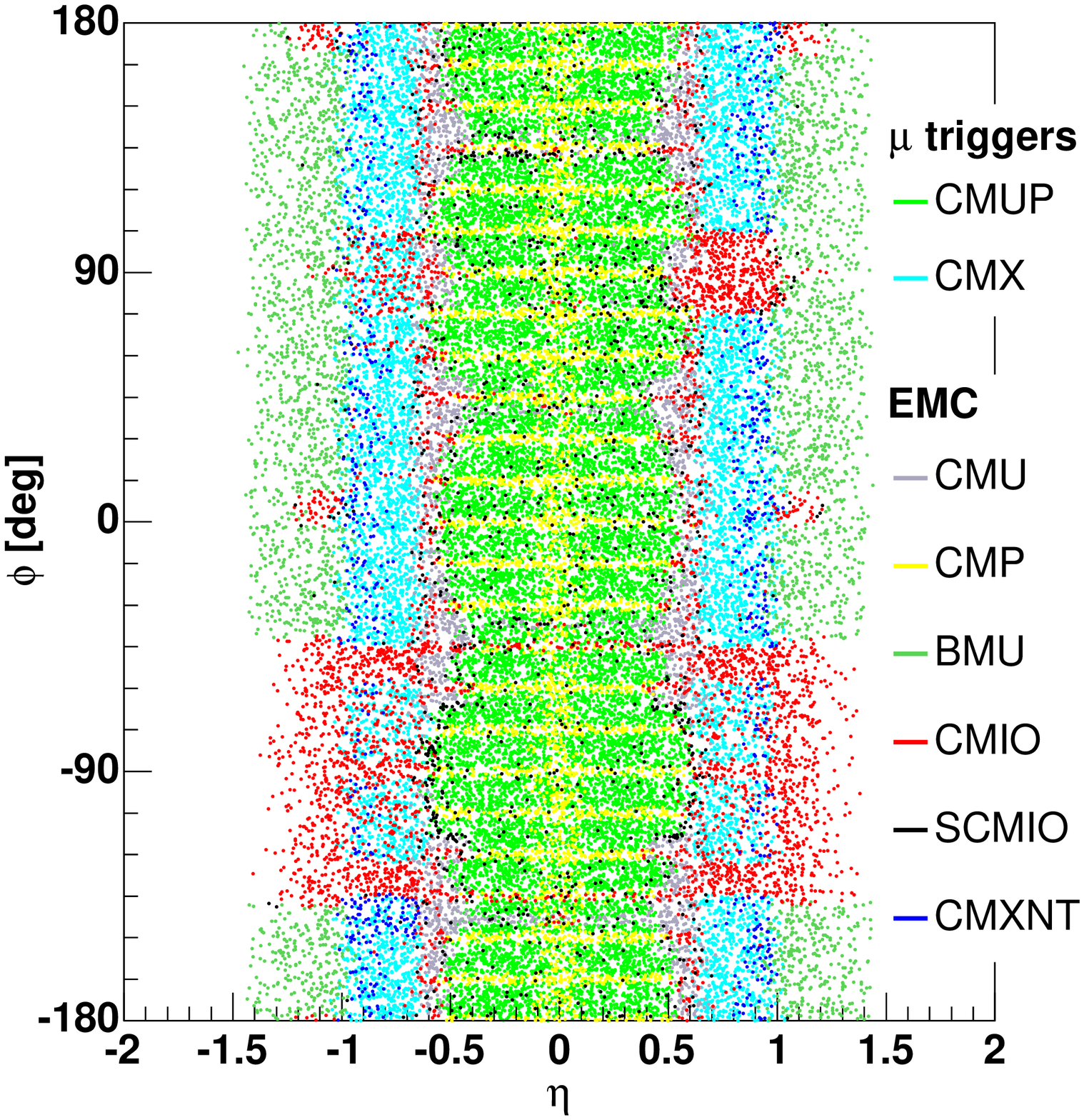}
\label{fig:mutypes}}
\end{center}
\caption{\label{fig:Triggertypes}Distributions in ($\phi-\eta$) space of
the electron (a) and muon (b) selection categories, showing the
  coverage of the detector that each lepton category provides.   The muon categories are
  more complicated due to the geometrical limitations of the several different muon
  detectors of CDF.}
\end{figure*}

Electron candidates in the CEM and PHX are rejected if an additional
high-\pt\ track is found which forms a common vertex with the track of
the electron candidate and has the opposite sign of the
curvature. These events are likely to stem from the conversion of a
photon.  Figure~\ref{fig:Triggertypes}(a) shows the $(\eta,\phi)$ distributions of
CEM and PHX electron candidates. 

Muon candidates are identified by requiring the presence of a
COT track with \pt$>$ 20~GeV$/c$ that extrapolates to a track
segment in one or more muon chambers.   The muon trigger may be
satisfied by two types of muon candidates, called CMUP and CMX.
A CMUP muon candidate is one in which
track segments matched to the COT track are found in both the CMU and the CMP chambers.
A CMX muon is one in which the track segment is found in the CMX muon detector.  
In order to minimize background contamination, further requirements
are imposed. The energy deposition in the electromagnetic and hadronic
calorimeters has to correspond to that expected from a
minimum-ionizing particle.  To reject cosmic-ray muons and muons from
in-flight decays of long-lived particles such as $K_{S}^0,
K_{L}^0$, and $\Lambda$ particles, the distance of closest approach of the track
to the beam line in the transverse plane is required to be less than 0.2~cm if there
are no silicon hits on the muon candidate's track, and less than 0.02~cm if there
are silicon hits.  The
remaining cosmic rays are reduced to a negligible level by taking 
advantage of their characteristic track
timing and topology.  

In order to add acceptance for events containing muons that cannot be 
triggered on directly, several additional muon types are taken from the extended
muon coverage (EMC) of the $\EtMiss$+jets trigger path: 
a track segment only in the CMU and a COT track not pointing to
CMP(CMU), a track segment only in the CMP and COT track not pointing to CMU
(CMP), a track segment in the BMU (BMU), an isolated track not fiducial to any
muon chambers (CMIO), an isolated track matched to a muon segment that is
not considered fiducial to a muon detector (SCMIO), and a track segment only in the CMX but in a
region that can not be used in the trigger due to tracking limitations of the
trigger (CMXNT).  Figure~\ref{fig:Triggertypes}(b) shows the $(\eta,\phi)$ distributions of
muon candidates in each of these categories. 

We require exactly one isolated charged lepton
candidate with $|\eta | < 1.6$.  A candidate is considered isolated if the \ET\ not
assigned to the lepton inside a cone defined by 
$R\equiv \sqrt{(\Delta\eta)^2+(\Delta\phi)^2} < 0.4$ centered around the lepton is less
than 10~\% of the lepton \ET\ (\pt) for electrons (muons).  This lepton is
called a tight lepton.  
Loose charged lepton candidates pass all of
the lepton selection criteria except for the isolation requirement.
We reject events which have an additional
tight or loose lepton candidate in order to reduce the $Z/\gamma^*$+jets and diboson
background rates.  

Jets are reconstructed using a cone algorithm by summing the transverse
calorimeter energy \ET\ in a cone of radius $R \le 0.4$.  
The energy deposition of an identified electron candidate, if present,
is not included in the jet energy sum. The \ET\ of a cluster is calculated with respect to the $z$
coordinate of the primary vertex of the event.  The energy of each jet is
corrected~\cite{Bhatti:2005ai} for the $\eta$ dependence and the nonlinearity of the calorimeter
response.  Routine calibrations of the calorimeter response are performed and these calibrations
are included in the jet energy corrections.
The jet energies are also adjusted by subtracting
the extra deposition of energy from additional inelastic \ppbar\
collisions on the same bunch crossing as the triggered event.

Reconstructed jets in events with identified charged lepton candidates
must have corrected \ET$> 20 \;\mbox{GeV}$ and
detector $|\eta| < 2.8$.  Detector $\eta$ is defined as the pseudorapidity of the jet
calculated with respect to the center of the detector. Only events
with exactly two or three jets are accepted.  At least one of
the jets must be tagged as containing a $B$~hadron by requiring a
displaced secondary vertex within the jet, using the {\sc secvtx} algorithm~\cite{Acosta:2004hw}. 
Secondary vertices are accepted if the
transverse decay length significance ($\Delta L_{xy} / \sigma_{xy}$)
is greater than or equal to 7.5.

Events passing the $\EtMiss$+jets trigger path and the EMC muon segment
requirements described above
are also required to have two sufficiently separated jets:
$\Delta R_{\rm jj}>1$.  Furthermore,
one of the jets must be central, with $|\eta_{\rm jet}|<0.9$, and both jets
are required to have transverse energies above 25~GeV.  These offline selection requirements
ensure full efficiency of the $\EtMiss$+jets trigger path.

The vector missing \ET\ ($\EtMissVec$) is defined by
\begin{eqnarray}
\EtMissVec & = & - \sum_{i} E_{\rm T}^i \hat{n}_i, \\
~i& = & \rm calorimeter~tower~number~with~|\eta| < 3.6, 
\end{eqnarray}
where $\hat{n}_i$ is a unit vector perpendicular to the beam axis and
pointing at the $i^{\rm th}$ calorimeter tower. We also define $\EtMiss=
|\EtMissVec|$.  Since this calculation is based on calorimeter
towers, $\not\!\! E_{\rm T}$ is adjusted for the effect of the
jet corrections for all jets.

A correction is applied to $\EtMissVec$ for muons since they traverse
the calorimeters without showering.  The transverse momenta of all
identified muons are added to the measured transverse energy sum and
the average ionization energy is removed from the measured calorimeter energy
deposits.   We require the corrected $\EtMiss$ to be greater than 25~GeV in order to purify a sample
containing leptonic $W$ boson decays.

A portion of the background consists of multijet events which do not contain $W$~bosons.
We call these ``non-$W$'' events below.  We select against the non-$W$ background by
applying additional selection requirements which are based on the assumption that
these events do not have a large $\EtMiss$ from an escaping neutrino,
but rather the $\EtMiss$ that is observed comes from lost or
mismeasured jets.  In events lacking a $W$ boson, one would expect
small values of the transverse mass, defined as
\begin{equation}
M_{\rm T}^W =
\sqrt{2 \left(p^\ell_{\rm T}{\mbox{$\EtMiss$}} - p^\ell_{x}\mbox{$\EtMiss$}^x -
p^\ell_y\mbox{$\EtMiss$}^y \right)}.
\end{equation}
Because the $\EtMiss$ in events that do not contain $W$ bosons often
comes from jets which are erroneously identified as charged leptons, $\EtMissVec$ often
points close to the lepton candidate's direction, giving the event a low transverse
mass.  Thus, the transverse mass is required to be above~10 GeV for
muons and 20~GeV for electrons, which have more of these events.

\begin{figure*}
\begin{center}
\includegraphics[width=1.5\columnwidth]{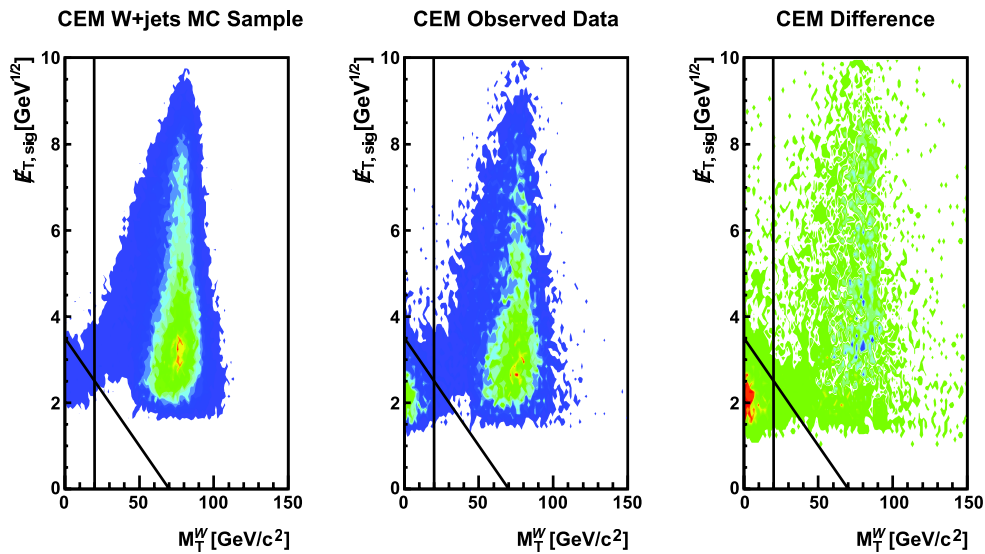}
\end{center}
\caption{\label{fig:QCDveto} Plots of $\protect\EtMissSig$ vs. $M_{\rm T}^W$
for $W+$jets Monte Carlo, the selected data in the $\ell+\EtMiss$+2 jets sample, 
and the two distributions subtracted for all CEM candidates. The black
lines indicate the requirements which are applied. 
Events with lower $\protect\EtMissSig$ or $M_{\rm T}^W$ are not selected.
}
\end{figure*}

Further removal of non-$W$ events is performed
with a variable called $\EtMiss$ significance ($\EtMissSig$), defined as
\begin{widetext}
\begin{equation}
\EtMissSig =
\frac{\EtMiss}{\sqrt{\sum_{\rm{jets}}C_{\rm{JES}}^2\cos^2\left(\Delta\phi_{\rm{jet},\EtMissVec}\right)E_{\rm{T,jet}}^{\rm{raw}}+
\cos^2\left(\Delta\phi_{{\vec{E}}_{\rm{T,uncl}},\EtMissVec}\right)\sum E_{\rm{T,uncl}}}},\label{eq:metsig}
\end{equation}
\end{widetext}
where $C_{\rm{JES}}$ is the jet energy correction factor~\cite{Bhatti:2005ai}, 
$E_{\rm{T,jet}}^{\rm{raw}}$ is a jet's energy before corrections are applied,
${\vec{E}}_{\rm{T,uncl}}$ refers to the vector sum of the
transverse components of calorimeter energy deposits not included in any
reconstructed jets, and $\sum E_{\rm{T,uncl}}$ is the sum of the magnitudes of these
unclustered energies.  The angle between the projections in the $r\phi$ plane of a jet
and $\EtMissVec$ is denoted $\Delta\phi_{{\rm{jet}},{\vec{E}}_{\rm{T,uncl}}}$, and
the angle between the projections in the $r\phi$ plane of $\sum E_{\rm{T,uncl}}$ and $\EtMissVec$
is denoted $\Delta\phi_{{\vec{E}}_{\rm{T,uncl}},\EtMissVec}$.
When the energies in Equation~\ref{eq:metsig} are measured in GeV, $\EtMissSig$ is an approximate significance,
as the dispersion in the measured $\EtMiss$ in events with no true $\EtMiss$ is approximated by the denominator.
Central electron events are required to have $\EtMissSig > 3.5 - 0.05 M_{\rm T}$ and
$\EtMissSig > 2.5 - 3.125 \Delta \phi_{{\rm jet2},{\scriptsize\EtMiss}}$, where jet~2 is the jet with the
second-largest $E_{\rm T}$, and all energies are measured
in GeV.  Plug electron events
must have $\EtMissSig > 2$ and $\EtMiss>45-30\Delta\phi_{{\rm jet},{\scriptsize\EtMissVec}}$ 
for all jets in the event.  These requirements reduce the amount of
contamination from non-$W$ events substantially, as shown in the plots
in Fig.~\ref{fig:QCDveto}.

  To remove events containing $Z$ bosons, we reject events in which the trigger
lepton candidate can be paired with an oppositely-signed track such that the invariant mass of the
pair is within the range 76~GeV/$c^2\le m_{\ell,{\rm{track}}}\le 106$~GeV/$c^2$.
Additionally, if the trigger lepton candidate is identified as an electron, 
the event is rejected if a cluster is found in the electromagnetic calorimeter 
that, when paired with the trigger lepton candidate, forms an invariant mass in the same range.

\section{\label{sec:SignalModel} Signal Model}

In order to perform a search for a previously undetected signal such
as single top quark production, accurate models predicting the
characteristics of expected data are needed for both the signal being
tested and the SM background processes.  This analysis uses Monte Carlo
programs to generate simulated events for each signal and background
process, except for non-$W$ QCD multijet events for which events
in data control samples are used.

\subsection{{\it s}-channel Single Top Quark Model}

The matrix element generator {\sc madevent}~\cite{Maltoni:2002qb} is
used to produce simulated events for the signal samples. The generator is
interfaced to the CTEQ5L~\cite{Lai:1999wy} parameterization of the parton distribution
functions (PDFs). The {\sc pythia}~\cite{Sjostrand:2000wi,Sjostrand:2006za} program is used to
perform the parton shower and hadronization.  Although {\sc madevent} uses only a leading-order
matrix element calculation, studies~\cite{Sullivan:2004ie,Sullivan:2005ar} indicate
that the kinematic distributions of $s$-channel events are only negligibly affected by
NLO corrections.  The parton shower simulates the higher-order effects of gluon radiation
and the splitting of gluons into quarks, and the Monte Carlo samples include contributions
from initial-state sea quarks via the proton PDFs.

\subsection{{\it t}-channel Single Top Quark Model} 

The $t$-channel process is more complicated. Several authors
point out~\cite{Boos:2006af,Sullivan:2004ie,Campbell:2009ss,Campbell:2009gj} that the leading-order
contribution to $t$-channel single top quark production as modeled in
parton-shower Monte Carlo programs does not adequately represent the
expected distributions of observable jets, which are better predicted
by NLO calculations.

\begin{figure}[t]
\begin{center}
\includegraphics[width=0.95\columnwidth]{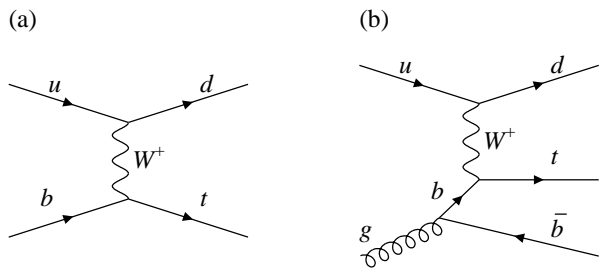}
\caption{The two different $t$-channel processes considered in our
signal model: (a) the $2\rightarrow 2$ process and (b) the $2\rightarrow 3$ process.}
\label{fig:tchan2-2_2-3}
\end{center}
\end{figure}

The leading-order process is a $2\rightarrow 2$ process with a $b$
quark in the initial state: $b+u\rightarrow d+t$, as shown 
in Fig.~\ref{fig:tchan2-2_2-3}(a).  For antitop quark
production, the charge conjugate processes are implied.  A parton distribution function
for the initial state $b$ quark is used for the calculation.
Since flavor is conserved in the strong interaction, a $\bar{b}$ quark
must be present in the event as well.  In what follows, this
$\bar{b}$ quark is called the spectator $b$ quark.  Leading-order
parton shower programs create the spectator $b$ quark through backward
evolution following the {\sc dglap}
scheme~\cite{Dokshitzer:1977sg,Gribov:1972ri,Altarelli:1977zs}.  Only
the low-$p_\mathrm{T}$ portion of the transverse momentum distribution
of the spectator $b$ quark is modeled well, while the high-$p_\mathrm{T}$
tail is not estimated adequately~\cite{Sullivan:2004ie}.  In addition,
the pseudorapidity distribution of the spectator $b$ quark, as simulated
by the leading-order process, is biased towards higher pseudorapidities
than predicted by NLO theoretical calculations.

We improve the modeling of the $t$-channel single top quark
process by using two samples: one for
the leading $2\rightarrow 2$ process $b+q\rightarrow q^\prime +t$, and
a second one for the $2\rightarrow 3$ process in which an initial-state
gluon splits into $b{\bar{b}}$, $g+q\rightarrow q^\prime + t + \bar{b}$.
In the second process the spectator $b$ quark is produced directly in the
hard scattering described by the matrix element
(Fig.~\ref{fig:tchan2-2_2-3}(b)). This sample describes the most
important NLO contribution to $t$-channel
production and is therefore suitable to describe the
high-$p_\mathrm{T}$ tail of the spectator $b$ quark
$p_\mathrm{T}$ distribution.  This sample, however, does not
adequately describe the low-$p_\mathrm{T}$ portion of the spectrum of
the spectator $b$ quark.  In order to construct a Monte Carlo sample
which closely follows NLO predictions, the $2\rightarrow 2$ process
and the $2\rightarrow 3$ process must be combined.

 A joint event sample was created by matching the $p_\mathrm{T}$
spectrum of the spectator $b$ quark to the differential cross
section predicted by the {\sc ztop} program~\cite{Sullivan:2004ie}
which operates at NLO.  The matched $t$-channel
sample consists of $2\rightarrow 2$ events for spectator $b$ quark
transverse momenta below a cutoff, called $K_\mathrm{T}$, and 
of $2\rightarrow 3$ events
for transverse momenta above $K_\mathrm{T}$.  The rates of
$2\rightarrow 2$ and $2\rightarrow 3$ Monte Carlo events are adjusted
to ensure the continuity of the spectator $b$ quark
$p_{\rm T}$ spectrum at $K_\mathrm{T}$.
The value of $K_\mathrm{T}$ is adjusted until the prediction of the
fraction of $t$-channel signal events with a detectable
spectator $b$ quark jet -- with $p_\mathrm{T} >
20\,\mathrm{GeV}/c$ and $|\eta|<2.8$ -- matches the prediction by
{\sc ztop}.   We obtain $K_\mathrm{T}=20\,\mathrm{GeV}/c$.   All
detectable spectator $b$ quarks with $p_\mathrm{T}>20\,\mathrm{GeV}/c$ of
the joint $t$-channel sample are simulated using the $2\rightarrow 3$
sample.

\begin{figure*}[t]  
\begin{center}
\subfigure[]{
\includegraphics[width=0.8\columnwidth]{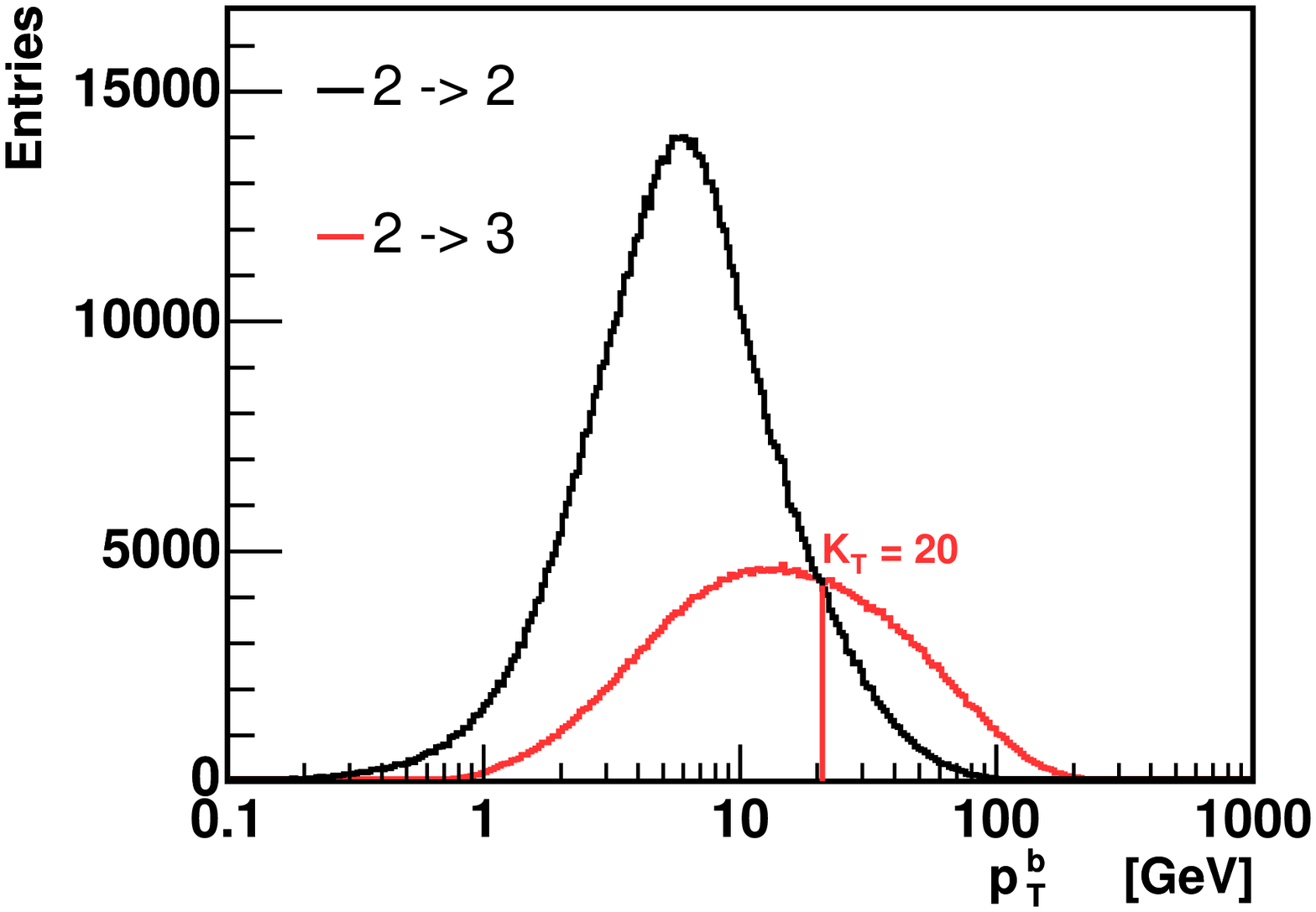} 
\label{fig:t-channel-matching_log}
}
\subfigure[]{
\includegraphics[width=0.8\columnwidth]{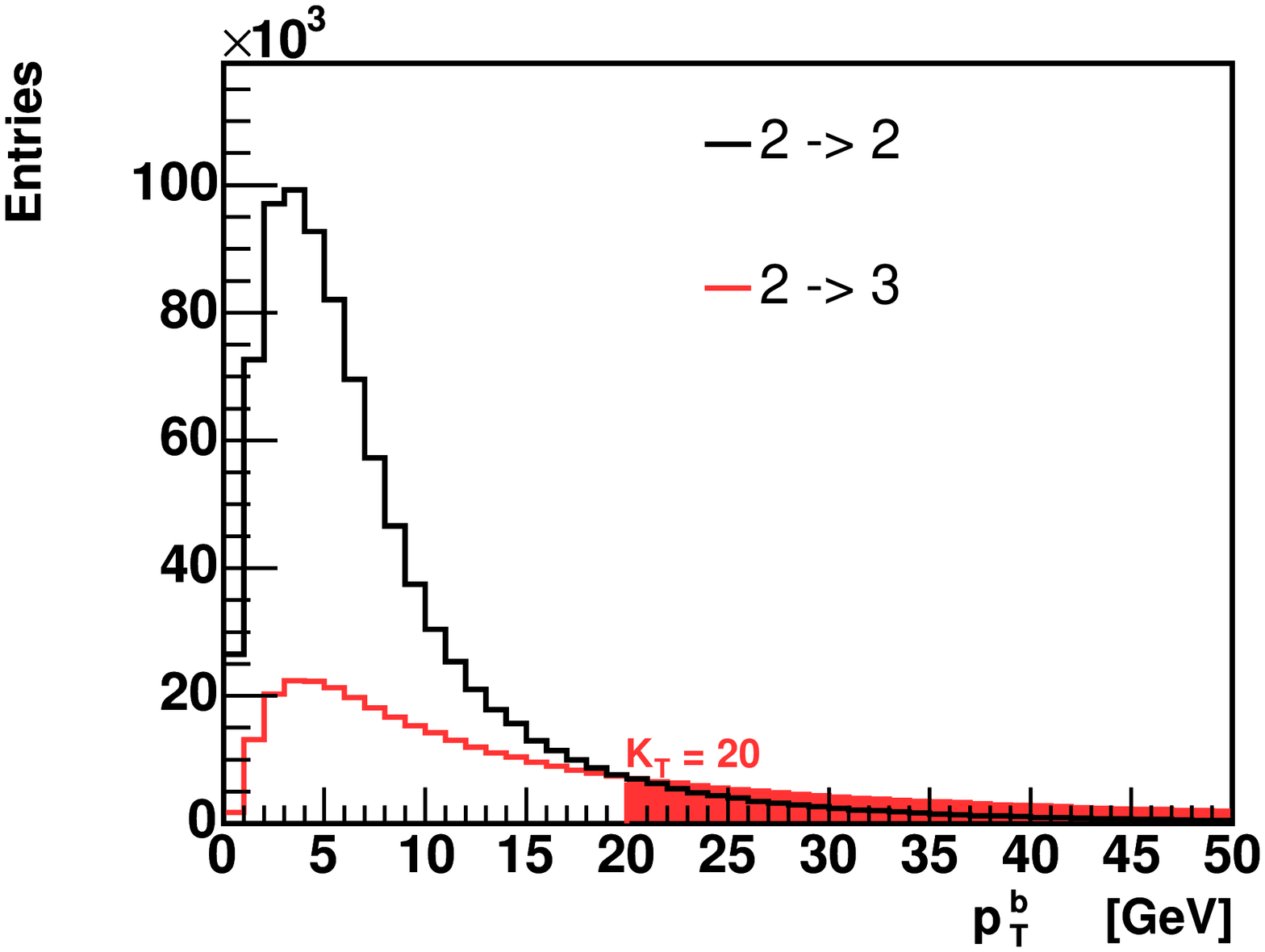}
\label{fig:t-channel-matching_lin}
}
\subfigure[]{
\includegraphics[width=0.8\columnwidth]{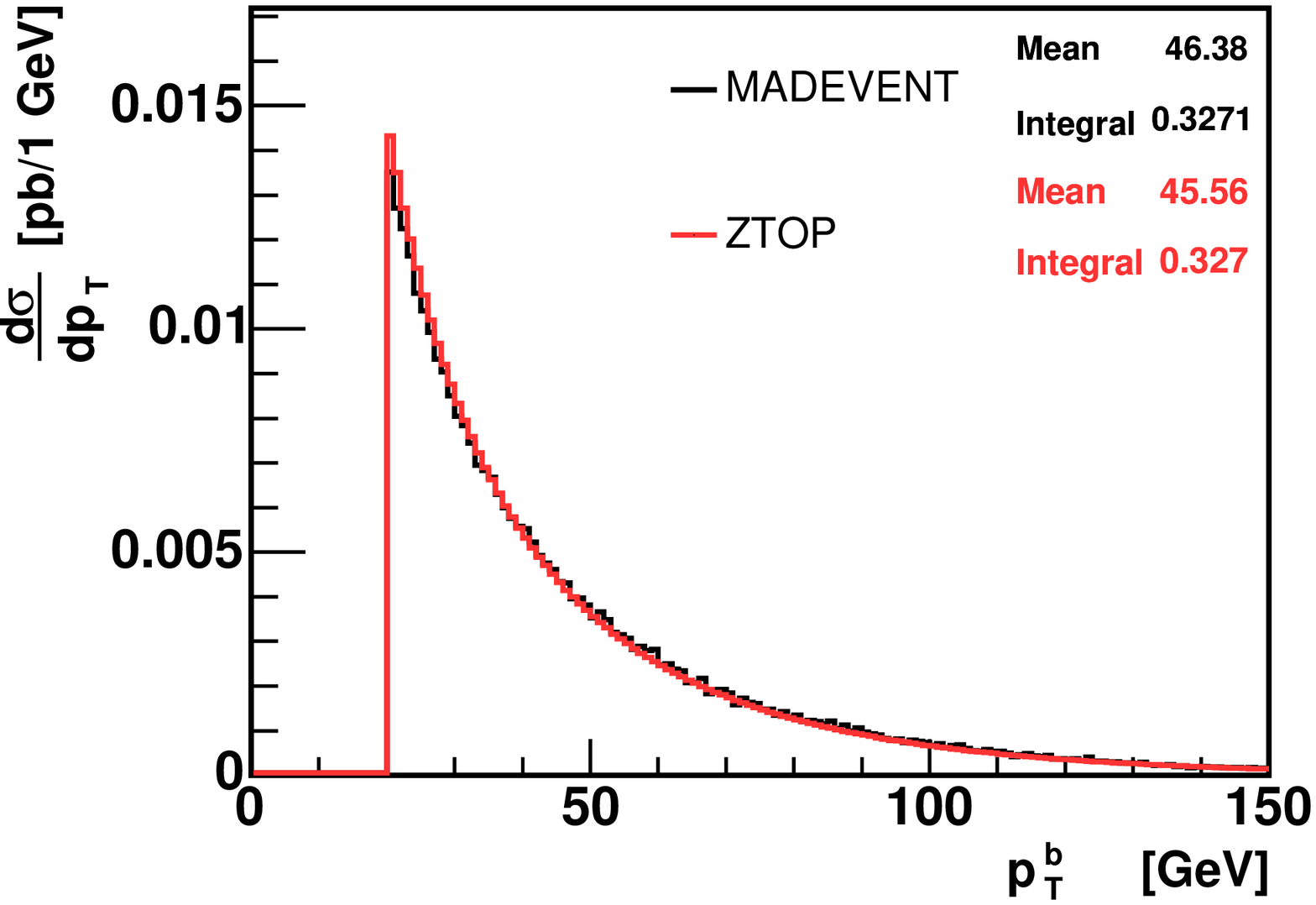}
\label{fig:t-channel-matching_ptb}
}
\subfigure[]{
\includegraphics[width=0.8\columnwidth]{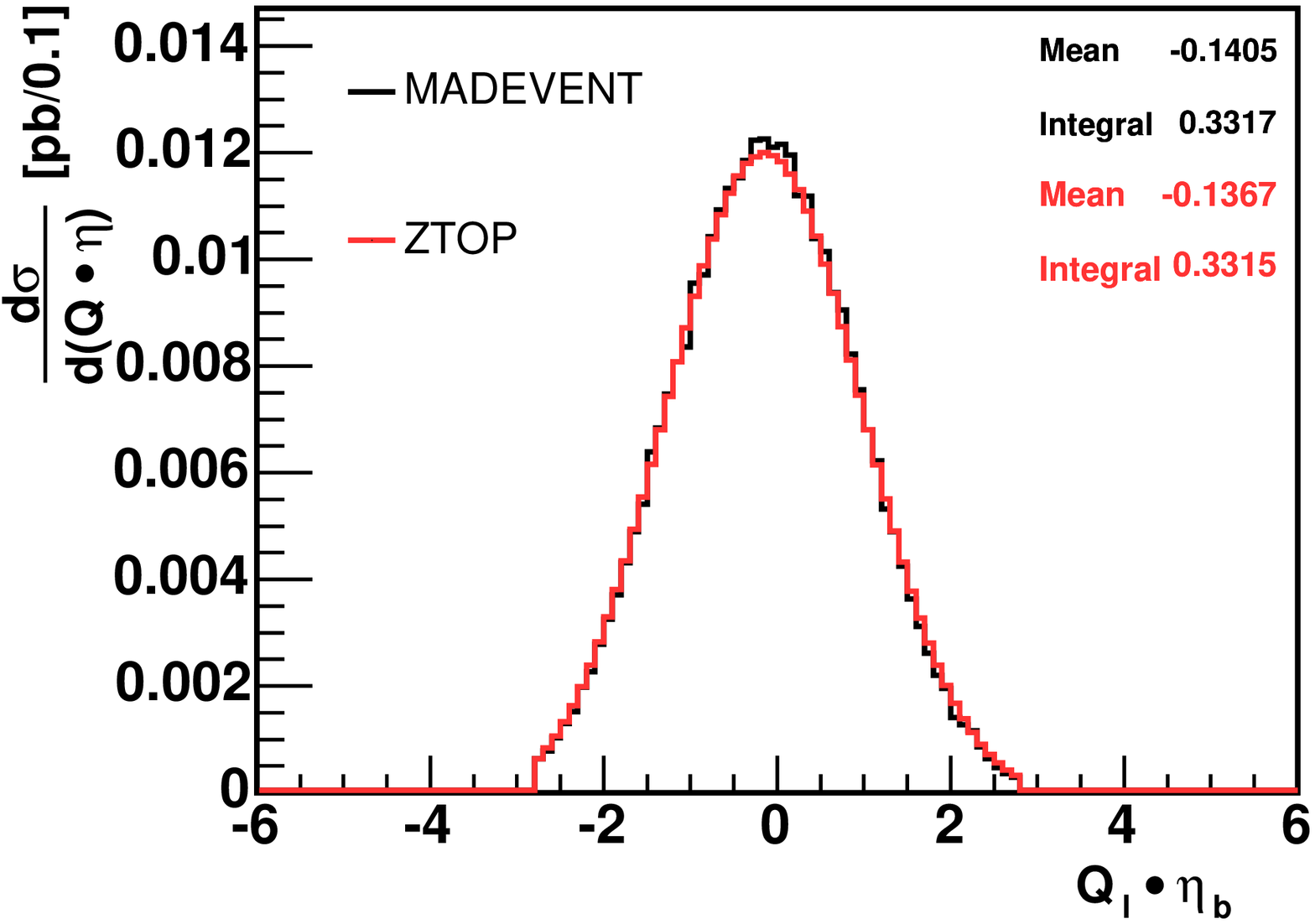} 
\label{fig:t-channel-matching_etab}
}
\vspace*{3mm}
\end{center}
\caption[tchannel-matching]{\label{fig:t-channel-matching} Matching of 
$t$-channel single top quark events of the $2\rightarrow 2$ and the
$2\rightarrow 3$ process.  The $p_\mathrm{T}$ distributions of the
spectator $b$ quark are shown,
\subref{fig:t-channel-matching_log} on a logarithmic $p_{\rm T}$ scale, 
and \subref{fig:t-channel-matching_lin} on a linear $p_{\rm T}$ scale. 
The ratio of $2\rightarrow 2$ to $2\rightarrow 3$ events is adjusted
such that the rate of spectator $b$ quarks with $p_\mathrm{T} >
20\,\mathrm{GeV}/c$ and $|\eta|<2.8$ matches the theoretical prediction. The
fraction of these events is illustrated in
\subref{fig:t-channel-matching_lin} by the shaded area. 
The matched {\sc madevent} sample reproduces both the rate
and the shape of the differential {\sc ztop} 
$p_\mathrm{T}$~\subref{fig:t-channel-matching_ptb} and $Q_\ell\cdot \eta$~\subref{fig:t-channel-matching_etab} cross section
distributions of the spectator $b$ quark.} 
\end{figure*}

Figure~\ref{fig:t-channel-matching} illustrates the matching
procedure and compares the outcome with the differential 
$p_\mathrm{T}$ and $Q_\ell \cdot \eta$ cross sections of the
spectator $b$ quark, where $Q_\ell$ is the charge of the lepton
from $W$ boson decay.  Both the falling $p_\mathrm{T}$ spectrum of
the spectator $b$ quark and the slightly
asymmetric shape of the $Q_\ell \cdot \eta$ distribution are
well modeled by the matched {\sc madevent} sample.
Figure~\ref{fig:t-channel-matching}\subref{fig:t-channel-matching_log}
shows the $p_\mathrm{T}$ distribution of the spectator $b$ quark on a
logarithmic scale.  The combined sample of $t$-channel events has a
much harder $p_\mathrm{T}$ spectrum of spectator $b$ quarks than the
$2\rightarrow 2$ sample alone provides.  The tail of the distribution
extends beyond $100\,\mathrm{GeV}/c$, while the $2\rightarrow 2$
sample predicts very few spectator $b$ quarks with $p_\mathrm{T}$ above
$50\,\mathrm{GeV}/c$.

\subsection{Validation}

It is important to evaluate quantitatively the modeling of single top quark
events.  We compare the kinematic distributions of the primary 
partons obtained from the $s$-channel and the matched $t$-channel {\sc madevent} 
samples to theoretical differential cross
sections calculated with {\sc ztop}~\cite{Sullivan:2004ie}.
We find, in general, very good agreement.
For the $t$-channel process in particular, the pseudorapidity
distributions of the spectator $b$ quark in the two predictions are nearly identical,
even though that variable was not used to match the two $t$-channel
samples.
 
One can quantify the remaining differences between the Monte Carlo
simulation and the theoretical calculation by assigning weights to simulated
events.  The weight is derived from a comparison of six kinematic
distributions: the $p_\mathrm{T}$ and the $\eta$ of the top quark and
of the two highest-$E_\mathrm{T}$ jets which do not originate from
the top-quark decay.  In case of $t$-channel production, we
distinguish between $b$-quark jets and light-quark jets. The
correlation between the different variables, parameterized by the
covariance matrix, is determined from the simulated events generated
by {\sc madevent}.  We apply the single top quark event selection to the
Monte Carlo events and add the weights. This provides an
estimate of the deviation of the acceptance in the simulation compared
to the NLO prediction. In the $W+2$ jets sample we find a fractional discrepancy
of $(-1.8\pm0.9)\%$~(MC~stat.) for the $t$-channel,
implying that the Monte Carlo estimate of the
acceptance is a little higher than the NLO prediction.  In the
$s$-channel we find excellent agreement:
$-0.3\%\pm0.7\%\,(\mathrm{MC\; stat.})$.  More details on the
$t$-channel matching procedure and the comparison to {\sc ztop} can be
found in references~\cite{Lueck:2006hz} and~\cite{Lueck:2009zz}.  The general conclusion from
our studies is that the {\sc madevent} Monte Carlo events represent
faithfully the NLO single top quark production predictions. The
matching procedure for the $t$-channel sample takes the main NLO
effects into account. The remaining difference is covered by
a systematic uncertainty of $\pm 1\%$ or $\pm 2\%$ on the
acceptance for $s$- and $t$-channel events, respectively.

Recently, an even higher-order calculation of the $t$-channel
production cross section and kinematic distributions has been
performed~\cite{Campbell:2009ss,Campbell:2009gj}, treating the $2\rightarrow 3$
process itself at NLO.  The production cross section in this
calculation remains unchanged, but a larger fraction of events have a
high-$p_\mathrm{T}$ spectator $b$ within the detector acceptance.
This calculation became available after the analyses described in this
paper were completed.  The net effect is to slightly decrease the
predicted $t$-channel signal rate in the dominant sample with two jets
and one $b$ tag, and to significantly raise the comparatively low
signal prediction in the double-tagged samples and the three-jet
samples, compensating each other. Thus, the expected as well as the
observed change of the outcome is insignificant for the combined and
the separate extraction of the signal cross section and significance.


\subsection{Expected Signal Yields} 
\label{sec:signalyields}
The number of expected events is given by
\begin{equation}
\label{eq:Nsig}
  \hat{\nu}=\sigma\cdot\varepsilon_\mathrm{evt} 
  \cdot \lumi 
\end{equation}
where $\sigma$ is the theoretically predicted cross section
of the respective process,
$\varepsilon_\mathrm{evt}$ is the event detection efficiency, and
$\mathcal{L}_\mathrm{int}$ is the integrated luminosity.
The predicted cross sections for $t$-channel and $s$-channel single top quark 
production are quoted in section~\ref{sec:Introduction}.
The integrated luminosity used for the analyses presented in this 
article is $\lumi=3.2$~\fb.

The event detection efficiency is estimated by performing the event selection
on the samples of simulated events.  Control samples in the data are used to
calibrate the efficiencies of the trigger, the lepton identification, and the 
$b$-tagging.  These calibrations are then applied to the Monte Carlo samples we use.

We do not use a simulation of the trigger efficiency in the Monte Carlo samples; instead
we calibrate the trigger efficiency using data collected
with alternate trigger paths and also $Z\rightarrow\ell^+\ell^-$ events
in which one lepton triggers the event and the other lepton is used to
calculate the fraction of the time it, too, triggers the event.  
We use these data samples to calculate the efficiency
of the trigger for charged leptons as a function of the lepton's $E_{\rm{T}}$ and $\eta$.
The uncorrected Monte Carlo-based efficiency prediction, $\varepsilon_\mathrm{MC}$ 
is reduced by the trigger efficiency $\varepsilon_\mathrm{trig}$. 
The efficiency of the selection
requirements imposed to identify charged leptons
is estimated with data samples with high-$p_{\rm T}$ triggered
leptons.  We seek in these events oppositely-signed tracks forming the $Z$ mass with
the triggered lepton.  The fraction of these tracks passing the lepton
selection requirements gives the lepton identification efficiency.
The $Z$ vetoes in the single top quark candidate selection requirements enforce the
orthogonality of our signal samples and these control samples we use to
estimate the trigger and identification efficiencies.

A similar strategy is adopted for using the data to calibrate the
$b$-tag efficiency.  At LEP, for example, single- and double-$b$-tagged
events were used~\cite{Abbiendi:1998eh} to extract the $b$-tag efficiency
and the $b$-quark fraction in $Z$ decay.  Jet formation 
in $p{\bar{p}}$ collisions involves many more processes, however, and the
precise rates are poorly predicted.  A jet originating from a $b$ quark produced
in a hard scattering process, for example, may recoil against another
$b$~jet, or it may recoil against a gluon jet.
The invariant mass requirement used in the lepton identification
procedure to purify a sample of $Z$ decays is not useful for separating a sample
of $Z\rightarrow b{\bar{b}}$ decays because of the low 
signal-to-background ratio~\cite{Donini:2008nt}.

We surmount these challenges and calibrate the $b$-tag efficiency in the data using the method described
in Ref.~\cite{Acosta:2004hw}, and which is briefly summarized here.
We select dijet events in which one jet is
tagged with the {\sc secvtx} algorithm, and the other jet has an identified electron candidate
with a large transverse momentum with respect to the jet axis
in it, to take advantage of the characteristic semileptonic decays of $B$ hadrons.
 The purity of $b{\bar b}$ events in this sample is nearly
unity.   We determine the flavor fractions in the jets containing electron candidates by
fitting the distribution of the invariant mass of the reconstructed displaced
vertices to templates for $b$~jets,
charm jets, and light-flavor jets, in order to account for the presence
of non-$b$ contamination.

The fraction of jets with electrons in them passing the
{\sc secvtx} tag is used to calibrate the {\sc secvtx} tagging
efficiency of $b$ jets which contain electrons.  This efficiency
is compared with that of $b$ jets passing the same selection requirements
in the Monte Carlo, and the ratio of the efficiencies is applied to
the Monte Carlo efficiency for all $b$ jets.  Systematic uncertainites
to cover differences in Monte Carlo mismodeling of semileptonic and inclusive
$B$ hadron jets are assessed.  The
$b$-tagging efficiency is approximately 45\% per $b$~jet from top
quark decay, for $b$~jets with at least two tracks and which
have $|\eta|<1$.  The ratio between the data-derived efficiency
and the Monte Carlo prediction does not show a noticeable dependence
on the $|\eta|$ of the jet or the jet's $E_{\rm T}$.  

The differences
in the lepton identification efficiency and the $b$-tagging between
the data and the simulation are accounted for by a correction factor
$\varepsilon_\mathrm{corr}$ on the single top quark event detection
efficiency.  Separate correction factors are applied to the single
$b$-tagged events and the double $b$-tagged events.  Systematic
uncertainties are assessed on the signal acceptance due to the
uncertainties on these correction factors.

The samples of simulated events are produced such
that the $W$ boson emerging from top quark decay is only allowed to
decay into leptons, that is $e\nu_e$, $\mu\nu_\mu$, and $\tau\nu_\tau$.
Tau lepton decay is simulated with {\sc tauola}~\cite{Jadach:1993hs}. 
The value of $\varepsilon_\mathrm{MC}$, the fraction of all signal MC events passing
our event selection requirements, is multiplied by the branching fraction of $W$ bosons into
leptons, $\varepsilon_\mathrm{BR}=0.324$.   The selection efficiencies for events
in which the $W$ boson decays to electrons and muons are similar, but the selection efficiency
for $W\rightarrow\tau\nu_\tau$ decays is less, because many tau decays do not contain leptons, and
also because the $p_{\rm T}$ spectrum of tau decay products is softer than those of electrons and muons.
In total, the event detection efficiency 
is given by
\begin{equation} 
  \label{eq:epsevt}
  \mathrm{\varepsilon_{evt} = \varepsilon_{MC}\cdot\varepsilon_{BR}\cdot
  \varepsilon_{corr}\cdot\varepsilon_{trig}}  
\end{equation} 
Including all trigger and identification efficiencies
we find $\varepsilon_\mathrm{evt}(t\text{-channel}) = (1.2 \pm 0.1)\%$
and $\varepsilon_\mathrm{evt}(s\text{-channel}) = (1.8 \pm 0.1)\%$.
The predicted signal yields for the selected two- and three-jet events with one and
two (or more) $b$-tagged jets are listed in Tables~\ref{tab:EventYield1tag} and~\ref{tab:EventYield2tag}.

\section{\label{sec:Background} Background Model}

The final state of a single top quark event --  a charged lepton,
missing transverse energy from the undetected neutrino, and two or three jets with one or more
$B$ hadrons, is also the final state of the $Wb{\bar{b}}$ process, which
has a much larger cross section.  Other processes which produce
similar final states, such as $Wc{\bar{c}}$ and $t{\bar{t}}$,
also mimic the single top quark signature because
of misreconstruction or because of the
loss of one or more components of the expected final state.
A detailed understanding of the rates and of the
kinematic properties of the background processes is necessary
in order to accurately measure the single top quark production cross section.

\begin{figure}[t]
\begin{center}
\includegraphics[width=0.95\columnwidth]{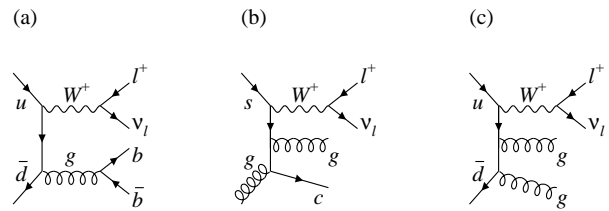}
\caption{Some representative diagrams of $W$+jets production. The
production cross sections of these processes are much larger than
that of single top quark production.}
\label{fig:Wjets}
\end{center}
\end{figure}

The largest background process is the associated production of a leptonically
decaying $W$ boson and two or more jets.
Representative Feynman diagrams are shown in Fig.~\ref{fig:Wjets}.
The cross section for $W$+jets production is much larger than that of the
single top quark signal, and the $W$+jets production cross sections
are difficult to calculate theoretically.  Furthermore, $W$+jets events
can be kinematically quite similar to the signal events we seek, and in the
case that the jets contain $b$ quarks, the final state can be identical
to that of single top quark production.  The narrow top quark
width, the lack of resonant structure in $W$+jets events, and color
suppression make the quantum-mechanical interference between the signal and
the background very small. 

\begin{figure}[t]
\begin{center}
\includegraphics[width=0.95\columnwidth]{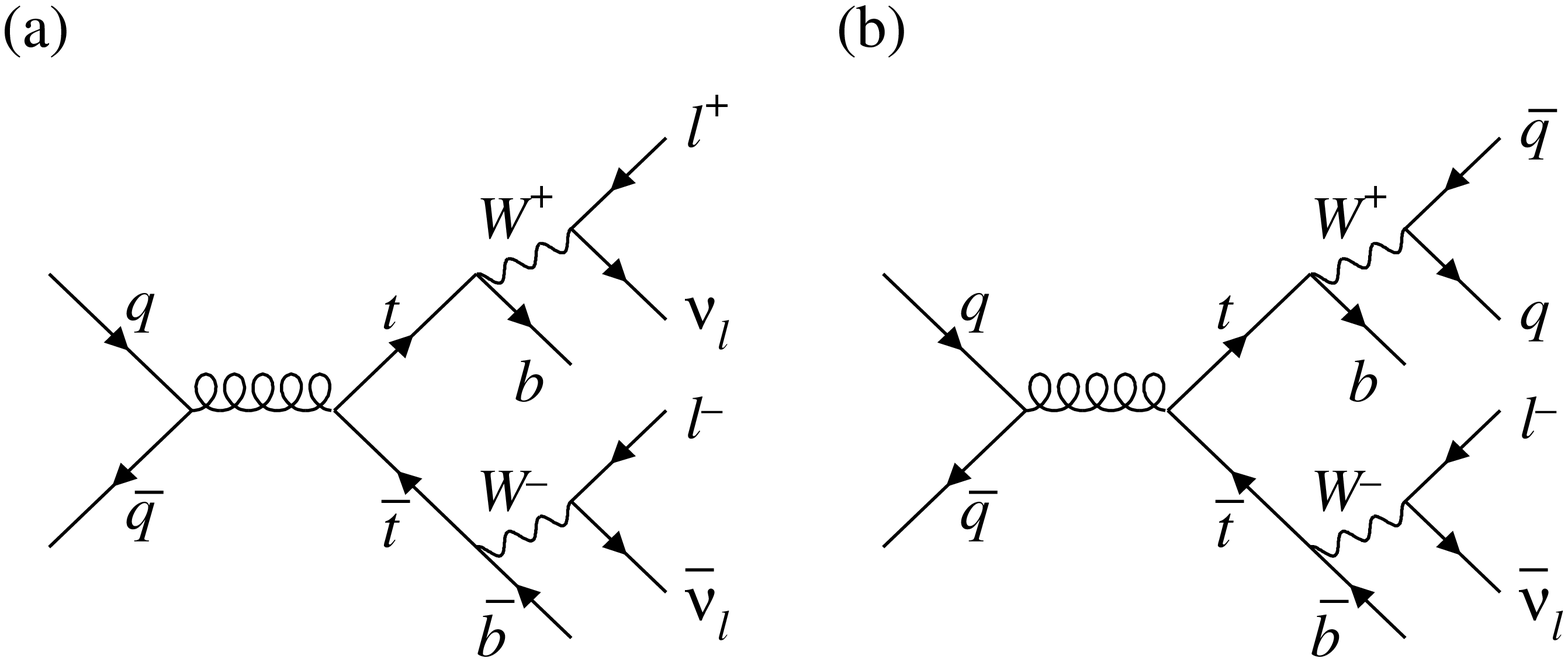}
\caption{Feynman diagrams of the $t\bar{t}$ background to single top quark
production.  To pass the event selection, these events must have one
charged lepton (a), or one or two hadronic jets (b), that go undetected.}
\label{fig:ttbar}
\end{center}
\end{figure}

\begin{figure}[t]
\begin{center}
\includegraphics[width=0.95\columnwidth]{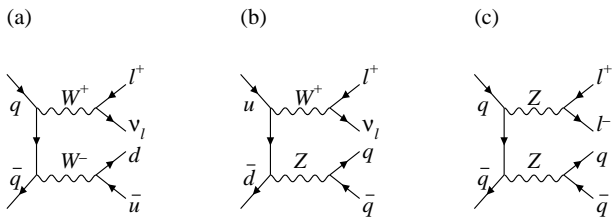}
\caption{Feynman diagrams for diboson production, which provides a
small background for single top quark production.}
\label{fig:diboson}
\end{center}
\end{figure}

\begin{figure}[t]
\begin{center}
\includegraphics[width=0.95\columnwidth]{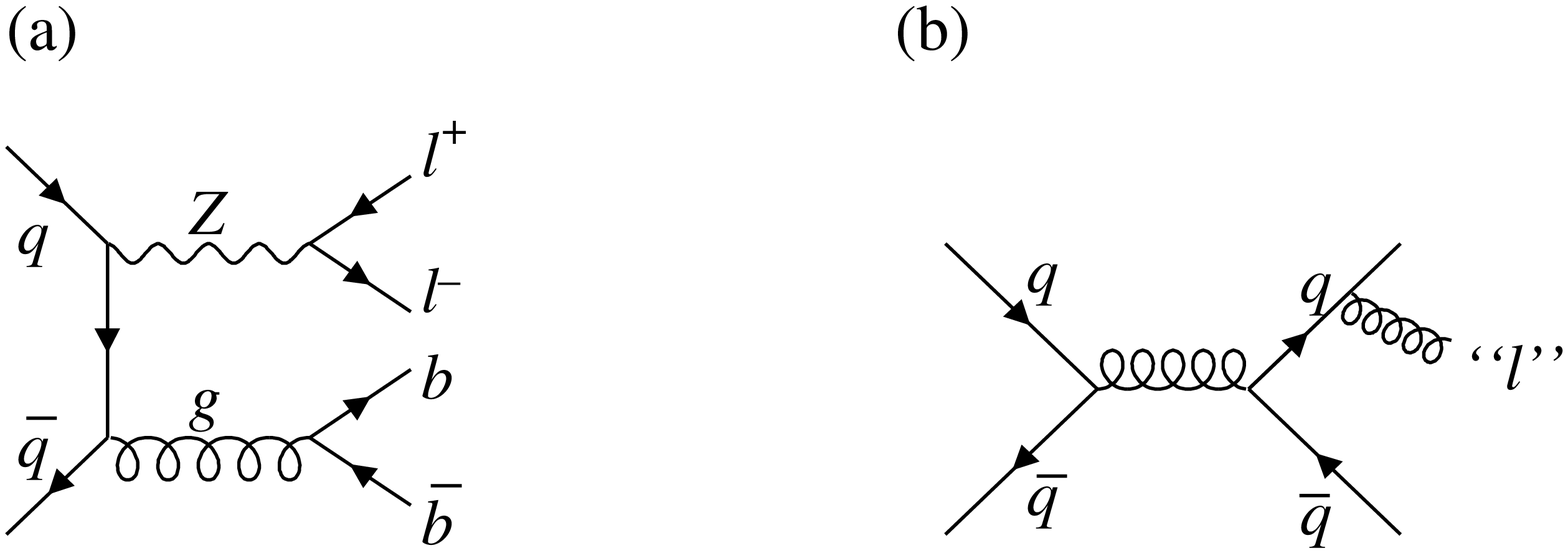}
\caption{Representative Feynman diagrams for (a) $Z/\gamma^*$+jets 
production and (b) non-$W$ events, in which a jet has to be
misidentified as a lepton and $\EtMiss$\ must be mismeasured to pass the
event selection.}
\label{fig:misc}
\end{center}
\end{figure}

Top quark pair production, in which one or two jets, or one charged lepton, has
been lost, also constitutes an important background process (Fig.~\ref{fig:ttbar}).
There are also contributions from the diboson
production processes $WW$, $WZ$, and $ZZ$, which are shown in
Fig.~\ref{fig:diboson}, $Z/\gamma^*$+jets processes in which one charged lepton 
from $Z$ boson decay is missed, (Fig.~\ref{fig:misc}(a)),
and QCD multijet events, which do not contain $W$ bosons but instead have
a fake lepton and mismeasured $\EtMiss$\ (Fig.~\ref{fig:misc}(b)).  
The rates and kinematic properties of these processes must be carefully
modeled and validated with data in order to make a precise measurement
of single top quark production.

Because there are many different background processes, we use a variety of methods
to predict the background rates.  Some are purely based on Monte Carlo simulations scaled
to high-order predictions of the cross section
(such as $t\bar{t}$); some are purely data-based (non-$W$); and some
require a combination of Monte Carlo and data ($W$+jets).

\subsection{Monte Carlo Based Background Processes} 

We use samples of simulated Monte Carlo events to estimate the
contributions of $t\bar{t}$, diboson, and $Z/\gamma^*$+jets production to the
$b$-tagged lepton+jets sample. The corresponding event detection
efficiencies $\varepsilon_\mathrm{evt}$ are calculated in the same way
as the single top quark processes described in Section~\ref{sec:SignalModel}
and Equation~\ref{eq:epsevt}. We apply Equation~\ref{eq:Nsig} to calculate
the final number of expected events. Therefore, it is essential that
the given physical process is theoretically well understood, {\it i.e.}, the
kinematics are well described in simulated events and the cross
section is well known.

To model the $t\bar{t}$ production contribution to our selected
samples, we use {\sc pythia}~\cite{Sjostrand:2006za} Monte Carlo samples,
scaled to the NLO theoretical cross section prediction~\cite{Bonciani:1998vc,Cacciari:2003fi}
of $\sigma_{t\bar{t}}=(6.70\pm 0.83)$~pb, assuming $m_t=175$~GeV$/c^2$.
The systematic uncertainty contains a component which covers the differences between
the calculation chosen and others~\cite{Berger:1997ed,Kidonakis:2003qe}.
The event selection efficiencies and the kinematic distributions
of $t{\bar{t}}$ events are predicted using these {\sc pythia} samples.
Because the Monte Carlo efficiencies for lepton identification and $b$ tagging
differ from those observed in the data, the $t\bar{t}$ efficiencies estimated from
the Monte Carlo are adjusted by factors $\epsilon_{\mathrm{corr}}$, which are functions of the numbers
of leptonically decaying $W$ bosons and $b$-tagged jets.

To estimate the expected number of diboson events in our selected data
sample we use the theoretical cross section predicted for a center of
mass energy of $\sqrt{s}=2.00$~TeV using the {\sc mcfm} program~\cite{PhysRevD.60.113006} 
and extrapolate
the values to $\sqrt{s}=1.96$~TeV. This leads to 
$\sigma_{WW}=(13.30\pm 0.80)$~pb, 
$\sigma_{WZ}=(3.96\pm 0.34)$~pb, and 
$\sigma_{ZZ}=(1.57\pm 0.21)$~pb.  The cross section uncertainties reported in~\cite{PhysRevD.60.113006}
are smaller than those obtained with {\sc mcfm} Version~5.4; we quote here the larger uncertainties.
The event selection efficiencies and the kinematic distributions
of diboson events are estimated with {\sc pythia} Monte Carlo samples, with corrections
applied to bring the lepton identification and $b$-tagging efficiency in line with those
estimated from data samples.

Events with $Z/\gamma^*$ boson production in association with jets are
simulated using {\sc alpgen}~\cite{Mangano:2002ea}, with {\sc pythia} used to model the parton shower
and hadronization.  The $Z/\gamma^*+$jets cross section is normalized to that
measured by CDF in the $Z/\gamma^*(\rightarrow e^+e^-)$+jets sample~\cite{Aaltonen:2007cp}, within the
kinematic range of the measurement, separately for the different numbers of jets.
Lepton universality is assumed in $Z$ decay.

\subsection{Non-{\it W} Multijet Events} 
\label{sec:nonw}

Estimating the non-$W$ multijet contribution to the sample is challenging because
of the difficulty of simulating these events. A variety of QCD
processes produce copious amounts of
multijet events, but only a tiny fraction of these events pass our 
selection requirements.  
In order for an event lacking a leptonic $W$ boson decay to be selected,
it must have a fake lepton or a real lepton from a heavy flavor quark decay.  In the same event,
the $\EtMiss$ must be mismeasured.  The rate at which fake leptons are reconstructed and the
amount of mismeasured  $\EtMiss$ are difficult to model
reliably in Monte Carlo.  

The non-$W$ background is modeled by selecting data samples which have
less stringent selection requirements than the signal sample.  These samples, which are described below,
are dominated by non-$W$ events with similar kinematic distributions as the
non-$W$ contribution to the signal sample.  The normalization of the non-$W$ prediction
is separately determined by fitting templates of the $\EtMiss$\ distribution to
the data sample.

We use three different data samples to model the non-$W$ multijet contributions. 
One sample is based
on the principle that non-$W$ events must have a jet which passes all
lepton identification requirements.  A data sample of inclusive jets is subjected
to all of our event selection requirements except the lepton identification requirements.
In lieu of an identified lepton, a jet is required
with $E_\mathrm{T} > 20$~GeV.  This jet must contain at least four tracks in order to reduce
contamination from real electrons from $W$ and $Z$ boson decay, and
80--95\% of the jet's total calorimetric energy must be 
in the electromagnetic calorimeter, in order
to simulate a misidentified electron.
The $b$-tagging requirement on other jets in the event is relaxed to requiring a taggable jet instead of a tagged
jet in order to increase the size of the selected sample.  A taggable jet is one
that is within the acceptance of the silicon tracking detector and which has at least
two tracks in it.  This sample is called the
jet-based sample.

The second sample takes advantage of the fact that fake leptons from
non-$W$ events have difficulty passing the lepton selection
requirements.  We look at lepton candidates in the central electron trigger 
that fail at least two of five identification requirements that do not depend
on the kinematic properties of the event, such as the fraction of
energy in the hadronic calorimeter.  These objects are treated as
leptons and all other selection requirements are applied.  This sample has the
advantage of having the same kinematic properties as the central
electron sample.  This sample is called the ID-based sample.

The two samples described above are designed to model events with misidentified
electron candidates.  Because of the similarities in the kinematic properties
of the ID-based and the jet-based events, we use the union of the jet-based and ID-based
samples as our non-$W$ model for triggered central electrons (the CEM sample).  
Remarkably, the same samples also simulate the kinematics of events with
misidentified triggered muon candidates; we use the samples again to model those events (the
CMUP and CMX samples).  The jet-based sample alone is used to model the non-$W$ background
in the PHX sample because the angular coverage is greater.

The kinematic distributions of the reconstructed objects
in the EMC sample are different from those in the CEM, PHX, CMUP, and CMX samples
due to the trigger requirements, and thus a separate sample must be used to model
the non-$W$ background in the EMC data.  This third sample
consists of events that are collected with the $\EtMiss$+jets trigger path
and which have a muon candidate passing all selection requirements except for the
isolation requirement.  It is called the non-isolated sample.

\begin{figure*}
\begin{center}
\subfigure[]{
\includegraphics[width=0.55\columnwidth]{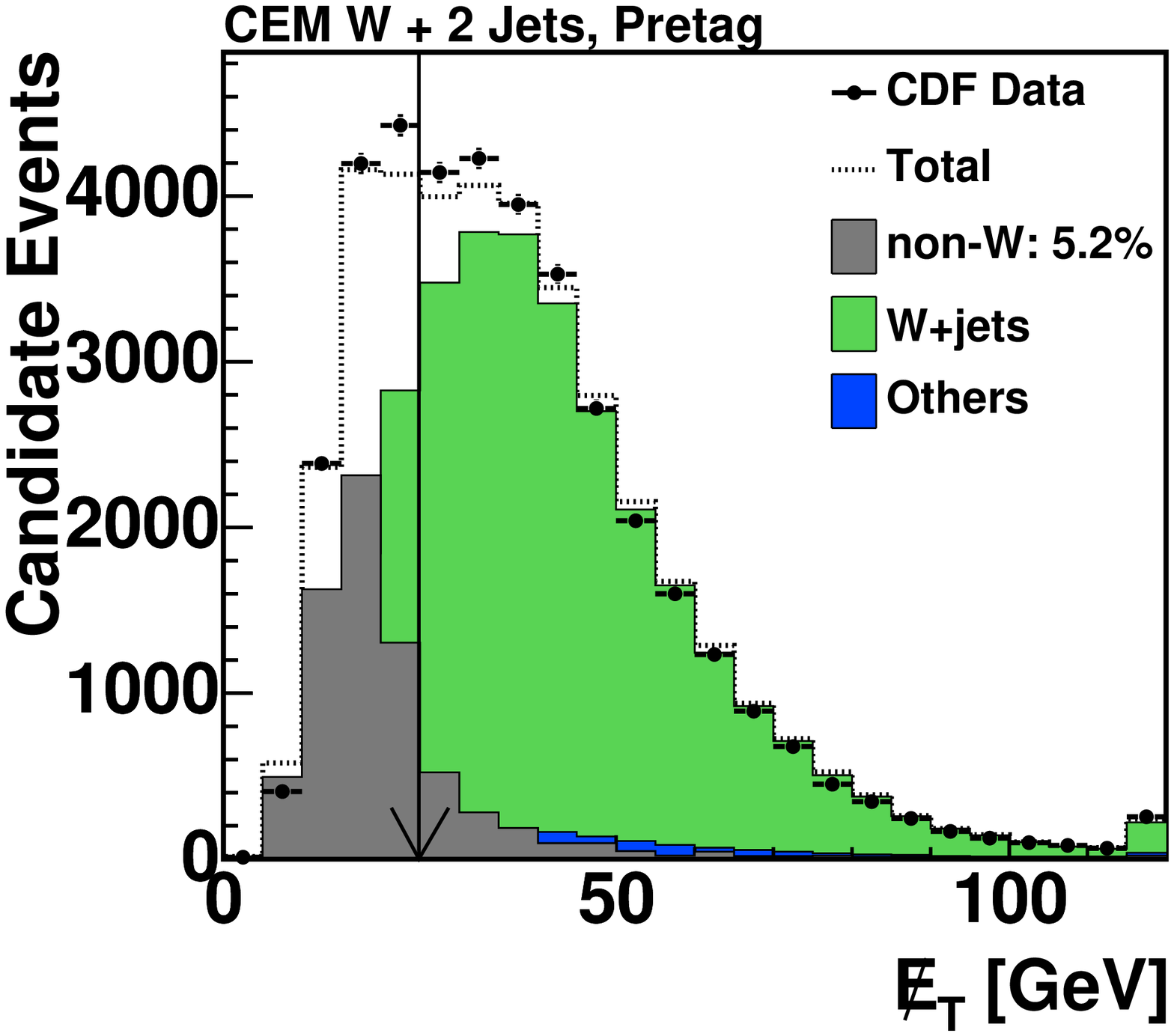}
\label{fig:pretagMetCEM}}
\subfigure[]{
\includegraphics[width=0.55\columnwidth]{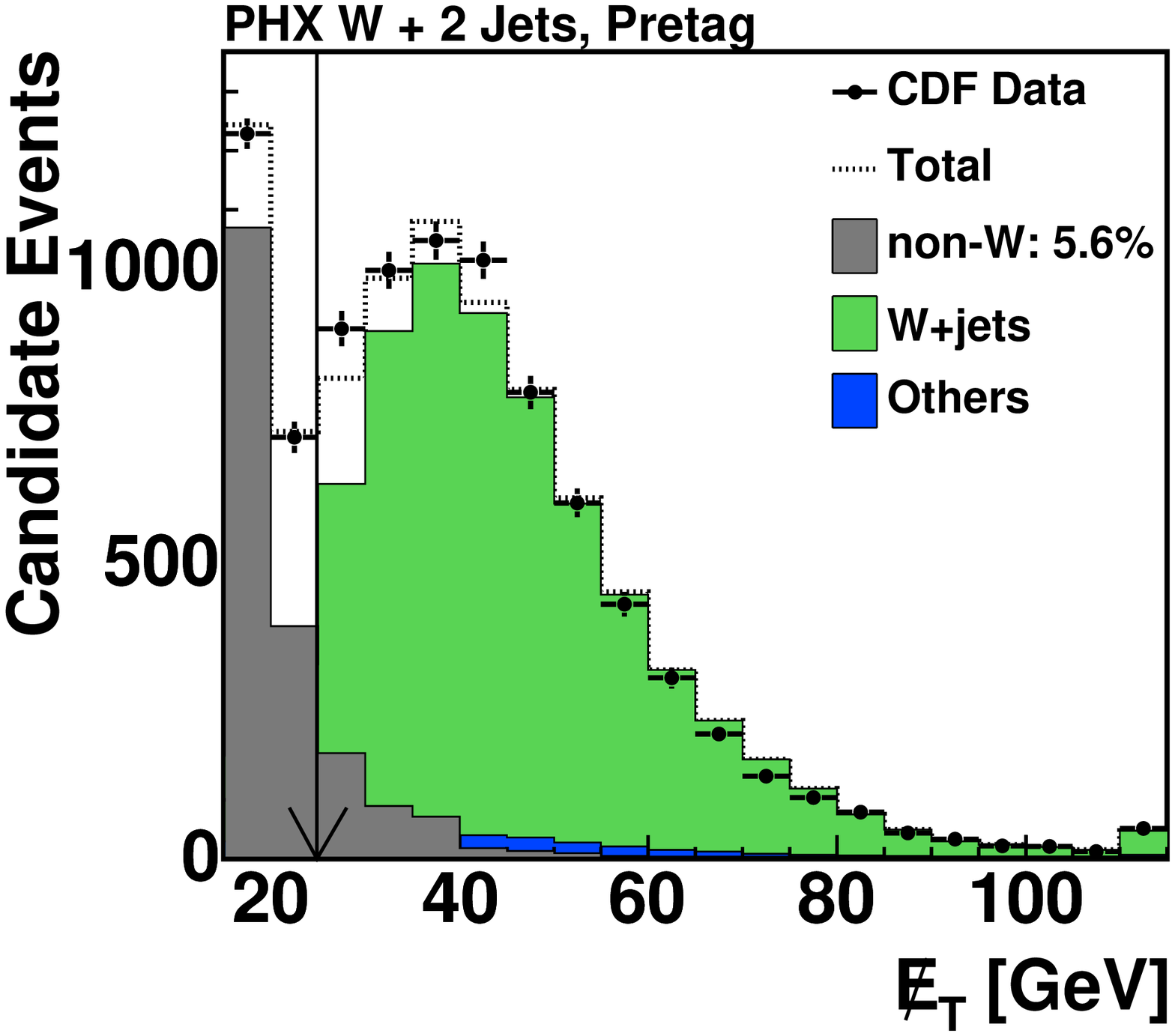}
\label{fig:pretagMetPHX}} 
\subfigure[]{
\includegraphics[width=0.55\columnwidth]{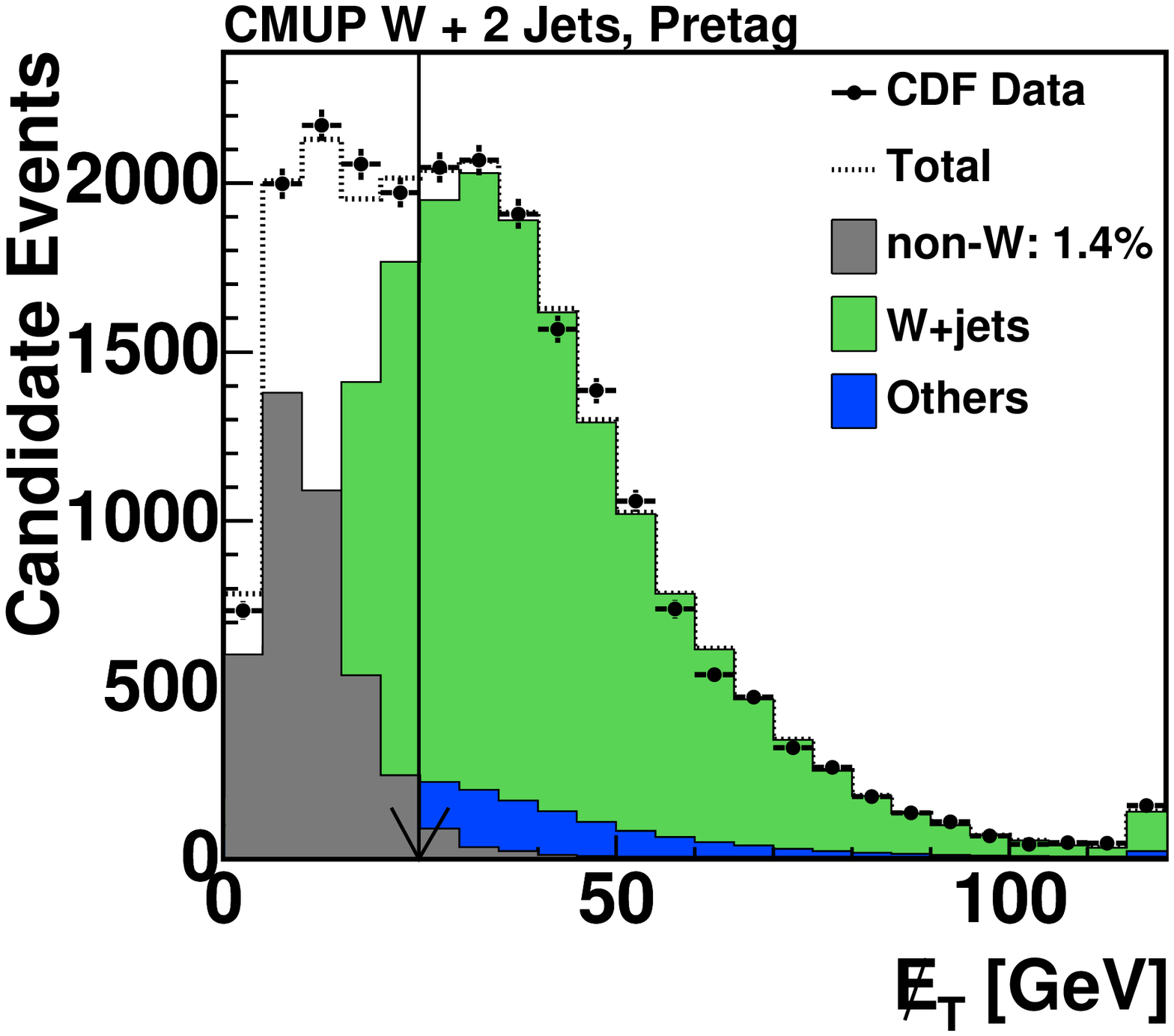}
\label{fig:pretagMetCMUP}} 
\subfigure[]{
\includegraphics[width=0.55\columnwidth]{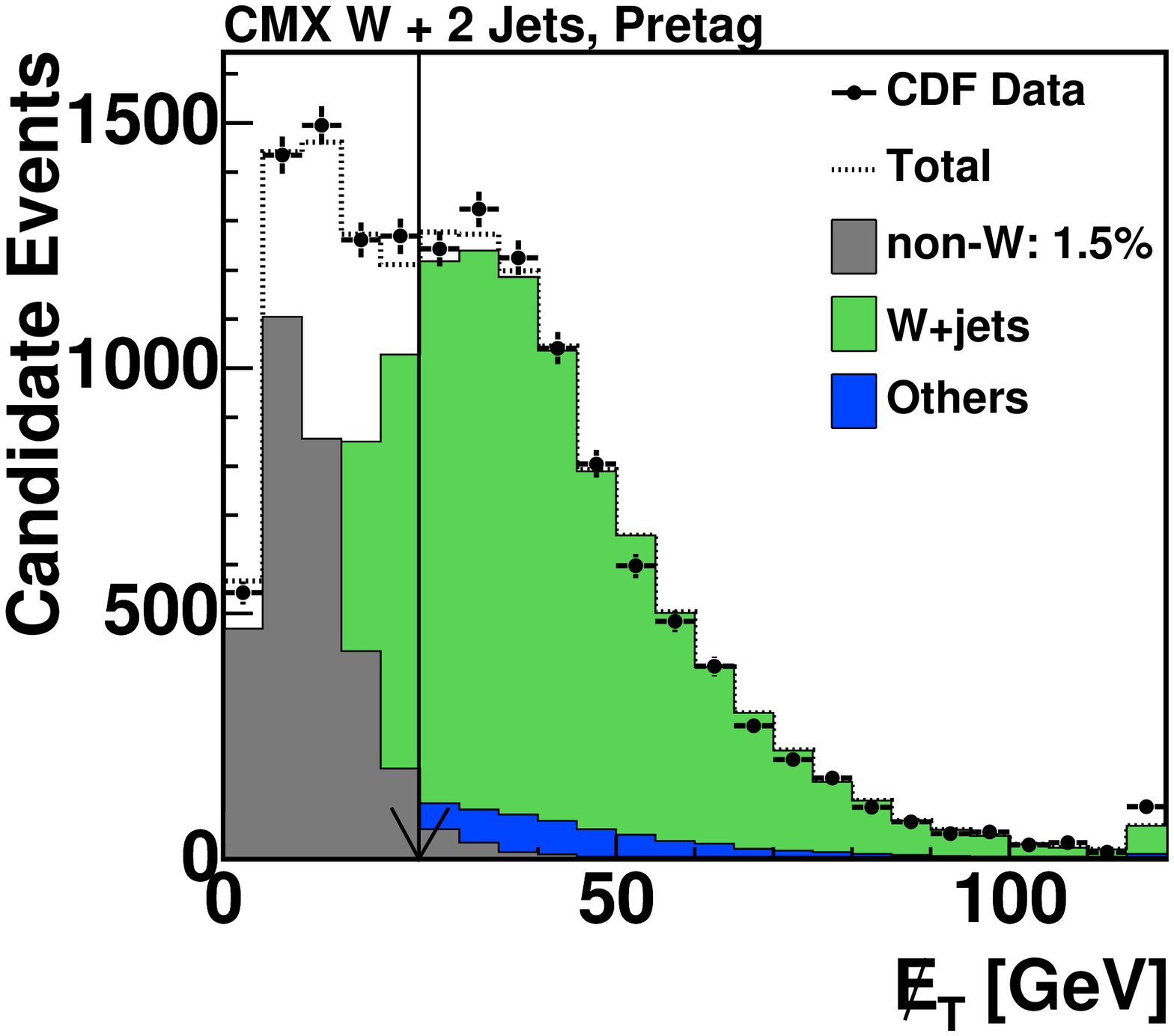}
\label{fig:pretagMetCMX}} 
\subfigure[]{
\includegraphics[width=0.55\columnwidth]{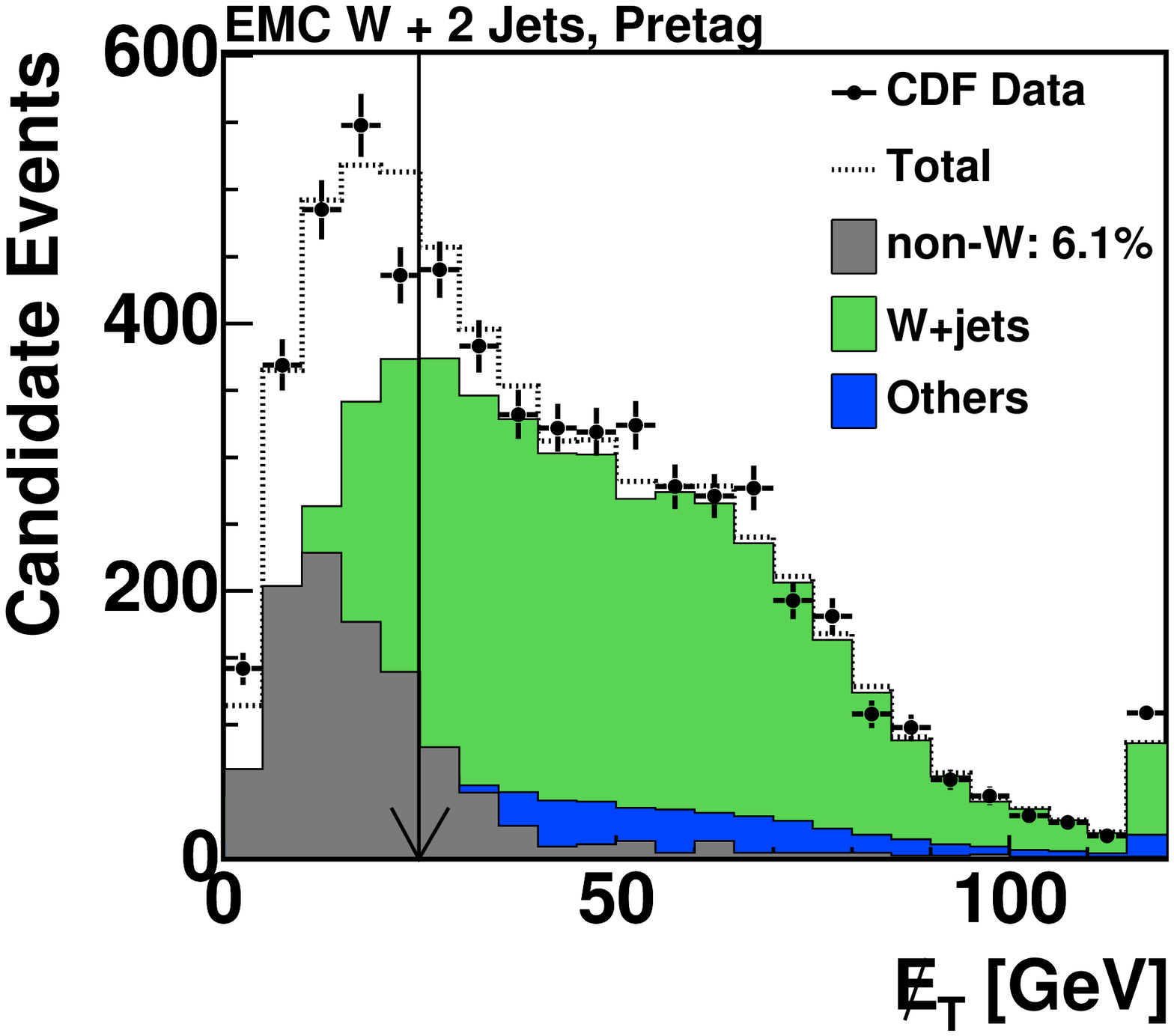}
\label{fig:pretagMetLOOSE}}
\end{center}
\caption{\label{fig:pretagMetFit}Fits to $\EtMiss$ distributions in the
pretag samples for the five different lepton categories (CEM, PHX, CMUP, CMX, EMC)  in $W$+two jet
events.  The fractions of non-$W$ events are estimated from the portions
of the templates above the $\EtMiss$ thresholds shown by the arrows.  Overflows are
collected in the highest bin of each histogram.  
The data are indicated with points
with error bars, and the shaded histograms show the best-fit predictions.  The non-$W$
templates are not shown stacked, but the $W$+jets and ``Others'' templates are stacked.
The unshaded histogram is the sum of the fitted shapes.
}

\begin{center}
\subfigure[]{
\includegraphics[width=0.55\columnwidth]{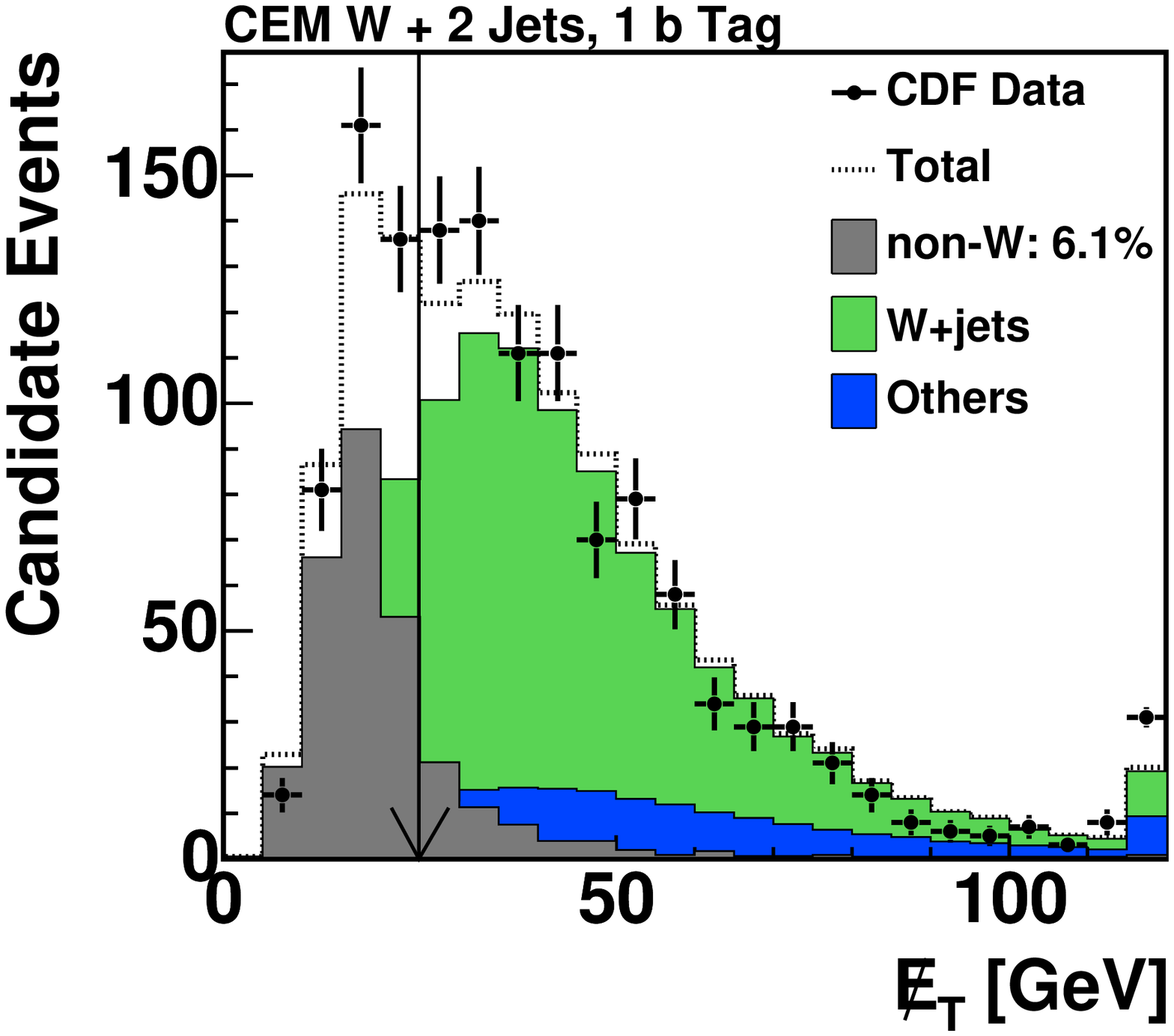}
\label{fig:1tagMetCEM}}
\subfigure[]{
\includegraphics[width=0.55\columnwidth]{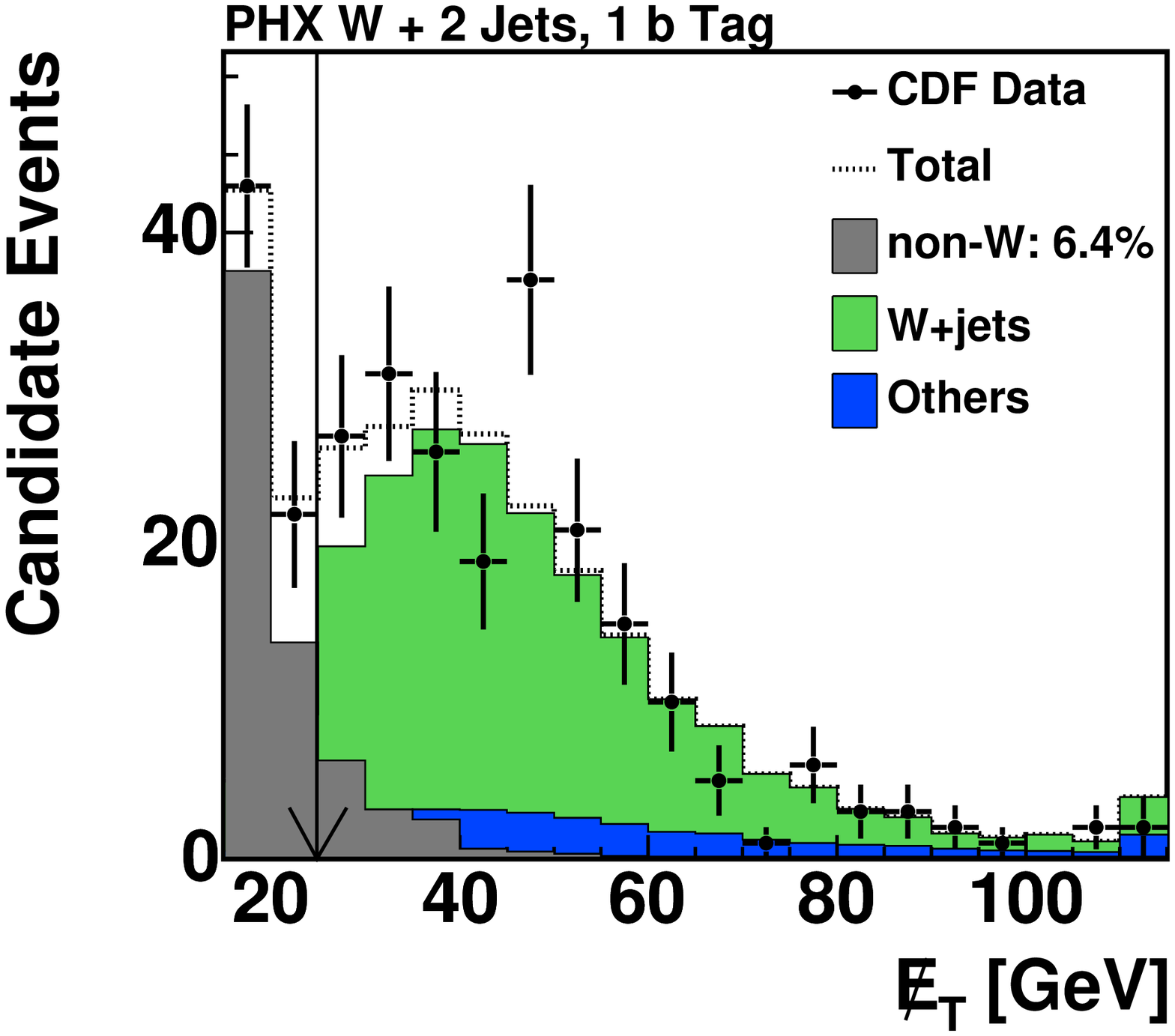}
\label{fig:1tagMetPHX}} 
\subfigure[]{
\includegraphics[width=0.55\columnwidth]{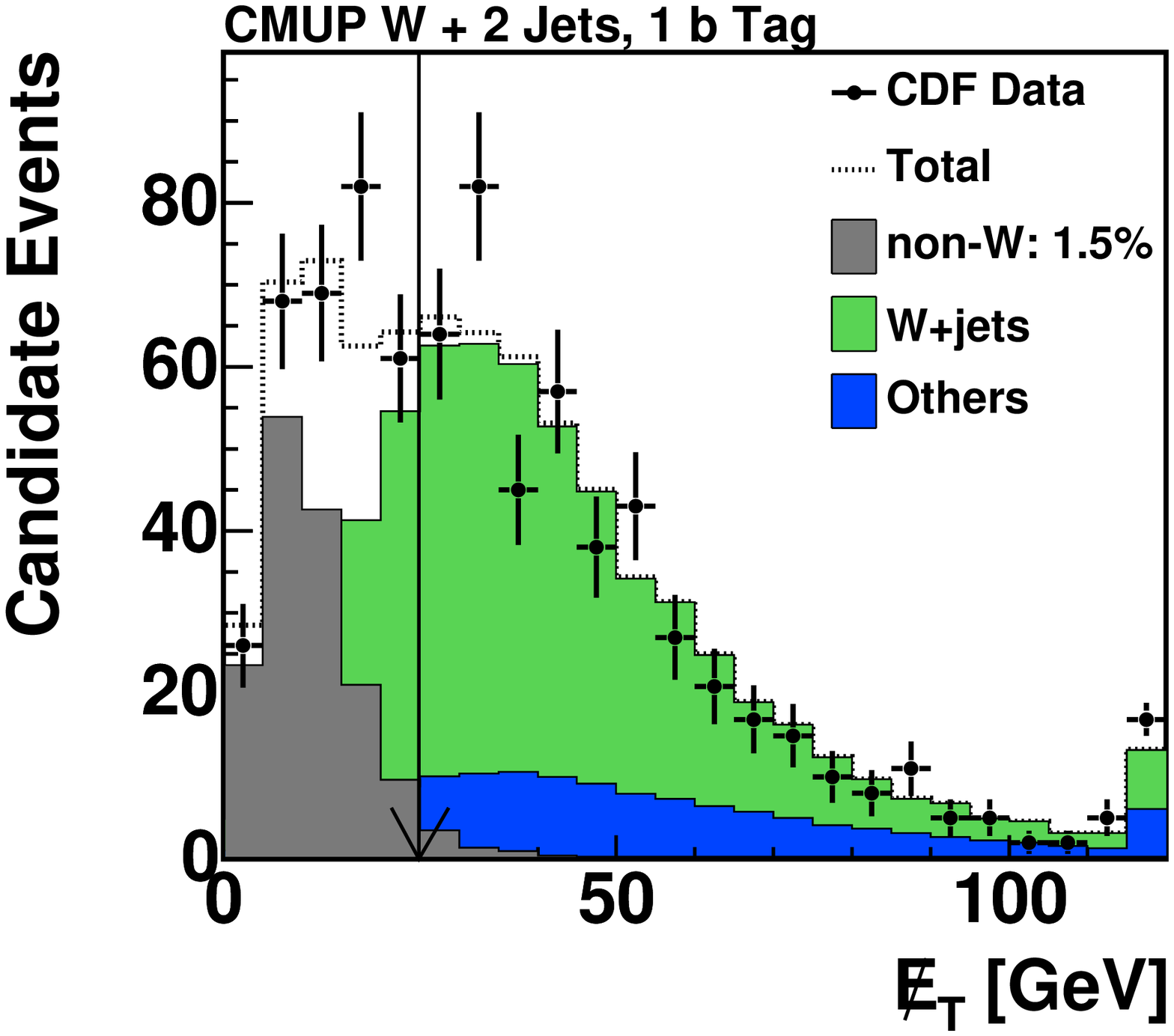}
\label{fig:1tagMetCMUP}}
\subfigure[]{
\includegraphics[width=0.55\columnwidth]{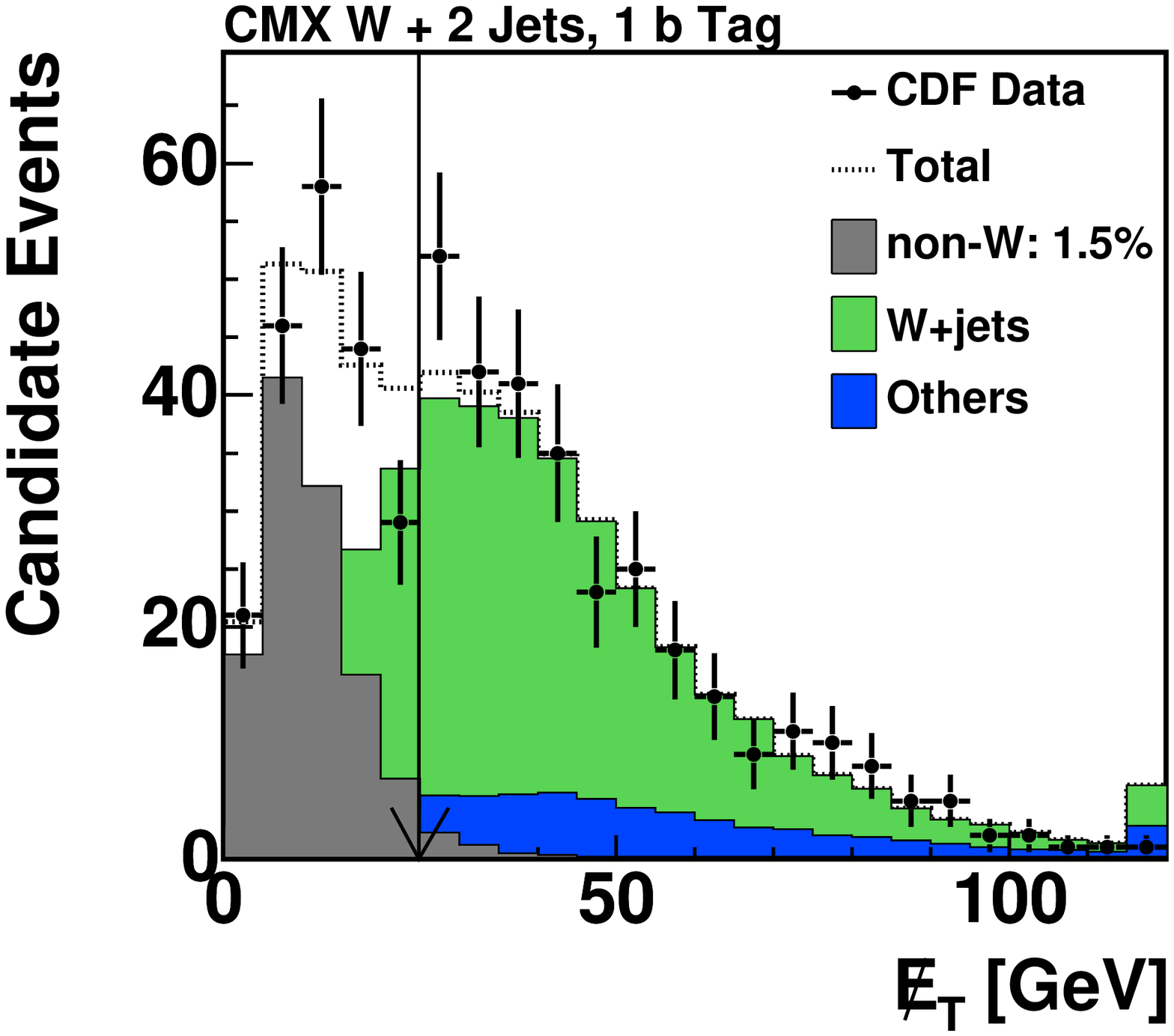}
\label{fig:1tagMetCMX}} 
\subfigure[]{
\includegraphics[width=0.55\columnwidth]{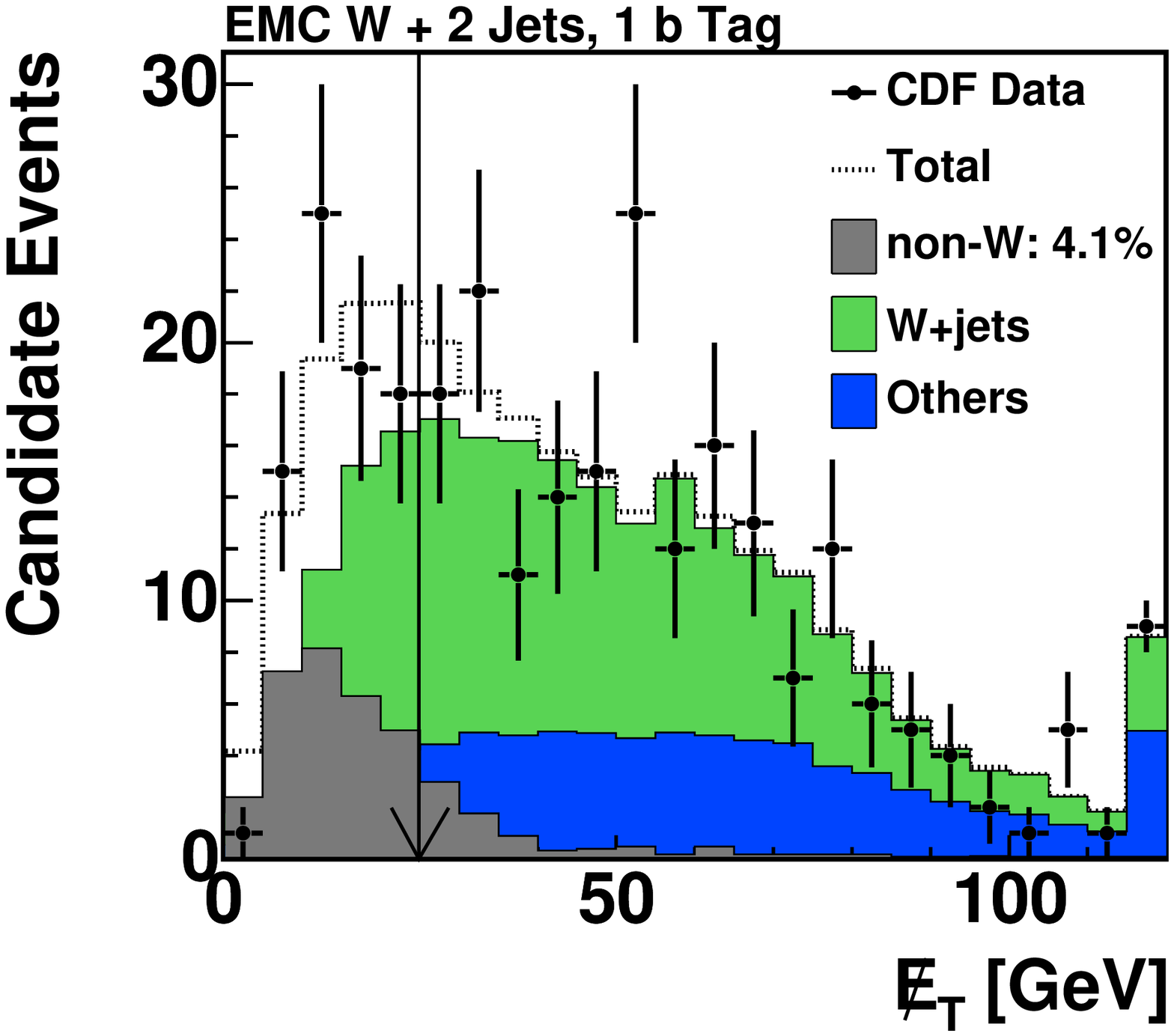}
\label{fig:1tagMetLOOSE}}
\end{center}
\caption{\label{fig:1tagMetFit}Fits to $\EtMiss$\ distributions in the
single-tagged sample for the five different lepton categories (CEM, PHX, CMUP, CMX, EMC) in $W$+2~jet
events.  The fraction of non-$W$ events is estimated from the fraction
of the template above the $\EtMiss$\ threshold shown by the arrows.  Overflows are
collected in the highest bin of each histogram.
The data are indicated with points
with error bars, and the shaded histograms show the best-fit predictions.  The non-$W$
template is not shown stacked, but the $W$+jets and ``Others'' templates are stacked.
The unshaded histogram is the sum of the fitted shapes.
}
\end{figure*}

The non-$W$ background must be determined not only for the data sample passing the event
selection requirements, but also for the control
samples which are used to determine the $W+$jets backgrounds, as described in Sections~\ref{sec:wplushf} 
and~\ref{sec:mistag}.
The expected numbers of non-$W$ events are estimated in
pretag events -- events in which all selection criteria are applied except the secondary vertex tag
requirement.  We require that at least one jet in a pretagged event is taggable.
In order to estimate the non-$W$ rates in this sample, we also remove the 
$\EtMiss$\ event selection requirement, but we retain all other non-$W$ rejection requirements.
We fit templates of the $\EtMiss$\ distributions
of the $W$+jets and the non-$W$ samples
to the $\EtMiss$\ spectra of the pretag data, holding constant
the normalizations of the additional templates needed to model the small diboson, $t{\bar{t}}$,
$Z+$jets, and single top backgrounds.  
The fractions of non-$W$ events are then calculated in
the sample with $\EtMiss > 25$ GeV.  The inclusion or omission of the single top
contribution to these fits has a negligible impact on the non-$W$ fractions that are fit.
These fits are performed separately for each lepton category (CEM, PHX, CMUP, CMX, and EMC) because the
instrumental fake lepton fractions are different for electrons and muons, and for the
different detector components.  In all lepton categories except PHX, the full 
$\EtMiss$\ spectrum is used in the fit.  For the PHX electron sample, we require 
$\EtMiss>15$~GeV in order to minimize sensitivity to the trigger.
The fits in the pretag region are also used to estimate the $W$+jets 
contribution in the pretag region, as described in Section~\ref{sec:wplushf}.
As Fig.~\ref{fig:pretagMetFit}
shows, the resulting fits describe the data quite well.

Estimates of the non-$W$ yields in the tagged samples used to search for the single top
signal are also needed.  These
samples are more difficult because the non-$W$ modeling samples are too
small to apply tagging directly -- only a few events pass the
secondary vertex requirement.  However, since the data show no dependence of
the $b$-tagging rate on $\EtMiss$, we use the untagged
non-$W$ templates in the fits to the $\EtMiss$ distributions in the tagged samples.
These fits are used
to extract the non-$W$ fractions in the signal samples.  As before, the
Monte Carlo predictions of diboson, $t{\bar{t}}$, $Z+$jets, and single top production
 are held constant and only the normalizations of the $W$+jets and the
non-$W$ templates are allowed to float. The resulting shapes are shown
in Fig.~\ref{fig:1tagMetFit} for the single-tagged sample, and these are used to derive the non-$W$ fractions
in the signal samples.  As before, the inclusion or omission of the single top
contributions in the fits has a negligible effect on the fitted non-$W$ fractions.
Because of the uncertainties in the tagging rates,
the template shapes, and the estimation methods, the estimated non-$W$ rates are
given systematic uncertainties of $\pm 40$\% in single-tagged events and $\pm 80$\% in
double-tagged events.  These uncertainties cover the differences in the results
obtained by fitting variables other than $\EtMiss$, as well as by changing
the histogram binning, varying the fit range, and using alternative samples
to model the non-$W$ background.  The uncertainty
in the double-tagged non-$W$ prediction is larger because of the larger
statistical uncertainty arising from the smaller size of the double-tagged sample.

\subsection{{\it W}+Heavy Flavor Contributions} 
\label{sec:wplushf}

Events with a $W$~boson accompanied by heavy flavor 
production constitute the majority of the
$b$-tagged lepton+jets sample.  These processes are $Wb{\bar{b}}$, shown in Fig.~\ref{fig:Wjets}(a),
$Wc{\bar{c}}$, which is the same process as $Wb{\bar{b}}$, but with charm quarks replacing the $b$ quarks, and $Wcj$, which is shown
in Fig.~\ref{fig:Wjets}(b).  Each process may be accompanied by more jets and pass the event selection
requirements for the $W$+3 jets signal sample.  Jets may fail to be detected, or they may fail to pass
our selection requirements, and such events may fall into
the $W$+1 jet control sample.  While these events can be simulated
using the {\sc alpgen} generator, the theory uncertainties on the
cross sections of these processes remain large compared with the size
of the single top quark 
signal~\cite{Bern:1996ka,Bern:1997sc,Giele:1996aa,Ellis:1998fv,FebresCordero:2006sj,Campbell:2006cu,Campbell:2008hh,Cordero:2009kv}.  
It is because of these large {\it a priori} uncertainties
on the background predictions and the small signal-to-background ratios
in the selected data samples that we must use advanced analysis techniques to
purify further the signal. We also use the data itself, both in control samples and
{\it in situ} in the samples passing all selection requirements, to constrain the
background rates, reducing their systematic uncertainties.  The {\it in situ} fits are
described in Section~\ref{sec:Interpretation}, and the control sample fits are described below.

The control samples used to estimate the $W+$ heavy flavor predictions and uncertainties
are the pretagged $W+n$~jets samples and the tagged $W+1$~jet sample.
We use the {\sc alpgen}+{\sc pythia} Monte Carlo model to extrapolate the measurements in the control samples to make
predictions of the $W+$heavy flavor background contributions in the data samples passing our signal
selection requirements.  The pretagged $W+n$~jets samples are used to scale the {\sc alpgen} predictions,
and the tagged $W+1$ jet sample is used to check and adjust {\sc alpgen}'s predictions of the 
fractions of $W$+jets events which are $Wb\bar{b}$, $Wc\bar{c}$, and $Wcj$ events.  A full description of the
method follows.

The number of pretag $W$+jets events is estimated by assuming that events not included in the
predictions based on Monte Carlo (these are the
$t{\bar t}$ and diboson predictions -- the single top quark signal
is a negligible component of the pretag sample) or non-$W$ multijet events, are
$W$+jets events.  That is:
\begin{equation}
  \label{eq:Wjets}
  N_{W\mathrm{+jets}}^{\mathrm{pretag}}=N_{\mathrm{data}}^{\mathrm{pretag}}\times (1-f_{\text{non-}W}^{\mathrm{pretag}})-N_{\mathrm{MC}}^{\mathrm{pretag}}
\end{equation}
where $N_{\mathrm{data}}^{\mathrm{pretag}}$ is the number of observed events in the pretag sample,
$f_{\text{non-}W}^{\mathrm{pretag}}$ is the fraction of non-$W$ events in the pretag
sample, as determined from the fits described in Section~\ref{sec:nonw},
and $N_{\mathrm{MC}}^{\mathrm{pretag}}$ is the expected number of pretag
$t{\bar t}$ and diboson events.   {\sc Alpgen} typically underestimates the
inclusive $W+$jets rates by a factor of roughly 1.4~\cite{Aaltonen:2007ip}.  
To estimate the yields of $Wb\bar{b}$, $Wc\bar{c}$, and $Wcj$ events, we multiply this data-driven estimate of the
$W$+jets yield by heavy flavor fractions.

The heavy flavor fractions in $W$+jets events are also not well predicted by our
{\sc alpgen}+{\sc pythia} model.   In order to improve the modeling of these
fractions, we perform fits to templates of flavor-separating variables in the
$b$-tagged $W$+1~jet data sample, 
which contains a vanishingly small component of single top quark signal events and is not otherwise
used in the final signal extraction procedure.
This sample is quite large and is almost
entirely composed of $W$+jets events.  We include Monte Carlo models of the
small contributions from $t{\bar{t}}$ and diboson
events as separate templates, normalized to their SM expected rates,
in the fits to the data.   Care must be exercised in the estimation of the $W$+heavy flavor
fractions, because fitting in the $W$+1~jet sample and using the fit values for the
$W$+2~jet and $W$+3~jet samples is an extrapolation.  We seek to estimate
the $b$ and charm fractions in these events with as many independent methods as possible and
we assign generous uncertainties that cover the differences between the several estimations
of the rates.

\begin{figure*}
\begin{center}
\subfigure[]{
\includegraphics[width=0.8\columnwidth]{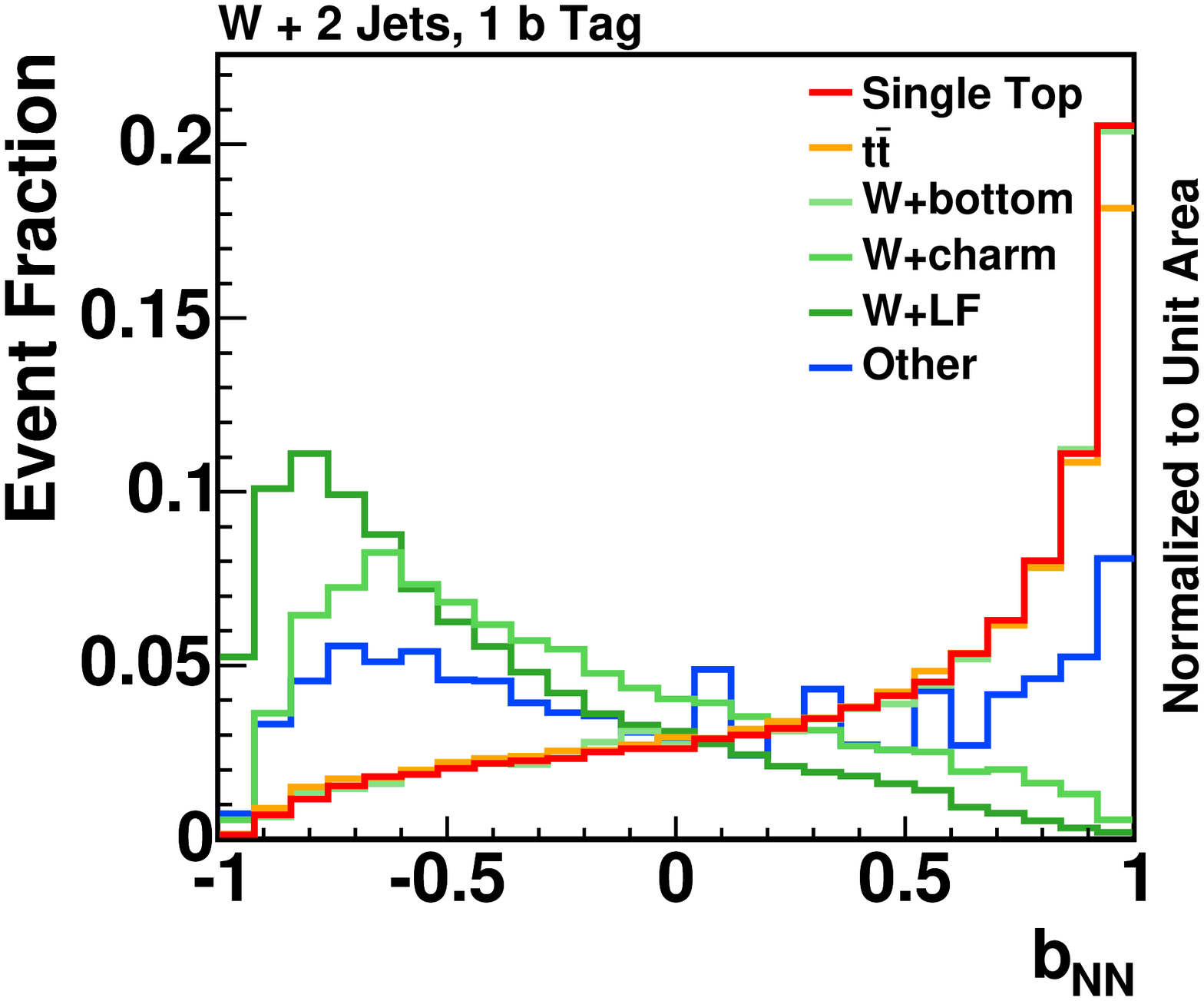}
\label{fig:nntagtemplates}}
\subfigure[]{
\includegraphics[width=0.8\columnwidth]{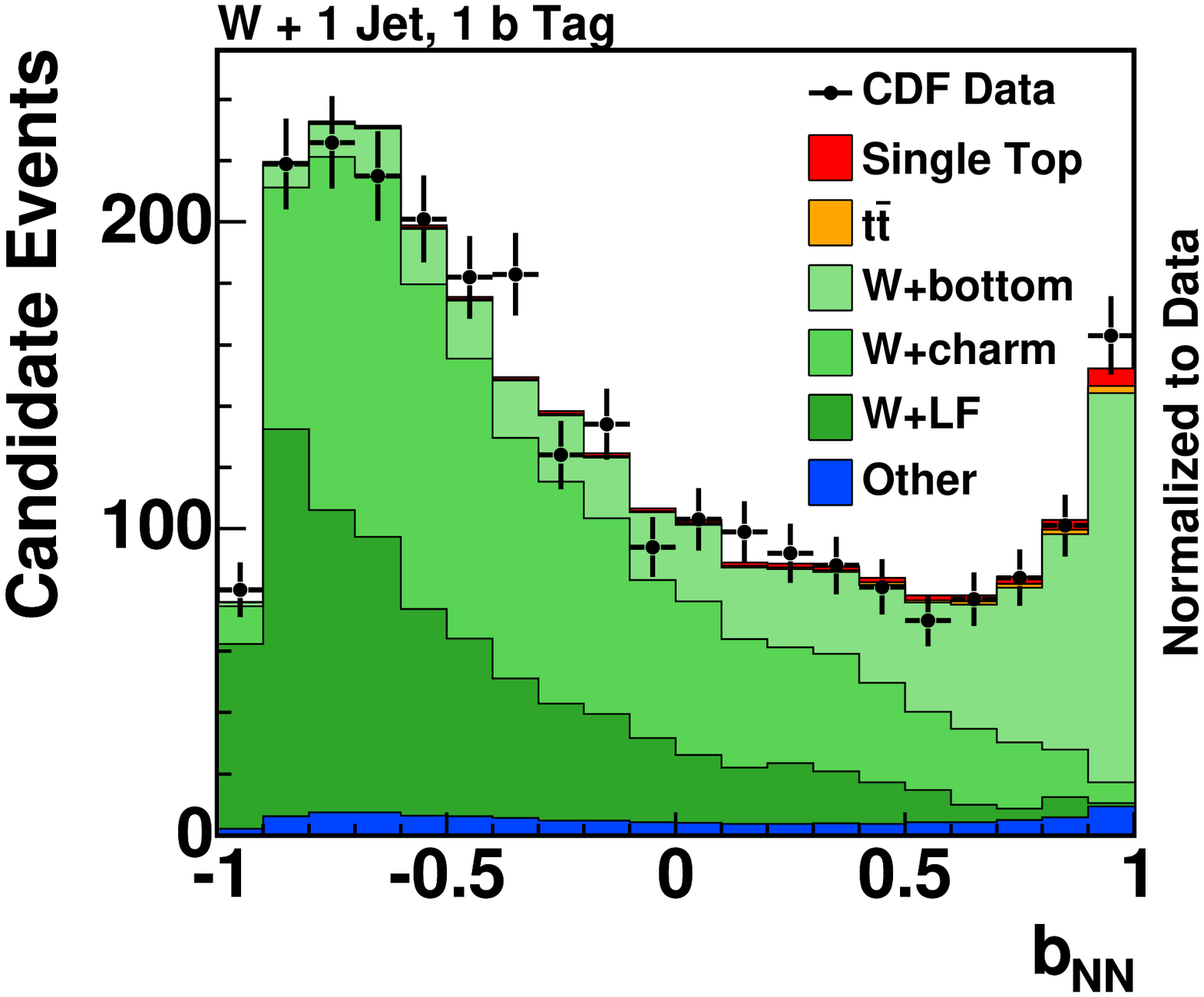}
\label{fig:bnnfit1j}}
\subfigure[]{
\includegraphics[width=0.8\columnwidth]{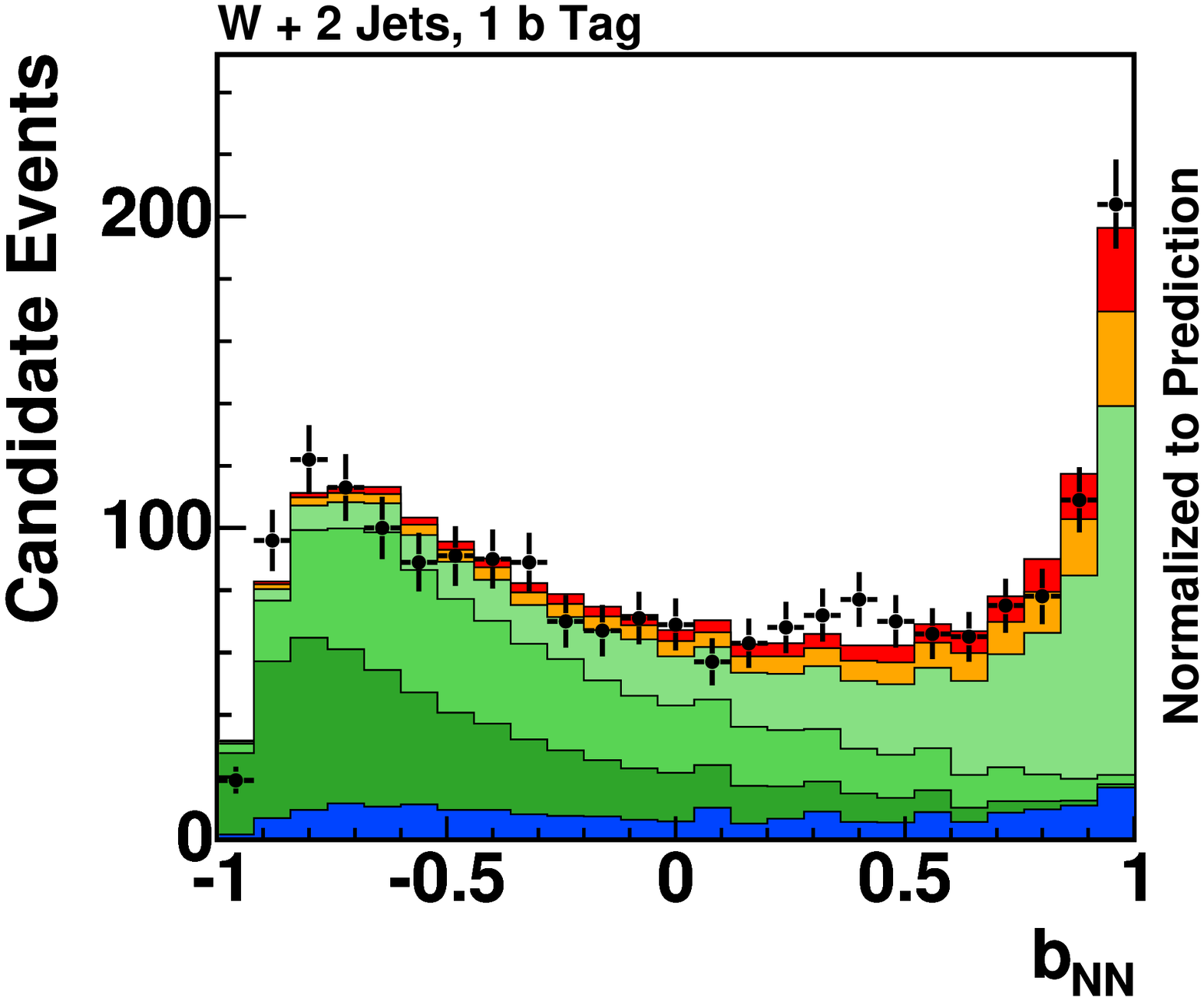}
\label{fig:bnn2j}}
\subfigure[]{
\includegraphics[width=0.8\columnwidth]{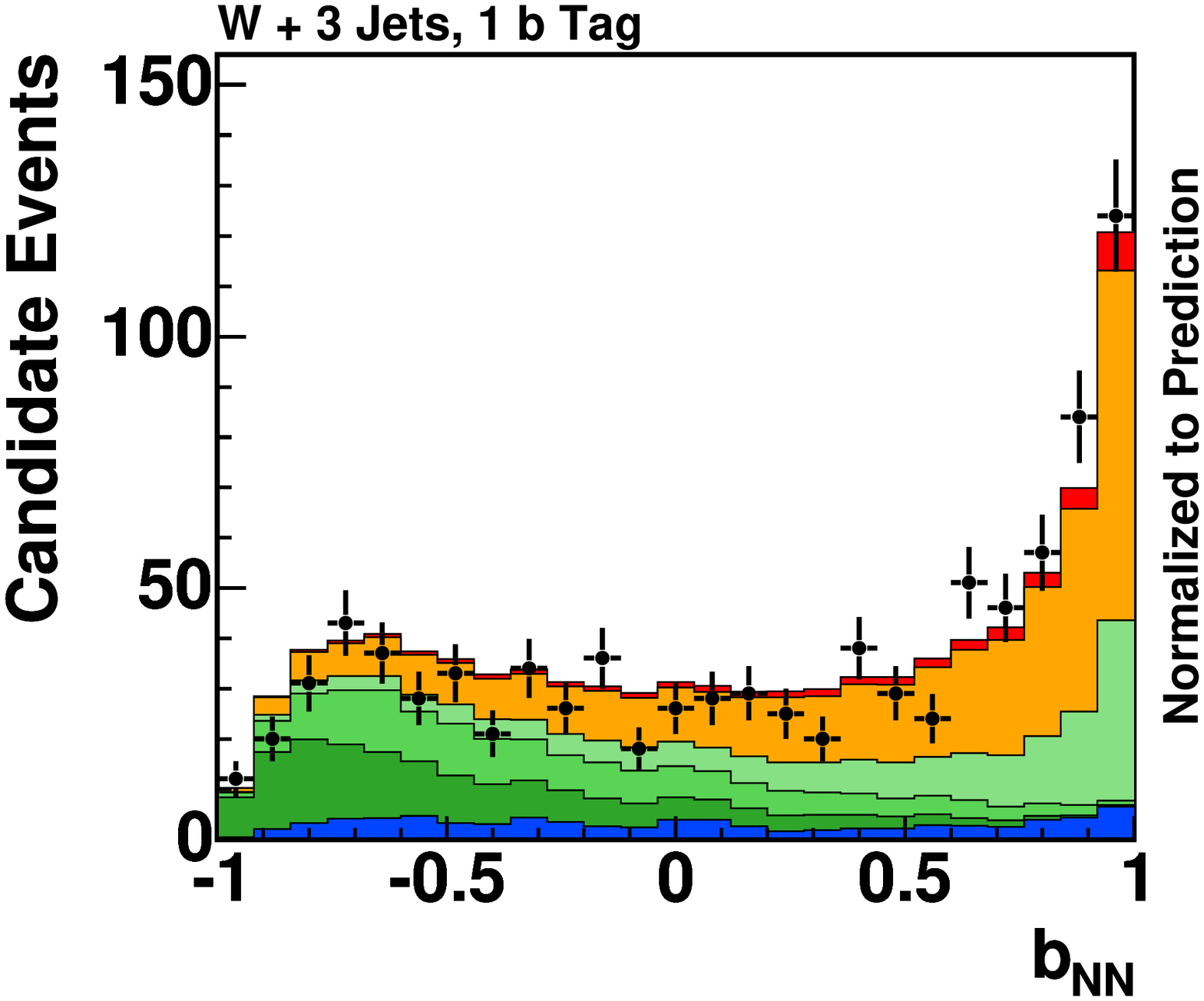}
\label{fig:bnn3j}}
\end{center}
\caption{\label{fig:bnntv} Templates (a) of the jet flavor
separator $b_{\mathrm{NN}}$ for $W$+light, $W$+charm (adding the $Wc{\bar{c}}$ and $Wcj$
contributions because of their similar shapes), and $W$+bottom events.  The template
labeled ``Other'' represents the diboson and $Z/\gamma^*$+jets contributions.  The strong
discrimination $b_{\rm{NN}}$ provides to separate jet flavors makes it a powerful variable in 
multivariate analyses.  Panel (b) shows the outcome of the fit to the $W$+1~jet data sample
allowing the $b$, $c$, and light-flavor components to 
float as described in Section~\protect\ref{sec:Background}.
Panels (c) and (d) compare the data and the corresponding predictions in the $W$+2~jet and $W$+3~jet samples.
In panels (b) through (d), the data are indicated with points with error bars, and the model predictions are
shown with shaded histograms, stacked in the same order as the legend.}
\end{figure*}

We fit the distribution of the
jet-flavor separator $b_{\mathrm{NN}}$ described in Section~\ref{sec:btagger}.   Template
distributions are created based on 
{\sc alpgen}+{\sc pythia} Monte Carlo
samples for the $W$+LF, $Wc{\bar{c}}$, $Wcj$, $Wb{\bar{b}}$, and $Z/\gamma^*$+jets processes,
where $W$+LF events are those in which none of the jets accompanying the leptonically decaying $W$
boson contains a $b$ or $c$ quark.  The template distributions for these five processes are shown
in Fig.~\ref{fig:bnntv}(a).
The $t{\bar{t}}$ and diboson templates are created using {\sc pythia} Monte Carlo
samples.  The non-$W$ model described in Section~\ref{sec:nonw} is also used.  The
$W$+LF template's rate is constrained by the data-derived mistag estimate, described in
Section~\ref{sec:mistag}, within its
uncertainty; the other $W+$jets templates' rates are not constrained.  The $t{\bar{t}}$, diboson,  
$Z/\gamma^*$+jets, and non-$W$ contributions are constrained within their uncertainties. 
The $Wb{\bar{b}}$ and $Wc{\bar{c}}$ components float in the fit but are scaled with the same
scaling factor, as the same diagrams, with
$b$ and $c$ quarks interchanged, contribute
in the {\sc alpgen} model, and we expect a similar correspondence for the leading
processes in the data.  We also let the $Wcj$ fraction float in the fit. 
The best fit in the $W+$1~jet sample is shown in Fig.~\ref{fig:bnntv}(b).

The fit indicates that the {\sc alpgen}-predicted $Wb{\bar{b}}+Wc{\bar{c}}$ fraction must be
multiplied by $1.4\pm 0.4$ in order for the templates to match the data, and the
best-fit value of the $Wcj$ fraction is also $1.4 \pm 0.4$ larger than that predicted
by {\sc alpgen}.  In addition to the fit to the $b_{\mathrm{NN}}$ distribution, we also 
fit the $W$+heavy flavor fractions in the $b$-tagged $W$+1-jet sample
with another variable, the reconstructed invariant mass of
the secondary vertex.  We perform this alternate fit in our standard $b$-tagged sample as well
as in one with loosened $b$-tag requirements.

We obtain additional information from~\cite{Wcharm:2007dm}, in which a direct measurement
of the $Wc$ fraction is made using lepton charge correlations.  The central value of this
measurement agrees well with the Monte Carlo predictions.  We thus set the multiplicative factor
of the $Wc$ component to $1.0\pm 0.3$ for use in the two- and three-jet bins.
 
The 30\% uncertainties assessed on the $Wb{\bar{b}}+Wc{\bar{c}}$ and $Wcj$ yields
cover the differences in the measured fit values and also approximates our uncertainty in
extrapolating this fraction to $W$+2 and 3 jet events.  
We check these extrapolations
in the $W+$2 and 3 jet events as shown in Figs.~\ref{fig:bnntv}(c) and~\ref{fig:bnntv}(d); 
no additional fit is performed
for this comparison.  The rates and flavor compositions match very well with the observed data
in these samples.
The uncertainties in the fit fractions arising from the uncertainties on the shapes of 
the $b_{\rm{NN}}$ templates discussed in Section~\ref{sec:btagger}
are a negligible component of the total uncertainty.

Since the yields of $W$+heavy flavor events are estimated from $b$-tagged data using the
same {\sc secvtx} algorithm as is used for the candidate event selection, the uncertainty
in the $b$-tagging efficiency does not factor into the prediction of these rates.

\subsection{Rates of Events with Mistagged Jets} 
\label{sec:mistag}

Some $W$+LF events pass our event selection
requirements due to the presence of mistagged jets.  A mistagged jet
is one which does not contain a weakly-decaying $B$ or charm hadron
but nonetheless passes all of the secondary vertex tagging
requirements of the {\sc secvtx} algorithm~\cite{Acosta:2004hw}.  Jets
are mistagged for several reasons: tracking errors such as hit
misassignment or resolution effects cause the reconstruction of false
secondary vertices, the multi-prong decays of long-lived particles
like the $K^0_s$ and the $\Lambda^0$ supply real secondary vertices,
and nuclear interactions with the detector material also provide a
real source of non-$b/c$ secondary vertices.

The estimation of the background yields from tracking resolution
related mistags is accomplished without the use of detector simulation.
The procedure is to measure the fractions of jets which have
negative decay lengths (defined below) to estimate the fraction of light-flavor
jets which have incorrect positive decay lengths.  This fraction is adjusted in order
to account for the asymmetry between the negative decay length
distribution and the positive decay length distribution, and to account for
the heavy-flavor contribution in the jet data, to obtain the
mistag probability.  This probability is multiplied by an estimate of $W$+LF
jet yield in each of our samples, separately for each lepton category and
jet-number category.  Each of these steps is described in detail below.

Events passing inclusive jet triggers with vertices with negative
two-dimensional (2D) decay lengths comprise the control sample used to
estimate the mistag rate.  The 2D decay
length $L_{xy}$ is the magnitude of the displacement from the primary
vertex to the reconstructed secondary vertex, projected first onto the
plane perpendicular to the beam axis, and then projected again onto
the jet axis's projection in the plane perpendicular to the beam axis.
The sign is given by the sign of the dot product of the 2D decay
length and the jet momentum.  Tracking resolution effects are expected
to produce a symmetric distribution of the 2D decay length of
light-flavor misreconstructed secondary vertices, centered on zero.  A
jet is said to be ``negatively tagged'' if the transverse decay length
significance $L_{xy} / \sigma_{L_{xy}} < -7.5$, while $L_{xy} /
\sigma_{L_{xy}}>7.5$ defines a ``positively tagged'' jet.  

The per-jet mistag rate is not a single number but rather it is parameterized as a function of
six kinematic variables: the $E_\mathrm{T}$ and $\eta$ of the jet, the
number of tracks in the jet, the scalar sum of transverse energy of
the tight jets, the number of reconstructed primary vertices, and the
$z$ coordinate of the primary vertex associated with the jet.
Since the negative tag rate does not fully reflect the positive mistags due
to the decays of long-lived particles and interactions with the detector material, a
correction factor $\alpha\beta$ for the mistag asymmetry is applied.
The factor $\alpha$ corrects for the asymmetry between the positive and negative
tag rates of light-flavor jets, and the factor
$\beta$ corrects for the presence of $b$ jets in the jet
samples used to derive the mistag rate.  These correction factors are extracted
from fits to distributions of the invariant mass of the reconstructed
secondary vertex in tagged jets in an inclusive jet sample.  A systematic uncertainty is derived
from fits to templates of pseudo-$c\tau$, which is defined as
$L_{xy}\frac{m}{p_\mathrm{T}}$~\cite{Acosta:2004hw}, where $m$ is the invariant mass of the
tracks in the displaced vertex, and $p_{\rm{T}}$ is the magnitude of the vector sum of the transverse
momenta of the tracks in the displaced vertex.  The systematic uncertainty on the asymmetry
factor $\alpha\beta$ is the largest component of the uncertainty on the mistag estimate.  Another
component is estimated from the differences in the negative tag rates computed with different
jet data samples with varying trigger requirements.  The average rate for jets to be mistagged
is approximately 1\%, although it depends strongly on the jet $E_{\rm T}$.

The per-jet mistag probabilities are multiplied by data-driven estimates of the $W$+LF yields, although
we must subtract the yields of the other components.  We subtract the pretagged $W$+heavy flavor
contributions from the pretagged $W$+jets yield of 
Equation~\ref{eq:Wjets} to estimate the $W$+LF yield:
\begin{equation}
\label{eq:WplusLF}
  N_{W\mathrm{+LF}}^{\mathrm{pretag}} =  N_{W\mathrm{+jets}}^{\mathrm{pretag}} 
- N_{Wb{\bar{b}}}^{\mathrm{pretag}}
- N_{Wc{\bar{c}}}^{\mathrm{pretag}}
- N_{Wcj}^{\mathrm{pretag}}
\end{equation} 
The pretagged $W$+heavy flavor contributions are estimated by dividing the tagged $W$+heavy flavor
contributions by the $b$-tagging efficiencies for each event category.  The mistag parameterization
is applied to each of the Monte Carlo and data samples used in Equations~\ref{eq:Wjets} and~\ref{eq:WplusLF},
in order for the total mistag yield prediction not to be biased by differences in the kinematics
of the several $W$+jets flavor categories.

We use {\sc alpgen}+{\sc pythia} Monte Carlo samples to predict the kinematics of $W$+LF events
for use in the analyses of this paper.  The mistag rate parameterization described above is applied 
to each jet in $W$+LF MC events, and these rates are used to weight the events 
to predict the yield of mistagged events in
each bin of each histogram of each variable.

\begin{figure}
\begin{center}
\includegraphics[width=0.8\columnwidth]{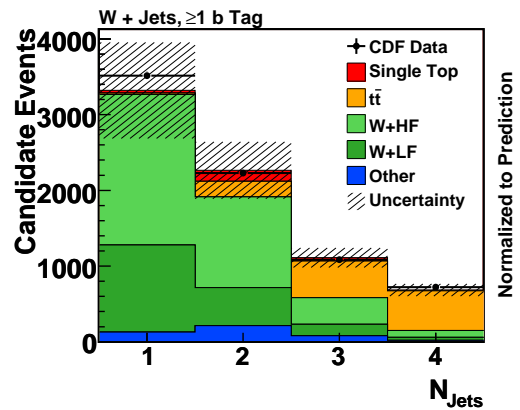}
\end{center}
\caption{\label{fig:Njets}The number of events predicted and observed
for $W$+jets events in which at least one jet is $b$-tagged.
The data are indicated with points, and the
shaded histograms show the signal and background predictions which are stacked
to form the total prediction. The stacking order is the same as the legend.
  The systematic uncertainty on the rates is far too large to
use a simple counting experiment to measure the single top quark cross section.}
\end{figure}

The predicted numbers of background events, signal events, and the
overall expected normalizations are given in
Tables~\ref{tab:EventYield1tag}, for events with exactly one $b$ tag, and in Table~\ref{tab:EventYield2tag}
for events with two or three $b$ tags.  Only two selected events in the data have three $b$ tags, consistent with
the expectation assuming that the third tag is a mistag.  The observed event counts and predicted yields
are summarized graphically as functions of jet multiplicity in Fig.~\ref{fig:Njets}.

\renewcommand{\arraystretch}{1.1}
\begin{table*}[tbh]
\caption{\label{tab:EventYield1tag} Summary of the predicted numbers of
signal and background events with exactly one $b$ tag, with systematic
uncertainties on the cross section and Monte Carlo efficiencies
included.   The total numbers of observed events passing the event selections are also
shown. 
The $W$ + 2 jets and $W$ + 3 jets samples are used to test for the signal, while the
$W$ + 1 jets and $W$ + 4 jets samples are used to check the background modeling. }
\begin{center}
\begin{tabular}{lcccc}\hline\hline
 & $W$ + 1 jet & $W$ + 2 jets & $W$ + 3 jets & $W$ + 4 jets \\
\hline

$Wb\bar{b}$	           & $823.7 \pm 249.6$    &  $581.1 \pm 175.1$  & $173.9 \pm 52.5$   & $44.8 \pm 13.7$  \\
$Wc\bar{c}$                & $454.7 \pm 141.7$    &  $288.5 \pm 89.0$   & $95.7  \pm 29.4$   & $27.2 \pm 8.5$  \\
$Wcj$                      & $709.6 \pm 221.1$    &  $247.3 \pm 76.2$   & $50.8  \pm 15.6$   & $10.2 \pm 3.2$  \\
Mistags                    & $1147.8 \pm 166.0$   &  $499.1 \pm 69.1$   & $150.3 \pm 21.0$   & $39.3 \pm 6.2$ \\
Non-$W$                    & $62.9  \pm 25.2$     &  $88.4  \pm 35.4$   & $35.4  \pm 14.1$   & $7.6  \pm 3.0$   \\
$t\bar{t}$ production      & $17.9 \pm 2.6$       &  $167.6 \pm 24.0$   & $377.3 \pm 54.8$   & $387.4 \pm 54.8$     \\
Diboson                    & $29.0  \pm 3.0$      &  $83.3  \pm 8.5$    & $28.1  \pm 2.9$    & $7.1  \pm 0.7$    \\
$Z/\gamma^*$+jets                   & $38.6  \pm 6.3$      &  $34.8  \pm 5.3$    & $14.6  \pm 2.2$    & $4.0  \pm 0.6$    \\
\hline			     		       	   		     	     		       		       				 		 
Total Background           & $3284.1\pm 633.8$    &  $1989.9\pm 349.6$  & $926.0 \pm 113.4$  & $527.7\pm 60.3$  \\
$s$-channel                & $10.7  \pm 1.6$      &  $45.3  \pm 6.4$    & $14.7  \pm 2.1$    & $3.3  \pm 0.5$    \\
$t$-channel                & $24.9  \pm 3.7$      &  $85.3  \pm 12.6$   & $22.7  \pm 3.3$    & $4.4  \pm 0.6$    \\
\hline			     		       	   		     	     		       		       				 		 
Total Prediction           & $3319.7 \pm 633.8$   &  $2120.4 \pm 350.1$ & $963.4 \pm 113.5$  & $535.4 \pm 60.3$ \\ 
\hline\hline		     		       	     			  		       		       	 
Observation                & 3516                 &  2090               & 920                & 567              \\ \hline 
\hline
\end{tabular}
\end{center}
\end{table*}

\begin{table*}[htb]
\caption{\label{tab:EventYield2tag} Summary of predicted numbers of
signal and background events with two or more $b$ tags, with systematic
uncertainties on the cross section and Monte Carlo efficiencies
included.  The total numbers of observed events passing the event selections are also
shown.
The $W$ + 2 jets and $W$ + 3 jets samples are used to test for the signal, while the
$W$ + 4 jets sample are used to check the background modeling. }
\begin{center}
\begin{tabular}{lccc}\hline\hline
 & $W$ + 2 jets & $W$ + 3 jets & $W$ + 4 jets \\
\hline

$Wb\bar{b}$	                    & $75.9 \pm 23.6$ & $27.4 \pm 8.5$  & $8.2  \pm 2.6$  \\
$Wc\bar{c}$                         & $3.7  \pm 1.2$  & $2.4  \pm 0.8$  & $1.1  \pm 0.4$  \\
$Wcj$                               & $3.2  \pm 1.0$  & $1.3  \pm 0.4$  & $0.4  \pm 0.1$  \\
Mistags                             & $2.2  \pm 0.6$  & $1.6  \pm 0.4$  & $0.7  \pm 0.2$  \\
Non-$W$                             & $2.3  \pm 0.9$  & $0.2  \pm 0.1$  & $2.4  \pm 1.0$  \\
$t\bar{t}$ production               & $36.4 \pm 6.0$  & $104.7\pm 17.3$ & $136.0\pm 22.4$ \\
Diboson                             & $5.0  \pm 0.6$  & $2.0  \pm 0.3$  & $0.6  \pm 0.1$  \\
$Z/\gamma^*$+jets                            & $1.7  \pm 0.3$  & $1.0  \pm 0.2$  & $0.3  \pm 0.1$  \\
\hline				    				       	  	       				     
Total Background                    & $130.4\pm 26.8$ & $140.6\pm 19.7$ & $149.8\pm 22.5$ \\
$s$-channel                         & $12.8 \pm 2.1$  & $4.5  \pm 0.7$  & $1.0  \pm 0.2$  \\
$t$-channel                         & $2.4  \pm 0.4$  & $3.5  \pm 0.6$  & $1.1  \pm 0.2$  \\
\hline				    				       	  	       				     
Total Prediction                    & $145.6\pm 26.9$ & $148.6\pm 19.7$ & $151.9\pm 22.5$ \\
\hline\hline			    			     	       	       	       			     
Observation                         & 139             & 166             & 154             \\ \hline  
\hline
\end{tabular}
\end{center}
\end{table*}

\subsection{Validation of Monte Carlo Simulation}
\label{sec:bgvalidation}

Because multivariate analyses depend so heavily on properly simulating
events, it is very important to validate the modeling of the
distributions in Monte Carlo by checking them with the data.  We do
this by comparing hundreds of data and Monte Carlo distributions.  We
make comparisons in control samples in which no jets have been
$b$-tagged to test the $W$+LF shapes, we test the modeling of
$W$+1~jet events to examine $W$+heavy flavor fraction and shapes, 
we compare the data and Monte Carlo distributions of kinematic
variables in the signal regions of tagged 2- and 3-jet events to check
the modeling of all of these variables, and we verify 
the modeling of the correlations between the discriminating variables.

\begin{figure*}
\begin{center}
\subfigure[]{
\includegraphics[width=0.55\columnwidth]{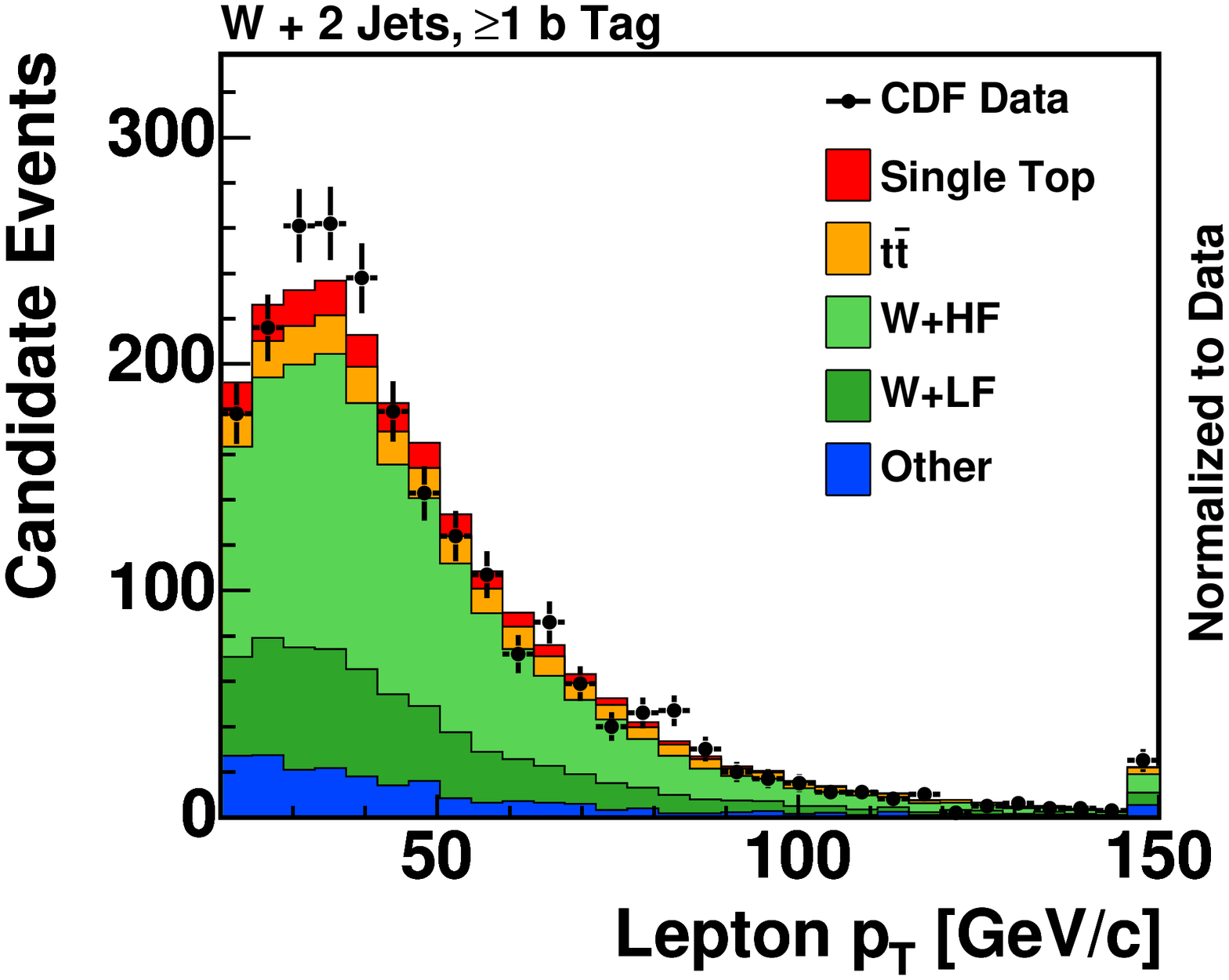}
\label{fig:LepPt2jet}}
\subfigure[]{
\includegraphics[width=0.55\columnwidth]{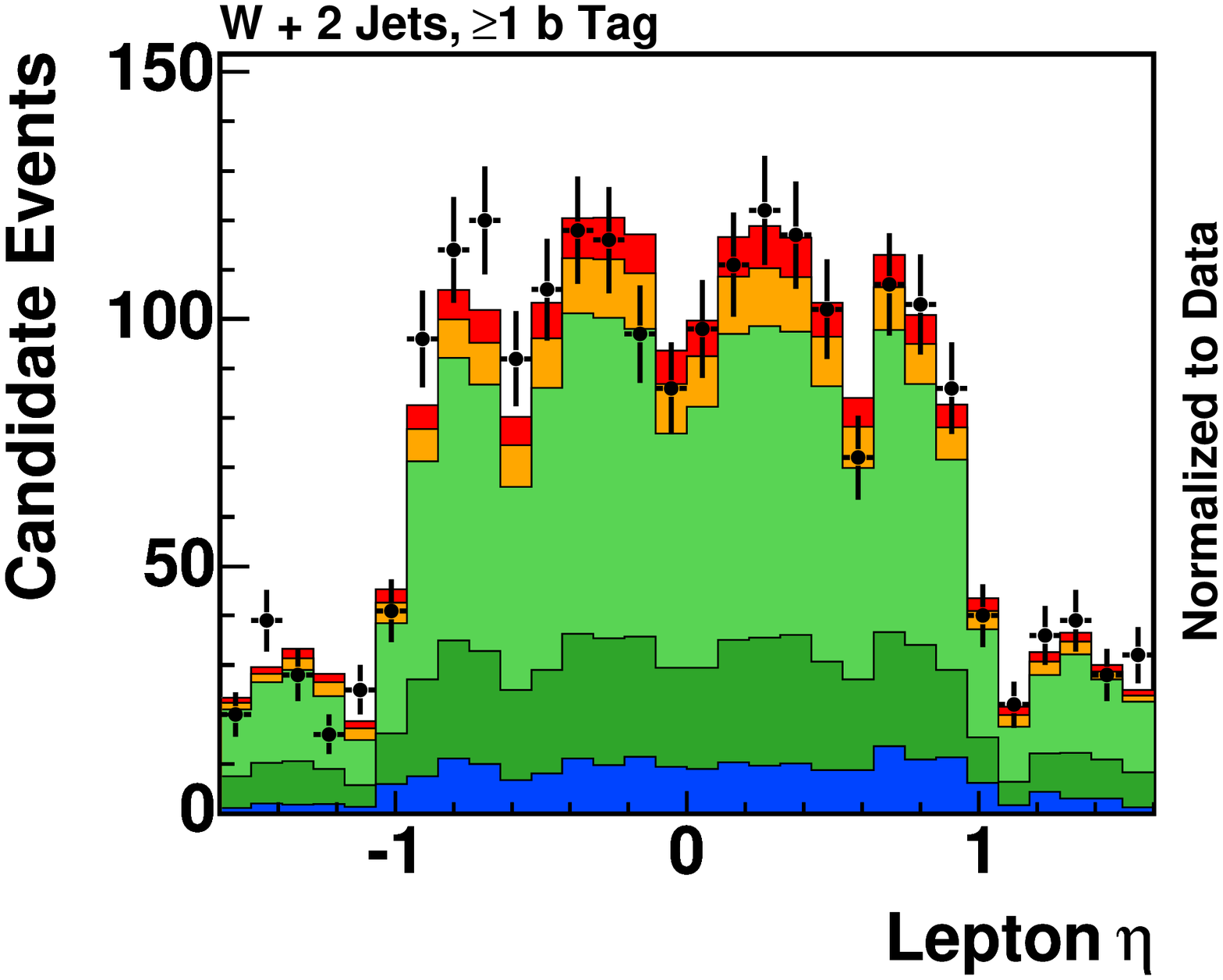}
\label{fig:LepEta2jet}}
\subfigure[]{
\includegraphics[width=0.55\columnwidth]{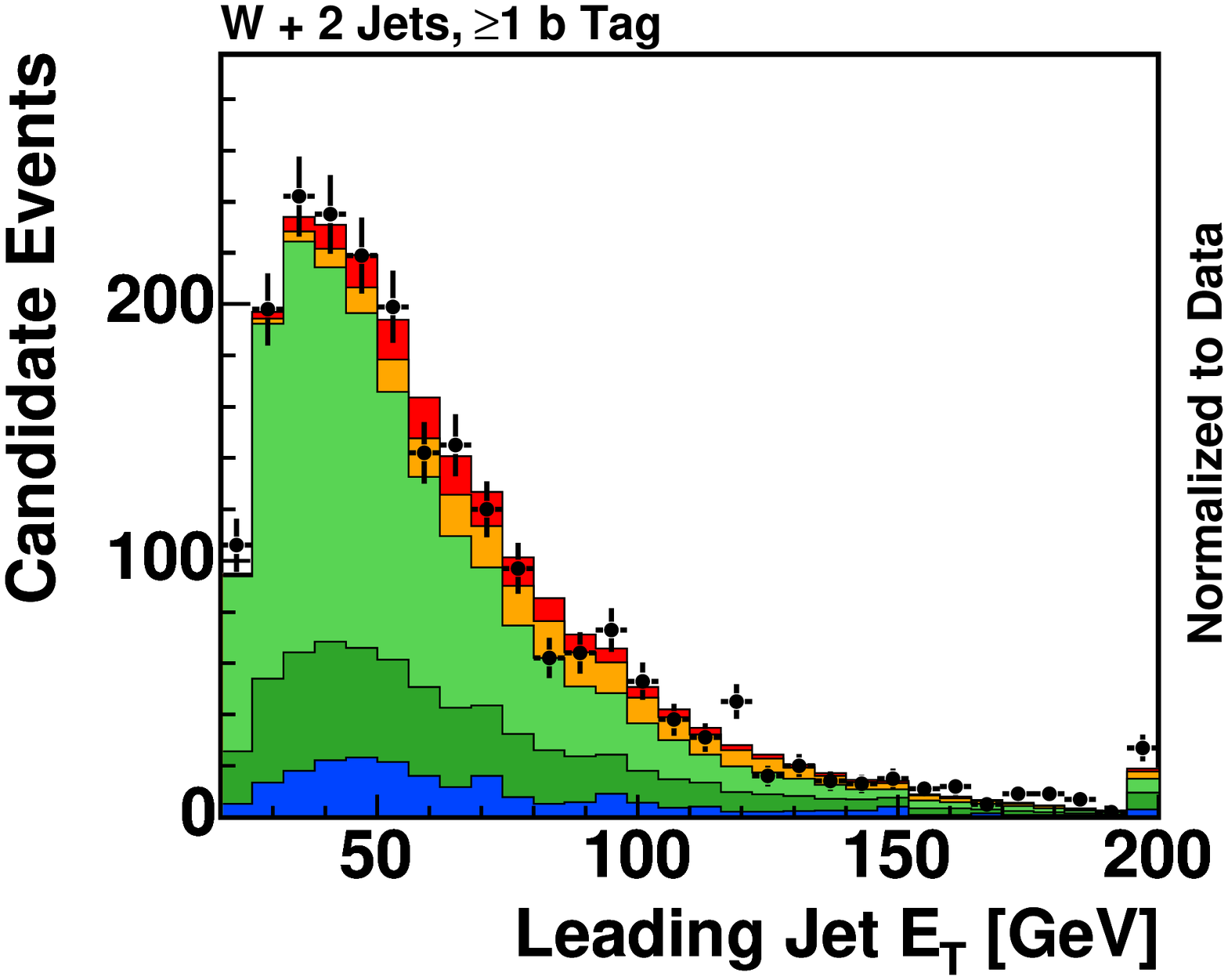}
\label{fig:J1Et2jet}}
\subfigure[]{
\includegraphics[width=0.55\columnwidth]{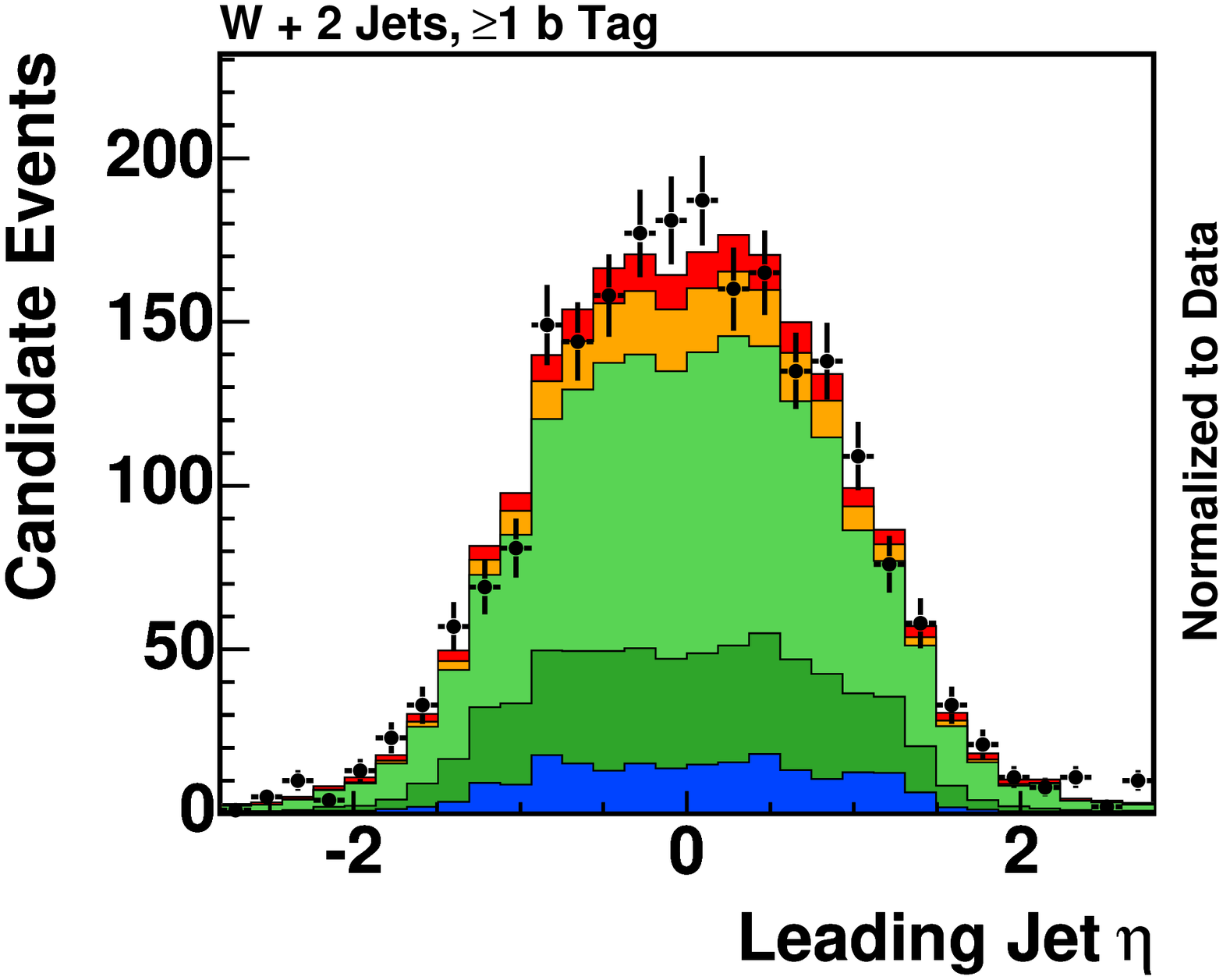}
\label{fig:J1Eta2jet}}
\subfigure[]{
\includegraphics[width=0.55\columnwidth]{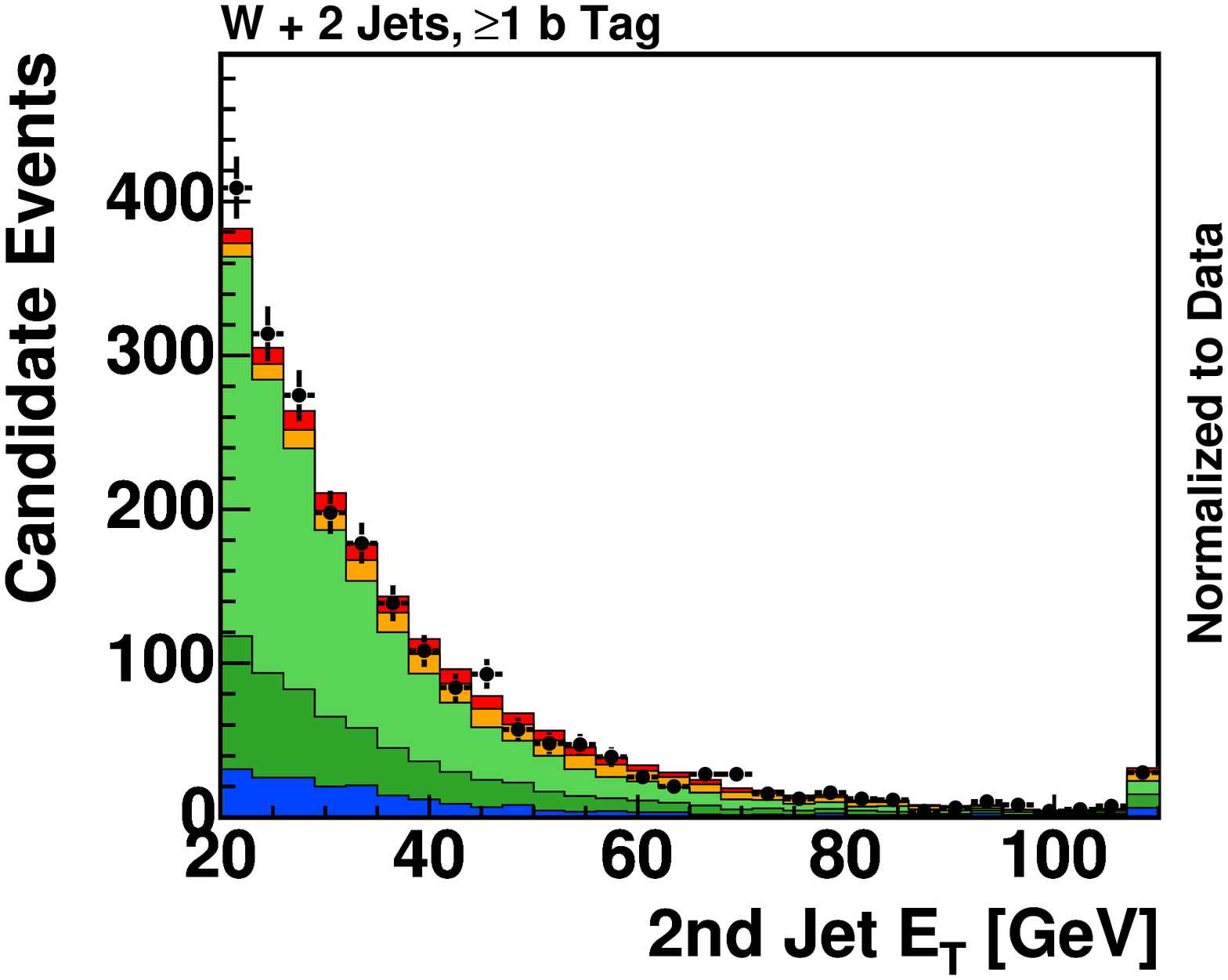}
\label{fig:J2ET2jet}}
\subfigure[]{
\includegraphics[width=0.55\columnwidth]{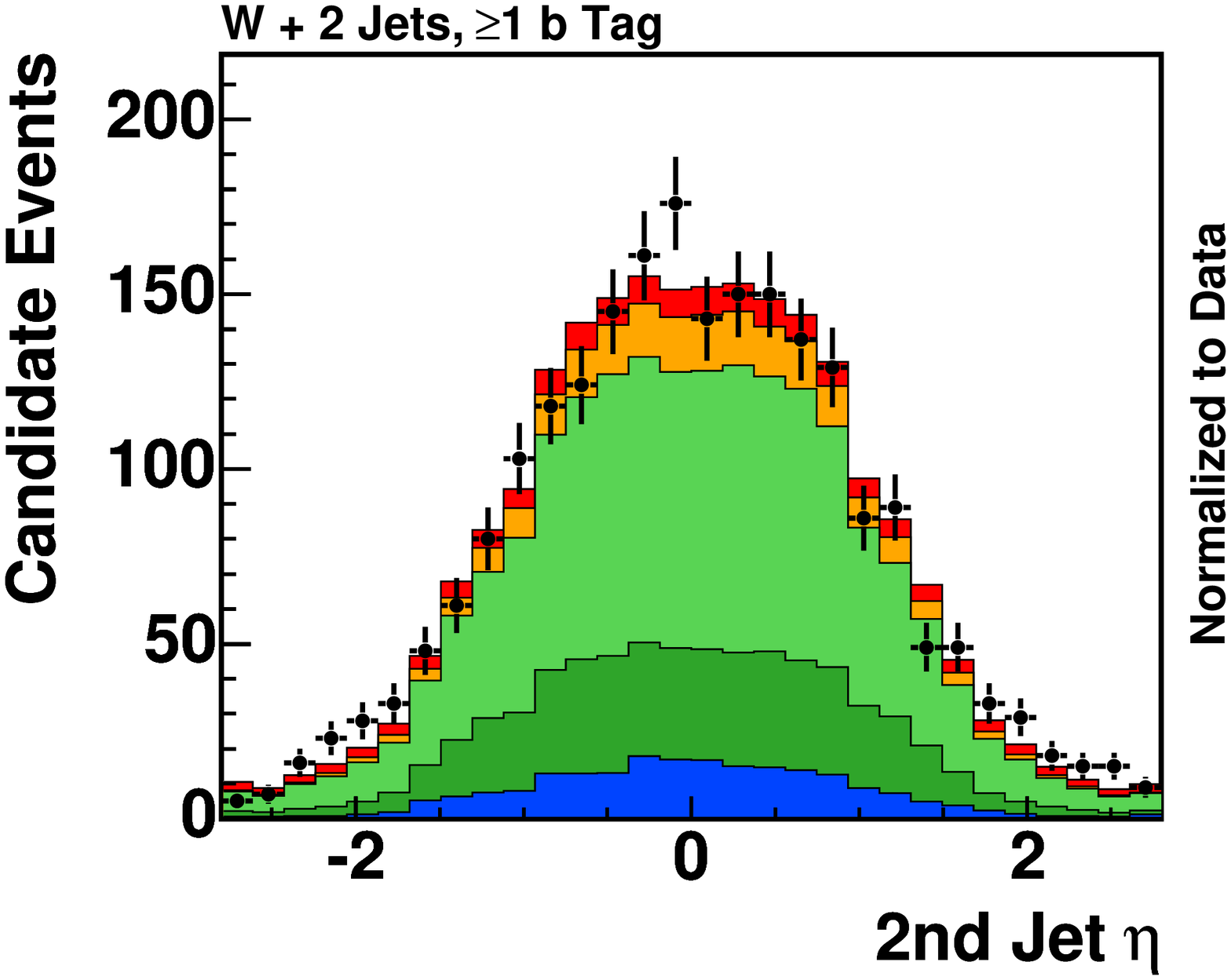}
\label{fig:J2Eta2jet}}
\end{center}
\caption{\label{fig:Input_validation_2jet}Validation plots comparing data
and Monte Carlo for basic kinematic quantities for events passing the event selection requirements
with two jets
and at least one $b$~tag.  
The data are indicated with points, and the
shaded histograms show the signal and background predictions which are stacked
to form the total prediction.  The stacking order follows that of the legend.}
\end{figure*}

\begin{figure*}
\begin{center}
\subfigure[]{
\includegraphics[width=0.55\columnwidth]{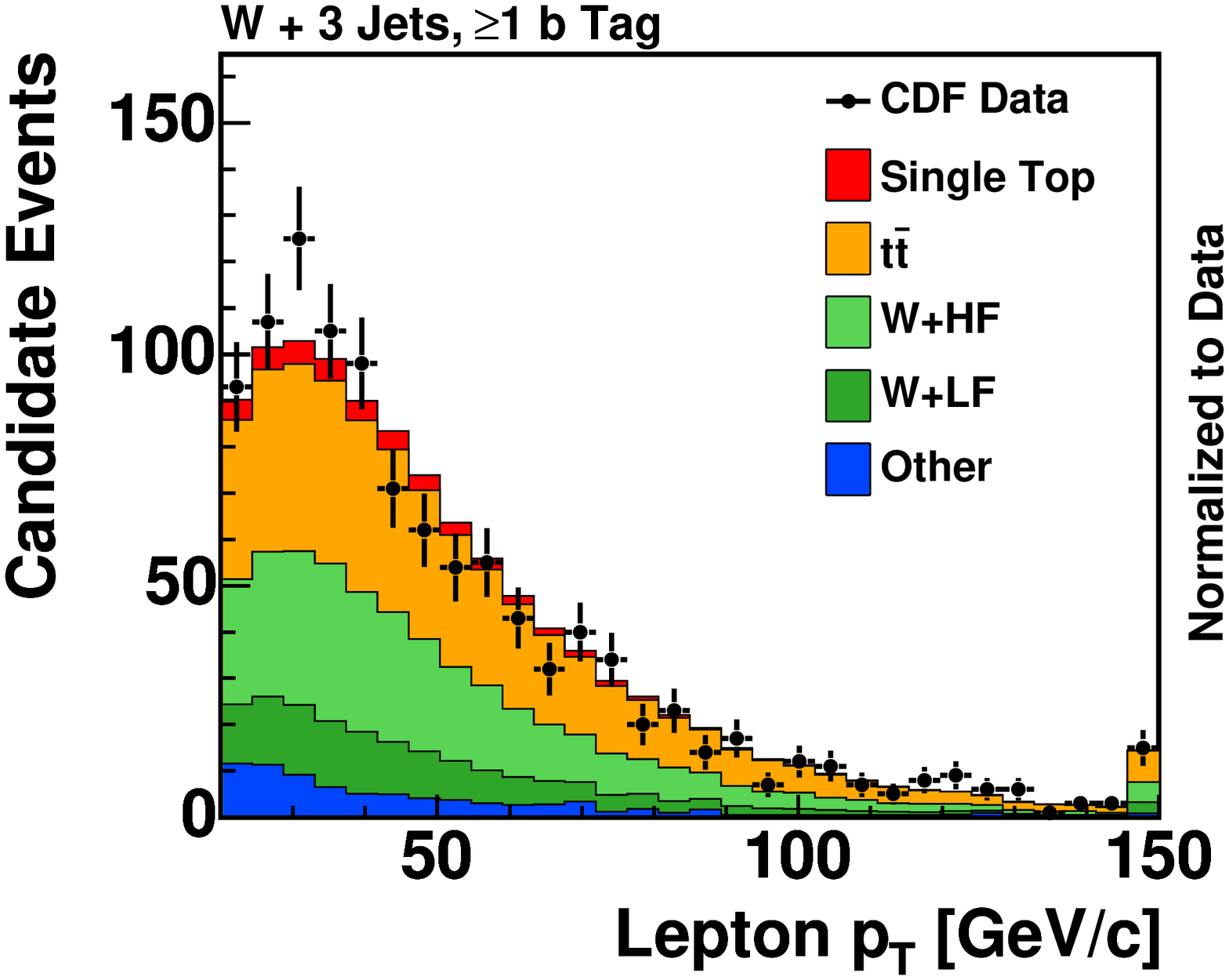}
\label{fig:LepPt3jet}}
\subfigure[]{
\includegraphics[width=0.55\columnwidth]{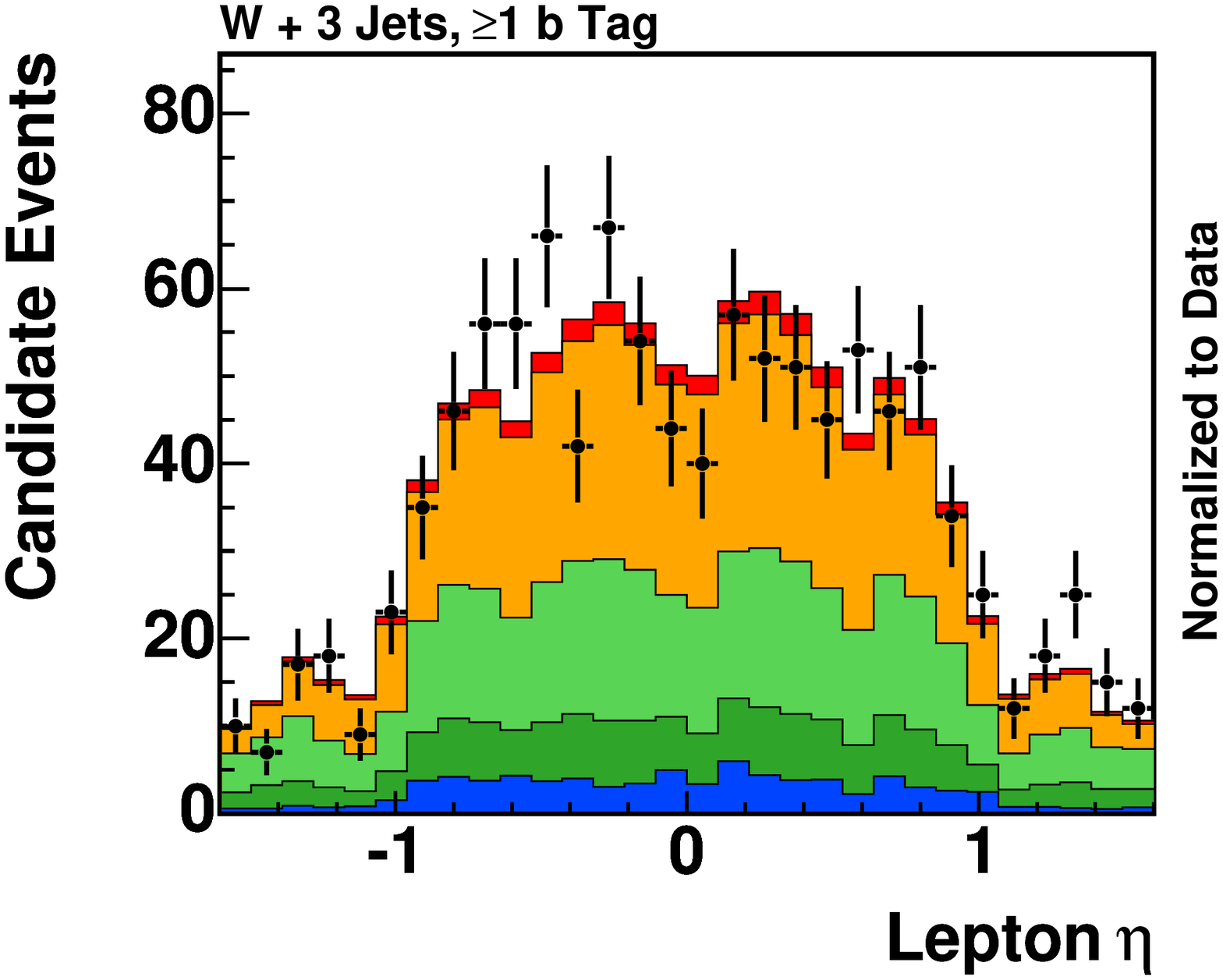}
\label{fig:LepEta3jet}}
\subfigure[]{
\includegraphics[width=0.55\columnwidth]{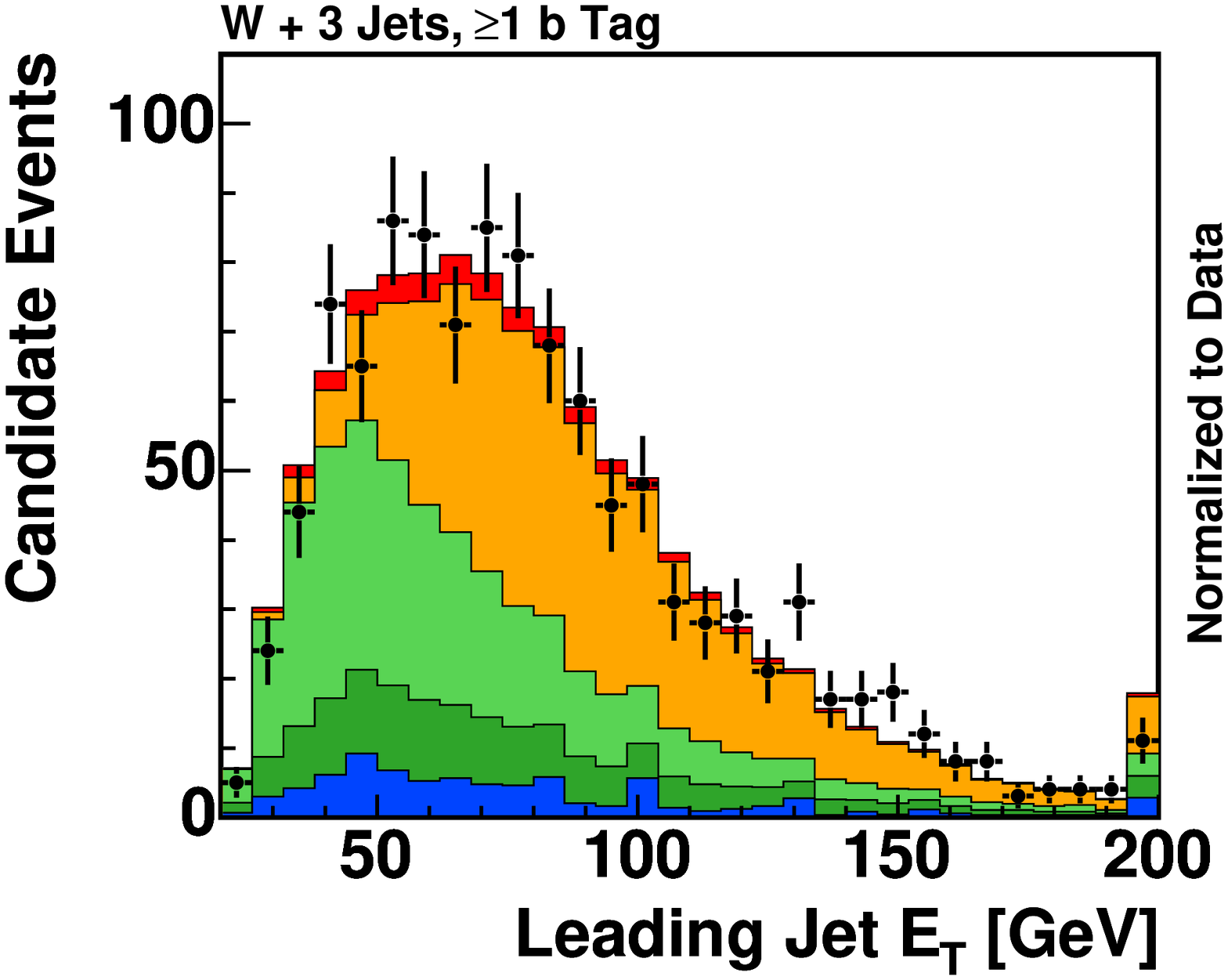}
\label{fig:J1Et3jet}}
\subfigure[]{
\includegraphics[width=0.55\columnwidth]{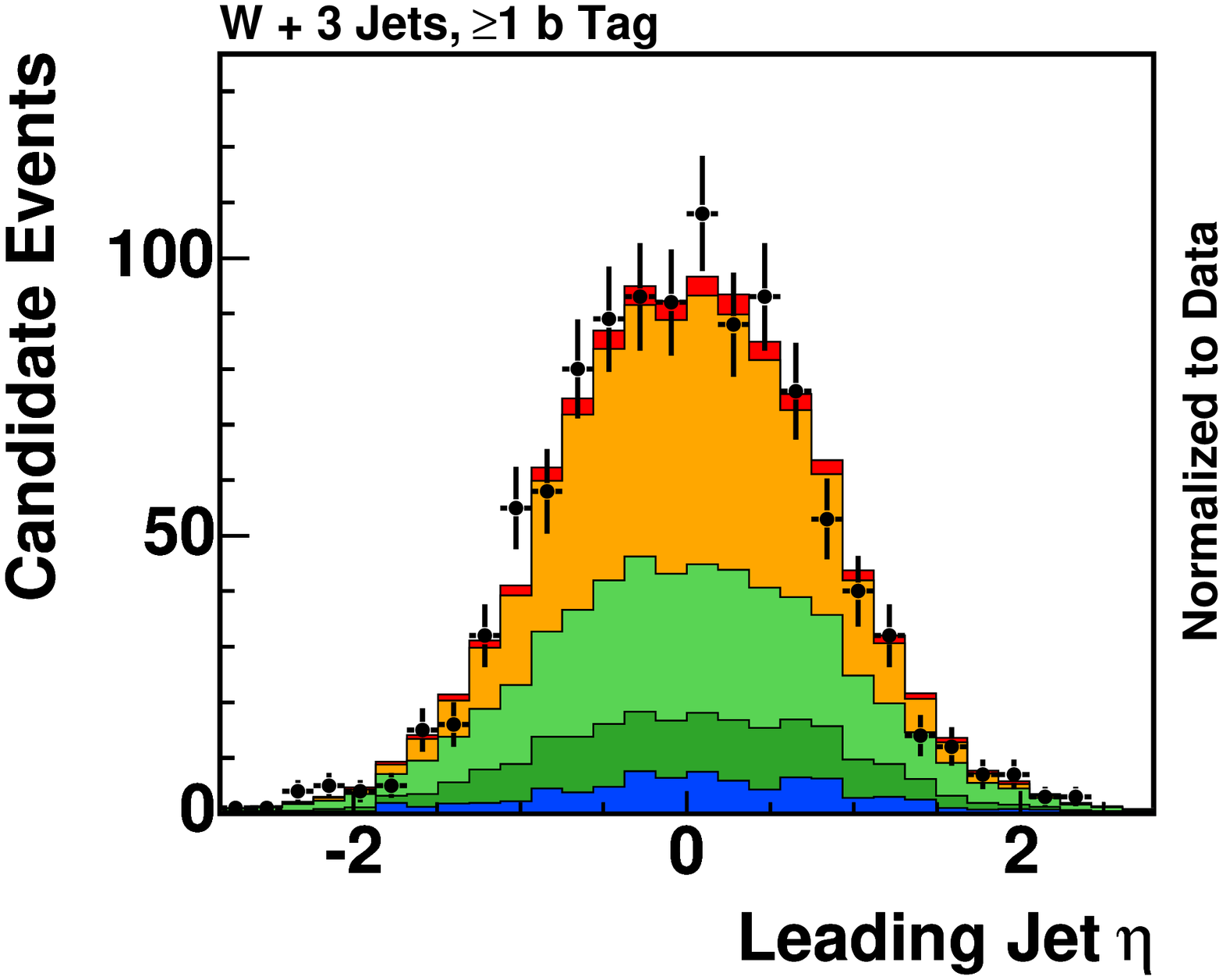}
\label{fig:J1Eta3jet}}
\subfigure[]{
\includegraphics[width=0.55\columnwidth]{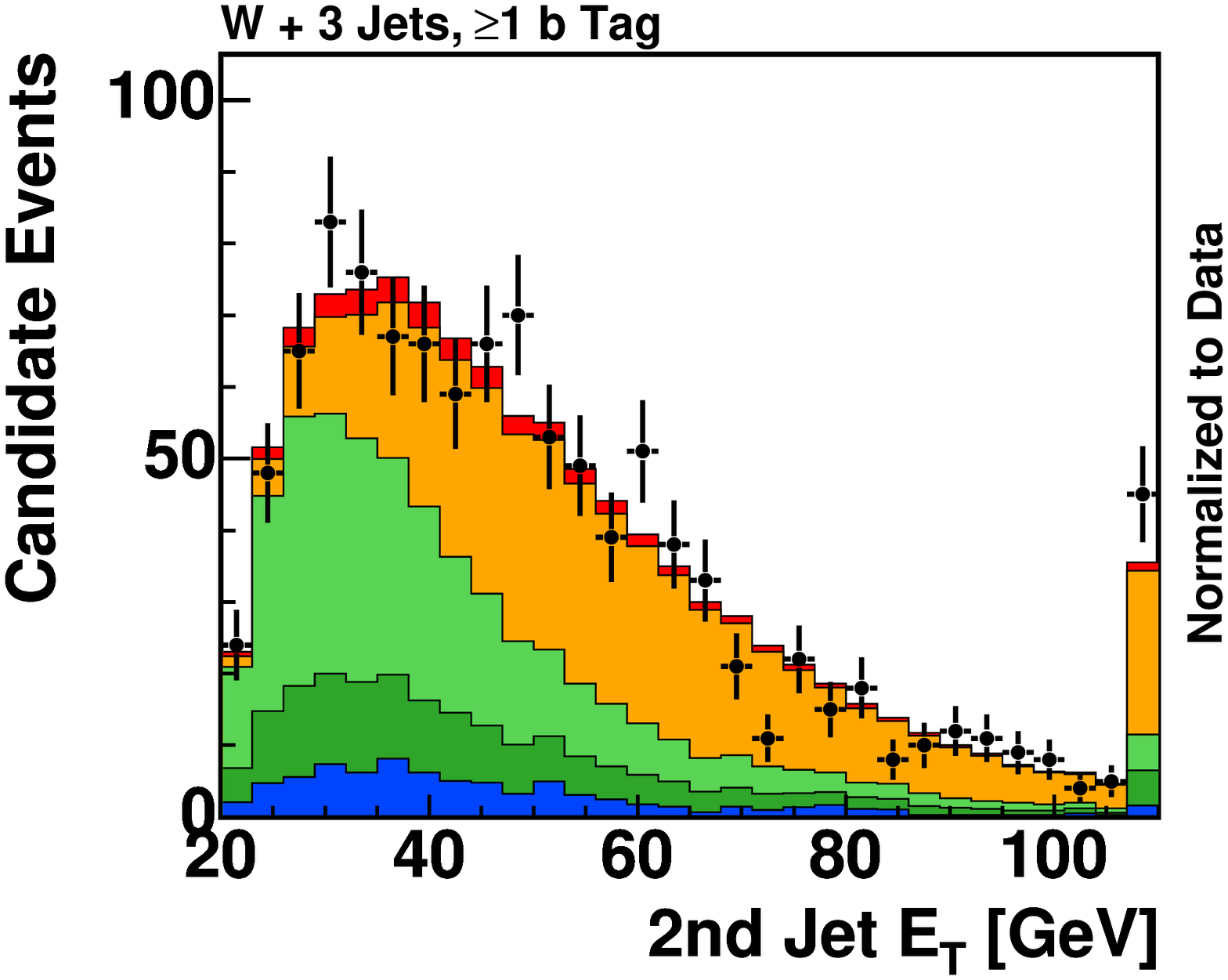}
\label{fig:J2ET3jet}}
\subfigure[]{
\includegraphics[width=0.55\columnwidth]{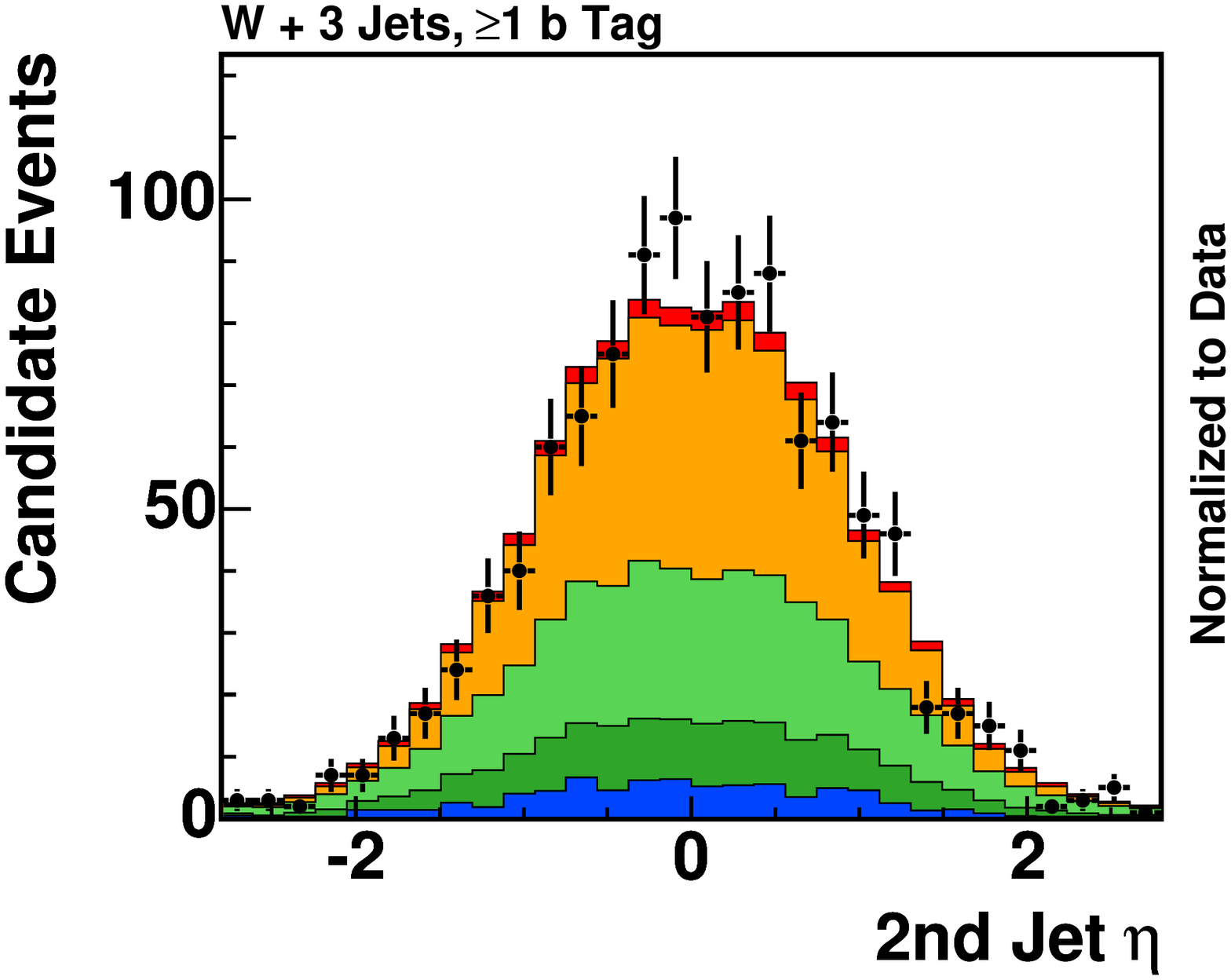}
\label{fig:J2Eta3jet}}
\end{center}
\caption{\label{fig:Input_validation_3jet}Validation plots comparing data
and Monte Carlo for basic kinematic quantities for events passing the event selection requirements
with three identified
jets and at least one $b$ tag.
The data are indicated with points, and the
shaded histograms show the signal and background predictions which are stacked
to form the total prediction.  The stacking order follows that of the legend.}
\end{figure*}

\begin{figure*}
\begin{center}
\subfigure[]{
\includegraphics[width=0.8\columnwidth]{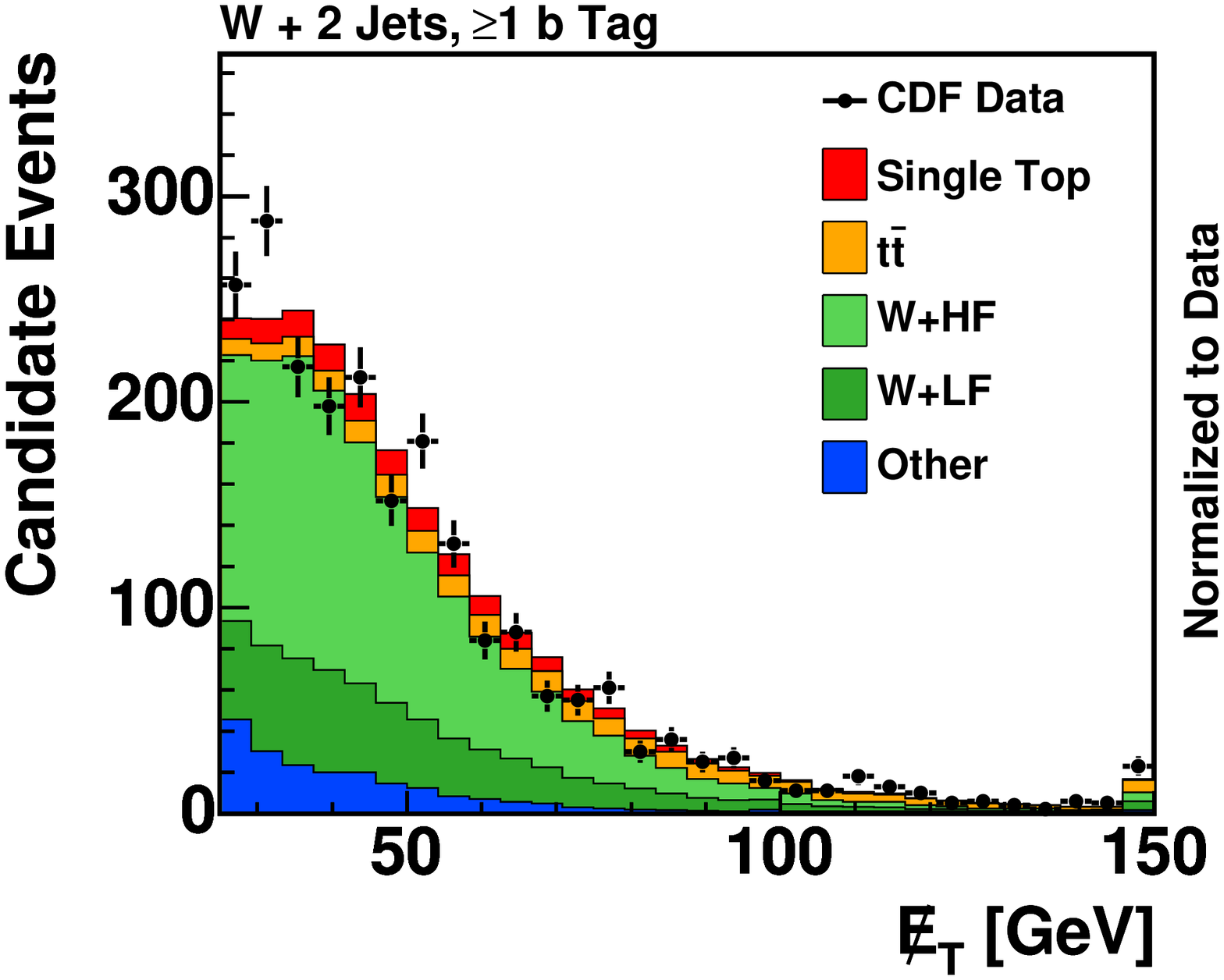}
\label{fig:MET2jet}}
\subfigure[]{
\includegraphics[width=0.8\columnwidth]{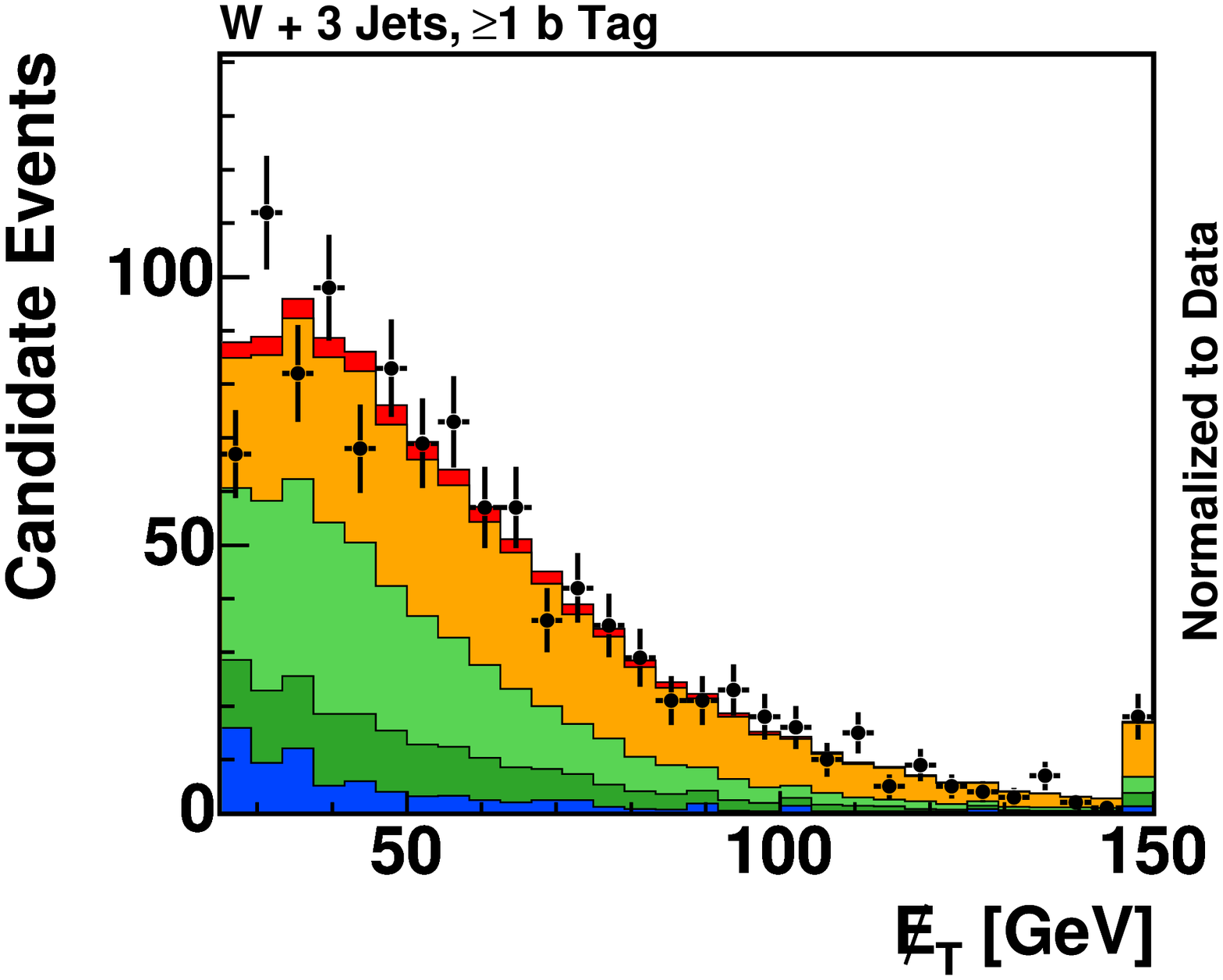}
\label{fig:MET3jet}}
\end{center}
\caption{\label{fig:Input_validation_met}Validation plots comparing data
and Monte Carlo for missing transverse energy for events passing our event
selection requirements with two jets (a)
and three jets (b), both with at least one $b$~tag.
The data are indicated with points, and the
shaded histograms show the signal and background predictions which are stacked
to form the total prediction.  The stacking order follows that of the legend.}
\end{figure*}

A sample of the validation plots we examine is shown in
Figures~\ref{fig:Input_validation_2jet},~\ref{fig:Input_validation_3jet},
and~\ref{fig:Input_validation_met}.  
The close match of the distributions gives
confidence in the results.
The validations of the modeling of other observable quantities
are shown later in this paper.

Out of the hundreds of distributions checked for discrepancies, only
two distributions in the untagged $W+$jets data were found to be
poorly simulated by our Monte Carlo model: the pseudorapidity of the
lowest-energy jet in both $W+2$~jet and $W$+3~jet events and the
distance between the two jets in $\phi - \eta$ space in $W+2$~jet
events.  These discrepancies are used to estimate systematic
uncertainties on the shapes of our final discriminant variables.
These distributions and the discussion of associated systematic
uncertainties are presented in Section~\ref{sec:Systematics}.

\section{\label{sec:btagger} Jet Flavor Separator}

In our event selection, we identify $b$-quark jets by requiring a
reconstructed secondary vertex.   A large fraction, 48\% of
the expected background events with $b$-tagged jets have no $B$ hadrons
in them at all.  This is due to the long lifetime and the mass of
charm hadrons, the false reconstruction of secondary vertices in
light jets, and the fact that the fraction of pretagged $W$+jets events containing
$B$ hadrons is small compared with the charm and light-flavored components.
Tagged jets without $B$ hadrons in them can be separated from those
containing $B$ hadrons by extending the vertex requirement using reconstructed quantities
that differentiate the two classes of jets.  These quantities take advantage of the long
lifetime ($\tau\approx 1.6$~ps) and the large mass ($m\approx 5$~GeV/$c^2$) of $B$ hadrons.
 
The invariant mass of the tracks in the reconstructed vertex is larger on average for 
vertices arising from a $B$ hadron decay than it is in vertices in jets that do not contain
$B$ hadrons.  The number of tracks in the
secondary vertex is also on average larger, and the significance of the transverse decay 
length ($\Delta L_{xy} / \sigma_{xy}$) is larger for $B$ hadron vertices.

In addition to the vertex properties, attributes of the tracks in the
jet are suitable to discriminate jets containing a $B$ hadron. Tracks
of charged particles originating from the decay of a $B$ hadron have
larger impact parameters and higher transverse momenta relative to the
jet axis.  The presence of semileptonic $B$ hadron decays increases the number and
transverse momenta relative to the jet axis 
of electrons and muons in $b$ jets as compared to non-$b$ jets.

To make full use of all discriminating quantities and their correlations,
the variables are used as inputs to a neural network which is
applied to jets selected by the {\sc secvtx} secondary vertex
tagger~\cite{Richter:2007zzc}.  This network is trained with simulated
events of single top quark production and the main background
processes, mixed according to the background estimation. 
Processes with secondary vertices due to
$B$ hadron decays are treated as signal events, namely single top quark,
\ttbar, and \Wbbbar\ production.  Physical processes containing no $b$
quarks but charm and light flavors are treated as background: \Wccbar,
$Wcj$, and $W$~+~light~jets.

The {\sc NeuroBayes} package~\cite{Feindt:2006pm} used for the
neural-network jet flavor separator combines a three-layer feed
forward neural network with a complex robust
preprocessing. Transforming the input variables to be distributed as unit-width Gaussians
reduces the influence of long tails; diagonalization and rotation
transform the covariance matrix of the variables into a unit matrix.
The neural network uses Bayesian regularization techniques for the training 
process. The network infrastructure consists of one input node for each 
input variable plus one bias node, ten hidden nodes, and one output node which gives a continuous output
variable $b_{\mathrm{NN}}$ in the interval 
[$-1,1$]. Jets with secondary vertices induced by the 
decay of a $B$ hadron tend to have $b_{\mathrm{NN}}$ values close to 1, while jets with falsely 
reconstructed vertices tend to have $b_{\mathrm{NN}}$ values near $-1$. 

The significances of the training variables are determined automatically 
during the preprocessing in {\sc NeuroBayes}. 
The correlation matrix of all preprocessed input variables is calculated, 
including the correlation of all variables to the target variable, which
is $+1$ for jets with $B$ hadron decays and $-1$ for all other jets.
The variables are omitted one at a time to determine the loss of total correlation to the target 
caused by their removal. The variable with the smallest loss of correlation is 
discarded leading to an ($n-1$)-dimensional correlation matrix. The same 
procedure is repeated with the reduced correlation matrix to find the least 
important of the ($n-1$) remaining variables.
The significance of each variable is calculated by dividing the loss of 
correlation induced by its removal by the square root of the sample size.
We investigated 50 candidate input variables but chose to include as inputs only those
with a significance larger than 3.0, of which there are 25.

Because the neural-network jet flavor separator is trained using
simulated events, it is essential to verify that the input and output distributions
are modeled well, and to assess systematic uncertainties where discrepancies are
seen.  The shapes of the input variable distributions in the data are found
to be reasonably well reproduced by the simulation.  We
also examine the distribution of $b_{\mathrm{NN}}$ for both $b$ signal and non-$b$
background.  The $b$ signal distribution is 
checked with double-{\sc secvtx}-tagged dijet events and compared against Monte Carlo
jets with $B$ hadron decays.  One jet in addition is required to have an electron with a large
transverse momentum with respect to the jet axis,
in order to purify further the $b$ content of the sample.
The jet opposite to the electron-tagged jet is probed for its distribution of the neural network output.
The distribution of $b_{\mathrm{NN}}$ in these jets is well simulated by that of $b$ jets in the
Monte Carlo~\cite{Richter:2007zzc}.

To test the response of the network
to light-flavored jets, negative-tagged jets were tested in data and Monte Carlo.  A
correction function was derived~\cite{Richter:2007zzc} 
to adjust for the small discrepancy observed in the output shape.
This correction function is 
parameterized in the sum of transverse energies in the event, the number of 
tracks per jet, and the transverse energy of the jet.
The correction function is applied to light-flavored and charm Monte Carlo jets
in the analyses presented in this paper, but not to $b$~jets.  The uncorrected
neural network outputs are used to evaluate systematic uncertainties on the shapes
of the final discriminant distributions.

The resulting network output $b_{\mathrm{NN}}$ distinguishes the $b$
signal from the charm and light-flavored background processes with a purity that
increases with increasing $b_{\mathrm{NN}}$, as can be seen in~Fig.~\ref{fig:bnntv}(a).  
Furthermore, the network gives very similar shapes for
different $b$-quark-producing processes, indicating that it is sensitive to
the properties of $b$-quark jets and does not depend on the underlying processes that produce them.

Not only is $b_{\mathrm{NN}}$ a valuable tool for separating the single top quark signal from background processes that
do not contain $b$~jets, it is also valuable for separating the different flavors of $W$+jets events,
which is crucial in estimating the background composition.  As described in Section~\ref{sec:Background},
the distribution of $b_{\mathrm{NN}}$ is fit in $b$-tagged $W$+1~jet events, and the heavy-flavor fractions
for $b$ and charm jets are extracted.  Using also a direct measurement of the 
$Wc$ rate~\cite{Wcharm:2007dm}, predictions are made of the $b$ and charm jet fractions
in the two- and three-jet bins.   These predictions are used to scale the {\sc alpgen} Monte Carlo samples,
which are then compared with the data in the
two- and three-jet $b$-tagged samples, without refitting the heavy-flavor composition, as shown
in Fig.~\ref{fig:bnntv}(c) and~(d).   The three-jet sample has a larger sample of $t{\bar{t}}$ events
which are enriched in $b$~jets.  The successful modeling of the changing flavor composition as a function
of the number of identified jets provides confidence in the correctness of the background simulation.

All multivariate methods described here use $b_{\mathrm{NN}}$ as an input
variable, and thus we need $b_{\mathrm{NN}}$ values for all Monte Carlo and data events
used to model the final distributions.  For the mistagged $W$+LF shape prediction, we use the
$W$+LF Monte Carlo sample, where the events are weighted by the data-based mistag prediction
for each taggable jet.  This procedure improves the modeling over what would be obtained
if Monte Carlo mistags were used, as the mistag probabilities are
based on the data, and it increases the sample size we use for the
mistag modeling.  An issue that arises is that parameterized mistagged events do
not have $b_{\mathrm{NN}}$ values and random values must be chosen for
them from the distribution in light-flavor events.  If a $W$+LF event has more than one
taggable jet, then random values are assigned to both jets.  These events are used
for both the single-mistag prediction and the double-mistag prediction with appropriate
weights.  The randomly chosen flavor-separator values must be the same event-by-event and jet-by-jet
for each of the four analyses in this paper in order for the super discriminant
combination method to be consistent.

The distributions of $b_{\mathrm{NN}}$ for 
non-$W$ multijet events are more difficult to predict because 
the flavor composition of the jets in these events is poorly known.
Specifically, since a non-$W$ event must have a fake lepton (or a lepton
from heavy-flavor decay), and also mismeasured $\EtMiss$, the flavor composition
of events passing the selection requirements depends on the details of the detector
response, particularly in the tails of distributions which are difficult to model.
It is necessary therefore to constrain these flavor fractions
with CDF data, and the flavor fractions thus estimated are specific to this analysis.
The non-$W$ event yields are constrained by the data as explained in Section~\ref{sec:nonw}.

The fraction of each flavor: $b$, charm, and light-flavored 
jets (originating from light quarks or gluons),
is estimated by applying the jet flavor
separator to $b$-tagged jets in the $15<\EtMiss<25$~GeV sideband of the data.  In this
sample, we find a flavor composition of 45\% $b$ quark jets, 40\% $c$
quark jets, and 15\% light-flavored jets.  Each  event in the non-$W$ modeling
samples (see Section~\ref{sec:nonw}) is randomly
assigned a flavor according to the fraction given above and then
assigned a jet flavor separator value chosen at random from the
appropriate flavor distribution.  The fractions of the non-$W$
events in the signal sample are uncertain both due to the uncertainties in the
sideband fit and the extrapolation to the signal sample.   
We take as an alternative flavor composition estimate 
60\% $b$ quark jets, 30\% $c$ quark jets, and 10\% light-flavored jets, 
which is the most $b$-like possibility of the errors on
the flavor measurement.  This alternative flavor composition affects the shapes of the
final discriminant distribution through the different flavor-separator neural network values.

\section{\label{sec:Multivariate} Multivariate Analysis}

The search for single top quark production and the measurement of its cross
section present substantial experimental challenges.  Compared with
the search for \ttbar\ production, the search for single top quarks suffers
from a lower SM production rate and a larger background. 
Single top quark events are also kinematically more similar to $W+$jets events
than \ttbar\ events are, since there is only one heavy top quark and thus
only one $W$ boson in the single top quark events, while there are two top quarks, each
decaying to $Wb$, in \ttbar\ events.  The most serious challenge
arises from the systematic uncertainty on the background prediction, which
 is approximately three times the size of the
expected signal.  Simply counting events which pass our selection requirements
will not yield a precise measurement of the single top quark 
cross section no matter how much data are accumulated because the
systematic uncertainty on the background is so large.  In fact,
in order to have sufficient sensitivity to expect to observe a signal
at the $5\,\sigma$ level, the systematic uncertainty on the background
must be less than one-fifth of the expected signal rate.

Further separation of the signal from the background 
is required.  Events that are classified as being more signal-like 
are used to test for the presence of single top quark production and measure
the cross section, and events that are classified as being more background-like 
improve our knowledge of the rates of background processes.  In order to optimize our
sensitivity, we construct discriminant functions based on kinematic
and $b$-tag properties of the events, and we classify the events on a continuous
spectrum that runs from very signal-like for high values of the discriminants to very
background-like for low values of the discriminants.  We fit the distributions of these
discriminants to the background and
signal+background predictions, allowing uncertain parameters, listed in 
Section~\ref{sec:Systematics}, to float, in a manner described in Section~\ref{sec:Interpretation}.

To separate signal events from background events, we look for properties
of the events that differ between signal and background.  Events from
single top quark production have distinctive energy and angular properties.  The backgrounds, too,
have distinctive features which can be exploited to help separate them.
Many of the variables we compute for each selected candidate event are motivated
by a specific interpretation of the event as a signal event or a background event.
It is not necessary that all variables used in a discriminant are motivated by 
the same interpretation of an event, nor do we rely on the correctness of the motivation
for the interpretation of any given event.  Indeed, each analysis is made more optimal
when it includes a mixture of variables that are based on different ways to interpret
the measured particles in the events.  We optimize our analyses by using
variables for which the distributions are maximally different between signal 
events and background events, and for which we have reliable modeling as verified by the data.

\begin{figure*}
\begin{center}
\subfigure[]{
\includegraphics[width=0.8\columnwidth]{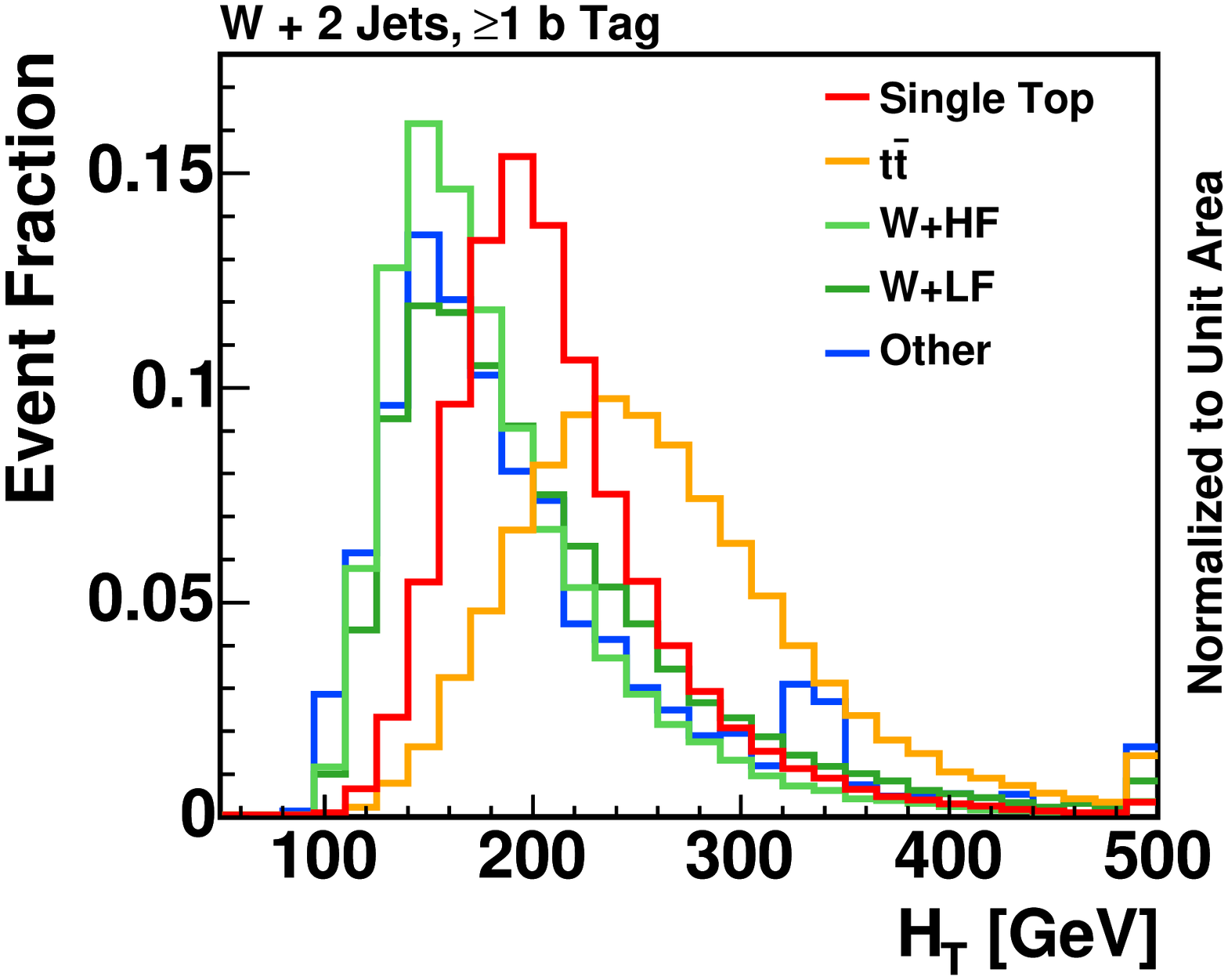}
\label{fig:HTshape}}
\subfigure[]{
\includegraphics[width=0.8\columnwidth]{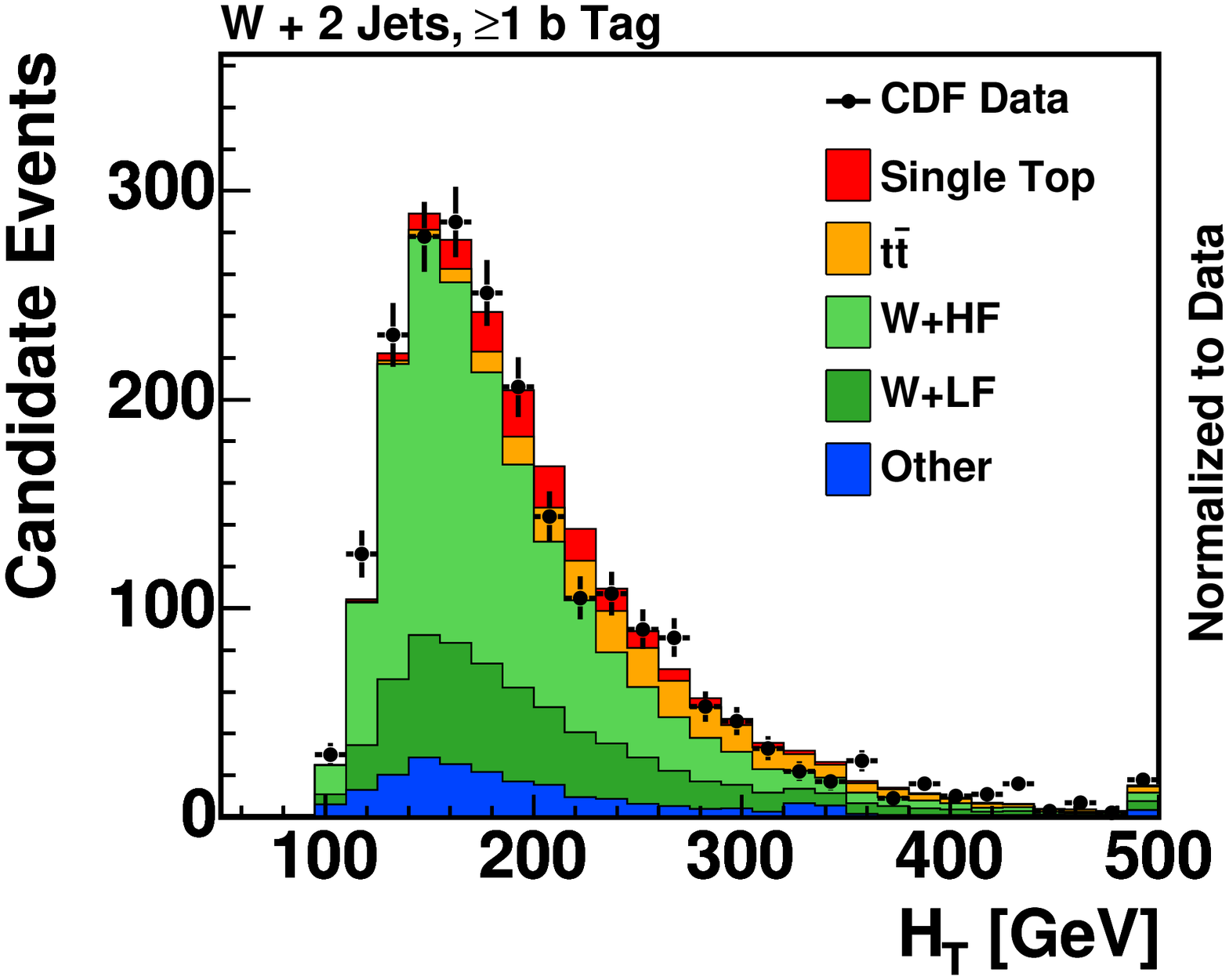}
\label{fig:HT}}
\subfigure[]{
\includegraphics[width=0.8\columnwidth]{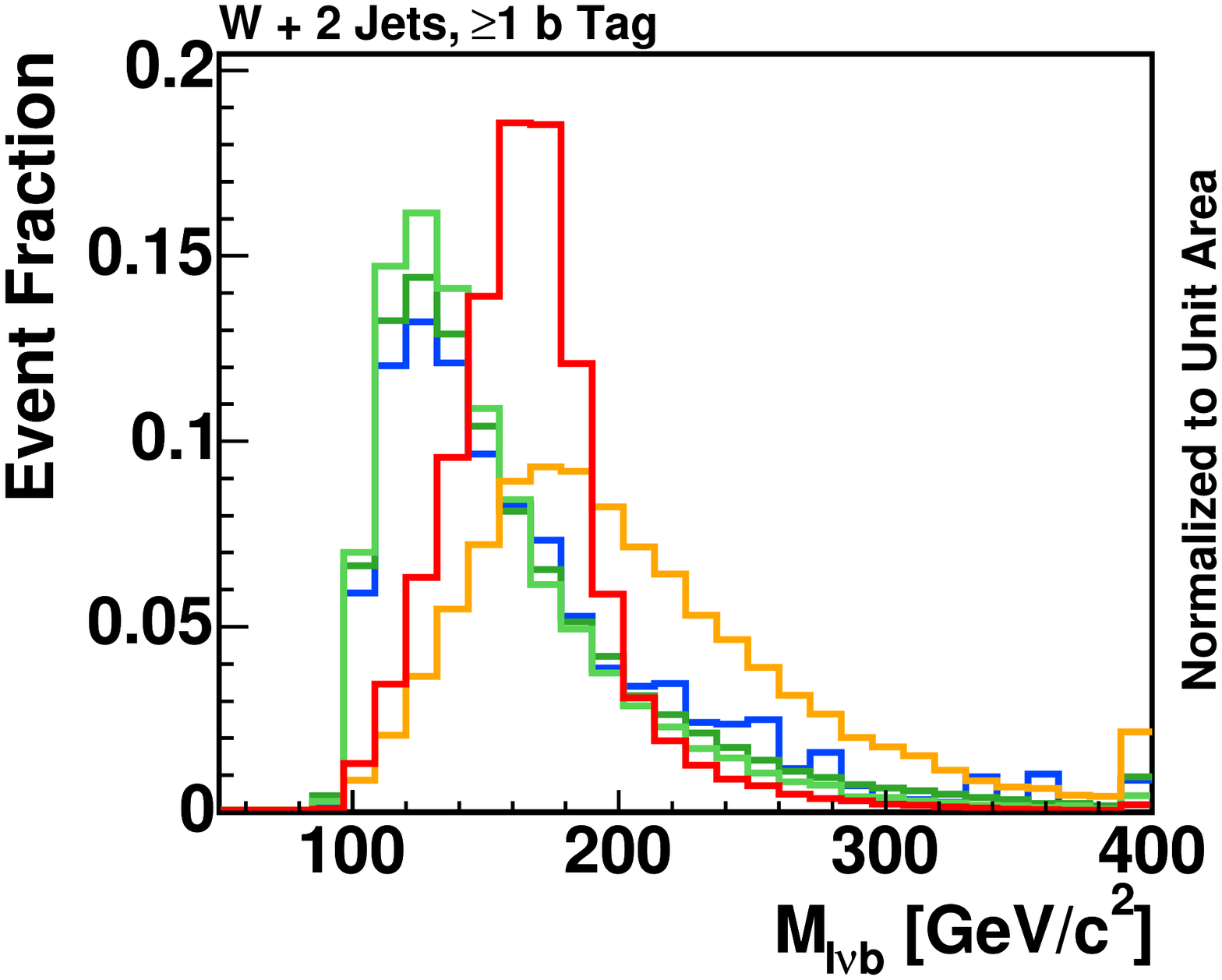}
\label{fig:Mlnubshape}}
\subfigure[]{
\includegraphics[width=0.8\columnwidth]{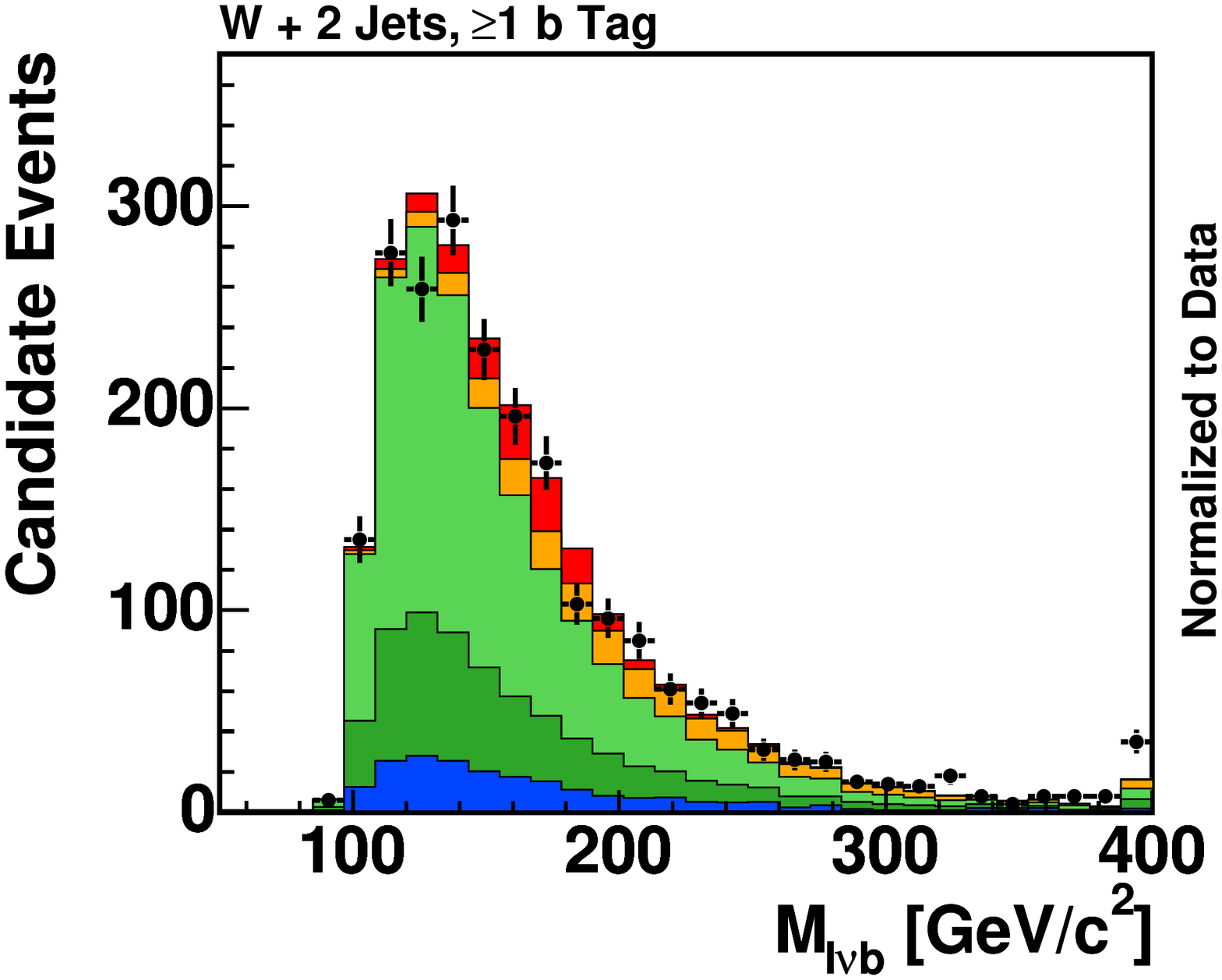}
\label{fig:Mlnub}}
\subfigure[]{
\includegraphics[width=0.8\columnwidth]{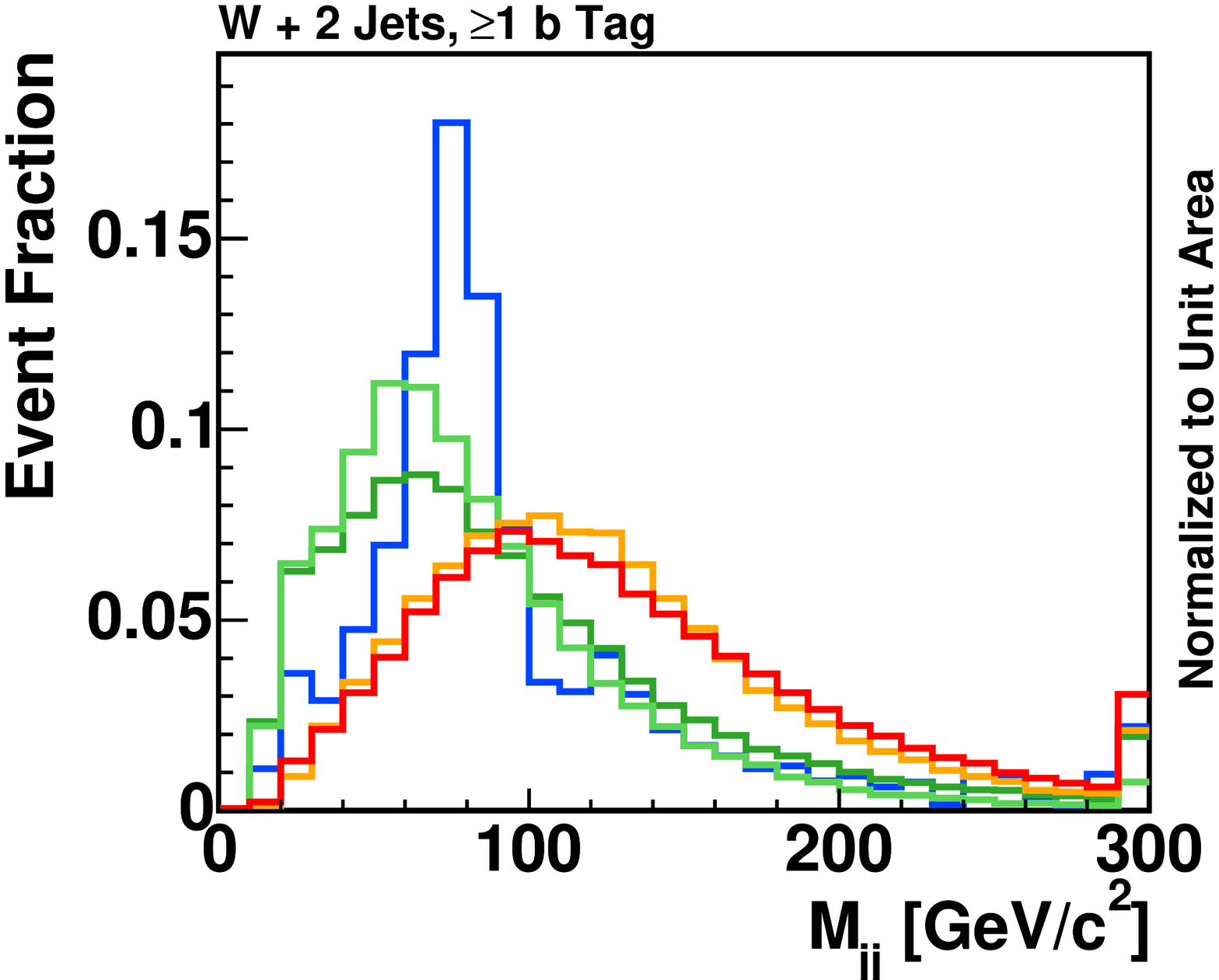}
\label{fig:mjjshape}}
\subfigure[]{
\includegraphics[width=0.8\columnwidth]{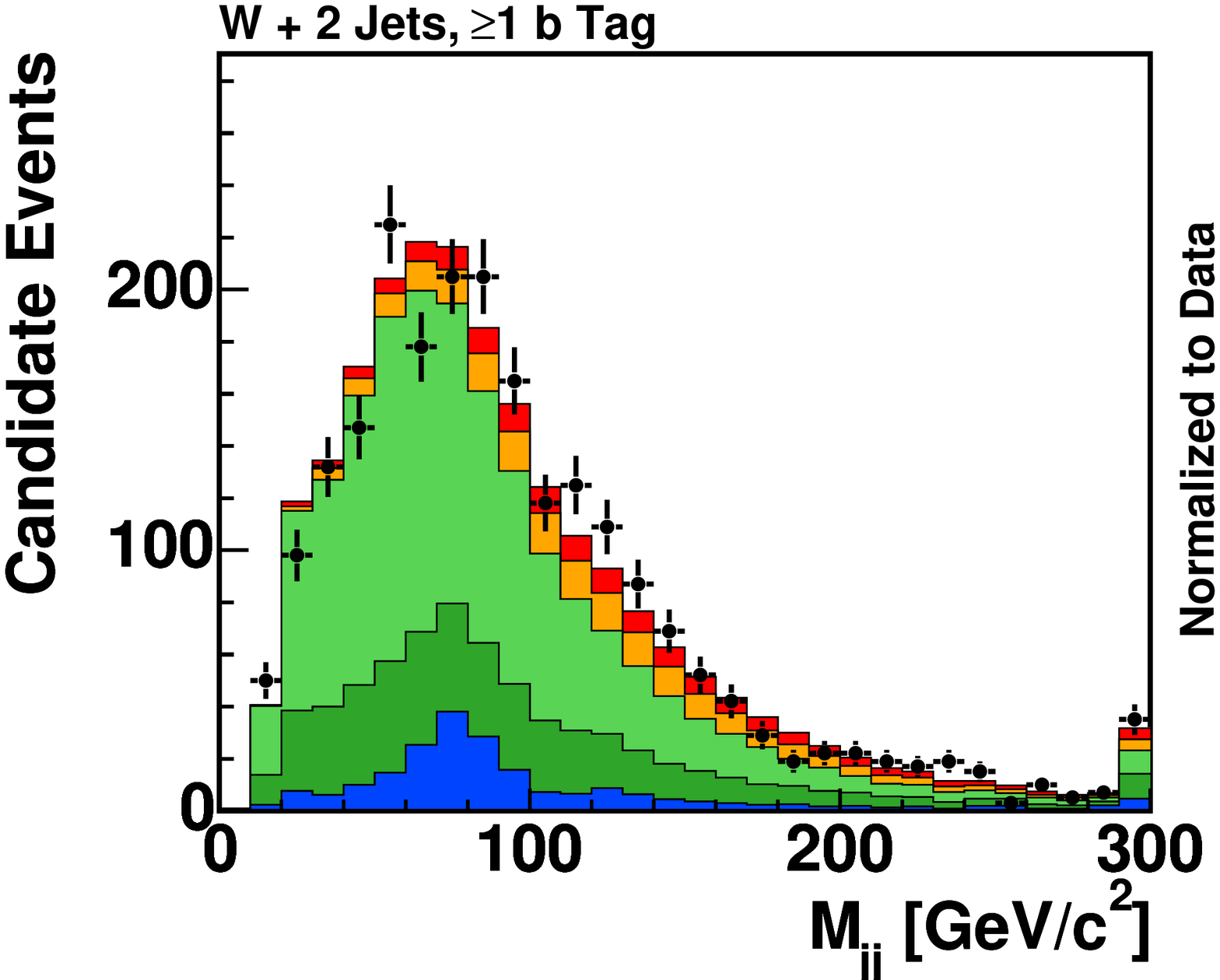}
\label{fig:mjj}}
\end{center}
\caption{\label{fig:Disc_vars_1}Monte Carlo templates (left) and
validation plots (right) comparing data and Monte Carlo for variables with
good discriminating power for events passing our selection requirements with two or three identified jets
and at least one $b$~tag.
The data are indicated with points, and the
shaded histograms show the signal and background predictions which are stacked
to form the total prediction.  The stacking order follows that of the legend.
Overflows are collected in the highest bin of each histogram.}
\end{figure*}

\begin{figure*}
\begin{center}
\subfigure[]{
\includegraphics[width=0.8\columnwidth]{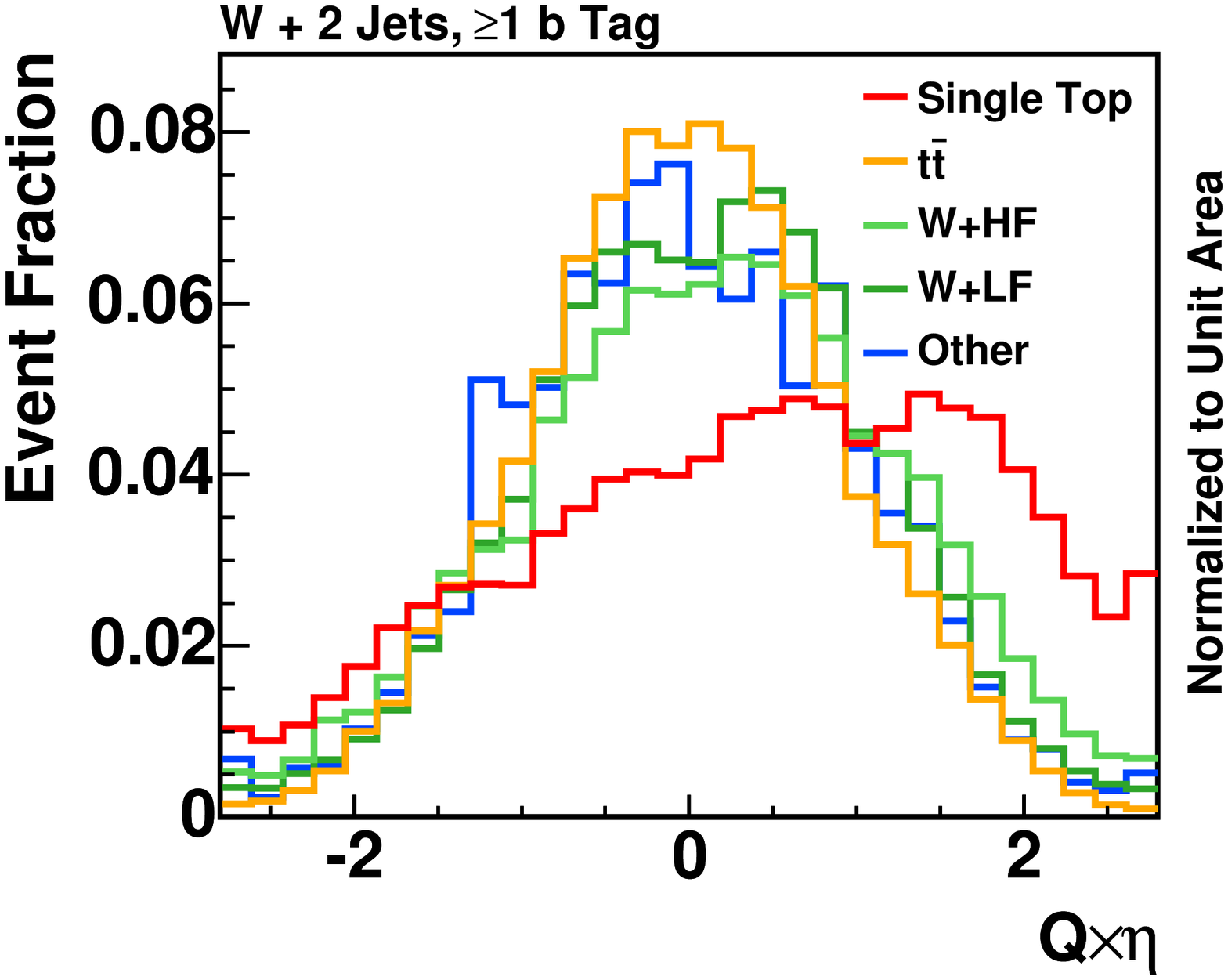}
\label{fig:Qetashape}}
\subfigure[]{
\includegraphics[width=0.8\columnwidth]{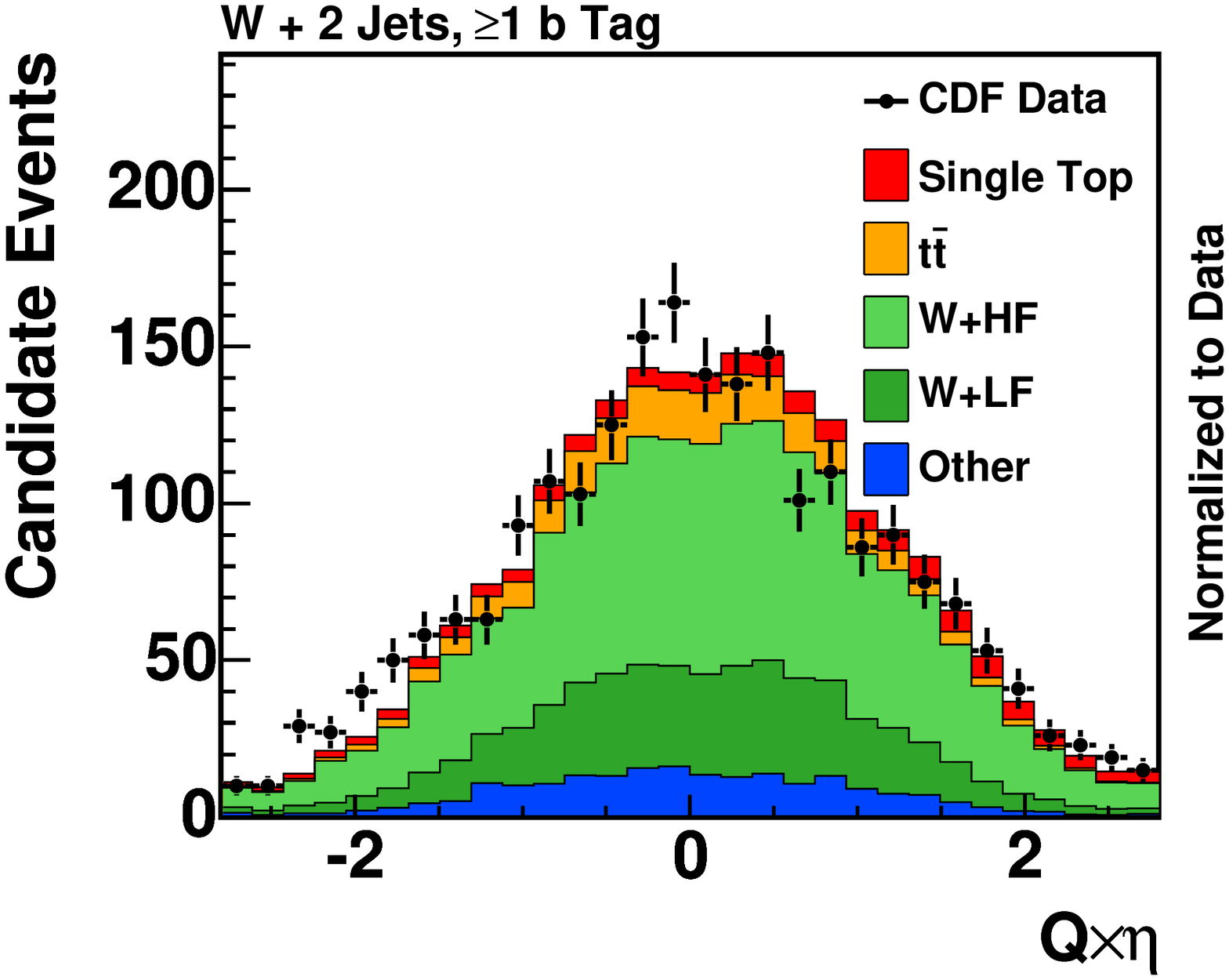}
\label{fig:Qeta}}
\subfigure[]{
\includegraphics[width=0.8\columnwidth]{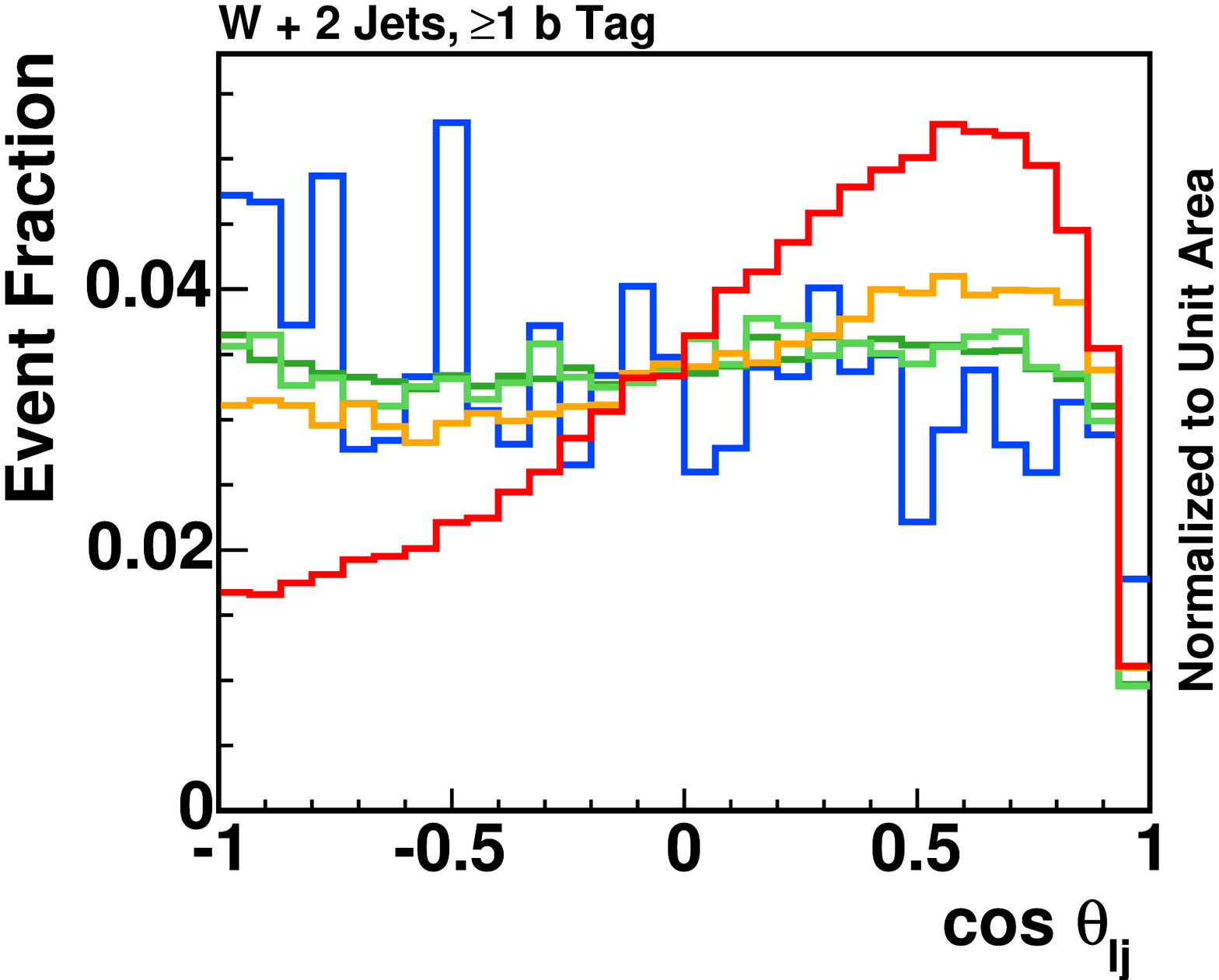}
\label{fig:cosljshape}}
\subfigure[]{
\includegraphics[width=0.8\columnwidth]{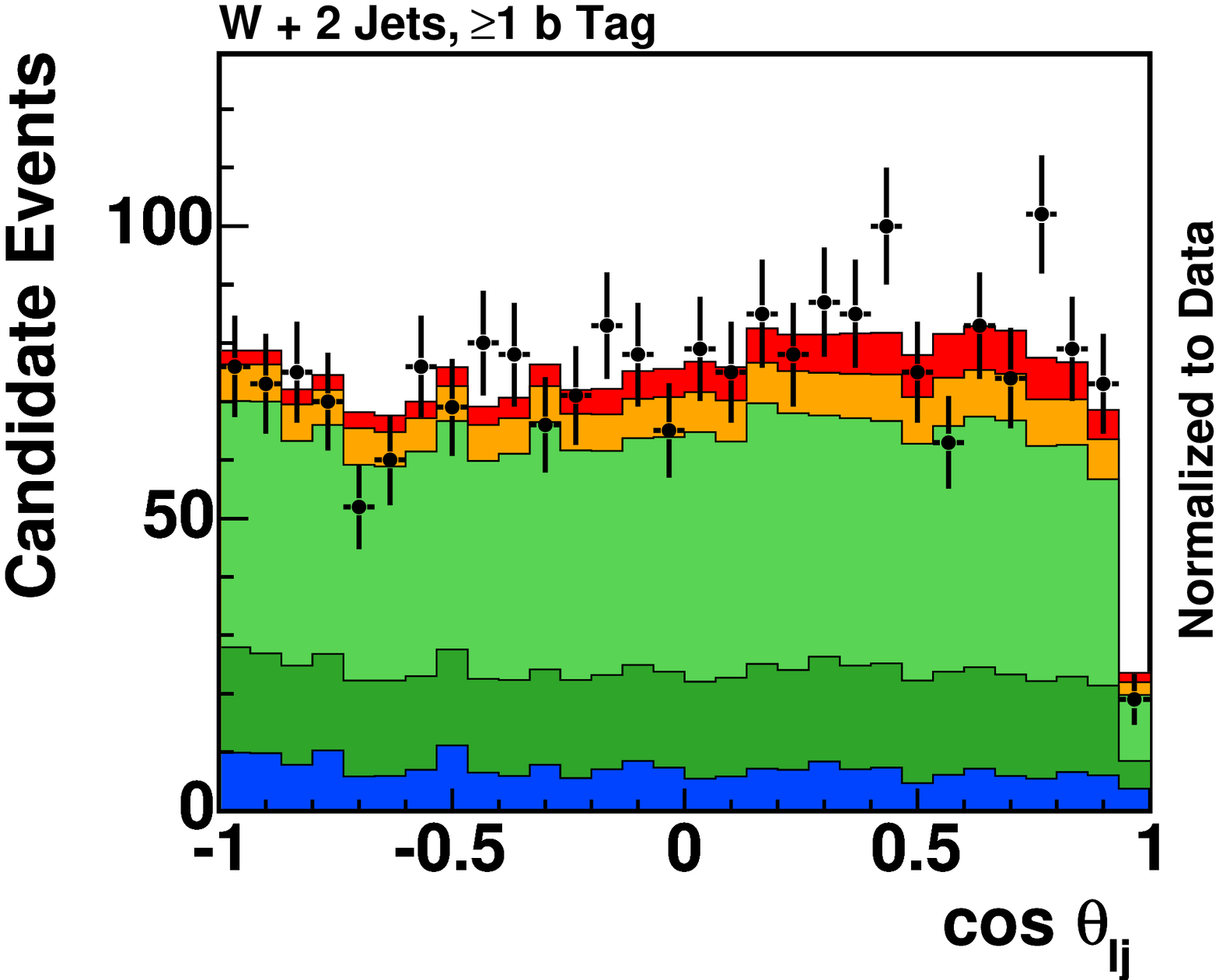}
\label{fig:coslj}}
\subfigure[]{
\includegraphics[width=0.8\columnwidth]{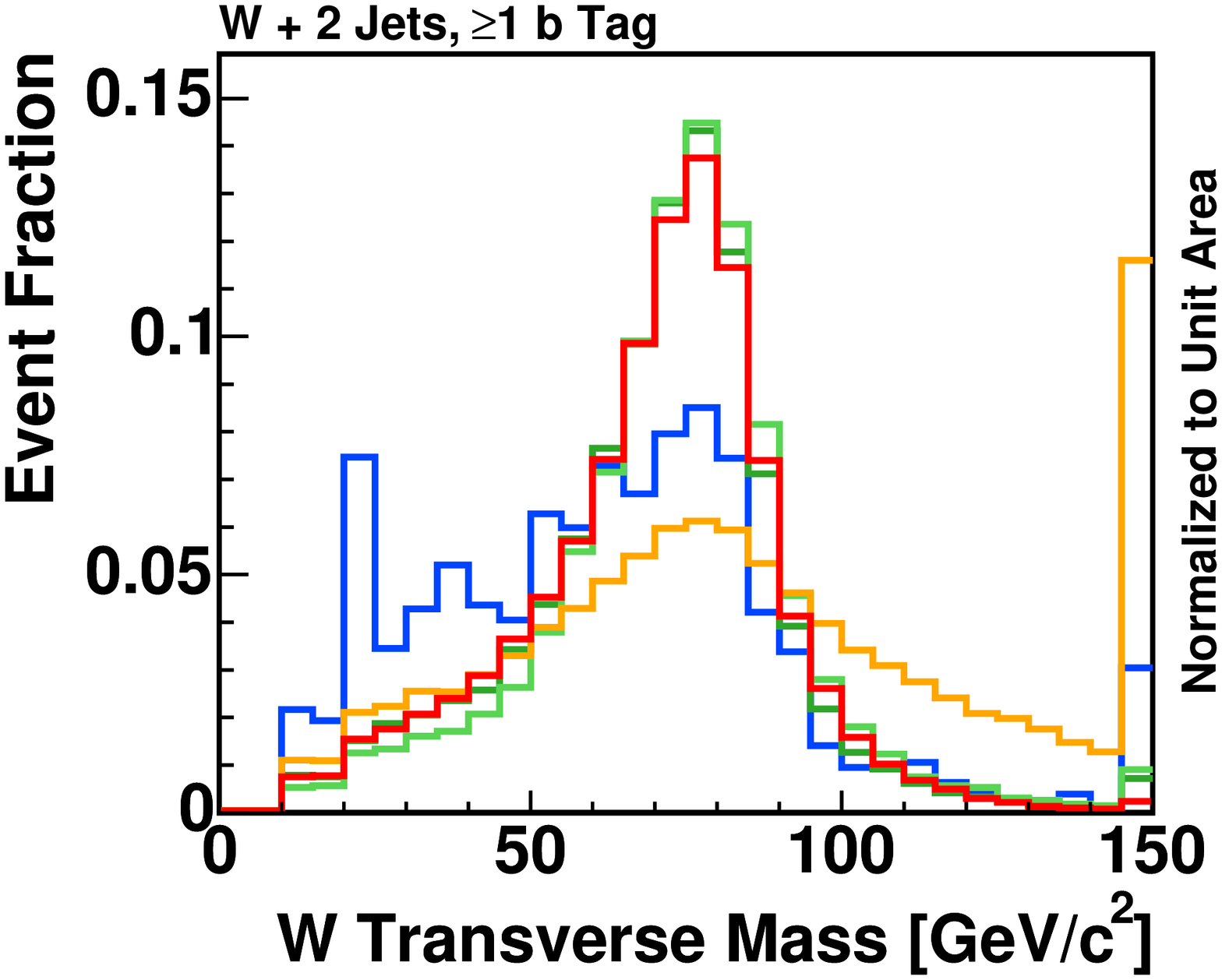}
\label{fig:mtwshape}}
\subfigure[]{
\includegraphics[width=0.8\columnwidth]{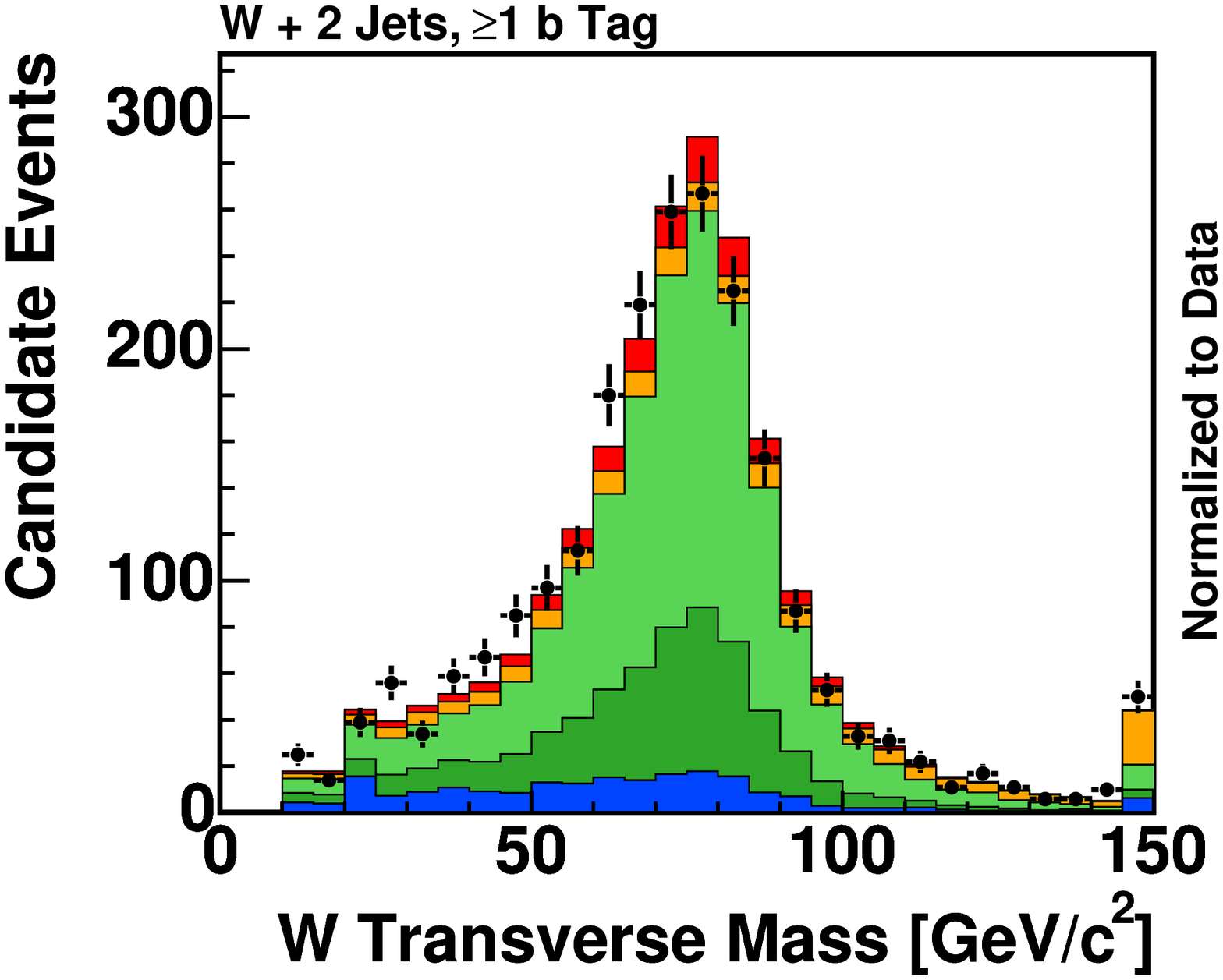}
\label{fig:mtw}}
\end{center}
\caption{\label{fig:Disc_vars_2}Monte Carlo templates (left) and
validation plots (right) comparing data and Monte Carlo for variables
with good discriminating power for events passing our selection requirements with two identified jets and
at least one $b$~tag.
The data are indicated with points, and the
shaded histograms show the signal and background predictions which are stacked
to form the total prediction. The stacking order follows that of the legend.
 Overflows are collected in the highest bin of each histogram.}
\end{figure*}

We list below some of the most sensitive variables, and explain why they are sensitive
in terms of the differences between the signal and background processes that they exploit.
The three multivariate discriminants, likelihood functions, neural networks, and boosted
decision trees, use these variables, or variations of them, as inputs; the analyses
also use other variables.  The matrix element
analysis uses all of these features implicitly, and it uses $b_{\rm NN}$ explicitly.
Normalized Monte Carlo predictions (``templates'') and modeling comparisons of these variables are shown in
Figs.~\ref{fig:Disc_vars_1} and~\ref{fig:Disc_vars_2}. 

\begin{itemize}

\item $M_{\ell \nu b}$: the invariant mass of the charged lepton,
the neutrino, and the $b$~jet from the top quark decay.  The $p_z$ of the
neutrino, which cannot be measured, is inferred by constraining $M_{\ell\nu}$ to
the $W$ boson mass, using the measured charged lepton candidate's momentum and 
setting $p_{\rm{T}}^\nu=$\met.  The neutrino's $p_z$ is the solution of a quadratic
equation, which may have two real solutions, one real solution, or two complex solutions.
For the case with two real
solutions, the one with the lower $|p_z|$ is chosen.  For the complex case, the real part of the
$p_z$ solution is chosen.  Some analyses use variations of this variable with different
treatment of the unmeasured $|p_z|$ of the neutrino.  The distribution of $M_{\ell\nu b}$ peaks
near $m_t$ for signal events, with broader spectra for background events from different processes.

\item $H_\mathrm{T}$: the scalar sum of the transverse energies of the
jets, the charged lepton, and $\EtMiss$ in the event. This quantity is much
larger for $t{\bar{t}}$ events than for $W$+jets events; single top quark events
populate the region in between $W$+jets events and $t{\bar{t}}$ events in this variable.

\item $M_{jj}$: the invariant dijet mass, which is
substantially higher on average for events containing top quarks than it is for events
with $W$+jets.

\item $Q \times \eta$: the sign of the charge of the lepton times the
pseudorapidity of the light quark jet~\cite{Yuan:1989tc}.  Large $Q \times \eta$ is
characteristic of $t$-channel single top quark events, because the light
quark recoiling from the single top quark often retains much of the momentum
component along the $z$ axis it had before radiating the $W$ boson.  It therefore
often produces a jet which is found at high $|\eta|$.  Multiplying $\eta$ by the
sign of the lepton's charge $Q$ improves the separation power of this variable
since 2/3 of single top quark production in the $t$-channel is initiated by
a $u$ quark in the proton or a ($\bar{u}$) quark in the antiproton, and the
sign of the lepton's charge determines the sign of the top quark's charge and is correlated with
the sign of the $\eta$ of the recoiling light-flavored jet.  The other 1/3 of
single top quark production is initiated by down-type quarks and has the opposite
charge-$\eta$ correlation.  $W$+jets and $t{\bar{t}}$ events lack this correlation,
and also have fewer jets passing our $E_{\rm T}$ requirement at large $|\eta|$ than 
the single top quark signal.

\item $\cos\theta_{\ell j}$: the cosine of the angle between the charged lepton
and the light quark jet~\cite{Mahlon:1996pn}.  For $t$-channel events, this tends to be
positive because of the $V-A$ angular dependence of the $W$ boson vertex.  This variable
is most powerful when computed in the rest frame of the top quark.

\item $b_{\mathrm{NN}}$: the jet flavor separator described in Section~\ref{sec:btagger}.  This
variable is a powerful tool to separate the signal from $W$+LF and $W$+charm events.

\item $M_{\rm T}^W$: the ``transverse mass'' of the charged lepton candidate and the $\EtMissVec$ vector.
The transverse mass is defined to be the invariant mass of the projections of the three-momentum
components in the plane perpendicular to the beam axis, and is so defined as to be independent
of the unmeasured $p_z$ of the neutrino.  Events without $W$ bosons in them (but with fake
leptons and mismeasured $\EtMiss$) have lower
$M_{\rm T}^W$ on average than $W$+jets events, signal events, and $t{\bar{t}}$ events.   Events
with two leptonically decaying $W$ bosons -- some diboson and $t{\bar{t}}$ events -- have even higher
average values of $M_{\rm T}^W$.  The distribution of $M_{\rm T}^W$ is an important cross-check
of the non-$W$ background rate and shape modeling.

\end{itemize}

While there are many distinctive properties of a single top quark
signal, no single variable is sufficiently sensitive to extract the
signal with the present data sample.  We must therefore use
techniques that combine the discrimination power of many
variables.   We use four such techniques in the $W$+jets sample,
a multivariate likelihood function, a matrix element method,
an artificial neural network, and a boosted decision tree.  These are described in detail
in the following sections.  Each of these techniques makes use of the most sensitive variables
described above in different ways, and in combination with other variables.  The measurements
using the separate techniques are highly correlated because the same events are analyzed with
each technique and because many of the same features are used, but the differences between
the techniques provide more discrimination power in combination as 
well as the ability to cross-check each
result with the others separately.

The measured single top quark cross section and the significance of the result
depend on the proper modeling of the input variable distributions for the
signals and the background processes.  We examine the distributions of all input variables
in the selected candidate events, comparing the data to the sum of the background and SM signal
predictions, and we also compare the distributions in a sample of events with no
$b$ tags but which pass all other event selection requirements.  The untagged event sample is
much larger than the tagged data sample and has no overlap with it, providing very precise
checks of the Monte Carlo's modeling of the data.  We do not
limit the investigation to input variables but also check the distributions of other
kinematic variables not used in the discriminants.
We also check the distributions of each discriminant output variable
in events with no $b$ tags.  Each of these investigations is done for each
technique, for 2-jet and 3-jet events separately, and for each category of charged lepton candidates,
requiring the examination of thousands of histograms.

\subsection{\label{sec:likelihood} Multivariate Likelihood Function}

A multivariate likelihood function (LF)~\cite{Ackerstaff:1997cz} is one
method for combining several sensitive variables.  This method makes use
of the relative probabilities of finding an event in histograms of each
input variable, compared between the signal and the background.

The likelihood function $L_k$ for event class $k$ is constructed using 
binned probability density
functions for each input variable.  The probability that an event from
sample $k$ will populate bin $j$ of input variable $i$ is defined to
be $f_{ijk}$.  The probabilities are normalized so that
$\sum_jf_{ijk}=1$ for all variables $i$ and all samples $k$.
For the signal, $k=1$, and in this paper, four background classes are
used to construct the likelihood function: $Wb{\bar{b}}$,
$t{\bar{t}}$, $Wc{\bar{c}}/Wc$, and $W+$LF, which are event classes
$k=$ 2, 3, 4, and 5, respectively.
Histogram underflows and overflows are properly accounted for.
The likelihood function for an
event is computed in two steps.  First, for each reconstructed
variable $i$, the bin $j$ in which the event falls is obtained, and
the quantities
\begin{equation}
p_{ik} = \frac{f_{ijk}}{\sum_{m=1}^5f_{ijm}},
\end{equation}
are computed for each variable $i$ and each event class $k$.  The $p_{ik}$ are used to compute
\begin{equation}
{\cal L}_k = \frac{\prod_{i=1}^{n_{{\mathrm{var}}}}p_{ik}}{\sum_{m=1}^5\prod_{i=1}^{n_{\mathrm{var}}}p_{im}},
\label{eq:lfdef}
\end{equation}
where $n_{\mathrm{var}}$ is the number of input variables.
The signal likelihood function, referred to as LF discriminant in the following, is the one which corresponds to the
signal class of events, ${\cal L}_1$.  
This method does not take advantage of the correlations between input variables,
which may be different between the signal and the background processes.
The predicted distributions of the likelihood functions are made 
from fully simulated Monte Carlo
and data sets where appropriate, with all correlations in them, and so while 
correlations are not taken advantage of, they are included in the necessary modeling.
The reduced dependence on the correlations makes the LF analysis
an important cross-check on the other analyses, which make use of the correlations. 
More detailed information on
this method can be found in~\cite{Budd:2008zz} and~\cite{Nakamura:2009zzd}.

Three likelihood functions are computed for use in the search for
single top quark production.  The first, $L_t$, is optimized for the $t$-channel signal; it is used for
events with two jets and one $b$ tag.  Another, $L_s$, is optimized for the
$s$-channel signal; it is applied to events with two jets and two $b$ tags.  The $L_s$-based analysis
was separately labeled the LFS analysis in~\cite{Aaltonen:2009jj}.
The third, $L_{3j}$, is optimized for the sum of both $s$- and $t$-channel single top quark production;
it is applied to events with three jets.
The inputs to these three likelihood functions are described in
Sections~\ref{sec:Likeli_t}, \ref{sec:Likeli_s}, and \ref{sec:Likeli_3jet}, respectively.

\subsubsection{\label{sec:kinematicConstraints}Kinematic Constraints}

The likelihood function input variables include the squares of the quantum-mechanical
matrix elements, using {\sc madgraph}~\cite{Maltoni:2002qb}, 
computed with the measured four-vectors.  These calculations 
depend very strongly on the invariant masses of the $\ell\nu$
system and the $\ell\nu b$ system, which result from the $W$ boson and
top quark decay, respectively.  The neutrino leaves no trace in the detector; 
$\EtMiss$\ is an approximation to its transverse momentum, and $p_z^\nu$ is not measured.
  The $b$ quark is also imperfectly
reconstructed; a $b$-tagged jet's energy is an approximation to the $b$~quark's momentum.
We solve for the $p_z$ of the neutrino
and the energy of the $b$ quark while requiring that $M_{\ell\nu}=M_W$ and
$M_{\ell\nu b}=m_t$.  The $W$ boson mass constraint results in two solutions.  If both are
real, the one with the smaller $|p_{z}|$ is used.  If both are
complex, a minimal amount of
additional $\EtMiss$\ is added parallel to the jet axis assigned to be the $b$
from the top quark's decay until a real solution for $|p_z^\nu|$ can be obtained.
In rare cases in which this procedure still fails to produce a real
$|p_z^\nu|$, additional $\EtMiss$\ is added along the $b$-jet
axis to minimize the imaginary part of $|p_z^\nu|$, and then a
minimal amount of $\EtMiss$\ is added perpendicular to the $b$-jet axis
until a real $|p_z^\nu|$ is obtained.

The top quark mass constraint can be satisfied by scaling the $b$-jet's
energy, holding the direction fixed,
until $M_{\ell\nu b}=m_t$.  As the $b$-jet's energy is scaled,
the $\EtMiss$\ is adjusted to be consistent with the change.  We then
recalculate $p_z^\nu$ using the $M_W$ constraint described above, 
and the process is iterated until $M_{\ell\nu b} = m_t$.  The resulting
four-vectors of the $b$ quark and the neutrino are then used with the
measured four-vector of the charged lepton in the matrix element expressions to construct discriminant
variables that separate the signal from the background.

\subsubsection{\label{sec:Likeli_t} 2-Jet $t$-channel Likelihood Function}
The $t$-channel likelihood function ${\cal L}_{t}$ uses seven
variables, and assumes the $b$-tagged jet comes from top quark decay.  The
variables used are:

\begin{itemize}
\item $H_{\rm T}$, the scalar sum of the $E_{T}$'s of the two jets, the lepton $E_{\rm T}$, and $\EtMiss$.
\item $Q\times \eta$, the charge of the lepton times the pseudorapidity of the jet which is not $b$-tagged.
\item $\chi^2_{\mathrm{kin}}$,  the $\chi^2$ of the comparison of the measured
$b$ jet energy and the one the kinematic constraints require in order
to make $M_{\ell\nu b}=m_t$ and $M_{\ell\nu}=M_W$, using the nominal uncertainty in the
$b$ jet's energy.  Any additional
$\EtMiss$\ which is added to satisfy the $m_{\ell\nu}=M_W$ constraint is
added to $\chi^2_{\mathrm{kin}}$
using the nominal uncertainty in the $\EtMiss$ measurement.
\item $\cos\theta_{\ell j}$, the cosine of the angle between 
the charged lepton and the untagged jet in the top quark decay frame.
\item $M_{jj}$, the invariant mass of the two jets. 
\item ME$_{t-\textrm{chan}}$, the differential cross
section for the $t$-channel process, as computed by {\sc madgraph}
using the constrained four-vectors of the $b$, $\ell$, and
$\nu$.
\item The jet flavor separator output $b_{\mathrm{NN}}$ described in
Section \ref{sec:btagger}.
\end{itemize}

\subsubsection{\label{sec:Likeli_s} 2-Jet $s$-channel Likelihood Function}

The $s$-channel likelihood function $L_s$ uses nine
variables.  Because these events have exactly two jets, both of which
are required to be $b$-tagged, we decide
which jet comes from the top quark decay with a separate likelihood
function that includes the transverse momentum of the $b$ quark, the
invariant mass of the $b$ quark and the charged lepton, and the product of the
scattering angle of the $b$ jet in the initial quarks' rest frame and
the lepton charge.  To compute this last variable, the $p_z$ of the neutrino
has been solved for using the $m_W$ constraint.

The variables input to $L_s$ are:

\begin{itemize}
\item $M_{jj}$, the invariant mass of the two jets. 
\item $p_{{\rm T}}^{jj}$, the transverse momentum of the two-jet system.
\item $\Delta R_{jj}$, the separation between the two jets in
$\phi$--$\eta$ space.
\item $M_{\ell\nu b}$, the invariant mass of the charged lepton,
the neutrino, and the jet assigned to be the $b$~jet from the top quark decay.
\item $E_{\rm T}^{j_{1}}$, the transverse energy of the leading jet, that is,
the jet with the largest $E_{\rm T}$.
\item $\eta_{j_{2}}$, the pseudorapidity of the non-leading jet.
\item $p_{\rm T}^\ell$, the transverse momentum of the charged lepton.
\item $Q\times \eta$, the charge of the lepton times the 
pseudorapidity of the jet which is not assigned to have come
from the top quark decay.
\item The logarithm of the likelihood ratio constructed by matrix
elements computed by {\sc madgraph}, using the $p_{z}^{\nu}$
solution which maximizes the likelihood described in the next point.
This likelihood ratio is defined as $\frac{ME_{s} + ME_{t}}{ME_{s} +
ME_{t} + ME_{Wbb}}$.
\item The output of a kinematic fitter which chooses a solution of
$p_{z}^{\nu}$ that maximizes the likelihood of the solution by allowing the
values of $p_{x}^{\nu}$ and $p_{y}^{\nu}$ to vary within their
uncertainties.  This likelihood is multiplied by the likelihood used
to choose the $b$ jet that comes from the top quark, and their product
is used as a discriminating variable.
\end{itemize}

\subsubsection{\label{sec:Likeli_3jet} 3-Jet Likelihood Function}

Three-jet events have more ambiguity in the assignment of jets to
quarks than two-jet events.  A jet must be assigned to be the one originating from the $b$ quark
from top quark decay, and another jet must be assigned to be the recoiling
jet, which is a light-flavored quark in the $t$-channel case and a $b$ quark
in the $s$-channel case.  In all there are six possible assignments of jets
to quarks not allowing for grouping of jets together.
  The same procedure described in
Section~\ref{sec:kinematicConstraints} is used on all six possible
jet assignments.  If only one jet is $b$-tagged, it is assumed
to be the $b$ quark from top quark decay.  If two jets are $b$-tagged, the
jet with the highest $-\log \chi ^2 + 0.005 p_{\rm T}$ is chosen,
where $\chi^2$ is the smaller of the outputs of the kinematic fitter,
one for each $p_{z}^{\nu}$ solution.  This algorithm correctly assigns
the $b$ jet 75\% of the time.

There is still an ambiguity regarding the proper assignment of the
other jets.  If exactly one
of the remaining jets is $b$-tagged, it is assumed to be from a $b$
quark, and the untagged jet assigned to be the $t$-channel recoiling jet; 
otherwise, the jet with larger $E_{\rm T}$ is assigned to be the $t$-channel
recoiling jet.  In all cases, the smaller $|p_{z}^{\nu}|$ solution is used.

The likelihood function $L_{3j}$ is defined with the following input variables:

\begin{itemize}
\item $M_{\ell\nu b}$, the invariant mass of the charged lepton,
the neutrino, and the jet assigned to be the $b$~jet from from the top quark decay.
\item $b_{\mathrm{NN}}$: the output of the jet-flavor separator.
\item The number of $b$-tagged jets.
\item $Q\times \eta$: the charge of the lepton times the pseudorapidity of the 
jet assigned to be the $t$-channel recoiling jet.
\item The smallest $\Delta R$ between any two jets, where $\Delta R$
is the distance in the $\phi$--$\eta$ plane between a pair of jets.
\item The invariant mass of the two jets not assigned to have come from top quark
decay.
\item $\cos\theta_{\ell j}$: the cosine of the
angle between the charged lepton and the jet assigned to be the $t$-channel recoiling jet
in the top quark's rest frame.
\item The transverse momentum of the lowest-\ET\ jet.
\item The pseudorapidity of the reconstructed $W$ boson.
\item The transverse momentum of the $b$ jet from top quark decay.
\end{itemize}

\subsubsection{Distributions}

\begin{figure*}
\begin{center}
\subfigure[]{
\includegraphics[width=0.65\columnwidth]{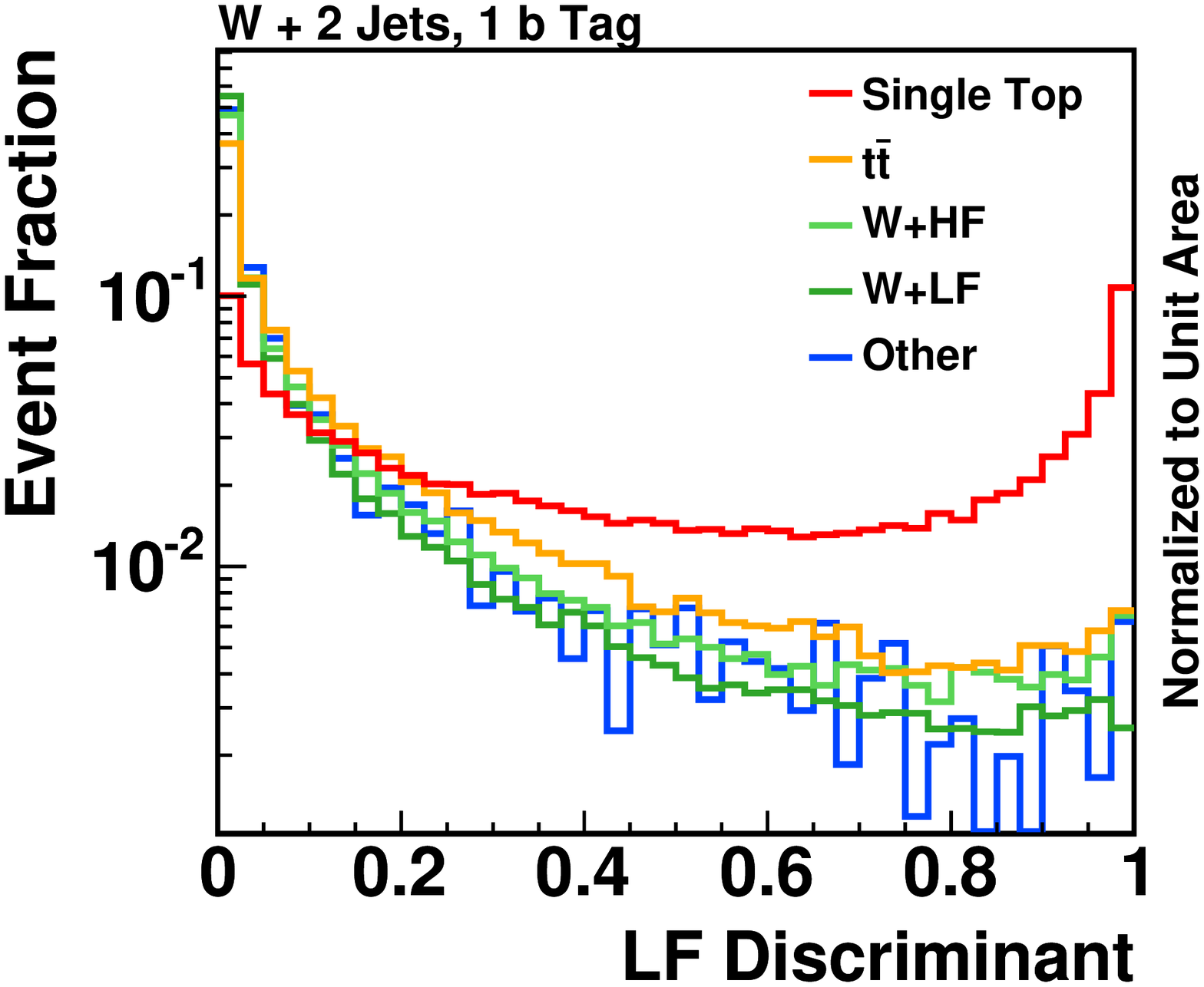}
\label{fig:LF2j1t_shape}}
\subfigure[]{
\includegraphics[width=0.65\columnwidth]{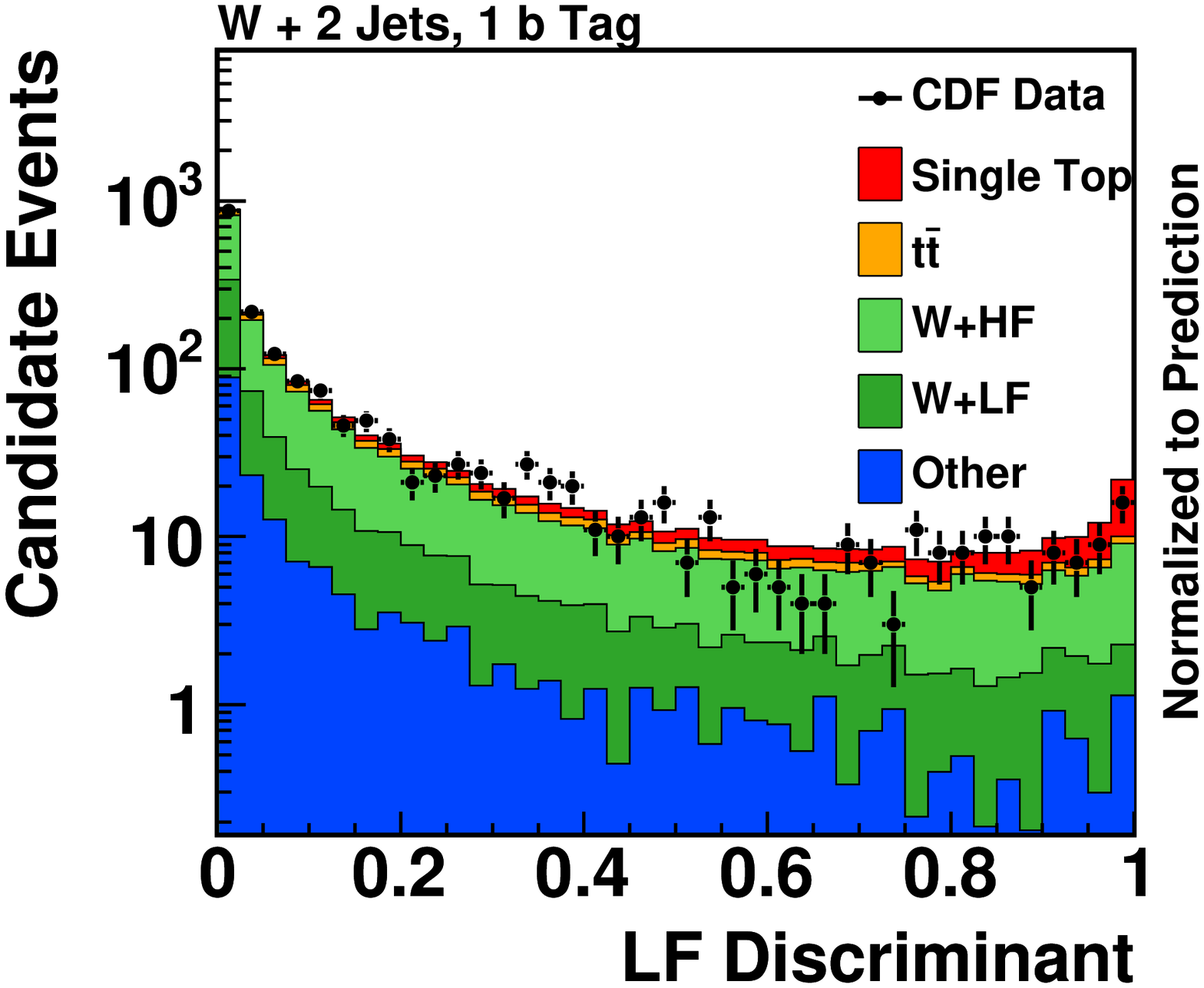}
\label{fig:LF2j1t}} \\
\subfigure[]{
\includegraphics[width=0.65\columnwidth]{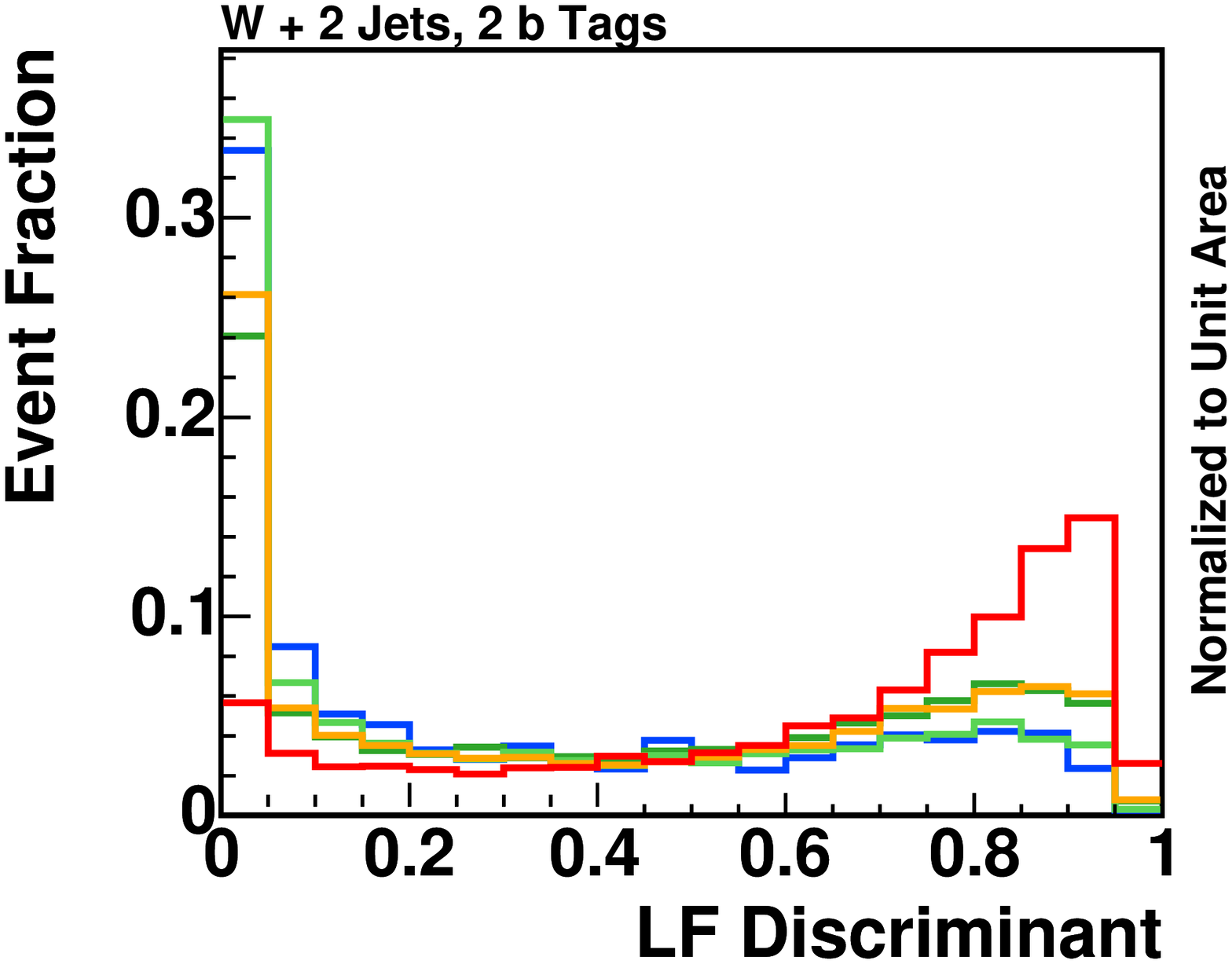}
\label{fig:LF2j2t_shape}}
\subfigure[]{
\includegraphics[width=0.65\columnwidth]{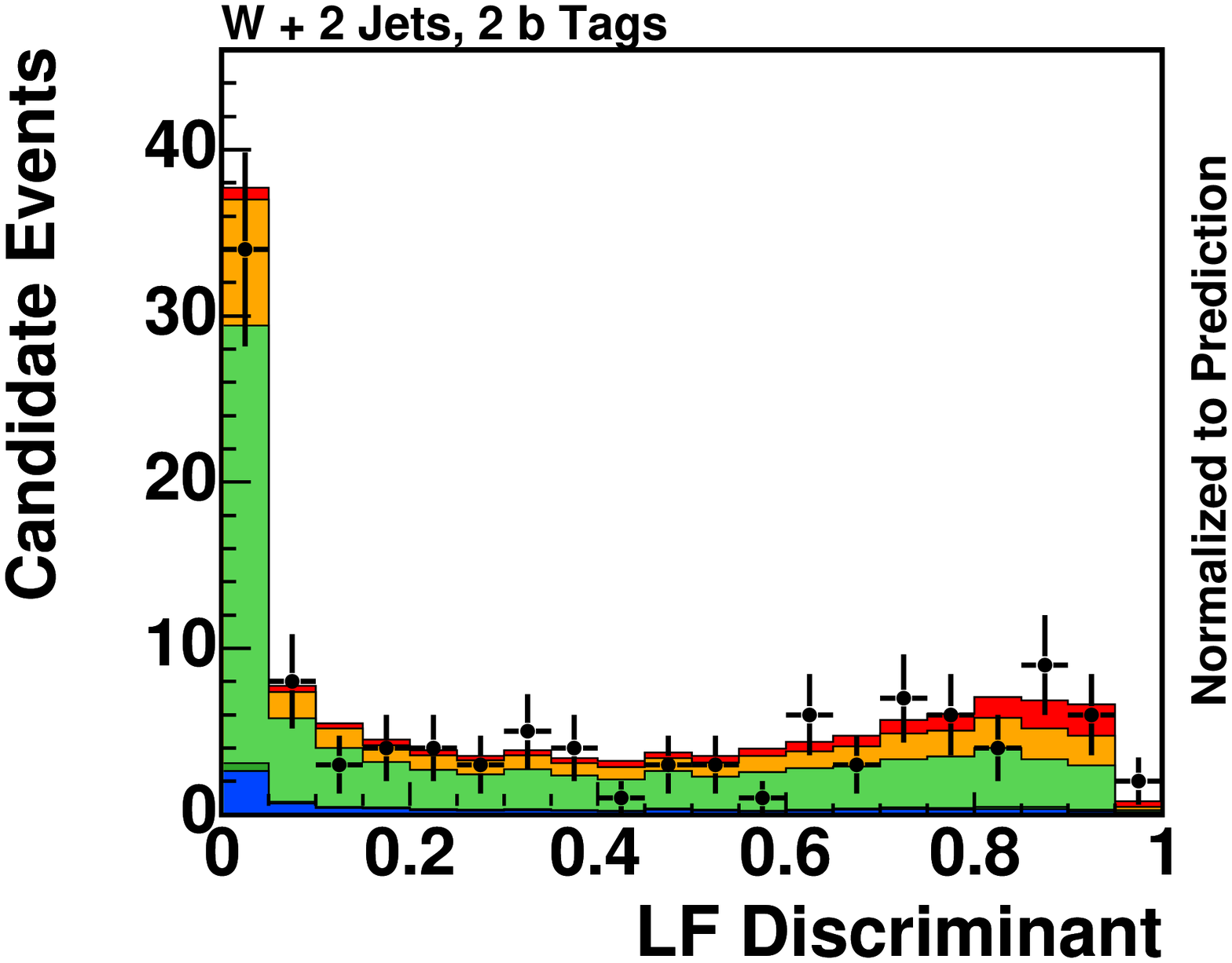}
\label{fig:LF2j2t}} \\
\subfigure[]{
\includegraphics[width=0.65\columnwidth]{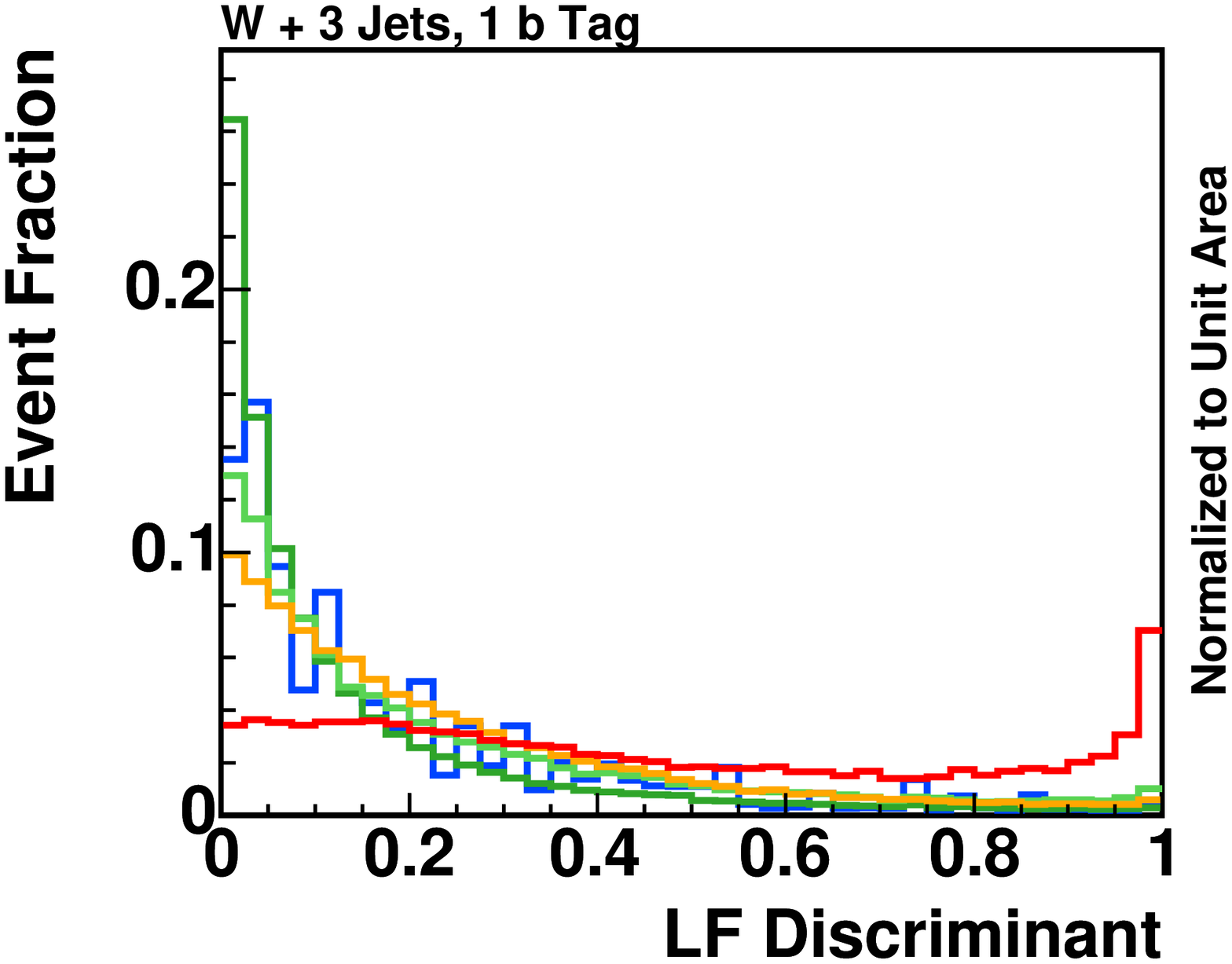}
\label{fig:LF3j1t_shape}}
\subfigure[]{
\includegraphics[width=0.65\columnwidth]{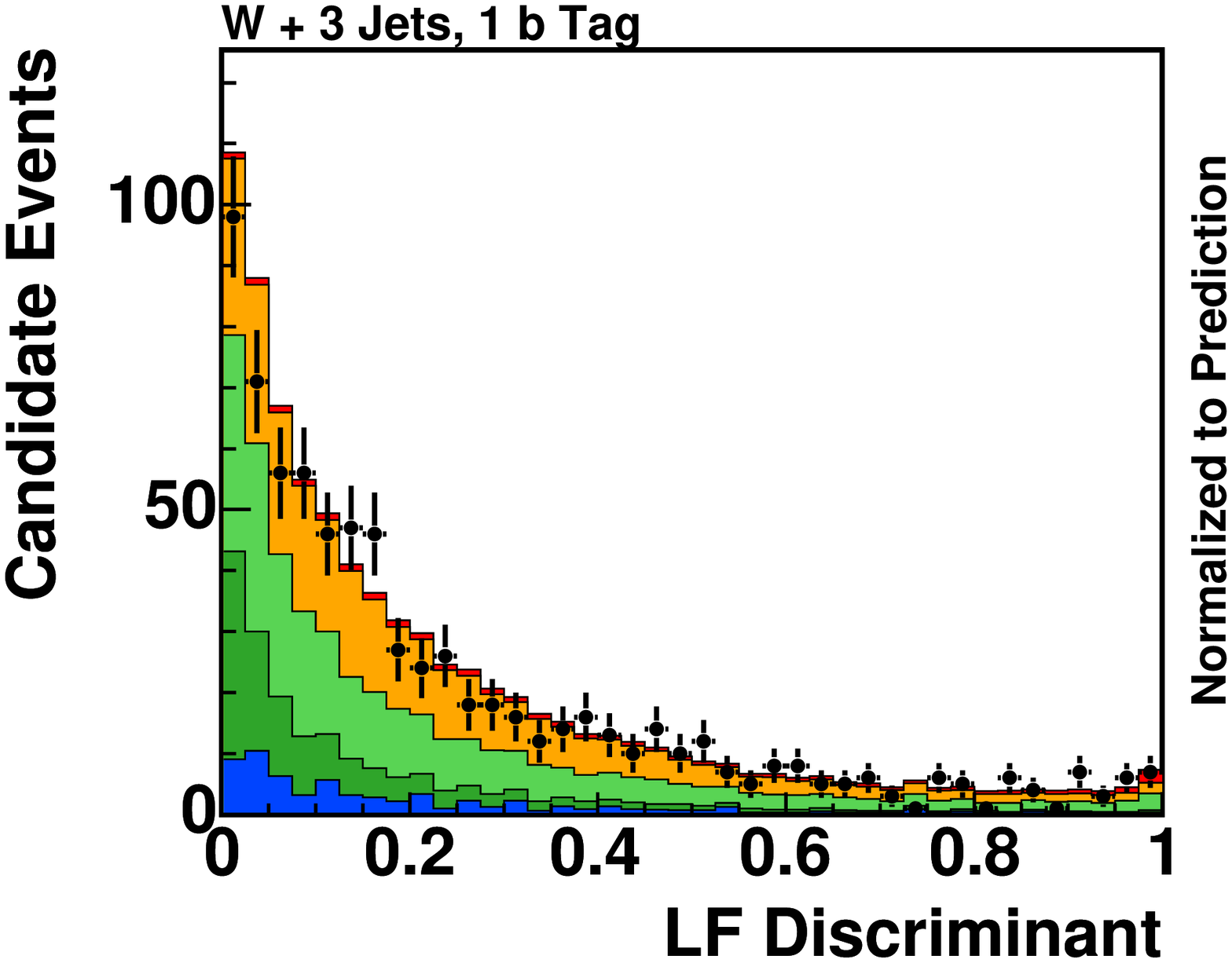}
\label{fig:LF3j1t}} \\
\subfigure[]{
\includegraphics[width=0.65\columnwidth]{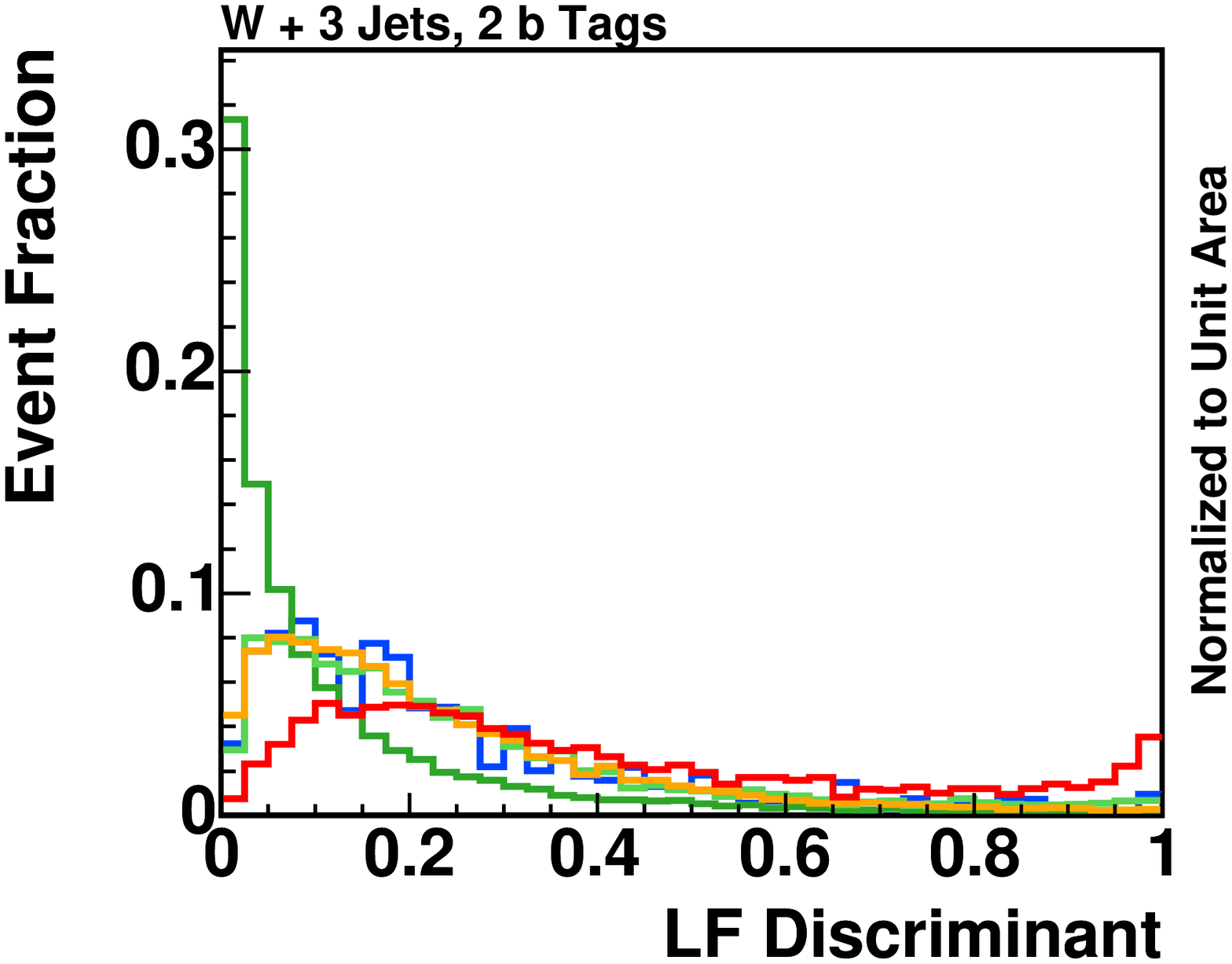}
\label{fig:LF3j2t_shape}}
\subfigure[]{
\includegraphics[width=0.65\columnwidth]{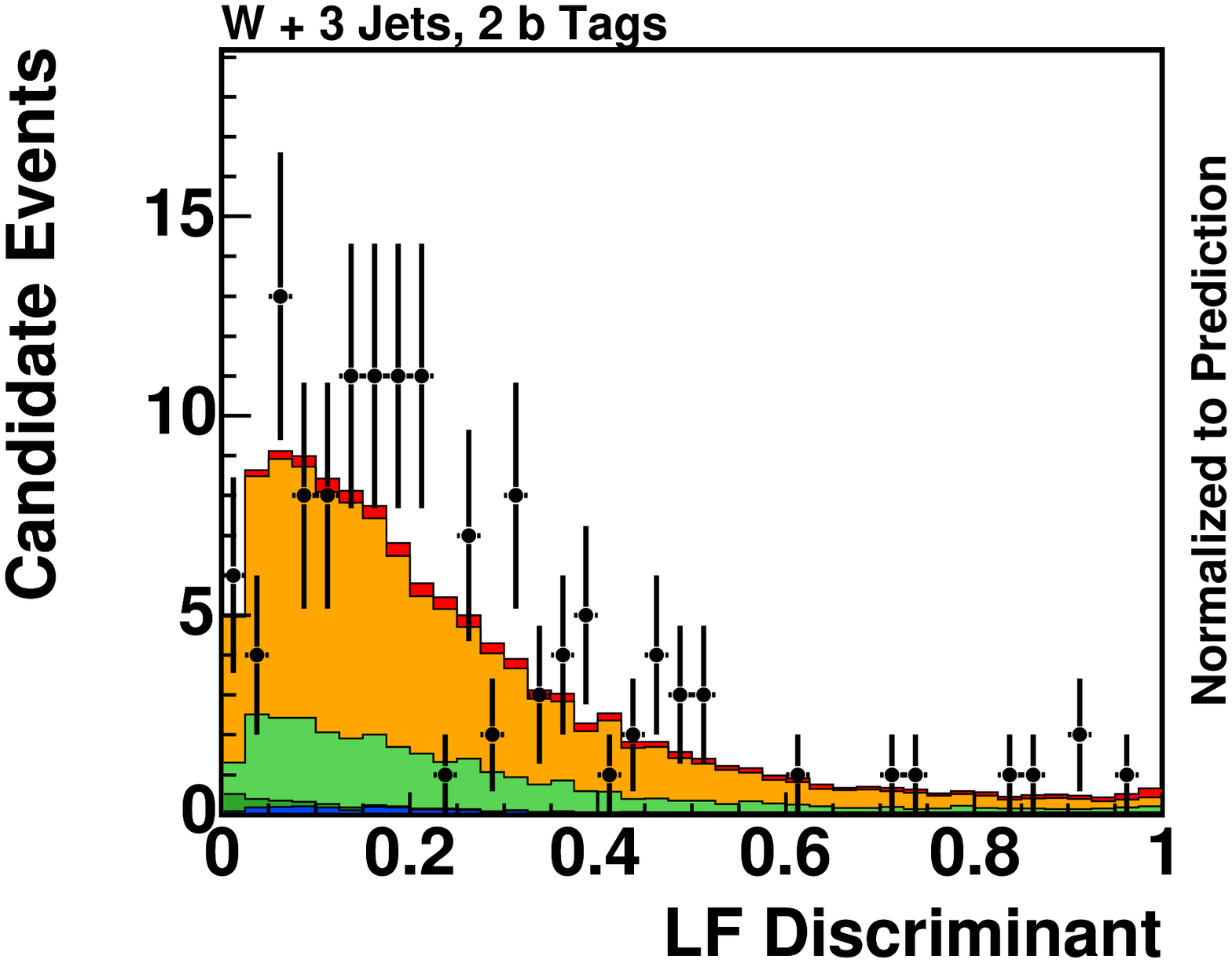}
\label{fig:LF3j2t}}
\end{center}
\caption{\label{fig:LF} 
Templates of predictions for the signal
and background processes, each scaled to unit area (left) and comparisons of
the data with the sum of the predictions (right)
of the likelihood function for each selected data sample. Single top quark events
are predominantly found on the right-hand sides of the histograms while
background events are mostly found on the left-hand sides. The two-jet, one-$b$-tag plots are
shown on a logarithmic vertical scale for clarity, while the others are shown on a linear scale.
The data are indicated by points with error bars, and 
the predictions are shown stacked, with the stacking order following that of the legend.
}
\end{figure*}

In each data sample, distinguished by the number of identified jets and the number
of $b$~tags, a likelihood function is constructed with the input variables described above.
The outputs lie between zero and one, where zero is background-like
and one is signal-like.  The predicted distributions of the signals and the
expected background processes are shown in~Fig.~\ref{fig:LF} for the four $b$-tag and jet
categories.  The templates, each normalized to unit area, are shown separately, indicating the separation
power for the small signal.  
The sums of predictions normalized to our signal and
background models, which are described in Sections~\ref{sec:Background}
and~\ref{sec:SignalModel}, respectively,
 are compared with the data.  
Figure~\ref{fig:allLF}\subref{fig:allLF_chan} 
shows the discriminant output distributions for the data and the predictions summed over all 
four $b$-tag and jet categories.

\subsubsection{Validation}

The distributions of the input variables to each likelihood function are checked in
the zero-, one-, and two-tag samples for two- and three-jet events.  Some of the most
important variables' validation plots are shown in Sections~\ref{sec:bgvalidation}
and~\ref{sec:Multivariate}.  The
good agreement seen between the predictions and the observations in both the input variables
and the output variables gives confidence in the validity of the technique.

\begin{figure*}
\begin{center}
\subfigure[]{
\includegraphics[width=0.8\columnwidth]{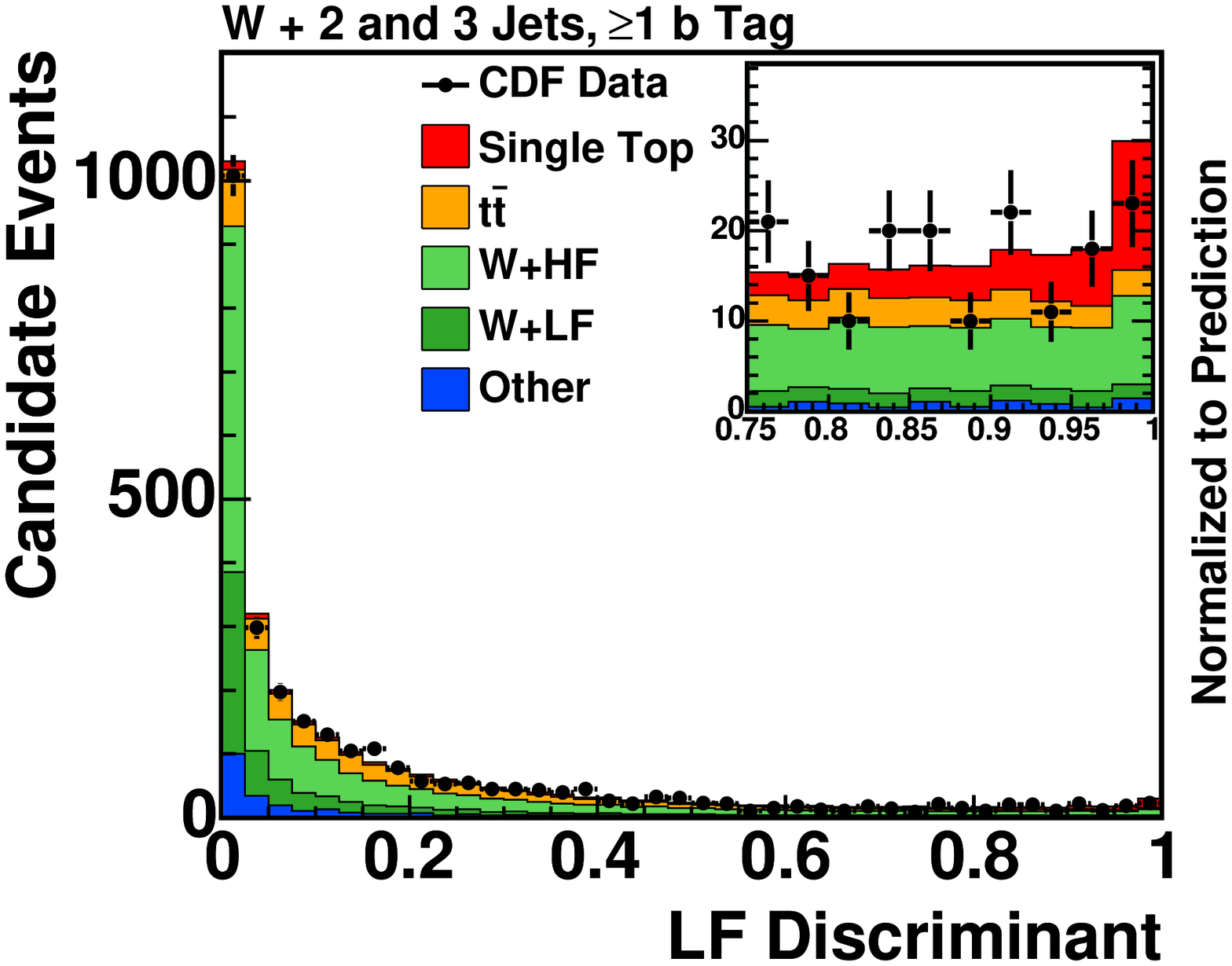}
\label{fig:allLF_chan}}
\subfigure[]{
\includegraphics[width=0.8\columnwidth]{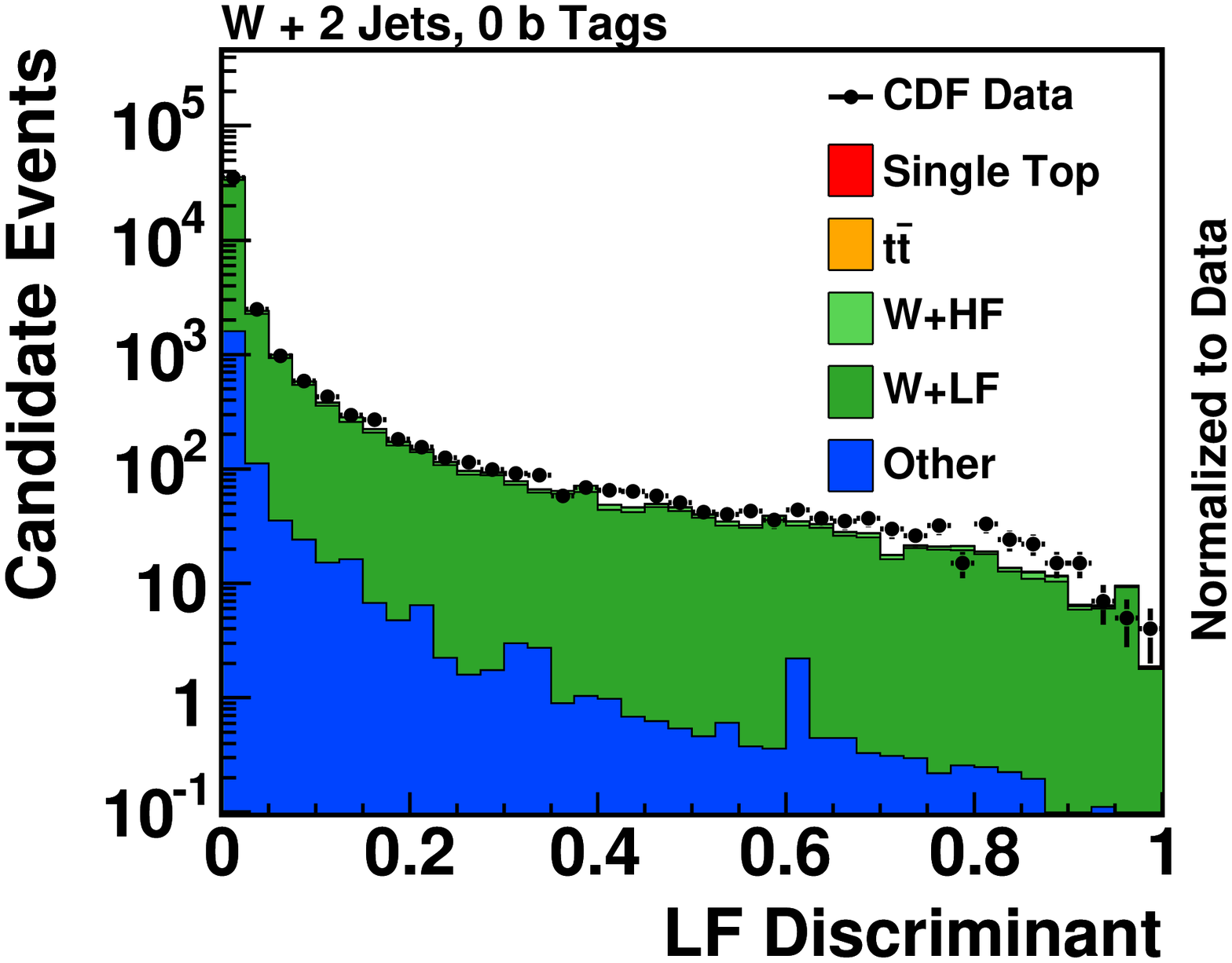}
\label{fig:allLF_0tag}}
\end{center}
\caption{\label{fig:allLF} Comparison of the data with the sum of the predictions 
of the likelihood function for the sum of all selected data samples (left) and for two-jet
one-tag events (right) applied to the untagged sideband, the latter with appropriate
modifications to variables that rely on $b$-tagging.
The stacking order follows that of the legend.  The discrepancies
between the prediction and the observation in the untagged sideband seen here
are covered by systematic uncertainties
on the $W+$jets background model.}

\begin{center}
\subfigure[]{
\includegraphics[width=0.8\columnwidth]{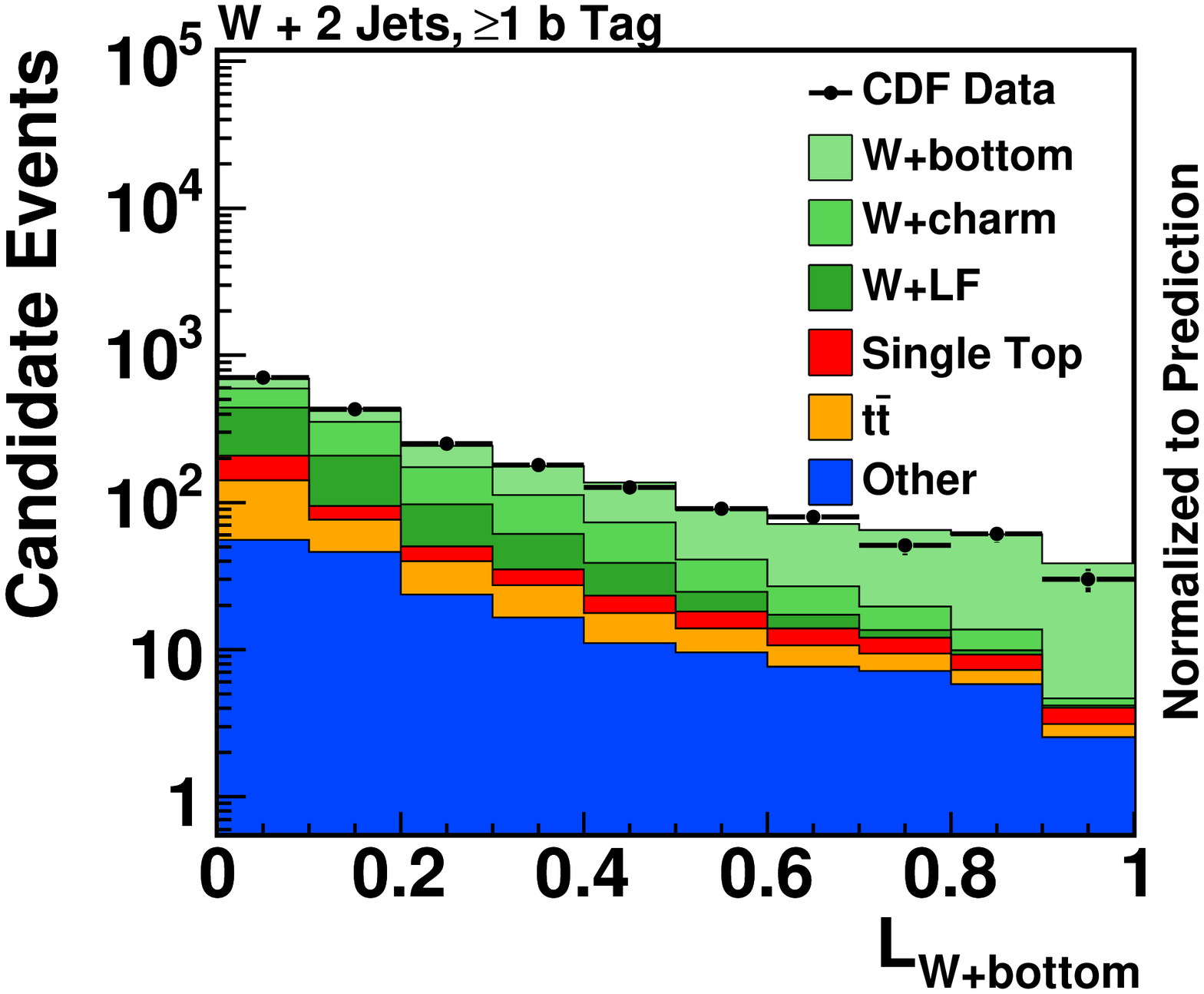}
\label{fig:wbblf}}
\subfigure[]{
\includegraphics[width=0.8\columnwidth]{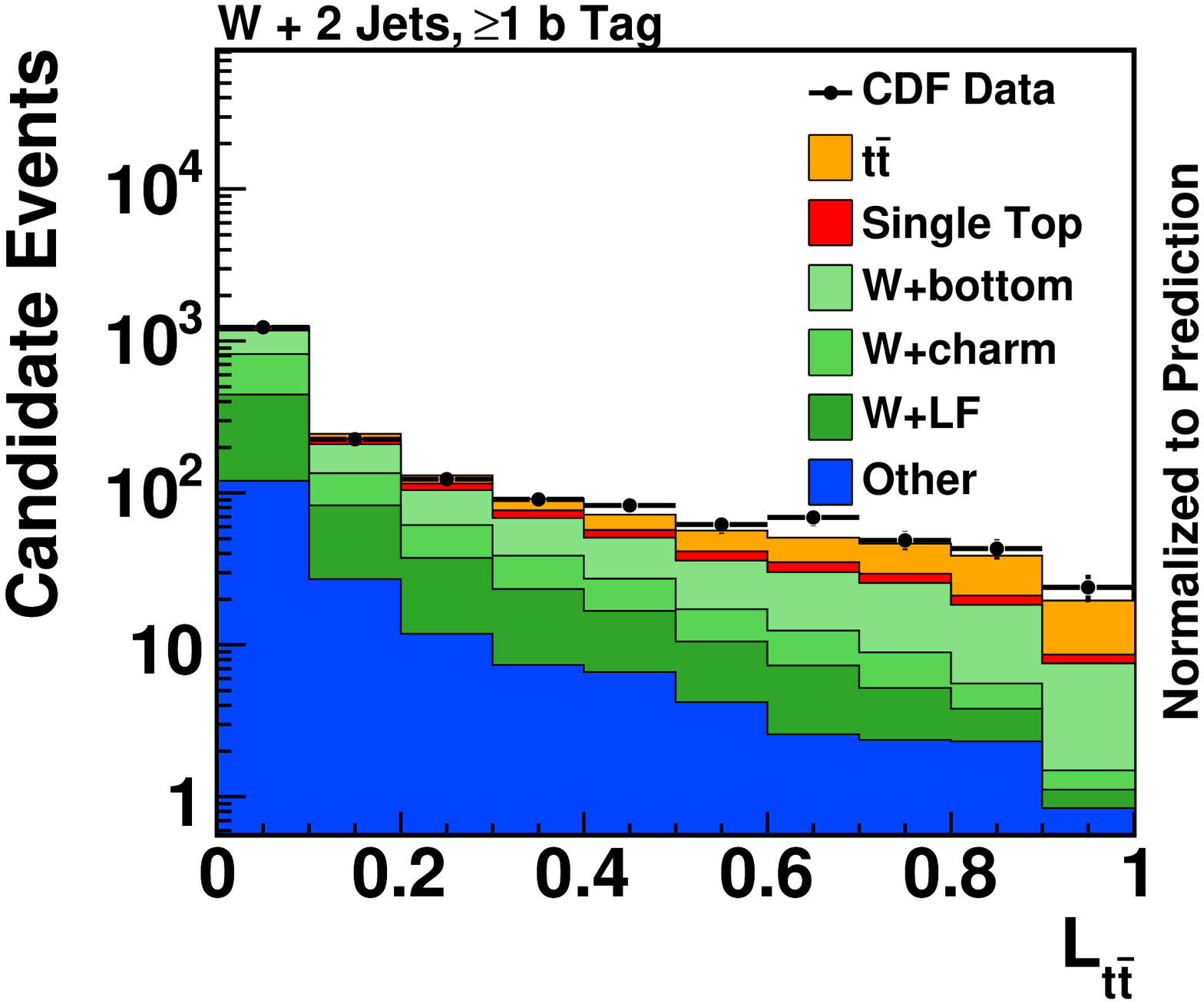}
\label{fig:ttlf}} \\
\subfigure[]{
\includegraphics[width=0.8\columnwidth]{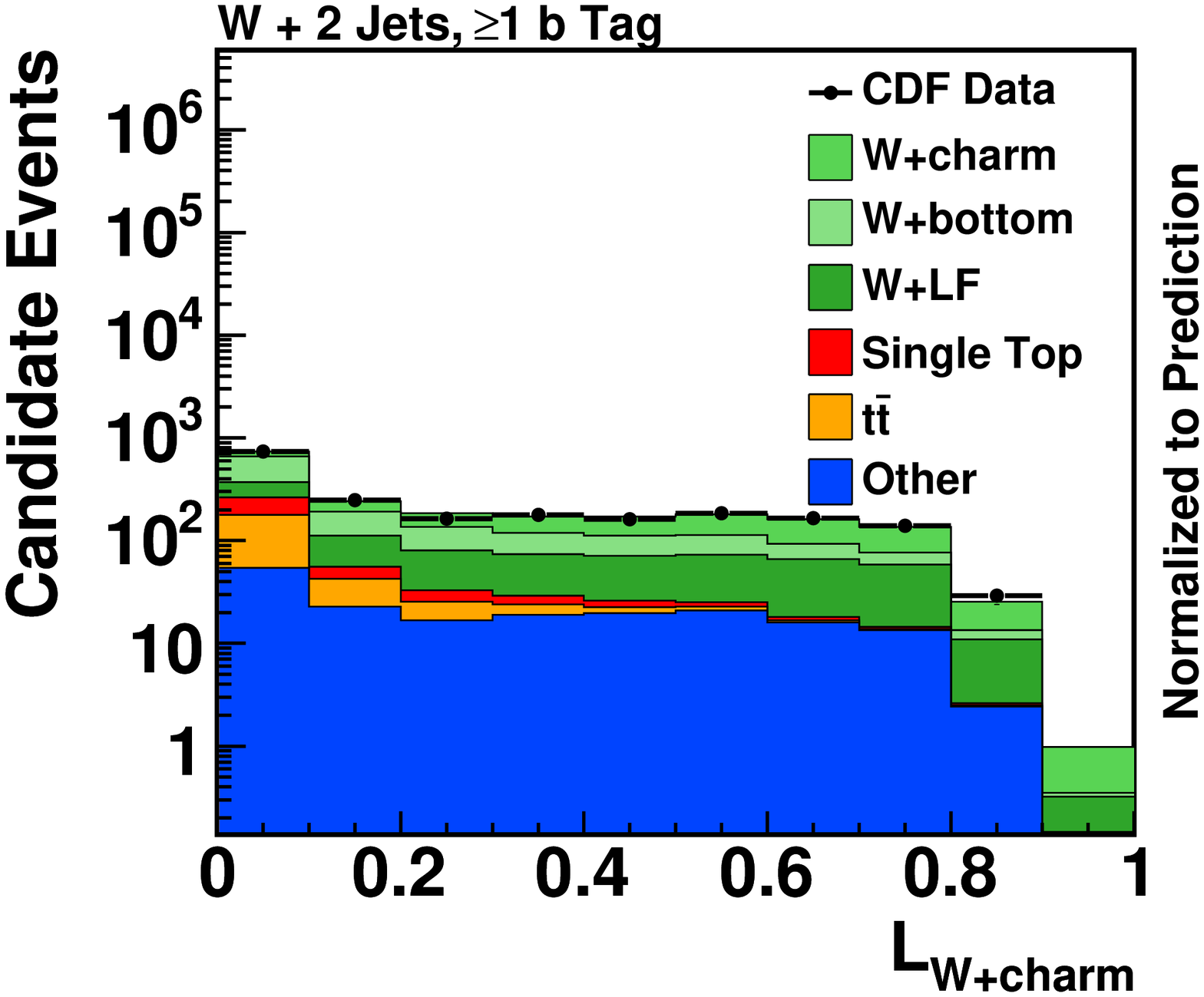}
\label{fig:wcharmlf}}
\subfigure[]{
\includegraphics[width=0.8\columnwidth]{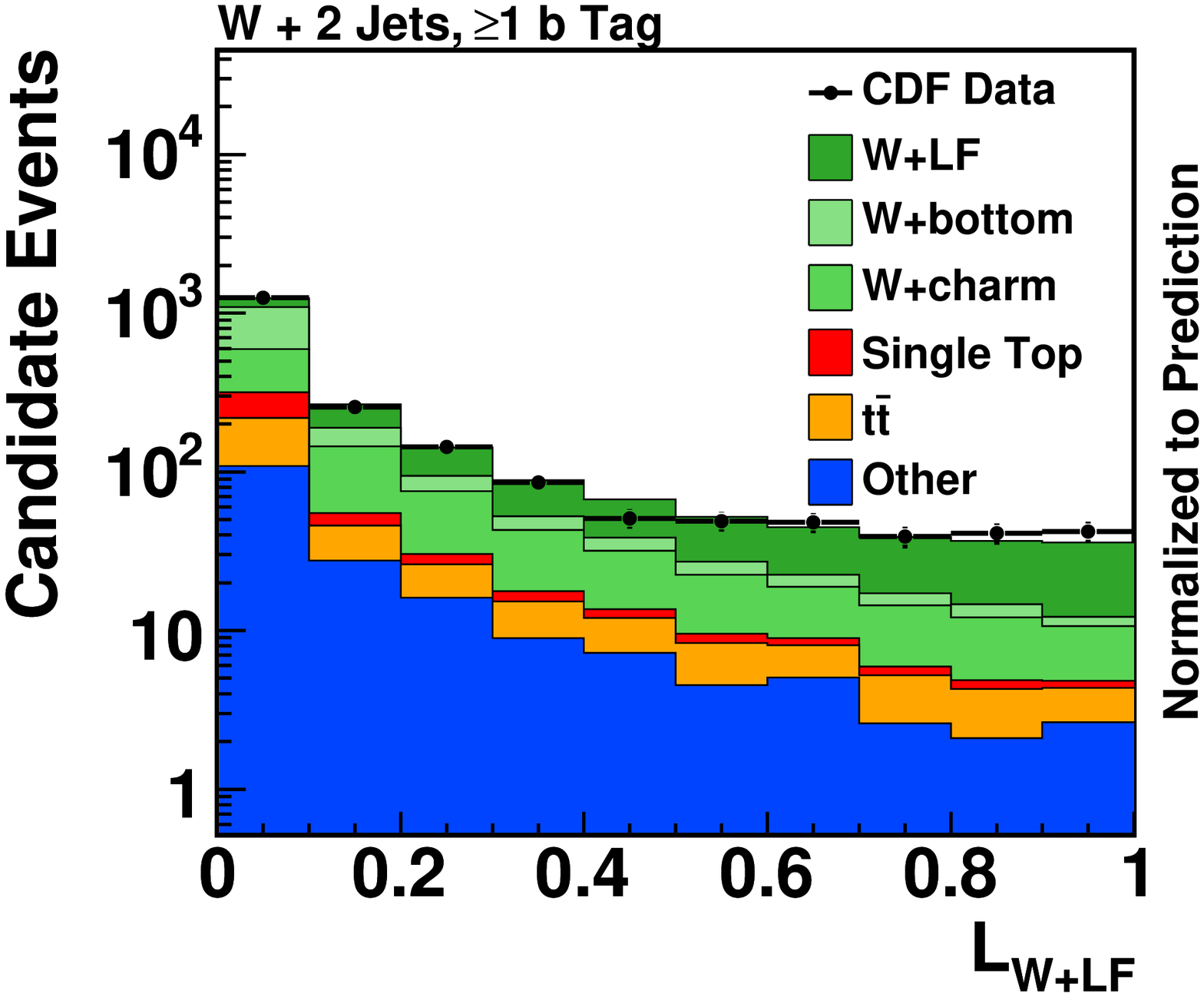}
\label{fig:mistaglf}}
\end{center}
\caption{\label{fig:lfbg} 
Distributions of 
$L_{W+{\rm{bottom}}}$,
$L_{t{\bar{t}}}$, 
$L_{W+{\rm{charm}}}$, 
and $L_{W+{\rm{LF}}}$
for $b$-tagged $W$+2~jet events
passing our event selection. The signal and
 background contributions are normalized to the same predicted rates that are used 
in the signal extraction histograms.  In each plot, the background process which the
discriminant treats as signal is stacked on top of the other background processes.
The stacking orderings follow those of the legends.
}
\end{figure*}

Each likelihood function is also tested in
the untagged sample, although the input variables which depend on $b$-tagging
are modified in order to make the test.  For example, $b_{\mathrm{NN}}$ is fixed to
$-1$ for untagged events, $Q\times\eta$ uses the jet with the largest $|\eta|$ instead
of the untagged jet, and the taggable jet with the highest $E_{\rm{T}}$ is used as the
$b$-tagged jet in variables which use the $b$-tagged jet as an input.
The modeling of the modified likelihood function in the untagged events is
not perfect, as can be seen in Fig.~\ref{fig:allLF}\subref{fig:allLF_0tag}.  This mismodeling
is covered by the systematic uncertainties on the {\sc alpgen} modeling of
$W$+jets events which constitute the bulk of the background.  Specifically,
using the untagged data as the model for mistagged $W$+jets events as well
as shape uncertainties on $\Delta R_{jj}$ and $\eta_{j2}$ cover the observed
discrepancy.

\subsubsection{Background Likelihood Functions}

Another validation of the Monte Carlo modeling and the likelihood function
discriminant technique is given by constructing discriminants that
treat each background contribution separately as a signal.  These discriminants then
can be used to check the modeling of the rates and distributions of the likelihood function
outputs for each background in turn by purifying samples of the targeted backgrounds and separating them
from the other components.
The same procedure of Equation~\ref{eq:lfdef}
is followed, except $k=2,$ 3, 4, or 5, corresponding to the $Wb{\bar{b}}$,
$t{\bar{t}}$, $Wc{\bar{c}}/Wc$, and the $W+$LF samples, respectively, changing only
the numerator of Equation~\ref{eq:lfdef}.  Each of these discriminants acts in the same
way as the signal discriminant, but instead it separates one category of background from
the other categories and also from the signals.  Distributions of 
$L_{W+{\rm{bottom}}}$,
$L_{t{\bar{t}}}$, 
$L_{W+{\rm{charm}}}$, 
and $L_{W+{\rm{LF}}}$ are shown in Fig.~\ref{fig:lfbg} for $b$-tagged $W$+2~jet events
passing our event selection.  The modeling of the rates and shapes of these distributions
gives us confidence that the individual background rates are well predicted and that
the input variables to the likelihood function are well modeled for the main
background processes, specifically in the way that they are used for the signal discriminant.

\begin{figure}
\end{figure}

\subsection{\label{sec:matrixelement} Matrix Element Method}

The matrix element (ME) method relies on the evaluation of event
probabilities for signal and background processes based on calculations
of the relevant SM differential cross sections. These probabilities are
calculated on an event-by-event basis for the signal and background
hypotheses and quantify how likely it is for 
the event to have originated from a given signal or
background process.  Rather than combine many complicated variables,
the matrix element method uses only the measured energy-momentum four-vectors
of each particle to perform its calculation.  The mechanics of the method
as it is used here are described below.  Further
information about this method can be found in \cite{Dong:2008zzc}.

\subsubsection{Event Probability}

If we could measure the four-vectors of the initial and final state
particles very precisely, the event probability for a specific process would be
\[P_{evt}\sim \frac{d\sigma}{\sigma}, \]
where the differential cross-section is given by \cite{Amsler:2008zz} and

\begin{equation}
\label{eqn:dsigma}
d\sigma=\frac{(2\pi)^4|{\cal M}|^2}{4\sqrt{(q_1\cdot q_2)^2 - m_{q_1}^2
m_{q_2}^2}}~d\Phi_n(q_1+q_2;p_1,..,p_n)
\end{equation}

\noindent where ${\cal M}$ is the Lorentz-invariant matrix element for the process
under consideration; $q_1$, $q_2$ and
$m_{q_1}$, $m_{q_2}$ are the four momenta and masses of the incident
particles; and $d\Phi_n$ is the $n$-body phase space given by
\cite{Amsler:2008zz}:

\begin{widetext}
\begin{equation}
\label{eqn:nphi}
d\Phi_n(q_1+q_2;p_1,..,p_n)=\delta^4\left(q_1+q_2-\sum_{i=1}^n p_i\right)\prod_{i=1}^n \frac{d^3p_i}{(2\pi)^32E_i}.
\end{equation}
\end{widetext}

However, several effects have to be considered: (1) the partons in the
initial state cannot be measured, (2) neutrinos in the final state are
not measured directly, and (3) the energy resolution of the detector
cannot be ignored. To address the first point, the differential cross
section is weighted by parton distribution functions. To address the
second and third points, we integrate over all particle momenta which
we do not measure (the momentum of the neutrino), or do not measure well, due
to resolution effects (the jet energies). The integration gives a weighted
sum over all possible parton-level variables $y$ leading to the observed
set of variables $x$ measured with the CDF detector. The mapping
between the particle variables $y$ and the measured variables $x$ is
established with the transfer function $W(y,x)$, which encodes the detector
resolution and is described in Section~\ref{sec:transferfunction}. Thus, the event
probability takes the form

\begin{widetext}
\begin{equation}
\label{eqn:evtprob}
P(x)=\frac{1}{\sigma}\int d\sigma(y)dq_1dq_2f\left(\left|q_1^z/p_{\rm{beam}}\right|\right)
f\left(\left|q_2^z/p_{\rm{beam}}\right|\right)W(y,x),
\end{equation}
\end{widetext}

\noindent where $d\sigma(y)$ is the differential cross section in 
terms of the particle variables; $f\left(q^z_i/p_{\rm{beam}}\right)$ are the PDFs, which are
functions of the fraction of the proton momentum $p_{\rm{beam}}$ carried by quark $i$.  The initial
quark momentum is assumed to be in the direction of the beam axis for purposes of this calculation.
Substituting Equations~\ref{eqn:dsigma} and \ref{eqn:nphi}
into Equation~\ref{eqn:evtprob} transforms the  event probability to
\begin{widetext}
\begin{equation}
\label{eqn:evtprob2}
P(x)=\frac{1}{\sigma}\int {2\pi^4}|{\cal{M}}|^2\frac{f\left(E_{q_1}/E_{\rm{beam}}\right)}{E_{q_1}}
\frac{f\left(E_{q_2}/E_{\rm{beam}}\right)}{E_{q_2}}W(y,x)d\Phi_4 dE_{q_1}dE_{q_2},
\end{equation}
\end{widetext}
where we have used the approximation
$\sqrt{(q_1\cdot q_2)^2 - m_{q_1}^2 m_{q_2}^2}\simeq 2E_{q_1}E_{q_2}$, 
neglecting the masses and transverse momenta of the initial partons.

We calculate the squared matrix element $|{\cal M}|^2$ for the event probability at
LO by using the {\sc helas} (HELicity
Amplitude Subroutines for Feynman Diagram Evaluations)
package~\cite{Murayama:1992}. The correct subroutine calls for a given process
are automatically generated by {\sc madgraph}~\cite{Maltoni:2002qb}. We calculate
event probabilities for all significant signal and background
processes that can be easily modeled to first order: $s$-channel and
$t$-channel single top quark production as well as the $Wb\bar{b}$,
$Wcg$, $Wgg$ (shown in Fig.~\ref{fig:Wjets}) and $t\bar{t}$
(Fig.~\ref{fig:ttbar}) processes.  The $Wcg$ and $Wgg$ processes are
only calculated for two-jet events because they have very little
contribution to three-jet background.

The matrix elements correspond to fixed-order tree-level calculations and thus are not
perfect representations of the probabilities for each process.  Since the integrated
matrix elements are not interpreted as probabilities but instead are used to form
functions that separate signal events from background events, the choice of the matrix
element calculation affects the sensitivity of the analysis but not its accuracy.
The fully simulated Monte Carlo
uses parton showers to approximate higher-order effects on kinematic distributions, and systematic
uncertainties are applied to the Monte Carlo modeling in this analysis in the same way as for
the other analyses.

While the matrix-element analysis does not directly use input variables that
are designed to separate signals from backgrounds based on specific kinematic properties
such as $M_{\ell\nu b}$, the information carried by these reconstructed variables is represented in
the matrix element probabilities.  For $M_{\ell\nu b}$ in particular, the pole in the
top quark propagator in ${\cal M}$ provides sensitivity to this reconstructed
quantity.  While the other multivariate analyses use the best-fit kinematics corresponding
to the measured quantities on each event, the matrix element analysis, by integrating
over the unknown parton momenta, extracts more information, also using the measurement
uncertainties.

\subsubsection{Transfer Functions}
\label{sec:transferfunction}

The transfer function, $W(y,x)$, is the probability of measuring the
set of observable variables $x$ given specific values of the parton
variables $y$.  In the case of well-measured quantities, $W(y,x)$ is taken as a
$\delta$-function (i.e. the measured momenta are used in the
differential cross section calculation). When the detector resolution
cannot be ignored, $W(y,x)$ is a parameterized resolution function
based on fully simulated Monte Carlo events.
For unmeasured quantities, such as the three components of the
momentum of the neutrino, the transfer
function is constant. Including a transfer function between the
neutrino's transverse momentum and $\EtMissVec$ would double-count the
transverse momentum sum constraint.  The choice of transfer function affects the
sensitivity of the analysis but not its accuracy, since the same
transfer function is applied to both the data and the Monte Carlo samples.

The energies of charged leptons are relatively well measured with the CDF detector and we
assume $\delta$-functions for their transfer functions. The angular resolution of the
calorimeter and the muon chambers is also good and we assume $\delta$-functions for the
transfer functions of the charged lepton and jet directions. The resolution of
jet energies, however, is broad and it is
described by a transfer function $W_{\rm jet}(E_{\rm parton},E_{\rm jet})$. 

The jet energy transfer functions map parton energies to measured jet
energies after correction for instrumental detector
effects~\cite{Bhatti:2005ai}. This mapping includes effects of radiation,
hadronization, measurement resolution, and energy outside the jet cone
not included in the reconstruction algorithm. The jet transfer
functions are obtained by parameterizing the jet response in fully
simulated Monte Carlo events. We parameterize the 
distribution of the difference between the parton and jet energies
as a sum of two Gaussian functions: one to account for
the sharp peak and one to account for the asymmetric tail. We
determine the parameters of the $W_{\rm jet}(E_{\rm parton},E_{\rm jet})$ by
performing a maximum likelihood fit to jets in events passing
the selection requirements.  The jets are required to be aligned within
a cone of $\Delta R<0.4$ with a quark or a gluon coming from the hard
scattering process.

We create three transfer functions: one for $b$ jets, which is
constructed from the $b$ quark from top quark decay in $s$-channel
single top quark events;
one for light jets, which is constructed from the light quark in
$t$-channel single top quark events; and one for gluons, which is
constructed from the radiated gluon in $Wcg$ events.  In each process,
the appropriate transfer function is used for each final-state parton.

\subsubsection{Integration}

To account for poorly measured variables, the differential cross
section must be integrated over all variables --- 14 variables for
two-jet events, corresponding to the momentum vectors of the four
final-state particles (12 variables) and the longitudinal momenta of
the initial state partons (2 variables). There are 11 delta functions
inside the integrals: four for total energy and momentum conservation
and seven in the transfer functions (three for the charged lepton's momentum
vector and four for the jet angles). The calculation of the event
probability therefore involves a three-dimensional integration. The
integration is performed numerically over the energies of the two quarks
and the longitudinal momentum of the neutrino ($p_{z}^{\nu}$).  For
three-jet events, the additional jet adds one more dimension to the
integral.

Because it is not possible to tell which parton resulted in a given
jet, we try all possible parton combinations, using the $b$-tagging
information when possible.  These probabilities are then added
together to create the final event probability.

Careful consideration must be given to $t\bar{t}$ events falling into the
$W+2$~jet and $W+3$~jet samples because these events have
final-state particles that are not observed.  In two-jet events, these missing
particles could be a charged lepton and a neutrino (in the case of 
$t\bar{t}\rightarrow\ell^+\nu_\ell\ell^{\prime-}{\bar{\nu}}_{\ell^\prime}b\bar{b}$ decays)
or two quarks (in the case of 
$t\bar{t}\rightarrow \ell^+\nu_\ell q\bar{q}^\prime b\bar{b} $ decays), 
and since both of these are decay products of a $W$ boson, we treat this matrix
element in either case as having a final-state $W$ boson that is missed in the
detector.  The particle assignment is not always correct, but the
purpose of the calculation is to construct variables that have maximal separation
power between signal and background events, and not that they produce a correct
assignment of particles in each event.  The choice of which particles are assumed to
have been missed is an issue of the optimization of the analysis and not of 
the validity of the result.
We integrate over the three components of the hypothetical
missing $W$ boson's momentum, resulting in a six-dimensional integral.  In the three-jet
case, we integrate over the momenta of one of the quarks from the $W$ boson decay.

The numerical integration for the simpler two-jet $s$- and $t$-channel
and $Wb\bar{b}$ diagrams is performed using an adaptation of the
CERNLIB routine {\sc radmul}~\cite{Genz:1980}.  This is a
deterministic adaptive quadrature method that performs well for
smaller integrations.  For the higher-dimensional integrations needed
for the three-jet and $t\bar{t}$ matrix elements, a faster integrator
is needed.  We use the {\sc divonne} algorithm implemented in the {\sc cuba}
library~\cite{Hahn:2004fe}, which uses a Monte-Carlo-based technique of
stratified sampling over quasi-random numbers to produce its answer.

\subsubsection{Event Probability Discriminant}
\label{sec:epd}

Event probabilities for all processes are calculated for each event
for both data events and Monte Carlo simulated events.  For each
event, we use the event probabilities as ingredients to build an event
probability discriminant ($EPD$), a variable for which the distributions
of signal events and background events are as different as possible.
Motivated by the Neyman-Pearson
lemma~\cite{Neyman:1933}, which states that a likelihood ratio is the
most sensitive variable for separating hypotheses, we define the $EPD$
to be $EPD=P_{\rm s}/(P_{\rm s}+P_{\rm b})$, where $P_{\rm s}$ and $P_{\rm b}$
are estimates of the
 signal and background probabilities, respectively.   This discriminant is close to zero if
$P_{\rm b} \gg P_{\rm s}$ and close to unity if $P_{\rm s} \gg P_{\rm b}$.  
There are four $EPD$
functions in all, for $W$+two- or three-jet events with one or two
$b$~tags.

\begin{figure*}
\begin{center}
\subfigure[]{
\includegraphics[width=0.65\columnwidth]{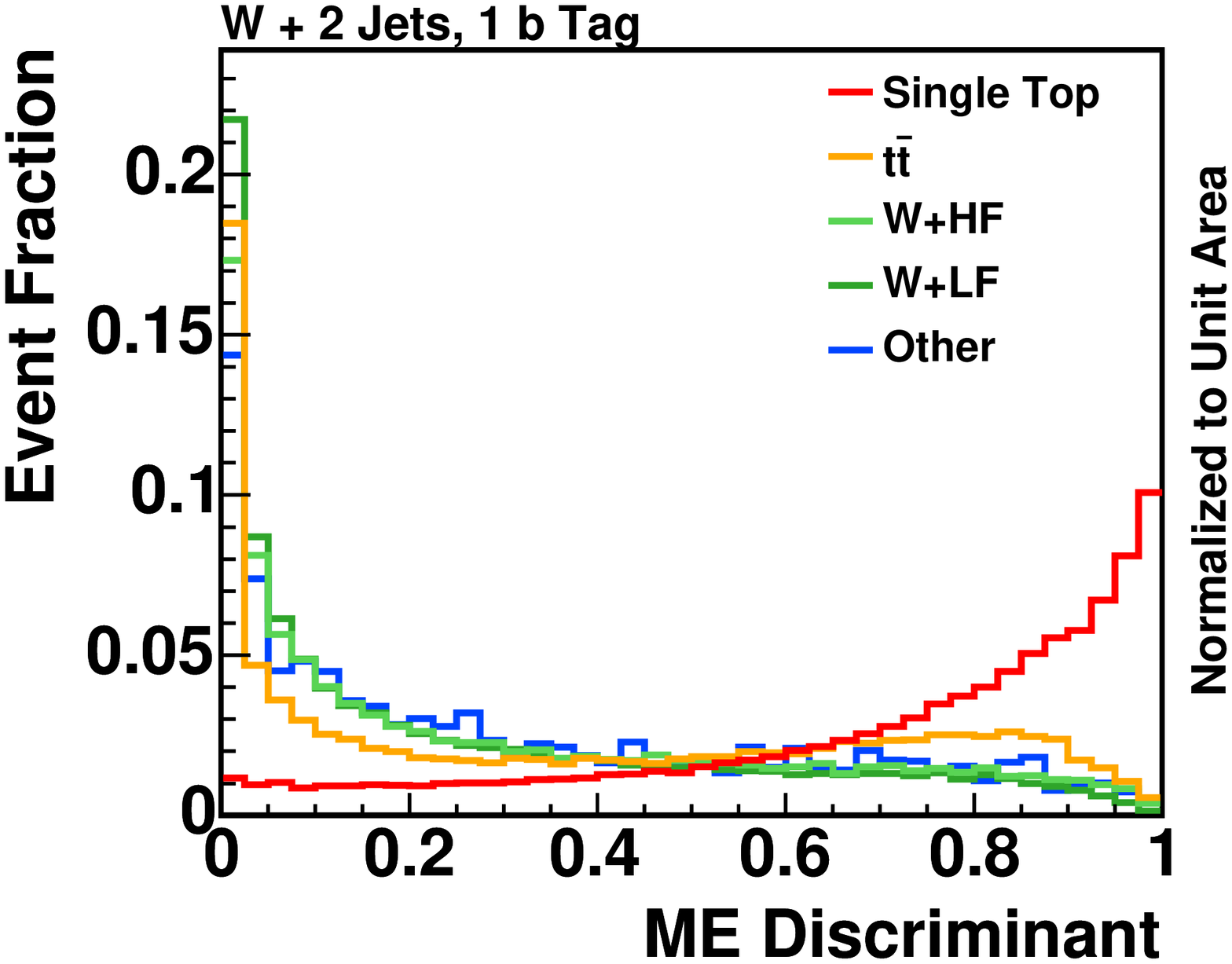}
\label{fig:EPD2j1t_shape}}
\subfigure[]{
\includegraphics[width=0.65\columnwidth]{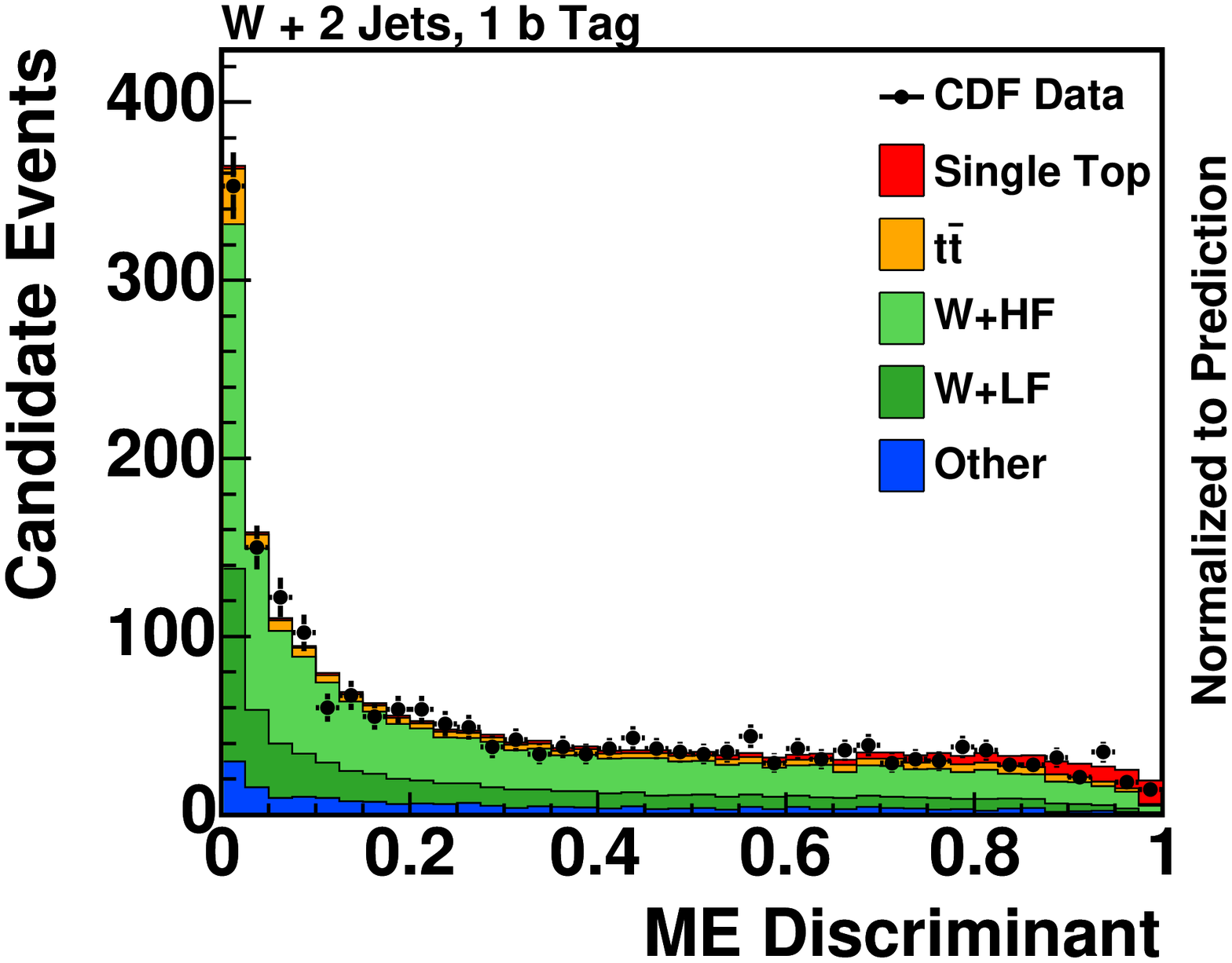}
\label{fig:EPD2j1t}} \\
\subfigure[]{
\includegraphics[width=0.65\columnwidth]{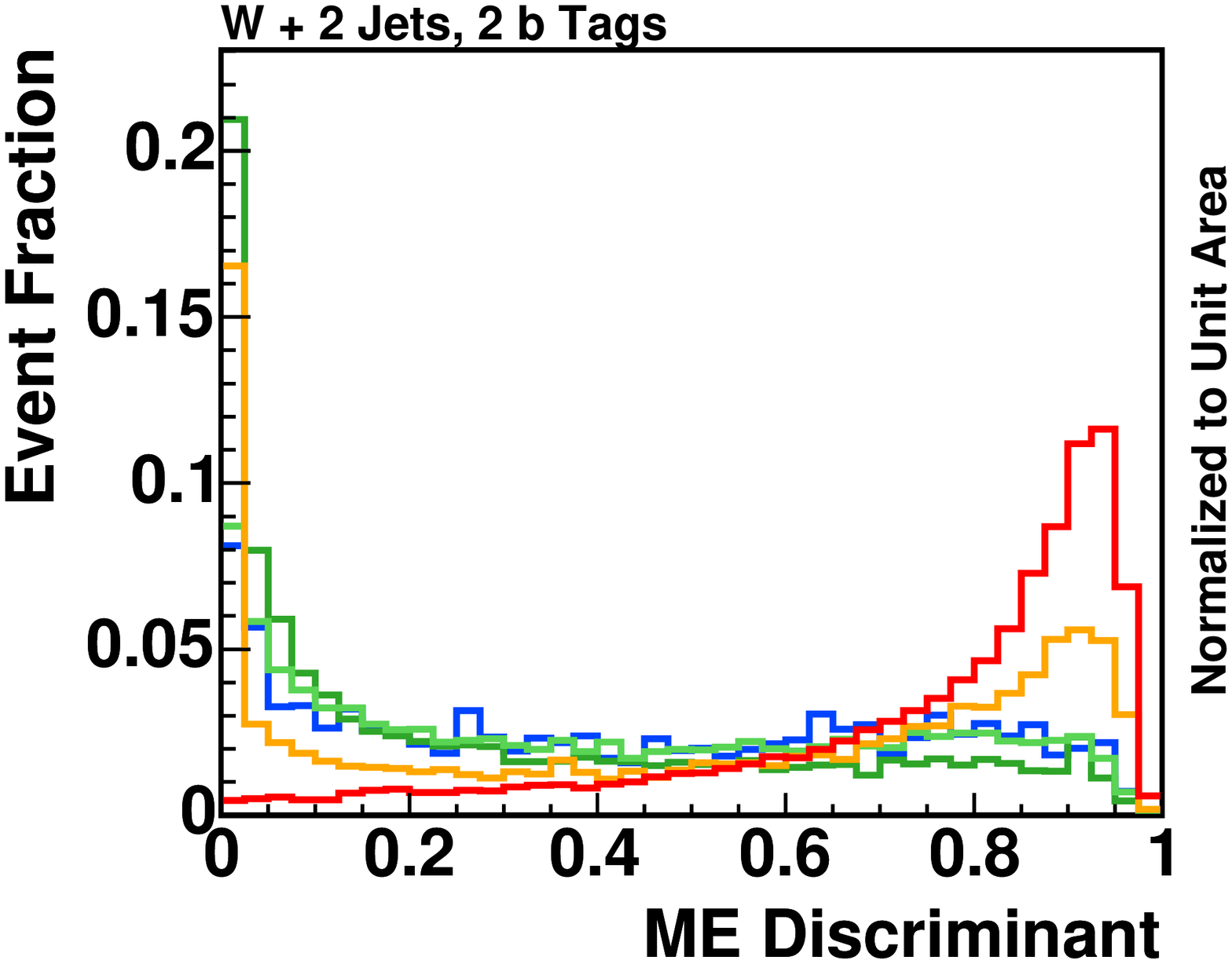}
\label{fig:EPD2j2t_shape}}
\subfigure[]{
\includegraphics[width=0.65\columnwidth]{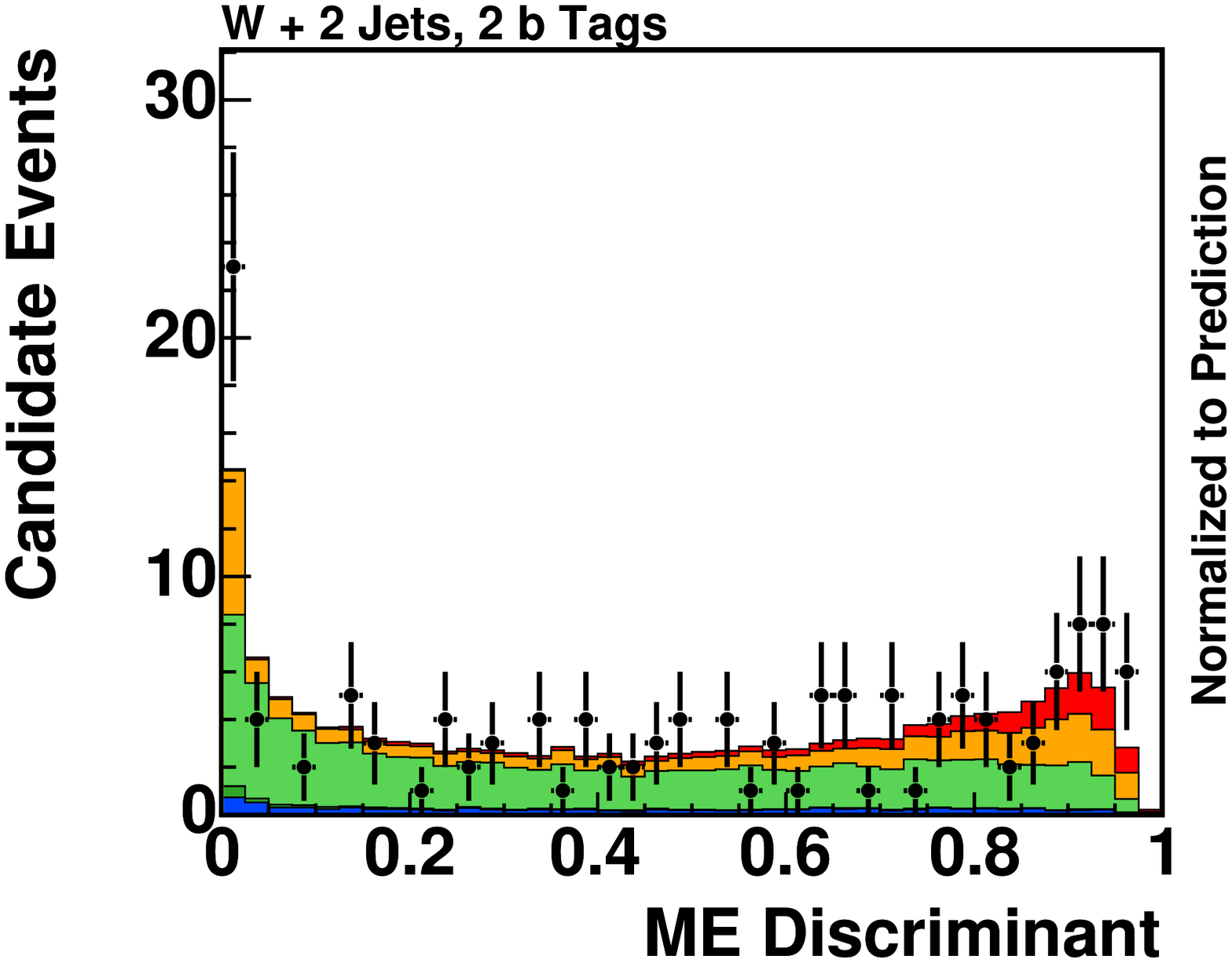}
\label{fig:EPD2j2t}} \\
\subfigure[]{
\includegraphics[width=0.65\columnwidth]{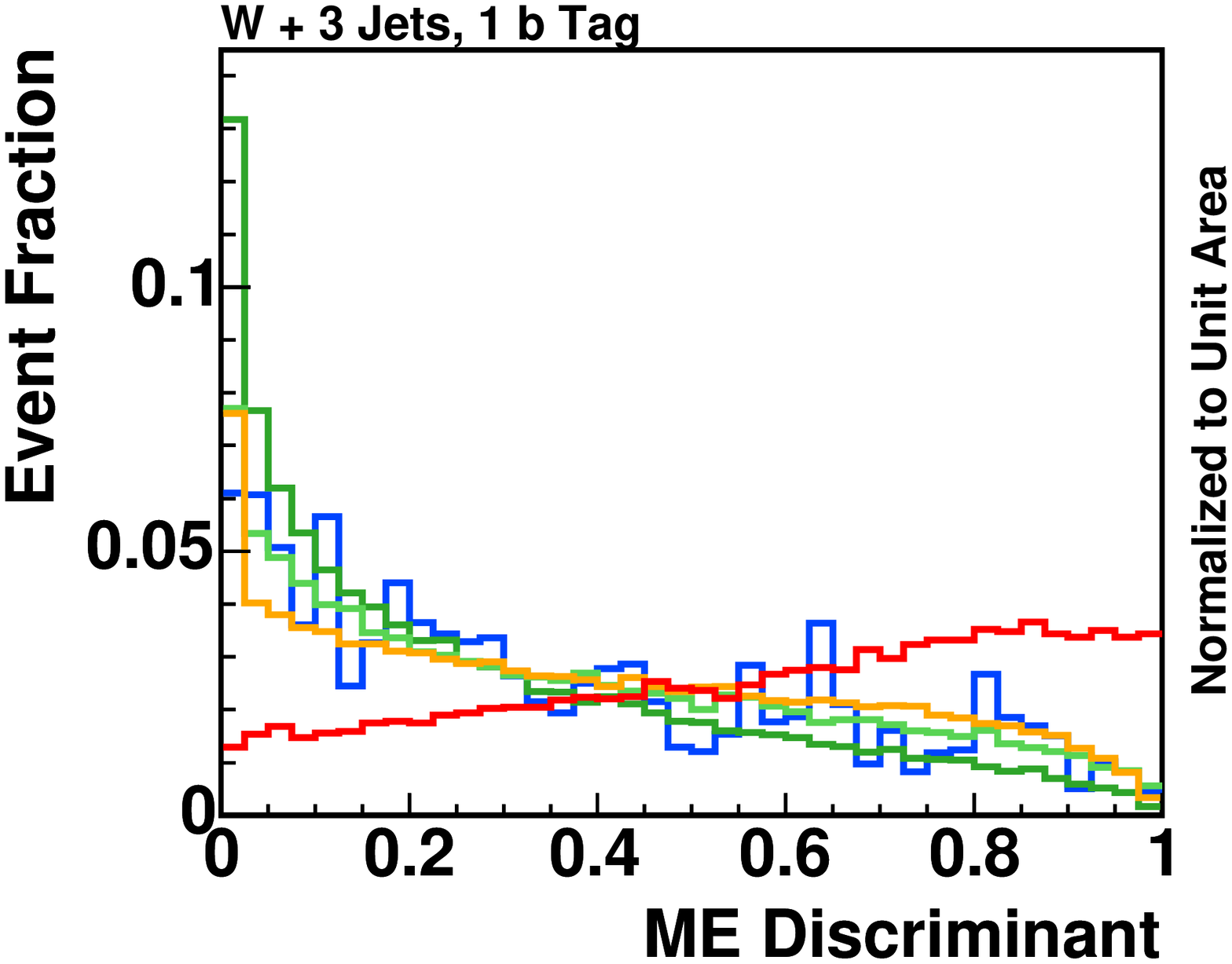}
\label{fig:EPD3j1t_shape}}
\subfigure[]{
\includegraphics[width=0.65\columnwidth]{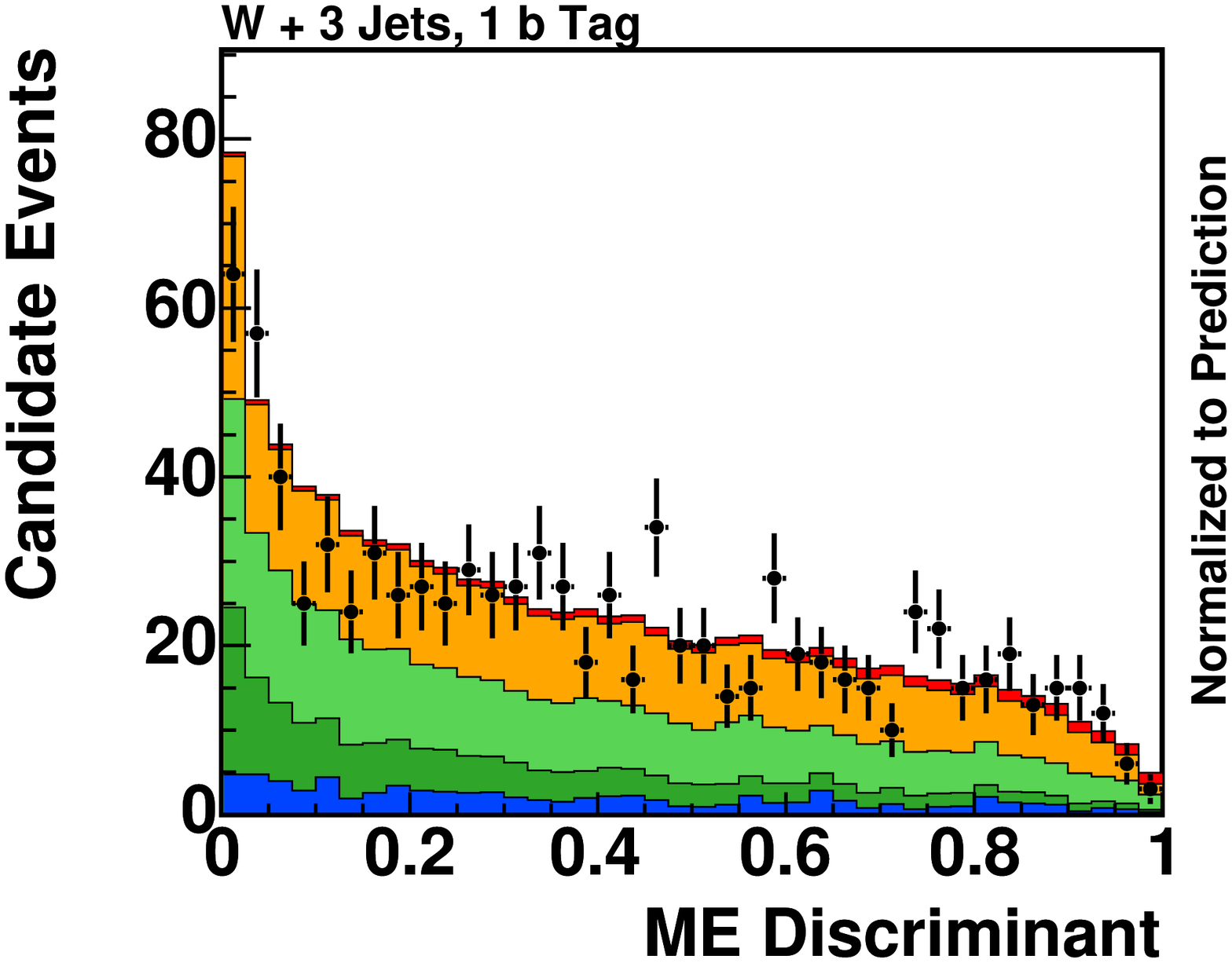}
\label{fig:EPD3j1t}} \\
\subfigure[]{
\includegraphics[width=0.65\columnwidth]{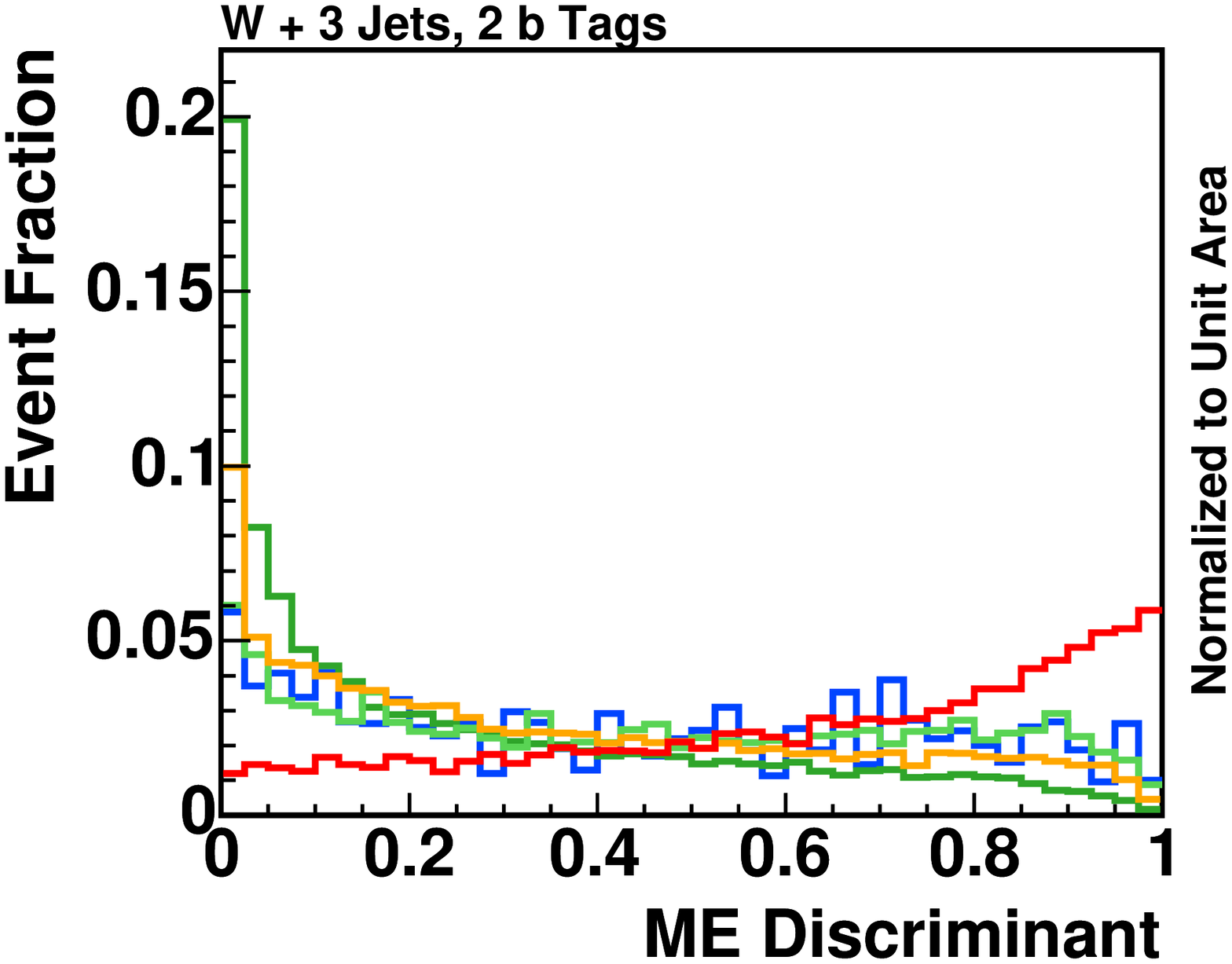}
\label{fig:EPD3j2t_shape}}
\subfigure[]{
\includegraphics[width=0.65\columnwidth]{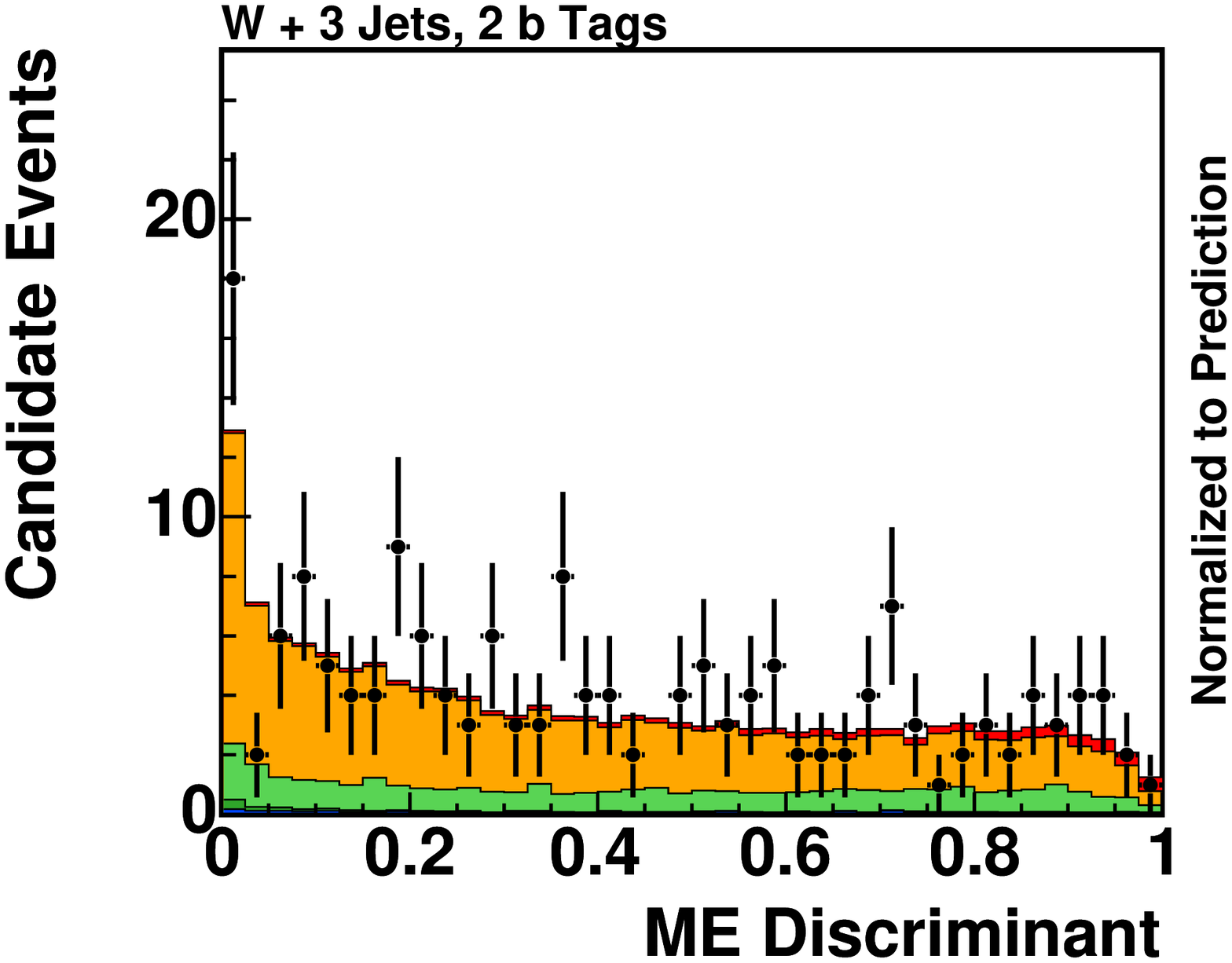}
\label{fig:EPD3j2t}}
\end{center}
\caption{\label{fig:EPD}
Templates of predictions for the signal
and background processes, each scaled to unit area (left) and comparisons of
the data with the sum of the predictions (right)
of the ME~discriminant $EPD$ for each selected data sample. Single top quark events
are predominantly found on the right-hand sides of the histograms while
background events are mostly found on the left-hand sides.
The data are indicated by points with error bars, and 
the predictions are shown stacked, with the stacking order following that of the legend.
}
\end{figure*}

Several background processes in this analysis have no $b$ jet in the
final state, and the matrix element probabilities do not
include detector-level discrimination between $b$~jets and non-$b$~jets.
In order to include this extra information,
we define the $b$-jet probability as
$b=(b_{\mathrm{NN}}+1)/2$ and use it to weight each matrix element
probability by the $b$ flavor probability of its jets. Since single top quark production always has a $b$
quark in the final state, we write the
event-probability-discriminant as:
\begin{widetext}
\begin{equation}
\label{eqn:epd_full}
EPD=\frac{b \cdot P_{\rm s}}{b \cdot \left( P_{\rm s} +
P_{Wb\bar{b}} + P_{t\bar{t}} \right) + (1-b) \cdot \left( P_{Wc\bar{c}} +
P_{Wcg} + P_{Wgg}\right)}
\end{equation}
\end{widetext}
where $P_{\rm s} =P_{s-{\rm channel}}+P_{t-{\rm channel}}$.  
Each probability
is multiplied by an arbitrary normalization factor, which is chosen
to maximize the expected sensitivity.  Different
values are chosen in each $b$-tag and jet category in order to
maximize the sensitivity separately in each.
The resulting templates and distributions are shown for all four $EPD$ functions
in their respective selected data samples in Fig.~\ref{fig:EPD}.  All of them provide good
separation between single top quark events and background events.
The sums of predictions normalized to our signal and
background models, which are described in Sections~\ref{sec:Background}
and~\ref{sec:SignalModel}, respectively,
 are compared with the data.  
Figure~\ref{fig:allME}\subref{fig:allME_chan} corresponds to the sum of all 
four $b$-tag and jet categories.

\subsubsection{Validation}

\begin{figure*}
\begin{center}
\subfigure[]{
\includegraphics[width=0.8\columnwidth]{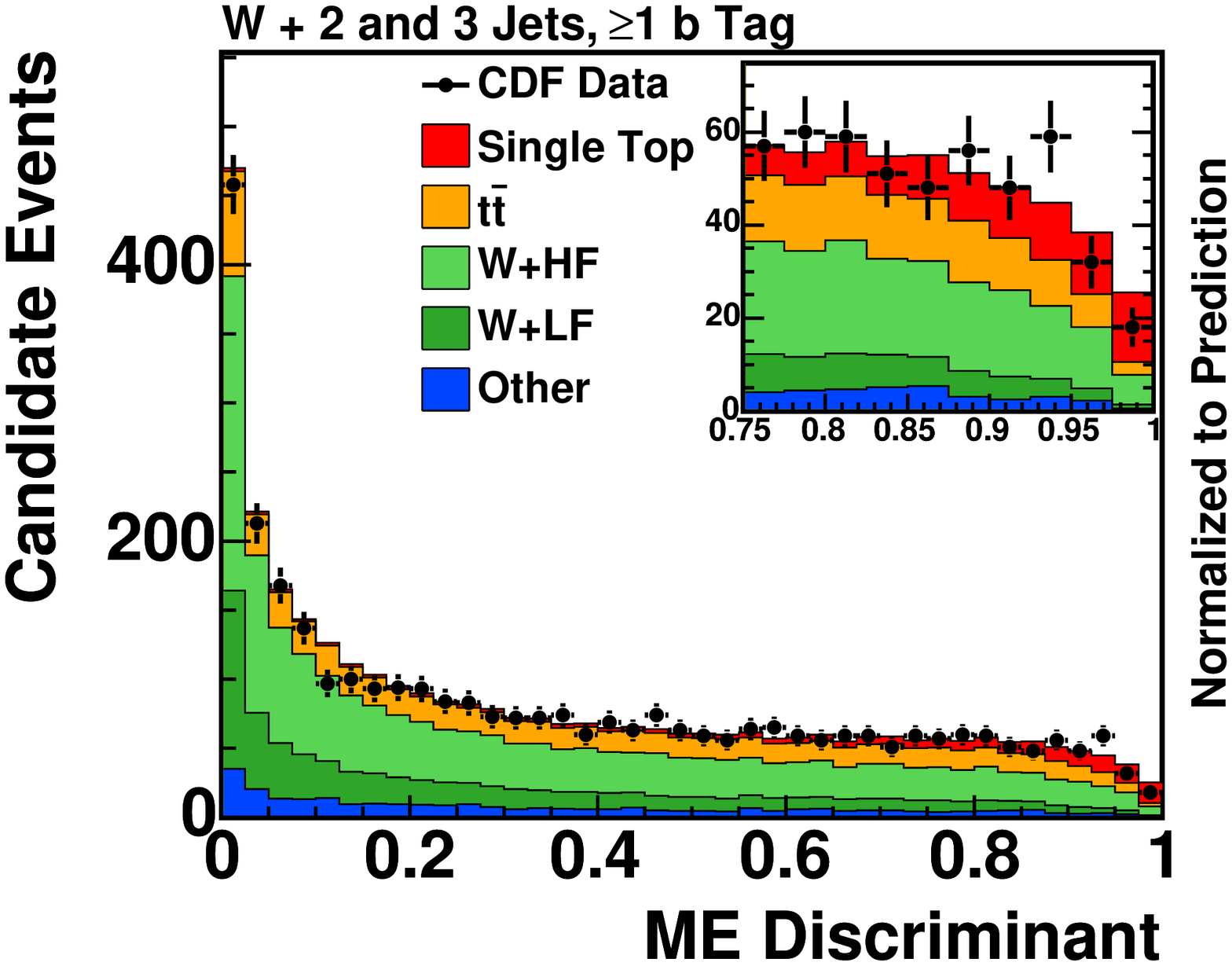}
\label{fig:allME_chan}}
\subfigure[]{
\includegraphics[width=0.8\columnwidth]{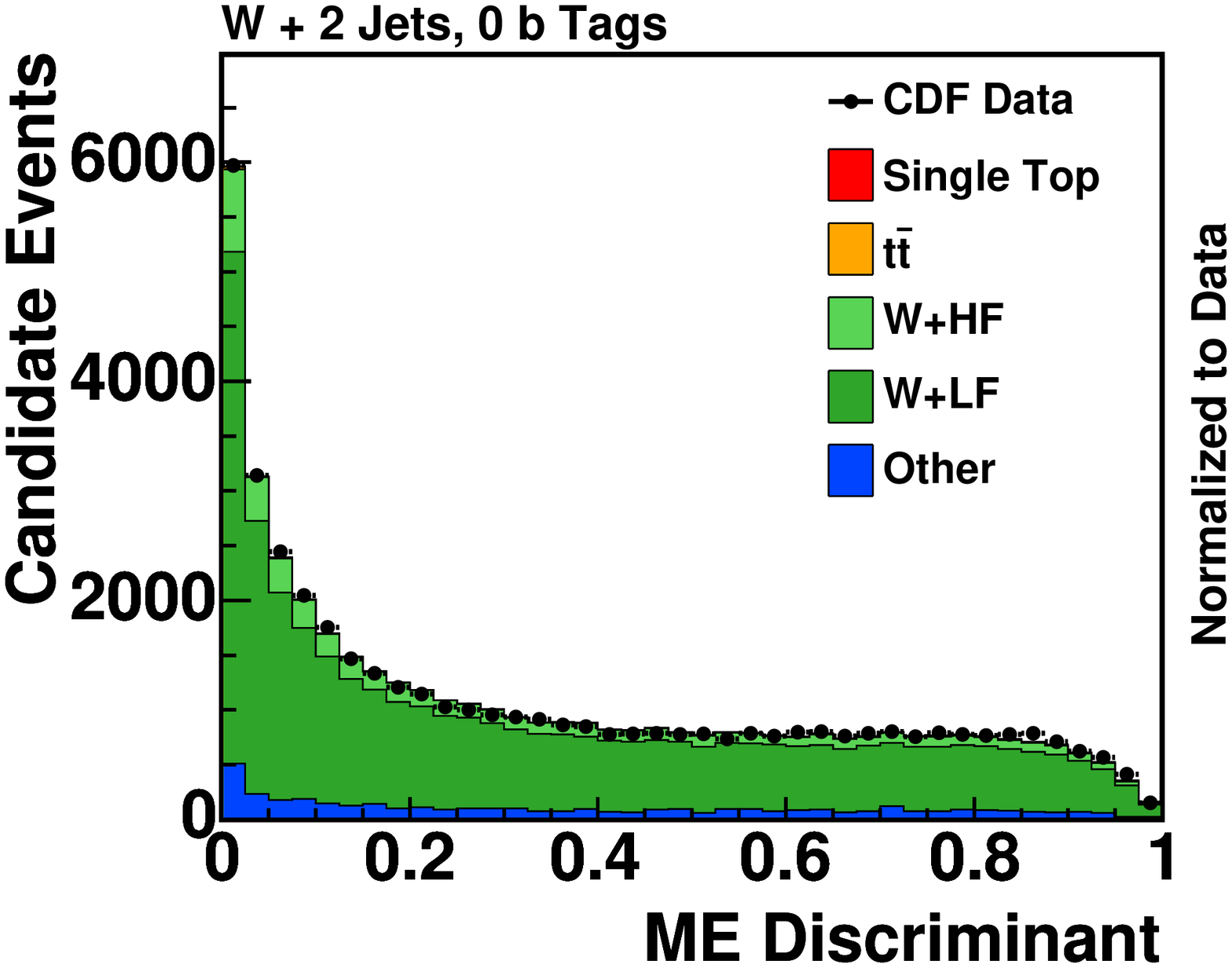}
\label{fig:allME_0tag}}
\end{center}
\caption{\label{fig:allME} Comparison of the data with the sum of the predictions 
of the matrix element discriminant for the sum of all selected data samples (left). The discriminant output 
for two-jet one-$b$-tag events applied to the untagged $W$+two jets control sample (right) shows 
that the Monte Carlo $W$+two jets samples model the ME distribution of the data well.
The data are indicated by points with error bars, and 
the predictions are shown stacked, with the stacking order following that of the legend.
}
\end{figure*}

We validate the performance of the Monte Carlo to predict
the distribution of each $EPD$ by checking the untagged
$W+$jets control samples, setting $b_{\mathrm{NN}}=0.5$ so
that it does not affect the $EPD$.  An example is shown in
Fig.~\ref{fig:allME}\subref{fig:allME_0tag} for $W$+two-jet events.  
The agreement in this control sample gives
us confidence that the information used in this analysis is well
modeled by the Monte Carlo simulation.

\begin{figure}
\begin{center}
\includegraphics[width=0.8\columnwidth]{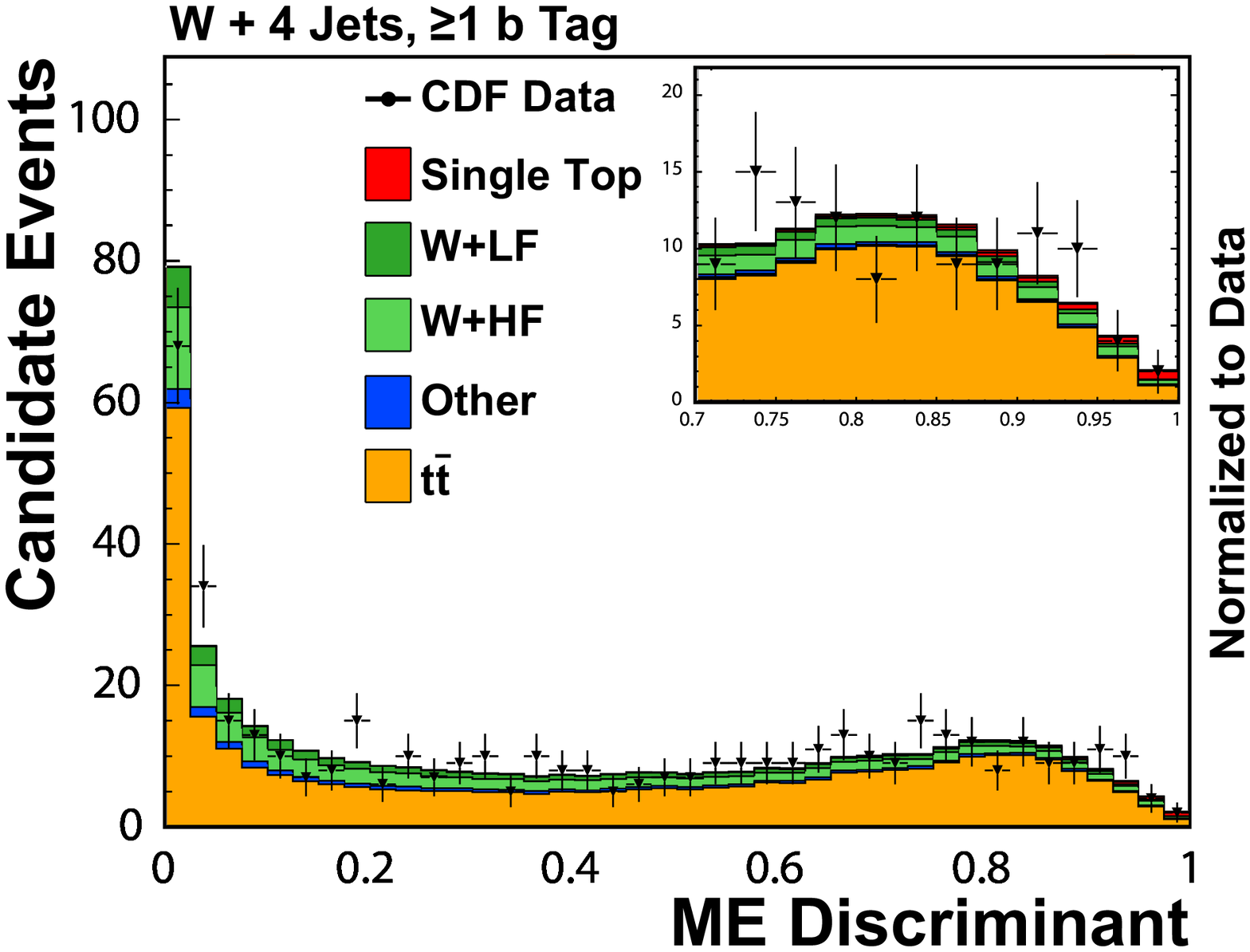}
\end{center}
\caption{\label{fig:ME4jet} The event probability discriminant for two-jet
one-$b$-tag events applied to the $b$-tagged $W$+four jets control sample, showing
that the Monte Carlo \ttbar\ samples model the $EPD$ distribution of the data well.
The data are indicated by points with error bars, and 
the predictions are shown stacked, with the stacking order following that of the legend.
}
\end{figure}

Because the \ttbar\ background is the most signal-like of the 
background contributions in this analysis, the matrix element distribution is specifically
checked in the $b$-tagged four-jet control sample, which is highly enriched
in \ttbar\ events.  Each EPD function is validated in this way, for two or three jets,
and one or two $b$~tags, using the highest-$E_{\rm{T}}$ jets in $W+$four-jet events with
the appropriate number of $b$ tags.  An example is shown in Fig.~\ref{fig:ME4jet}, for the
two-jet one-$b$-tag $EPD$ function.

\subsection{\label{sec:NN} Artificial Neural Network}

A different approach uses artificial neural networks (NN) to
combine sensitive variables to distinguish single top quark signal 
from background events. As with the
neural network flavor separator $b_{\mathrm{NN}}$ described in
Section~\ref{sec:btagger}, the
{\sc NeuroBayes}~\cite{Feindt:2006pm} package is
used to create the neural networks. 
We train a different neural network in each selected data sample -- indexed by
the number of jets, the number of $b$-tagged jets, and whether the charged lepton
candidate is a triggered lepton or an EMC lepton.
For all samples,  the $t$-channel Monte Carlo is used as the
signal training sample except for the two-jet two-$b$-tag
events, in which $s$-channel events are treated as signal. The
background training sample is a mix of Standard Model
processes in the ratios of the estimated yields given in
Tables~\ref{tab:EventYield1tag} and~\ref{tab:EventYield2tag}.

Each training starts with more than fifty variables, but the training
 procedure removes those with no significant discriminating power,
 reducing the number to 11--18 variables.  Each neural network has one
 hidden layer of 15 nodes and one output node.

As in other cases, the transverse momentum of the neutrino is inferred
from the $\EtMiss$\ of the event. The component of the momentum of the neutrino along the beam axis is
calculated from the assumed mass of the $W$ boson and the measured energy and momentum
of the charged 
lepton.  A quadratic equation in $p_z^\nu$ must be solved.  If there is one real solution, we use it.
If there are two real solutions,  we use the one with the smaller $|p_{z}^\nu|$.  
If the two solutions are complex, a kinematic fit which varies the 
transverse components of $\EtMissVec$ is performed to find a solution as close 
as possible to $\EtMissVec$~\cite{Papaikonomou:2009zz} which results in a real $p_z^\nu$.

If only one jet is $b$-tagged,
it is assumed to be from top quark decay.  If there is more 
than one $b$-tagged jet, the jet with the largest
$Q_{\ell} \times \eta$ is chosen. More detailed information about this method can 
be found in~\cite{Lueck:2009zz}.

\subsubsection{Input Variables}

The variables used in each network are summarized in
Table~\ref{tab:NNvars}.  Descriptions of the variables follow.

\begin{table}[h]
\caption{\label{tab:NNvars} Summary of variables used in the
different neural networks in this analysis. An explanation of the
variables is given in the text.}
\begin{center}
\begin{tabular}{lcccc}\hline\hline
 & \multicolumn{2}{c}{2-jet} & \multicolumn{2}{c}{3-jet} \\
Variable & 1-tag & 2-tag & 1-tag & 2-tag \\
\hline

$M_{\ell\nu b}$ & \Checkmark & \Checkmark & \Checkmark & \\
$M_{\ell\nu bb}$ & & \Checkmark & & \Checkmark \\
$M_{\rm T}^{\ell\nu b}$ & \Checkmark & \Checkmark & \Checkmark & \Checkmark \\
$M_{jj}$ & \Checkmark & \Checkmark & \Checkmark & \Checkmark \\
$M_{\rm T}^{W}$ & \Checkmark & \Checkmark & & \\
$E_{\rm T}^{b_{\rm top}}$ & & \Checkmark & \Checkmark & \\
$E_{\rm T}^{b_{\rm other}}$ & & & & \Checkmark\\
$\sum E_{\rm T}^{jj}$ & & & \Checkmark & \Checkmark \\
$E_{\rm T}^{\rm light}$ & \Checkmark & & & \Checkmark \\
$p_{\rm T}^{\ell}$ & \Checkmark & & & \\
$p_{\rm T}^{\ell\nu jj}$ & & & \Checkmark & \Checkmark \\
$H_{\rm T}$ & \Checkmark & & \Checkmark & \\
$\EtMiss$ & & \Checkmark & & \\
$\EtMissSig$ & & & \Checkmark & \\
$\cos \theta_{\ell j}$ & \Checkmark & & \Checkmark & \Checkmark \\
$\cos \theta_{\ell W}^{W}$ & \Checkmark & & & \\
$\cos \theta_{\ell W}^{t}$ & \Checkmark & & & \\
$\cos \theta_{jj}^{t}$ & & \Checkmark & & \Checkmark \\
$Q \times \eta$ & \Checkmark & & \Checkmark & \Checkmark \\
$\eta_{\ell}$ & & \Checkmark & & \\
$\eta_{W}$ & \Checkmark & \Checkmark & & \\
$\sum\eta_{j}$ & \Checkmark & & \Checkmark & \\
$\Delta \eta_{jj}$ & & & \Checkmark & \Checkmark \\
$\Delta \eta_{t,{\rm light}}$ & & & \Checkmark & \\
$\sqrt{\hat{s}}$ & & & & \Checkmark \\
Centrality & & & & \Checkmark \\
Jet flavor separator & \Checkmark & \Checkmark & \Checkmark & \\
\hline\hline
\end{tabular}
\end{center}
\end{table}

\begin{itemize}
\item $M_{\ell\nu b}$: The reconstructed top quark mass.
\item $M_{\ell\nu bb}$: The reconstructed mass of the charged lepton,
the neutrino, and the two $b$-tagged jets in the event.
\item $M_{\rm{T}}^{\ell\nu b}$: The transverse mass of the reconstructed top quark.
\item $M_{jj}$: The invariant mass of the two jets.  In the
three-jet networks, all combinations of jets are included as variables.
\item $M_{\rm T}^{W}$: The transverse mass of the reconstructed $W$
boson.
\item $E_{\rm T}^{b_{\rm top}}$: The transverse energy of the $b$
quark from top decay.
\item $E_{\rm T}^{b_{\rm other}}$: The transverse energy of the $b$
quark not from top decay.
\item $\sum E_{\rm T}^{jj}$: The sum of the transverse energies of the two most
energetic jets.  In the three-jet one-tag network, all combinations of two
jets are used to construct separate $\sum E_{\rm T}^{jj}$ input variables.
\item $E_{\rm T}^{light}$: The transverse energy of the untagged or
lowest-energy jet.
\item $p_{\rm T}^{\ell}$: The transverse momentum of the charged lepton.
\item $p_{\rm T}^{\ell\nu jj}$: The magnitude of the vector sum of the
transverse momentum of the charged lepton,
the neutrino, and all the jets in the event.
\item $H_{\rm T}$: The scalar sum of the transverse energies of the charged lepton, the
neutrino, and all the jets in the event.
\item $\EtMiss$: The missing transverse energy.
\item $\EtMissSig$: The significance of the missing transverse energy $\EtMiss$, as defined in
Equation~\ref{eq:metsig}.
\item $\cos\theta_{\ell j}$:
The cosine of the angle
between the charged lepton and the untagged or lowest-energy jet in the top
quark's reference frame.
\item $\cos \theta_{\ell W}^{W}$: The cosine of the angle between the
charged lepton and the reconstructed $W$ boson in the $W$ boson's reference frame.
\item $\cos \theta_{\ell W}^{t}$: The cosine of the angle between the
charged lepton and the reconstructed $W$ boson in the top quark's reference frame.
\item $\cos \theta_{jj}^{t}$: The cosine of the angle between the two
most energetic jets in the top quark's reference frame.
\item $Q \times \eta$: The charge of the
lepton multiplied by the pseudorapidity of the untagged jet.
\item $\eta_{\ell}$: The pseudorapidity of the charged lepton.
\item $\eta_{W}$: The pseudorapidity of the reconstructed $W$ boson.
\item $\sum\eta_{j}$: The sum of the pseudorapidities of all jets.
\item $\Delta \eta_{jj}$: The difference in pseudorapidity of the two
most energetic jets.  In the three-jet two-tag network, the difference
between the two least energetic jets is also used.
\item $\Delta \eta_{t,\rm light}$: The difference in
pseudorapidity between the untagged or lowest-energy jet and the
reconstructed top quark.
\item $\sqrt{\hat{s}}$: The energy of the center-of-mass system of the hard interaction, defined
as the $\ell\nu b$ system plus the recoiling jet.
\item Centrality: The sum of the transverse energies of the two leading jets
divided by $\sqrt{{\hat{s}}}$.
\item $b_{\mathrm{NN}}$: The jet flavor separator 
neural network output described in Section~\ref{sec:btagger}. 
For two-tag events, the sum of the two outputs is used.
\end{itemize}

\subsubsection{Distributions}

\begin{figure*}
\begin{center}
\subfigure[]{
\includegraphics[width=0.65\columnwidth]{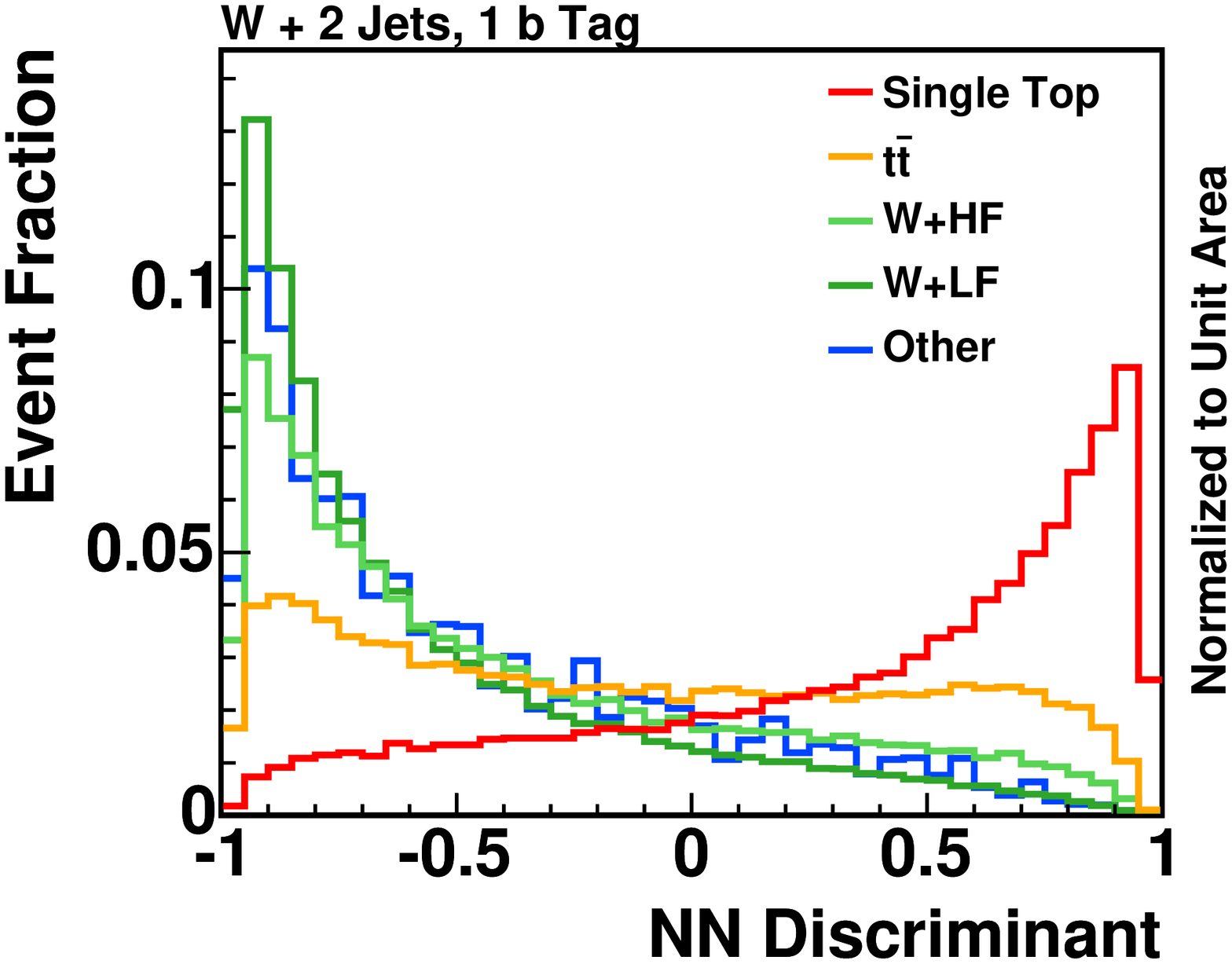}
\label{fig:NN2j1t_shape}}
\subfigure[]{
\includegraphics[width=0.65\columnwidth]{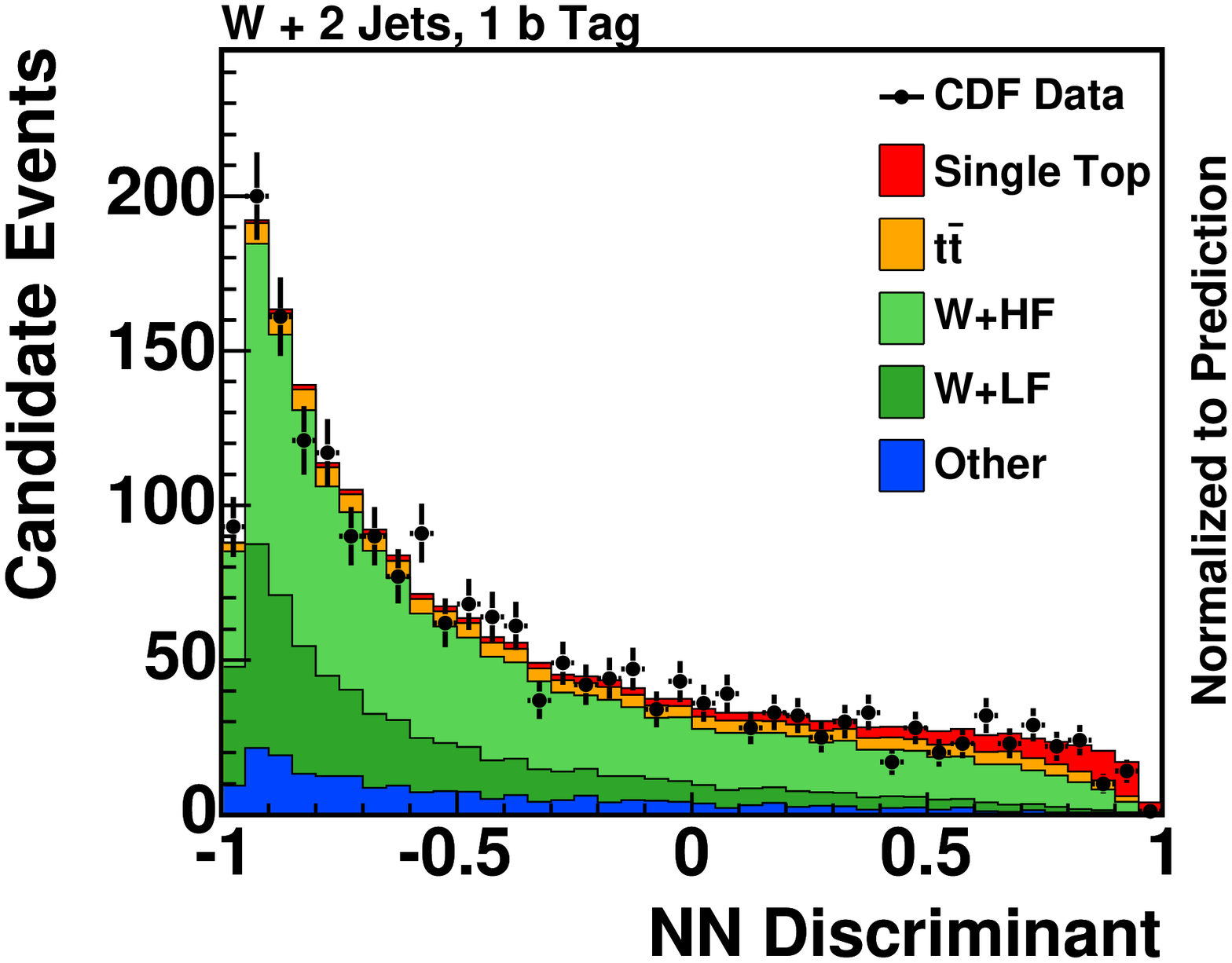}
\label{fig:NN2j1t}} \\
\subfigure[]{
\includegraphics[width=0.65\columnwidth]{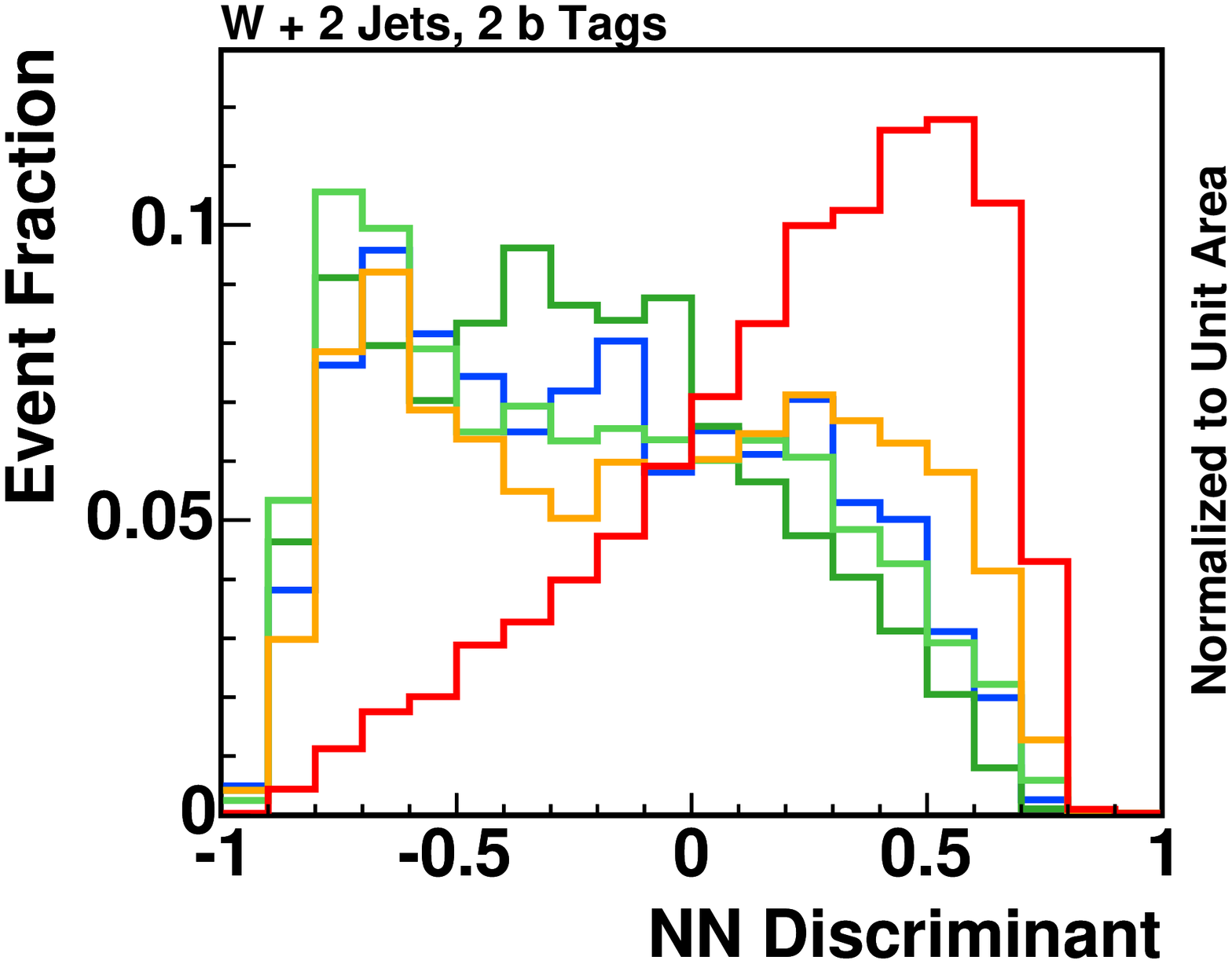}
\label{fig:NN2j2t_shape}}
\subfigure[]{
\includegraphics[width=0.65\columnwidth]{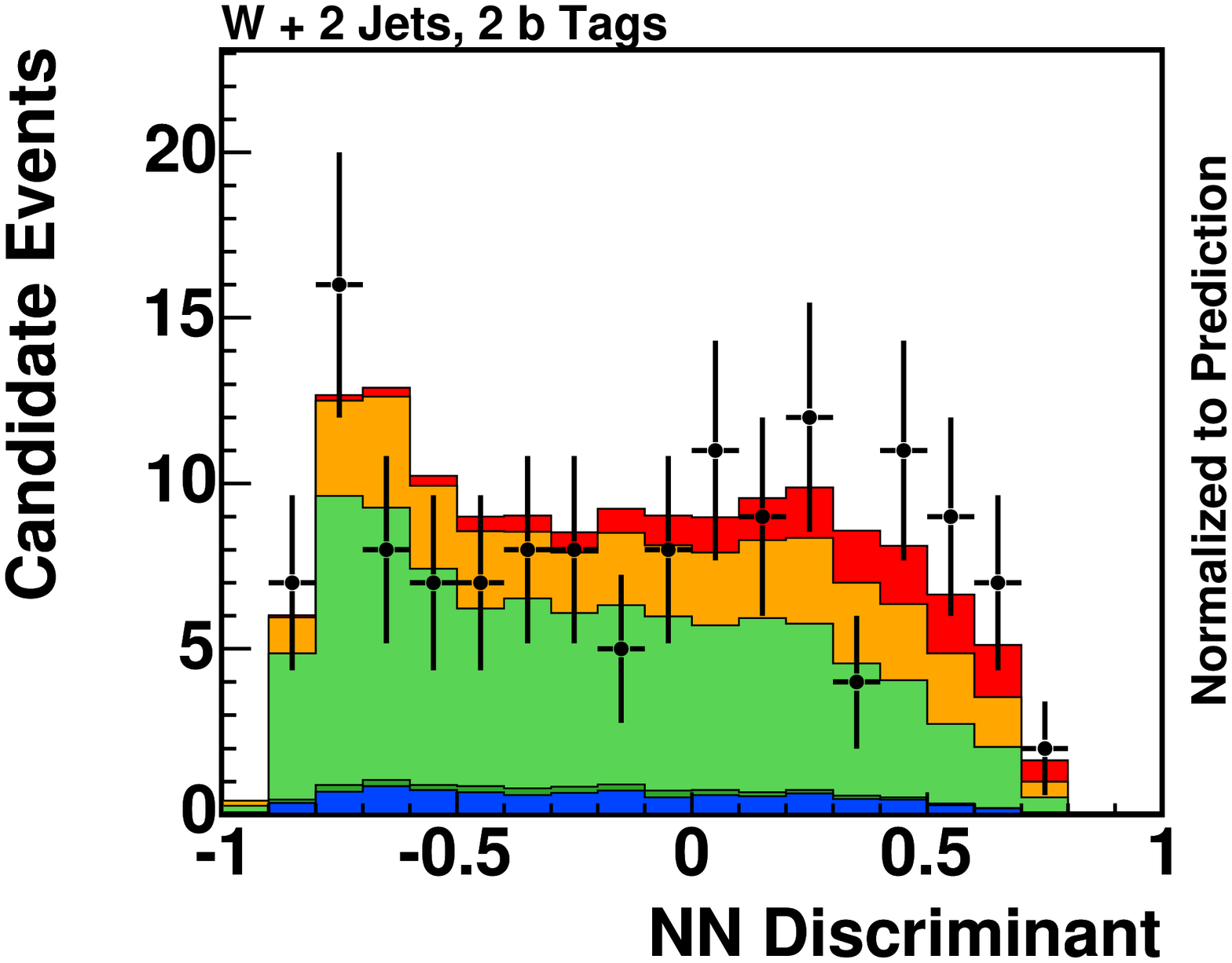}
\label{fig:NN2j2t}} \\
\subfigure[]{
\includegraphics[width=0.65\columnwidth]{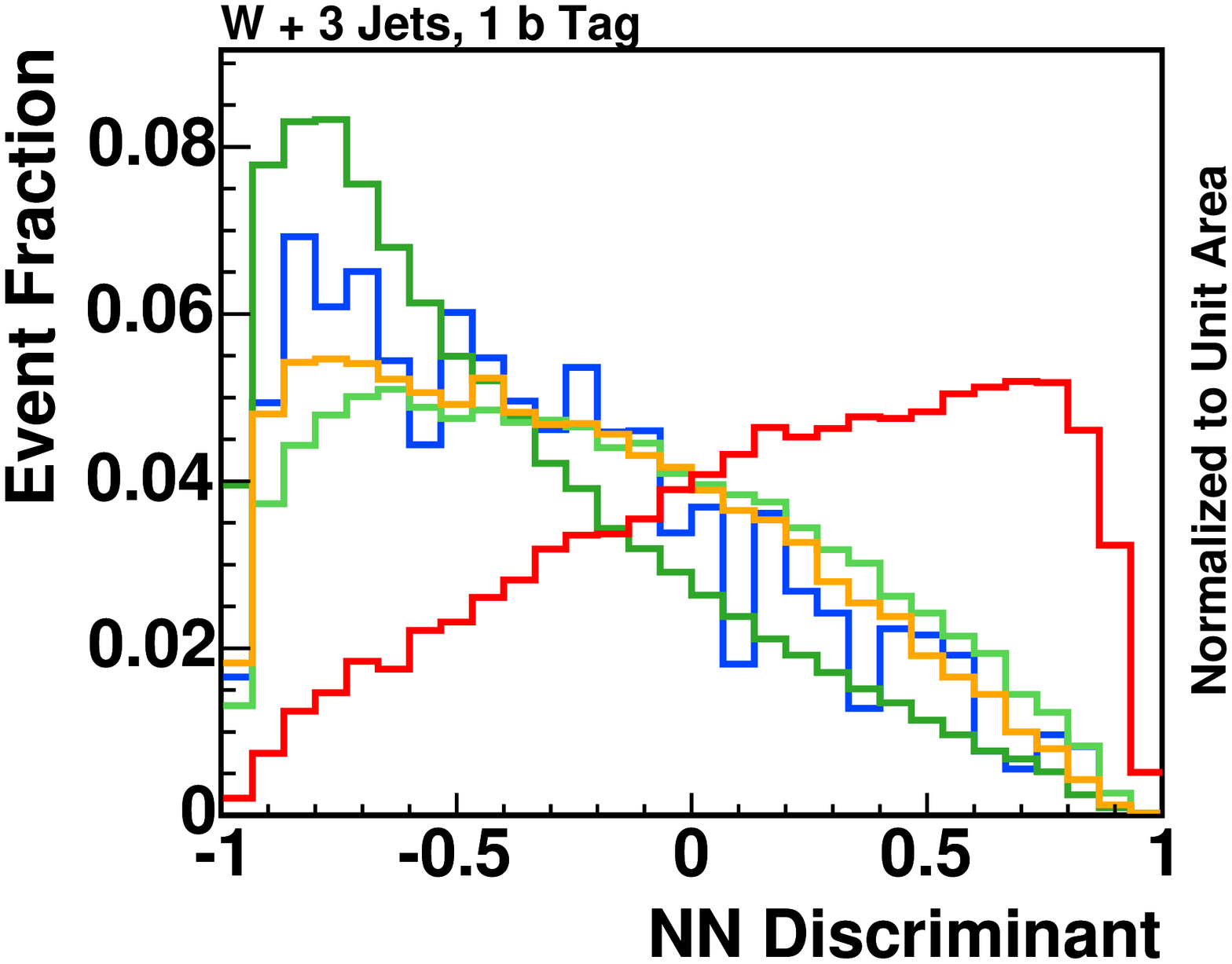}
\label{fig:NN3j1t_shape}}
\subfigure[]{
\includegraphics[width=0.65\columnwidth]{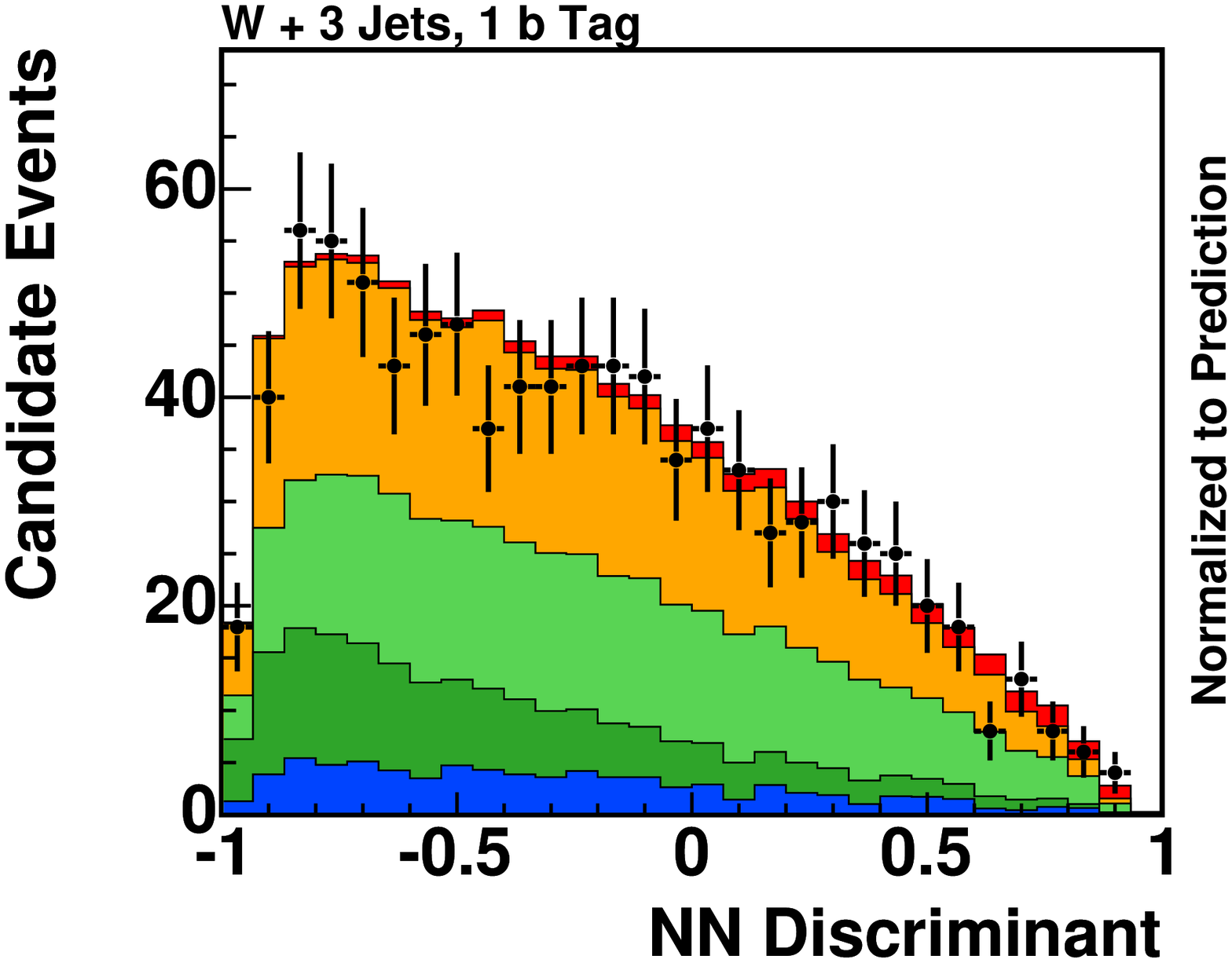}
\label{fig:NN3j1t}} \\
\subfigure[]{
\includegraphics[width=0.65\columnwidth]{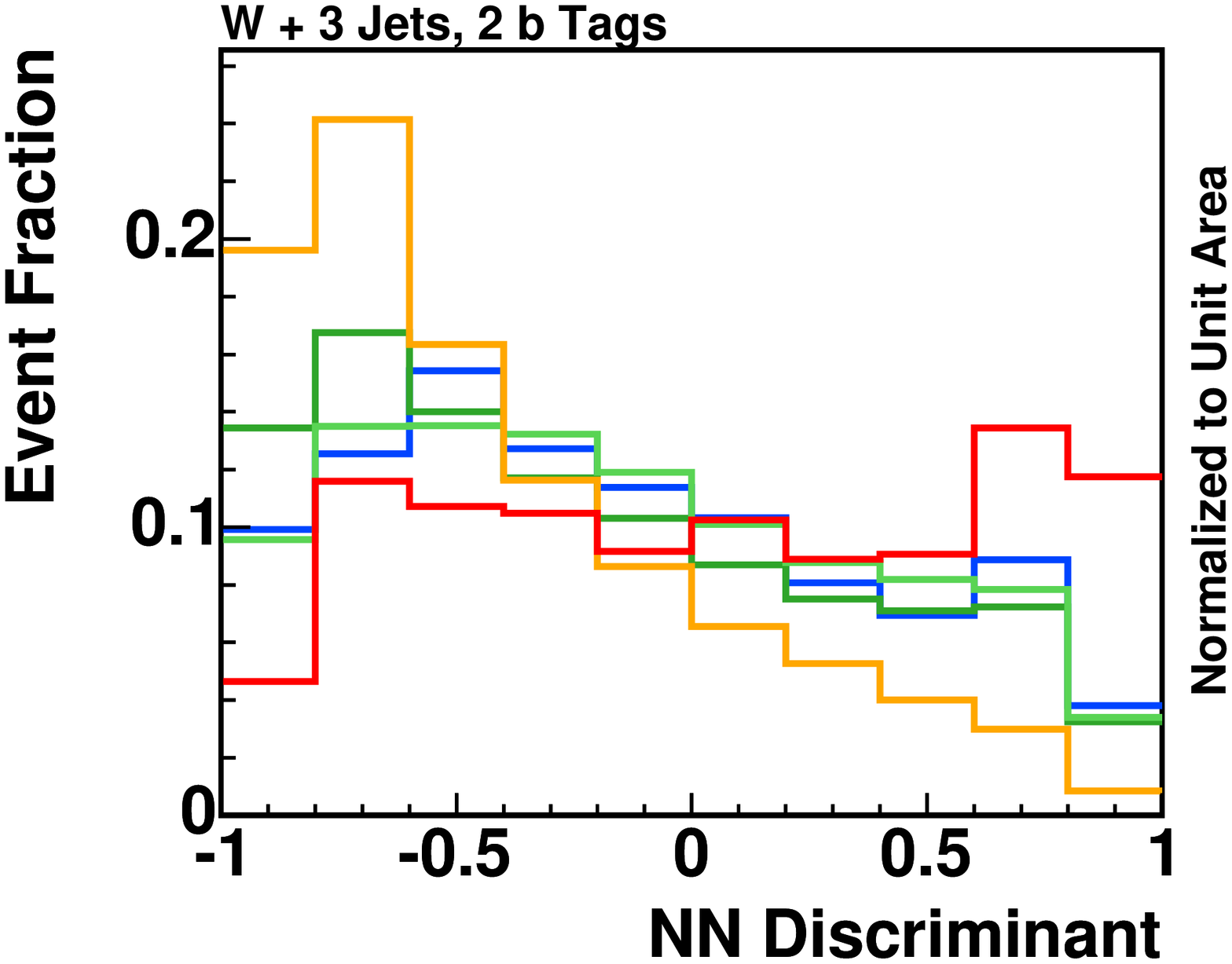}
\label{fig:NN3j2t_shape}}
\subfigure[]{
\includegraphics[width=0.65\columnwidth]{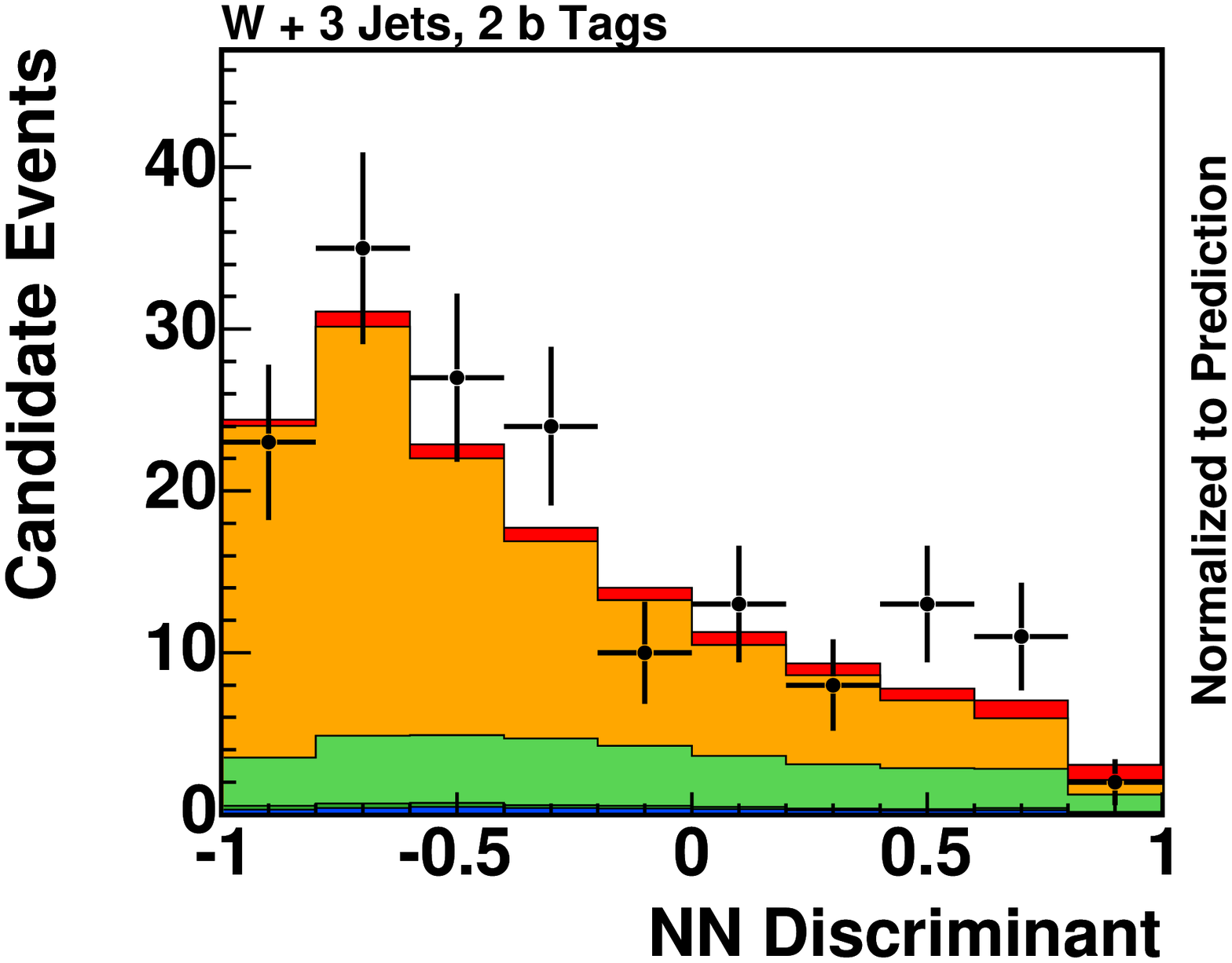}
\label{fig:NN3j2t}}
\end{center}
\caption{\label{fig:NN} 
Templates of predictions for the signal
and background processes, each scaled to unit area (left) and comparisons of
the data with the sum of the predictions (right)
of the neural network output for each signal region. Single top quark events
are predominantly found on the right-hand sides of the histograms while
background events are mostly found on the left-hand sides.
The data are indicated by points with error bars, and 
the predictions are shown stacked, with the stacking order following that of the legend.
}
\end{figure*}

In each data sample, distinguished by the number of identified jets and the number
of $b$~tags, a neural network is constructed with the input variables described above.
The outputs lie between $-1.0$ and $+1.0$, where $-1.0$ is background-like
and $+1.0$ is signal-like.  The predicted distributions of the signals and the
expected background processes are shown in~Fig.~\ref{fig:NN} for the four $b$-tag and jet
categories.  The templates, each normalized to unit area, are shown separately, indicating the separation
power for the small signal.  
The sums of predictions normalized to our signal and
background models, which are described in Sections~\ref{sec:Background}
and~\ref{sec:SignalModel}, respectively,
are compared with the data.   Figure~\ref{fig:allNN}\subref{fig:allNN_chan} 
corresponds to the sum of all four $b$-tag and jet categories.

\subsubsection{Validation}

The distributions of the input variables to each neural network are checked in
the zero, one, and two-tag samples for two- and three-jet events.  
Comparisons of the observed and predicted distributions of some of the variables
which confer the most sensitivity are shown in Sections~\ref{sec:bgvalidation} 
and~\ref{sec:Multivariate}.  The
good agreement seen between the predictions and the observations in both the input variables
and the output variables gives us confidence in the Monte Carlo modeling of the output discriminant
distributions.

\begin{figure*}
\begin{center}
\subfigure[]{
\includegraphics[width=0.8\columnwidth]{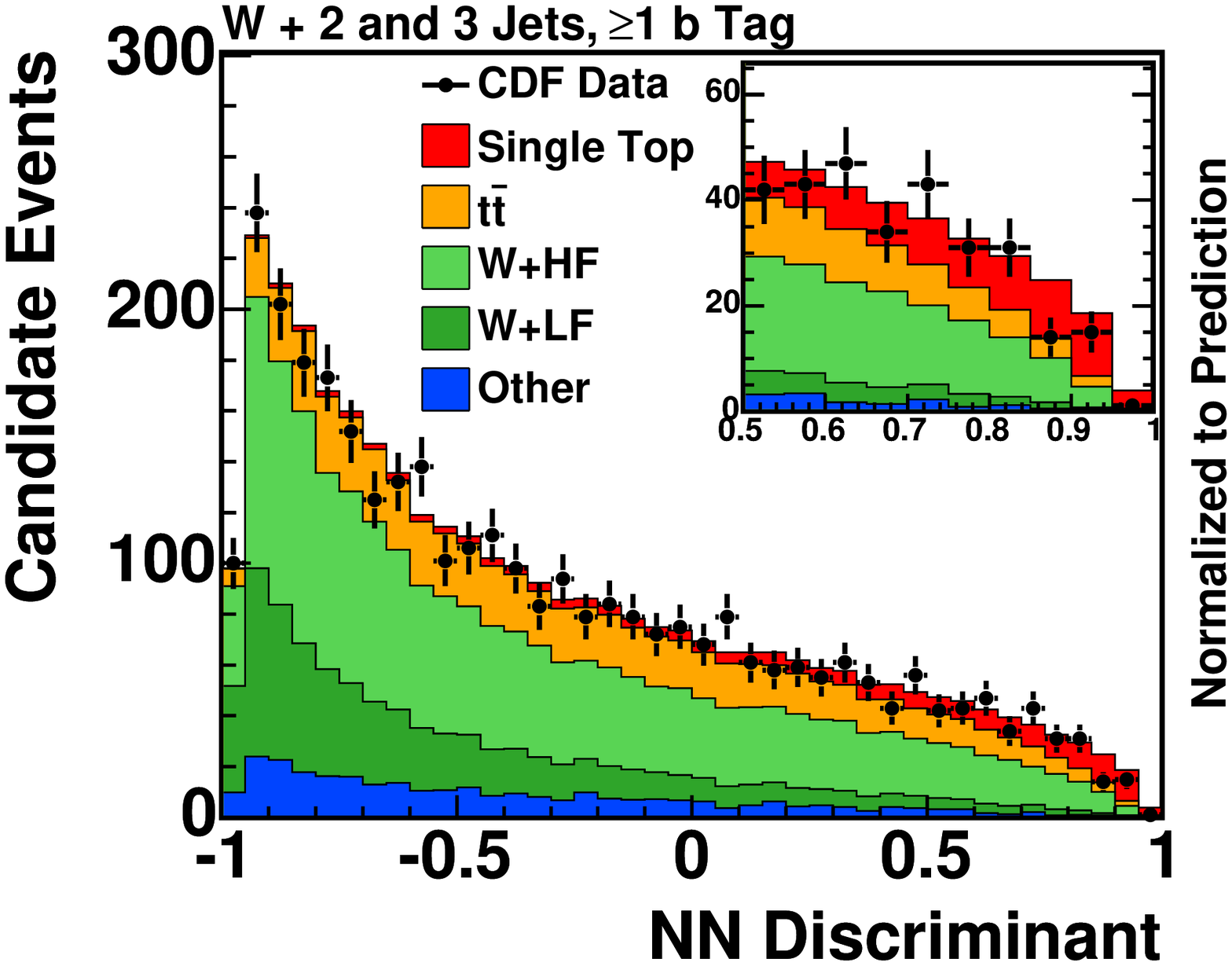}
\label{fig:allNN_chan}}
\subfigure[]{
\includegraphics[width=0.8\columnwidth]{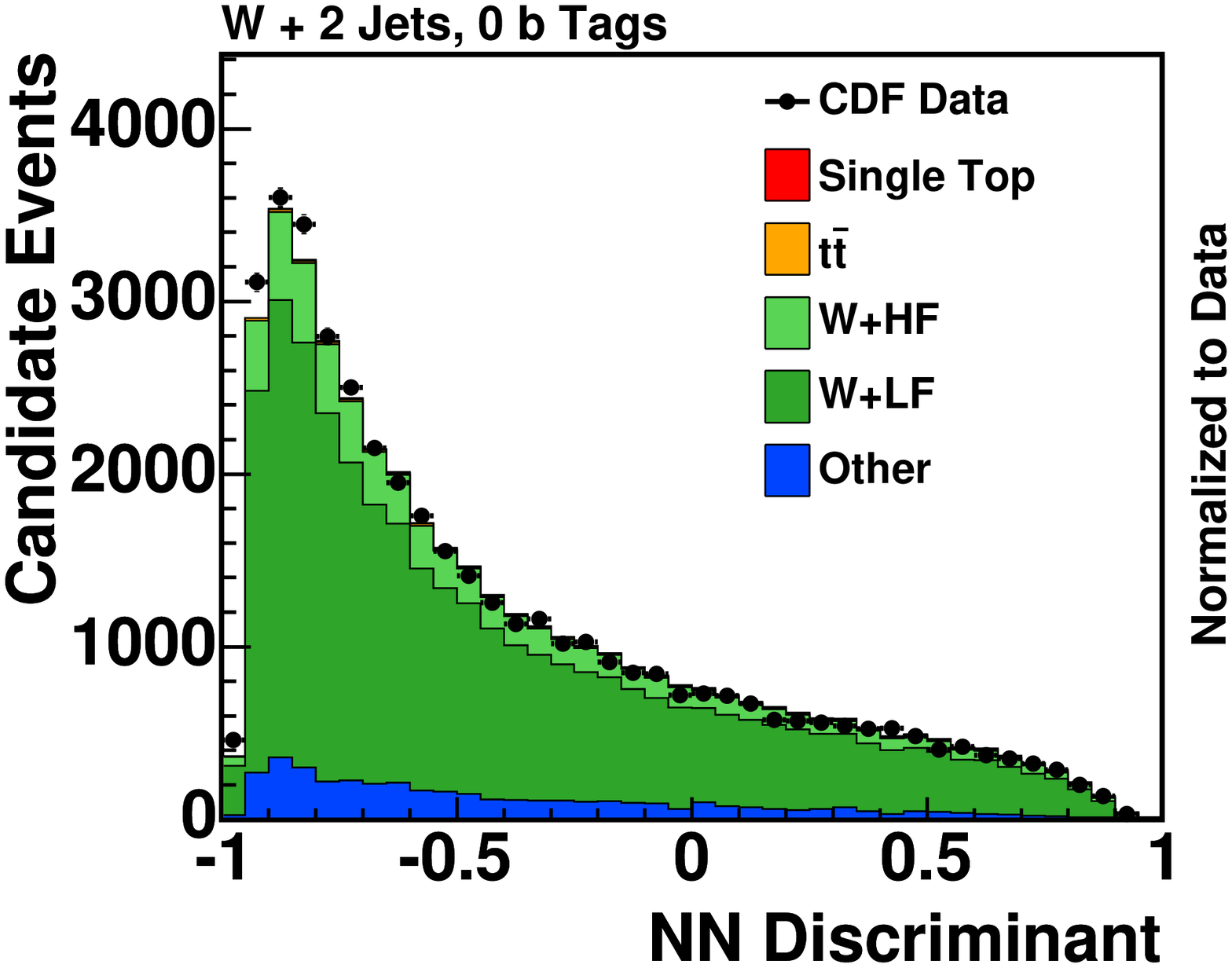}
\label{fig:allNN_0tag}}
\end{center}
\caption{\label{fig:allNN} Comparison of the data with the sum of the predictions 
of the neural network output for the sum of all selected signal data samples (left) and the neural 
network output for two-jet one-$b$-tag events applied to the untagged control sample, showing
close modeling of the data and good control over the $W$+light-flavor shape.
The data are indicated by points with error bars, and 
the predictions are shown stacked, with the stacking order following that of the legend.
}
\end{figure*}

We validate the performance of each network by checking it in the untagged
sideband, appropriately modifying variables that depend on tagging
information.  An example is shown in
Fig.~\ref{fig:allNN}\subref{fig:allNN_0tag}.  The agreement in this sideband gives
us confidence that the information used in this analysis is well
modeled by the Monte Carlo simulation.

\subsubsection{High NN Discriminant Output}

\begin{figure*}
\begin{center}
\subfigure[]{
\includegraphics[width=0.8\columnwidth]{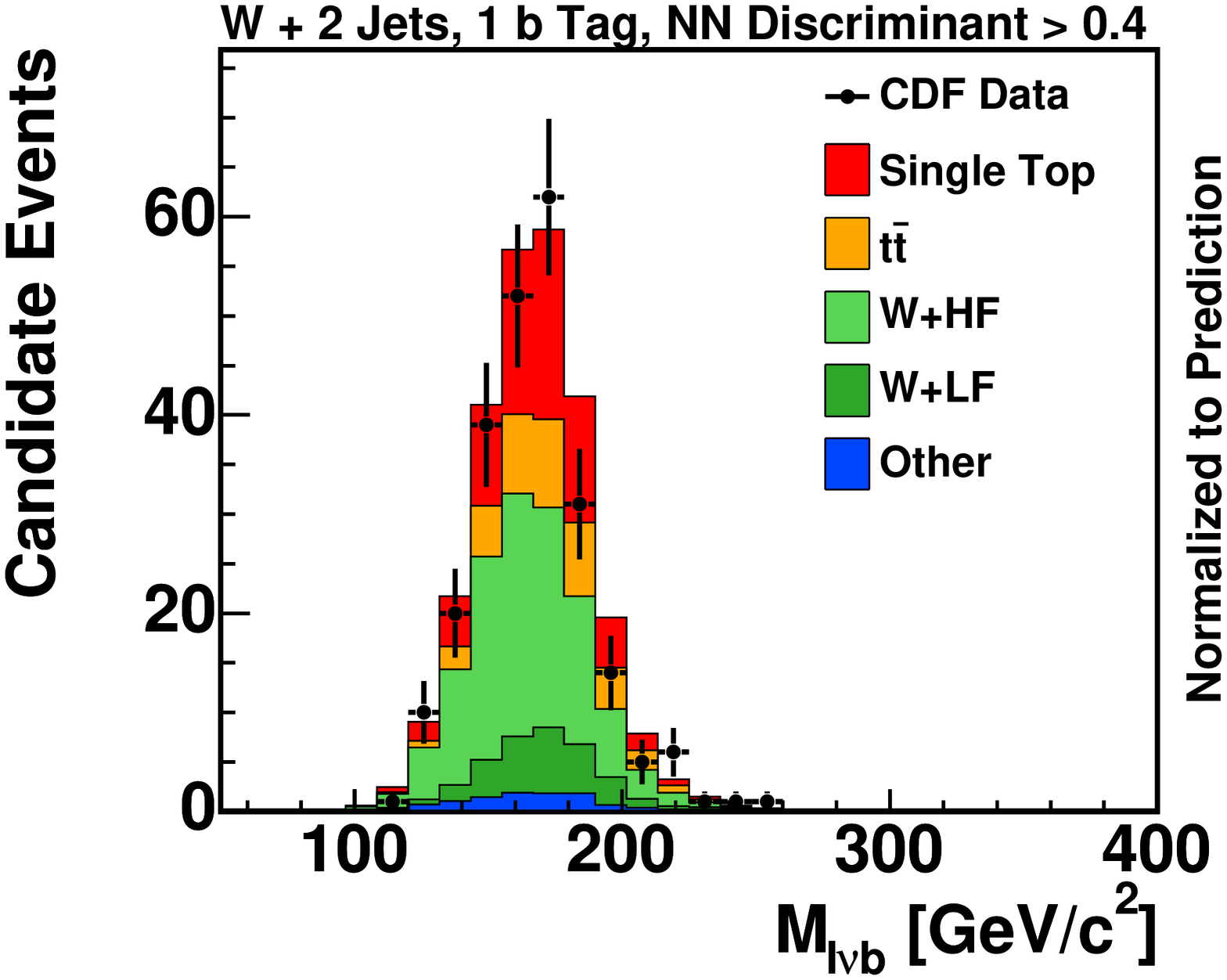}
\label{fig:NNhighNNout}}
\subfigure[]{
\includegraphics[width=0.8\columnwidth]{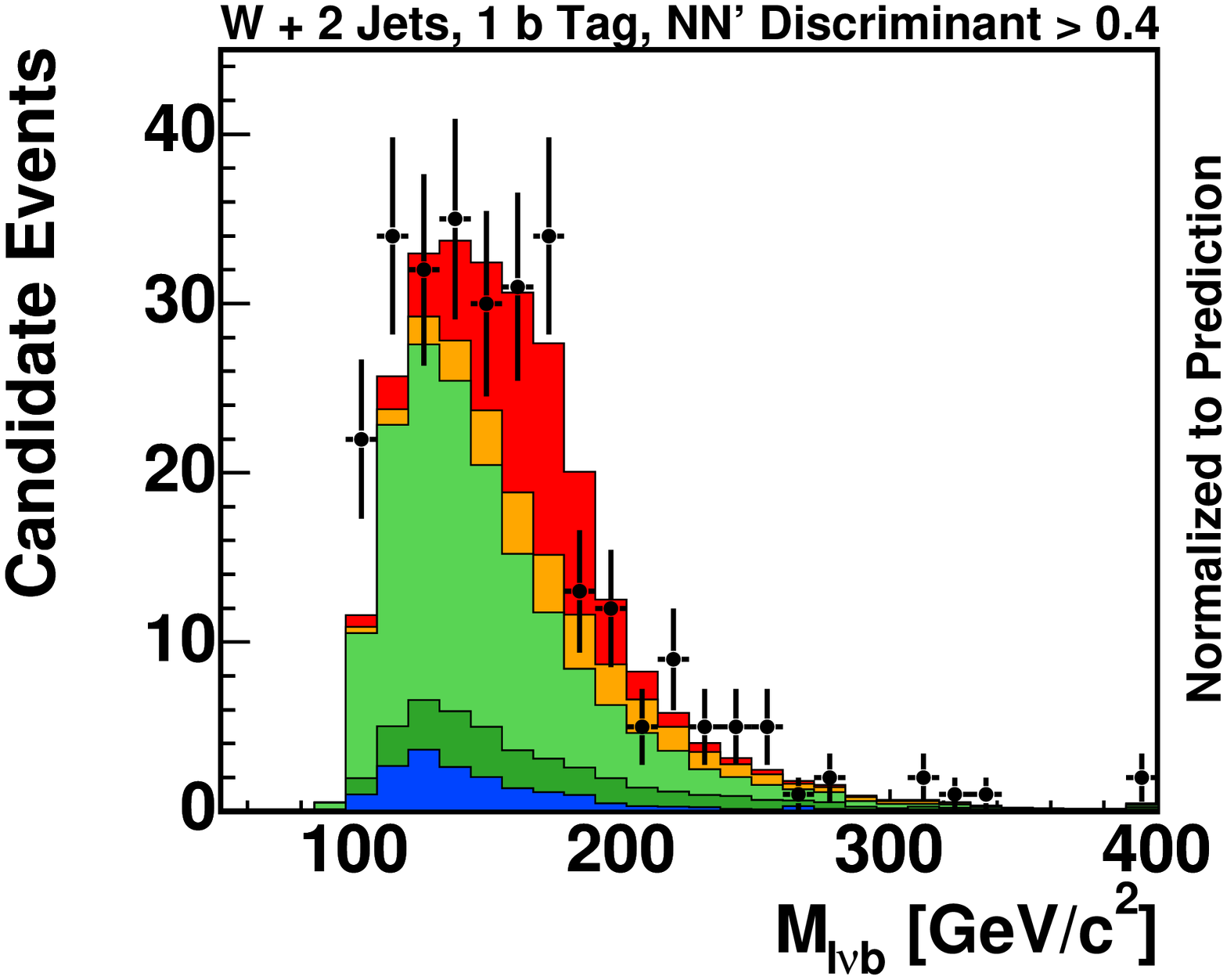}
\label{fig:NNhighNNPrimeout}} \\
\end{center}
\caption{\label{fig:NNhigh} 
Comparison of the predictions and the data for $M_{\ell \nu b}$ for events with an output above 0.4 of the original NN (left) and a specially trained NN$^{\prime}$ (right) discriminant.
The data are indicated by points with error bars, and 
the predictions are shown stacked, with the stacking order following that of the legend.
}
\end{figure*}

To achieve confidence in the quality of the signal contribution in the
highly signal-enriched region of the NN discriminant, further studies
have been conducted. By requiring a NN discriminant output above 0.4
in the event sample with 2 jets and 1 $b$ tag, a
signal-to-background ratio of about 1:3 is achieved. This subsample of
signal candidates is expected to be highly enriched with signal
candidates and is simultaneously sufficient in size to check the
Monte Carlo modeling of the data.  We compare the expectations
of the signal and background processes to the observed data of this
subsample in various highly discriminating variables. The agreement is
good, as is shown, for example, for the invariant mass of the charged lepton,
the neutrino, and the $b$-tagged jet $M_{\ell \nu b}$ in
Fig.~\ref{fig:NNhigh}(a). Since only very signal-like background
events are within this subsample, the background shapes are very
similar to the signal shapes.  This is because the $M_{\ell \nu b}$ is
one of the most important input variables of the NN discriminant,
leading to a signal-like sculpted shape for background events in this
subsample.  As a consequence, the shape of this distribution does not carry
information as to whether a signal is present or absent.

To overcome the similar shapes of signal and background events in the
signal-enriched subsample, a special neural network discriminant
(NN$^{\prime}$) is constructed in exactly the same way as the original,
but without $M_{\ell \nu b}$ as an input.  Since $M_{\ell \nu b}$ is
highly correlated with other original neural network input variables, such as 
$M_\mathrm{T}^{\ell \nu b}$ (with a correlation coefficient of 65\%), $H_\mathrm{T}$ (45\%), and
$M_{jj}$ (24\%), these variables are also omitted for the training of
the special NN$^{\prime}$ discriminant.  Despite the loss of
discrimination through the removal of some very important input
variables, the NN$^{\prime}$ discriminant is still powerful enough to
enrich a subsample of events with signal.  With the requirement NN$^{\prime}>0.4$,
the signal-to-background ratio is somewhat reduced compared with that of the original NN
discriminant. The benefit of this selection is that the
predicted distributions of the signal and background are now more different from each other.
We predict that background
events are dominant at lower values of $M_{\ell \nu b}$ while the
single top quark signal is concentrated around the reconstructed top
quark mass of 175 GeV/$c^2$, as shown in Fig.~\ref{fig:NNhigh}(b). Because of the
more distinct shapes of the signal and background expectations, the
observed shape of the in data distribution is no longer
explicable by the background prediction alone; a
substantial amount of signal events is needed to describe the observed
distribution.  The NN$^\prime$ network is used only for this cross-check; it is not
included in the main results of this paper.

\subsection{\label{sec:BDT} Boosted Decision Tree}

A decision tree classifies events with
a series of binary choices; each choice is based on a single variable.
Each node in the tree splits the sample into two subsamples, and a decision
tree is built using those two subsamples, continuing until the
number of events used to predict the signal and background 
in a node drops below a set minimum.  
In constructing a tree, for each node, the variable used to split the node's
data into subsamples and the value of the variable on the boundary of the two
subsamples are chosen to 
provide optimal separation between
signal and background events.  The same variable may be used in
multiple nodes, and some variables may not be used at all. 
 This procedure
results in a series of final nodes with maximally different signal-to-background
ratios.

Decision trees allow many input variables to be combined into
a single output variable with powerful discrimination between signal
and background. Additionally, decision trees are insensitive to the
inclusion of poorly discriminating input variables because the
training algorithm will not use non-discriminating variables when
constructing its nodes.  In this analysis, we train a different
boosted decision tree (BDT) in each data sample.  We use the
TMVA~\cite{Hocker:2007ht} package to perform this analysis~\cite{CasalLarana:2010zz}.  
The boosting procedure is described below.

The criterion used to choose the variable used to split each node's data
and to set the value of the variable on the boundary is to optimize the Gini index~\cite{gini}
$p(1-p) = sb/(s+b)^2$, where $p=s/(s+b)$ is the purity and
$s$ and $b$ are the number of signal and background events in the
node, respectively.

A shortcoming of decision trees is their instability with respect to
statistical fluctuations in the training sample from which the tree
structure is derived.  For example, if two input variables exhibit
similar separation power, a fluctuation in the training sample may
cause the algorithm to decide to use one variable early in the
decision chain, while a slightly different training sample may
result in a tree which uses the other variable in its place,
resulting in a substantially different tree.

This problem is overcome by a boosting~\cite{Freund:1997} procedure that extends 
this concept from one tree to several trees which form a
``forest'' of decision trees. The trees are derived from the same training
ensemble by reweighting events, and are 
finally combined into a single classifier which
is given by a weighted average of the individual decision 
trees. Boosting stabilizes
the response of the decision trees with respect to 
fluctuations in the training sample and is able to
considerably enhance the performance with respect to a single tree.

This analysis uses the {\sc adaboost}~\cite{Freund:1997} 
(adaptive boost) algorithm, in which the events that were
misclassified in one tree are multiplied by a common
boost weight $\alpha$ in the training of the next tree. The boost weight is derived from the
fraction of misclassified events, $r$, of the previous tree,
\begin{equation}
 \alpha = \frac{1-r}{r}.
\end{equation}

The resulting event classification $y_{\rm{BDT}}(x)$ for the 
boosted tree is given by
\begin{equation} \label{eq:boost-weights}
y_{\rm{BDT}}(x) = \sum_{i\in {\rm forest}}\ln(\alpha_i)\cdot h_i(x),
\end{equation}
where the sum is over all trees in the forest. Large (small) values of
$y_{\rm{BDT}}(x)$ indicate a signal-like (background-like) event.
The result $h_i(x)$ of an individual tree can either be defined to be
$+1$ ($-1$) for events ending up in a signal-like (background-like) leaf 
node according to the majority of training events in that
leaf, or $h_i(x)$
can be defined as the purity of the leaf node in which the event is found.
We found that the latter option performs better
for single-tag samples, while the double tag samples--which have
fewer events--perform better when trained with the former option.

While non-overlapping samples of Monte Carlo events are used to train
the trees and to produce predictions of the distributions of their outputs,
there is the possibility of ``over-training'' the trees.  If insufficient Monte Carlo
events are classified in a node of a tree, then the training procedure can falsely
optimize to separate the few events it has in the training sample and perform worse
on a statistically independent testing sample.
In order to remove statistically insignificant nodes from each tree we
employ the cost complexity~\cite{Breiman:1984} pruning algorithm.  
Pruning is the process of cutting back a tree from the bottom up after
it has been built to its maximum size. Its purpose is to remove
statistically insignificant nodes and thus reduce the over-training of
the tree.

The background processes included in the training are $t\bar t$ and
$Wb\bar{b}$ for double-$b$-tag channels, and those as well as $Wc$ and
$W+$LF for the single-$b$-tag channels. Including the non-dominant background processes
is not found to significantly increase the performance of the analysis.

\subsubsection{Distributions}

\begin{figure*}
\begin{center}
\subfigure[]{
\includegraphics[width=0.65\columnwidth]{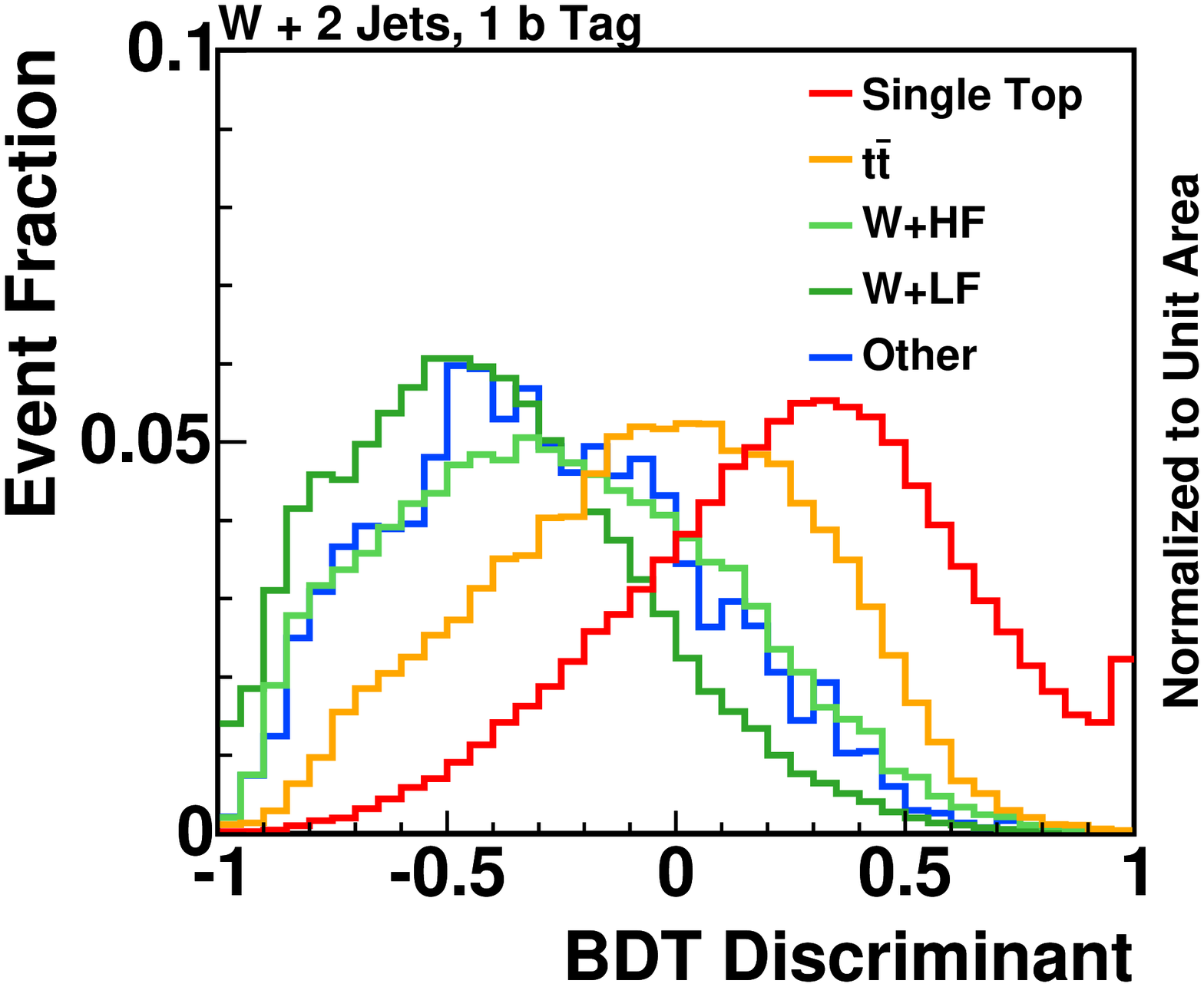}
\label{fig:BDT2j1t_shape}}
\subfigure[]{
\includegraphics[width=0.65\columnwidth]{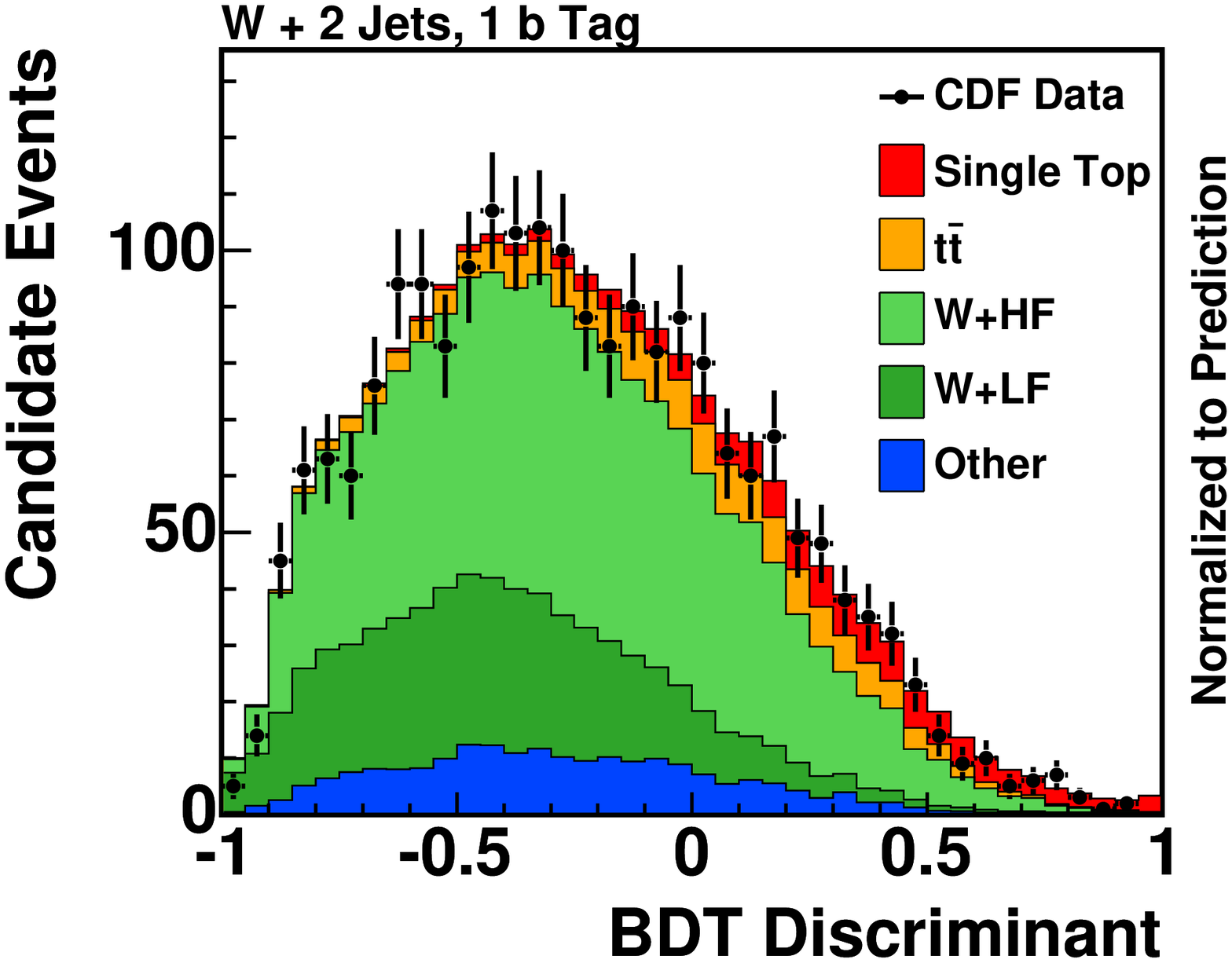}
\label{fig:BDT2j1t}} \\
\subfigure[]{
\includegraphics[width=0.65\columnwidth]{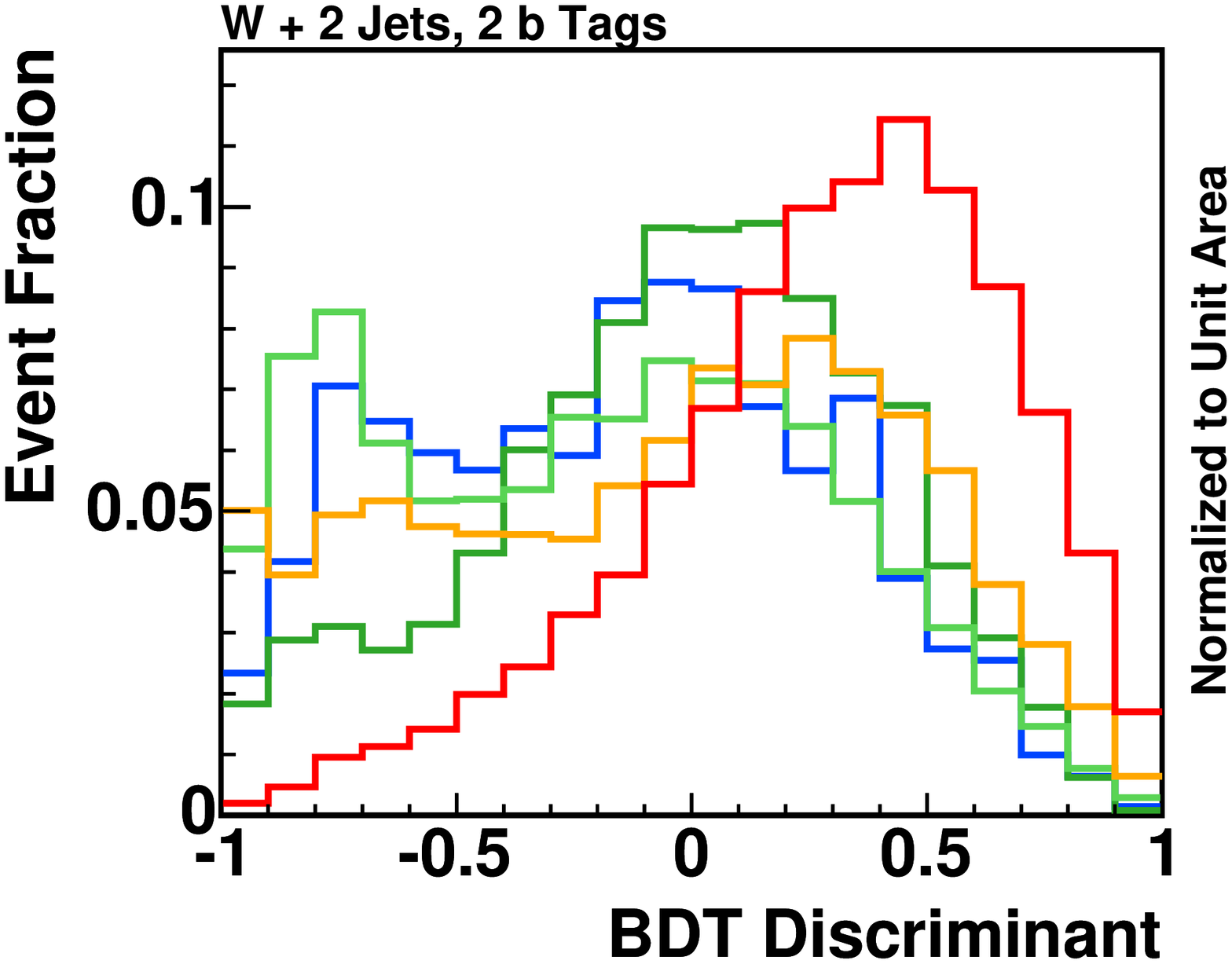}
\label{fig:BDT2j2t_shape}}
\subfigure[]{
\includegraphics[width=0.65\columnwidth]{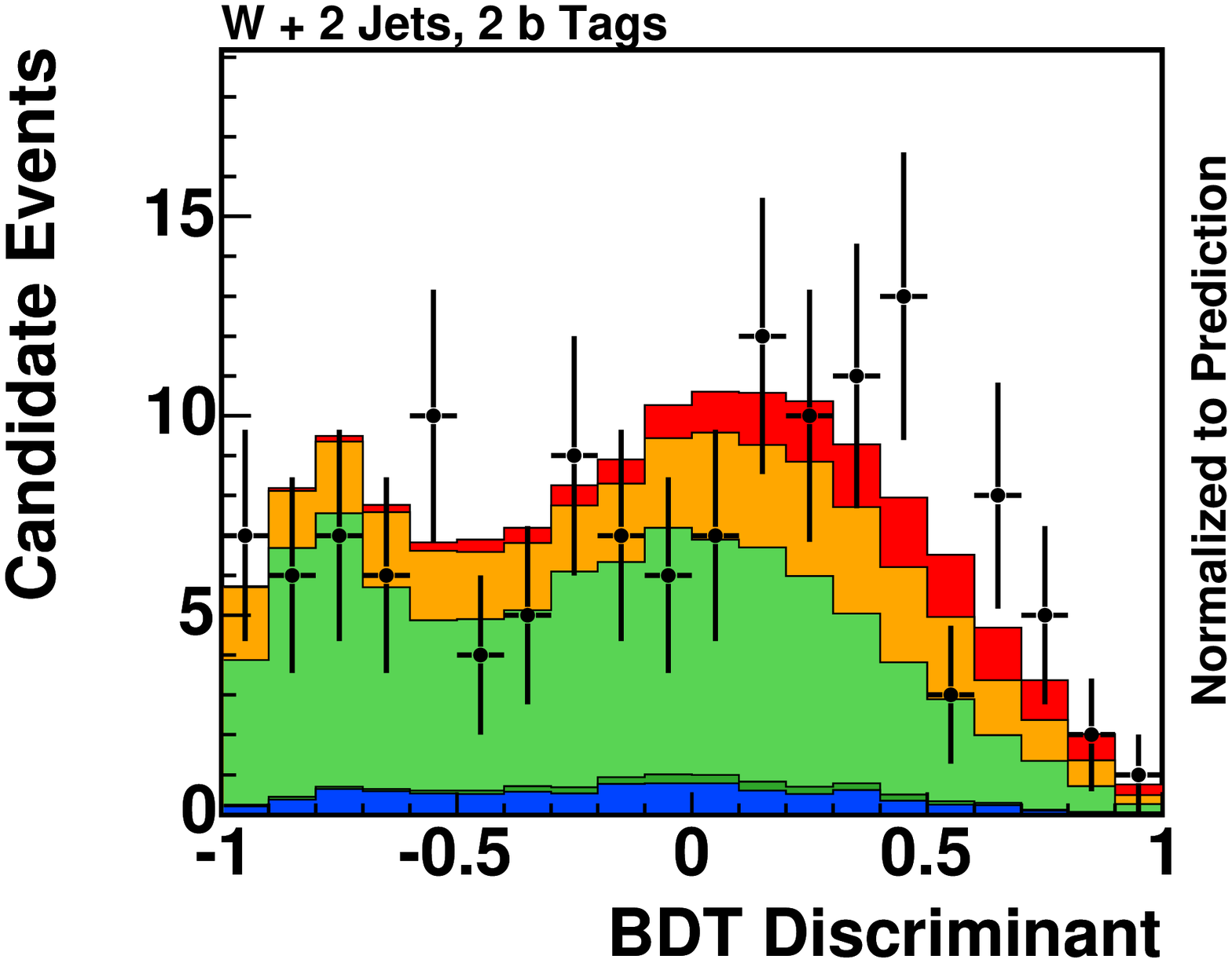}
\label{fig:BDT2j2t}} \\
\subfigure[]{
\includegraphics[width=0.65\columnwidth]{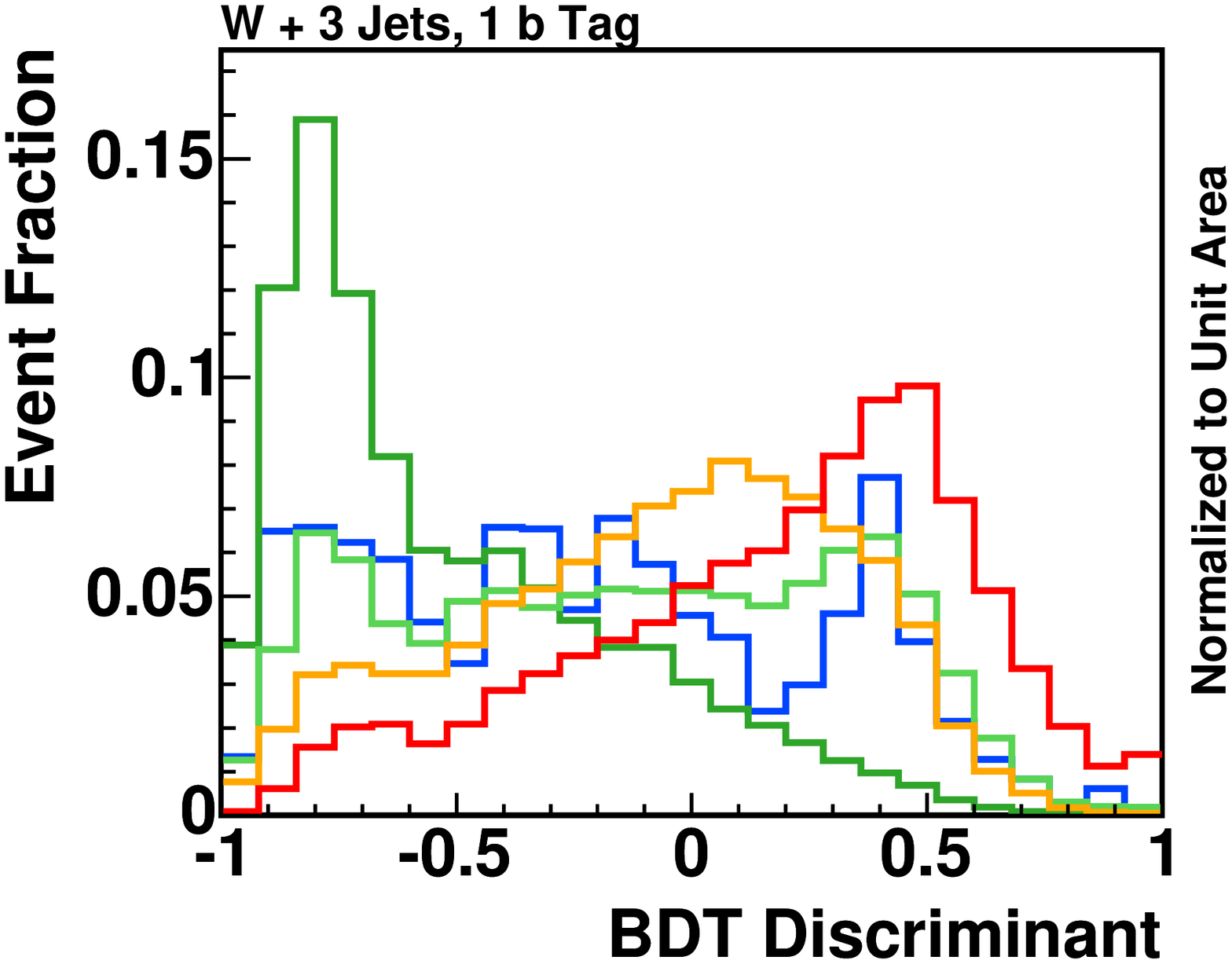}
\label{fig:BDT3j1t_shape}}
\subfigure[]{
\includegraphics[width=0.65\columnwidth]{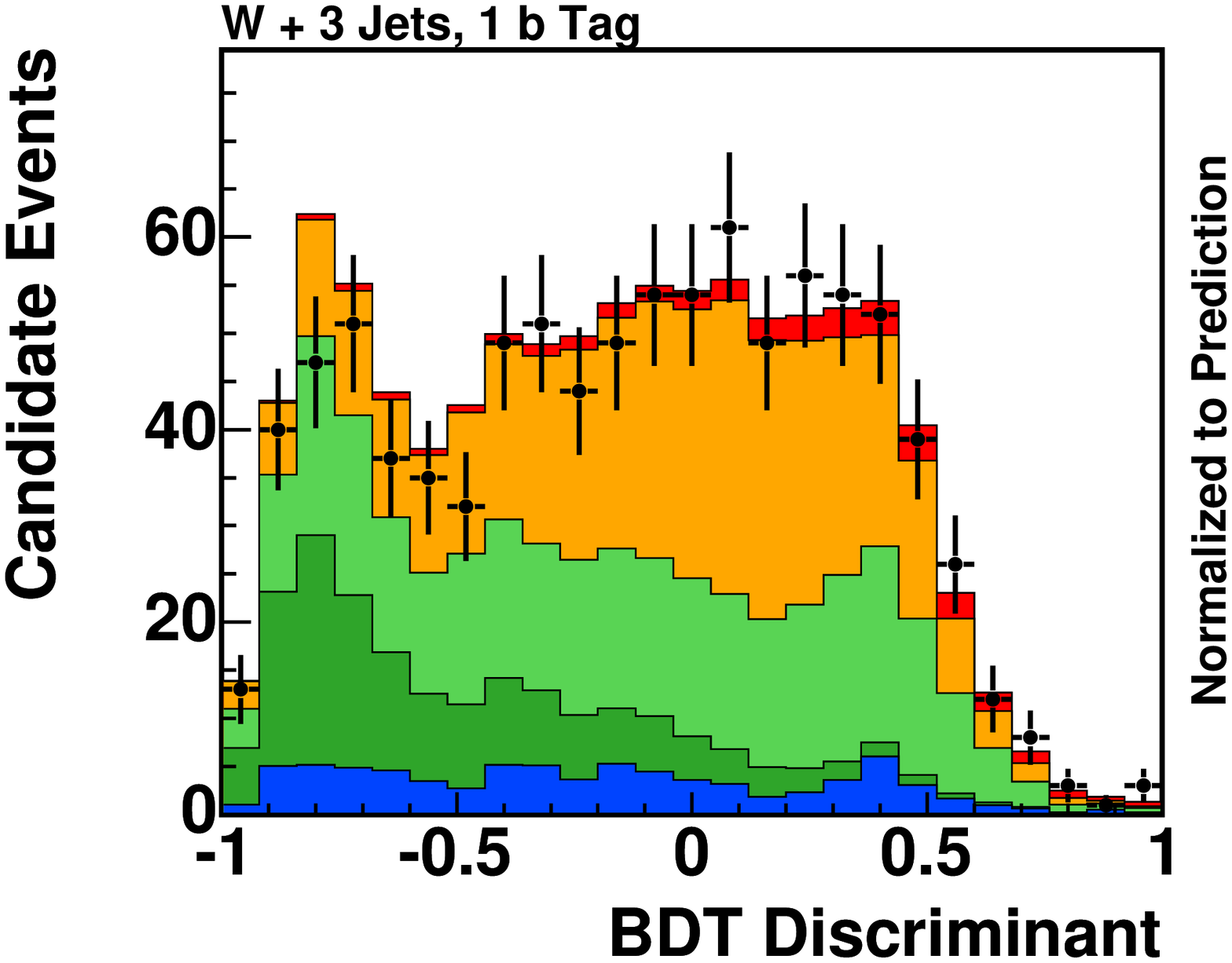}
\label{fig:BDT3j1t}} \\
\subfigure[]{
\includegraphics[width=0.65\columnwidth]{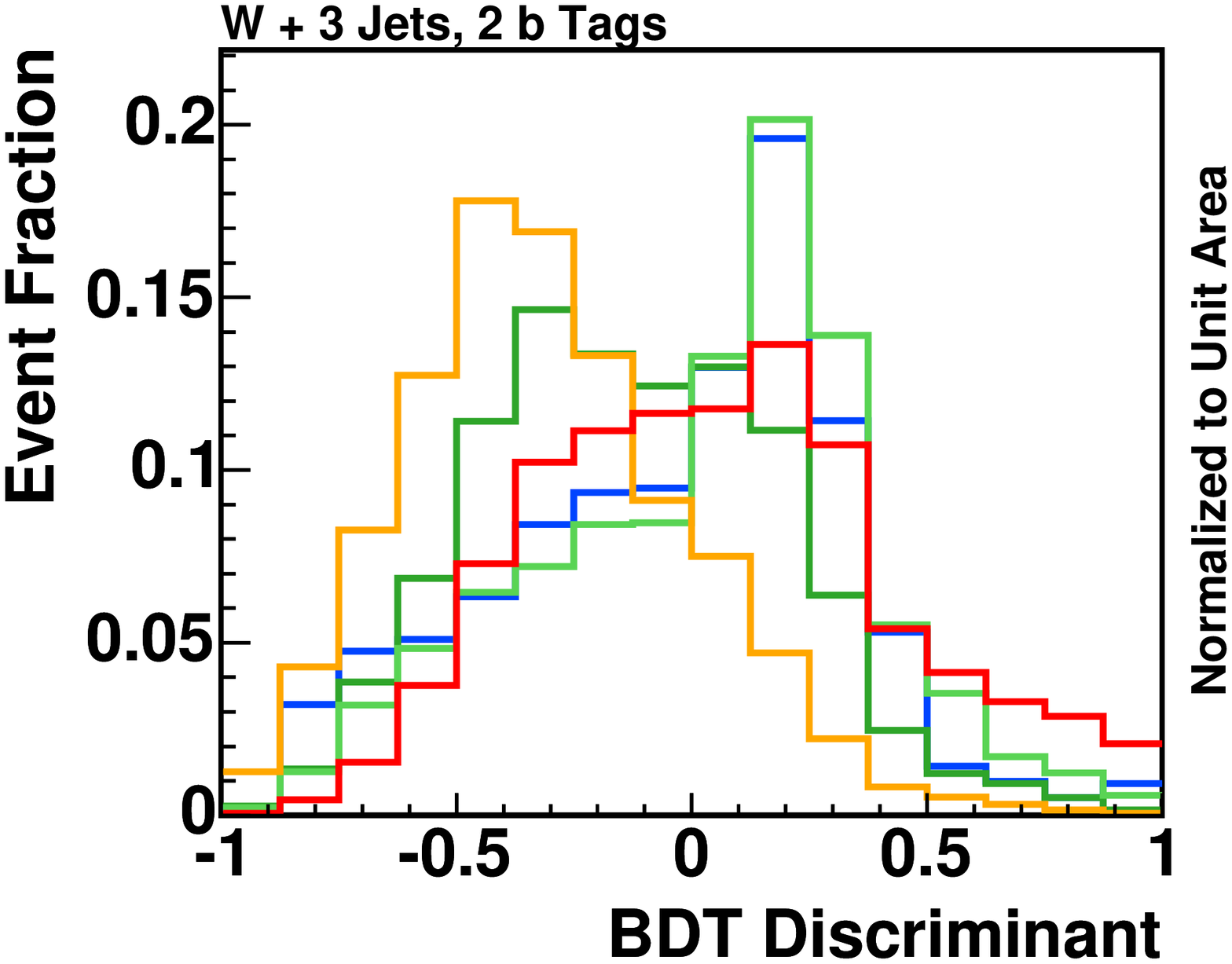}
\label{fig:BDT3j2t_shape}}
\subfigure[]{
\includegraphics[width=0.65\columnwidth]{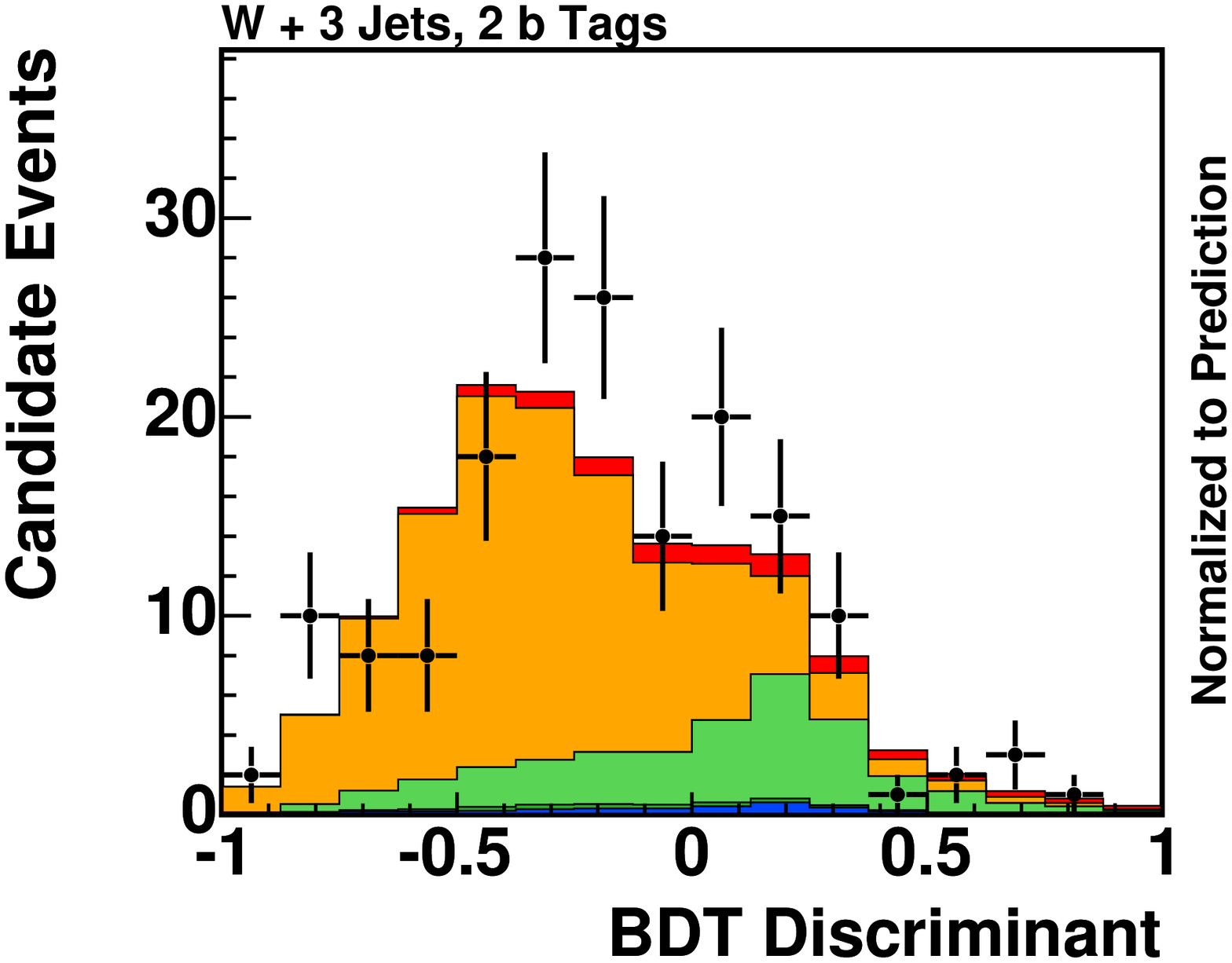}
\label{fig:BDT3j2t}}
\end{center}
\caption{\label{fig:BDT}
Templates of predictions for the signal
and background processes, each scaled to unit area (left) and comparisons of
the data with the sum of the predictions (right)
of the boosted decision tree output for each data sample. Single top quark events
are predominantly found on the right-hand sides of the histograms while
background events are mostly found on the left-hand sides.
The data are indicated by points with error bars, and 
the predictions are shown stacked, with the stacking order following that of the legend.
}
\end{figure*}

In each data sample, distinguished by the number of identified jets and the number
of $b$~tags, a BDT is constructed with the input variables described above.
The output for each event lies between $-1.0$ and 1.0, where $-1.0$ indicates the event has properties
that make it appear much more to be a background event than a signal event, 
and 1.0 indicates the event appears much more likely to have come from a single top
signal.  The predicted distributions of the signals and the
expected background processes are shown in~Fig.~\ref{fig:BDT} for the four $b$-tag and jet
categories.  The templates, each normalized to unit area, are shown separately, indicating the separation
power for the small signal.
The sums of predictions normalized to our signal and
background models, which are described in Sections~\ref{sec:Background}
and~\ref{sec:SignalModel}, respectively, are compared with the data.  
Figure~\ref{fig:allBDT}\subref{fig:allBDT_chan} 
corresponds to the sum of all four $b$-tag and jet categories.

\subsubsection{Validation}

The distributions of the input variables to each BDT are checked in
the zero, one, and two $b$-tag samples for two- and three-jet events, and also in
the four-jet sample containing events with at least one $b$~tag.  Some of the most
important variables' validation plots are shown in Sections~\ref{sec:bgvalidation} and~\ref{sec:Multivariate}.  The
good agreement seen between the predictions and the observations in both the input variables
and the output variables gives us confidence in the Monte Carlo
modeling of the distributions of the discriminant outputs.

\begin{figure*}
\begin{center}
\subfigure[]{
\includegraphics[width=0.8\columnwidth]{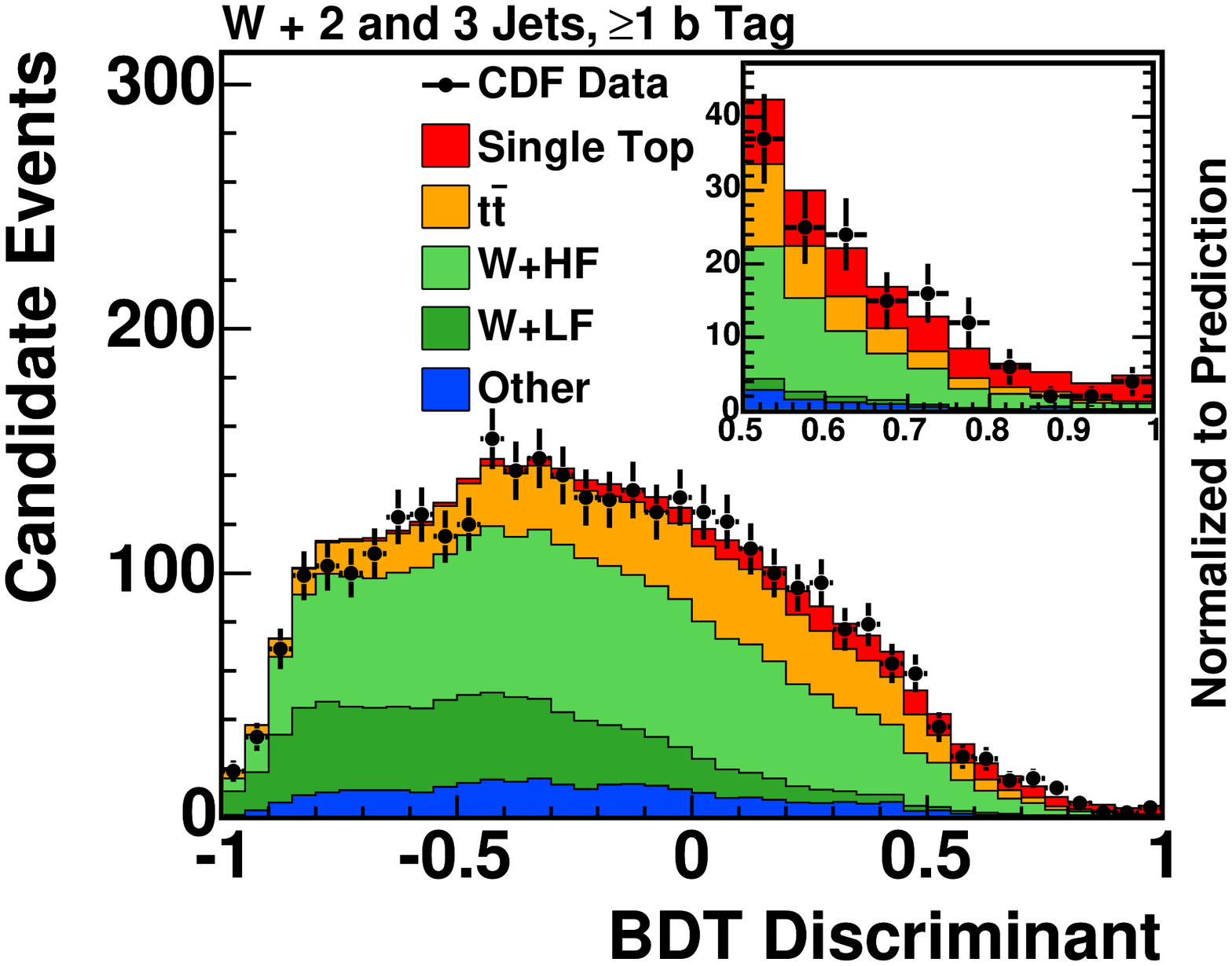}
\label{fig:allBDT_chan}}
\subfigure[]{
\includegraphics[width=0.8\columnwidth]{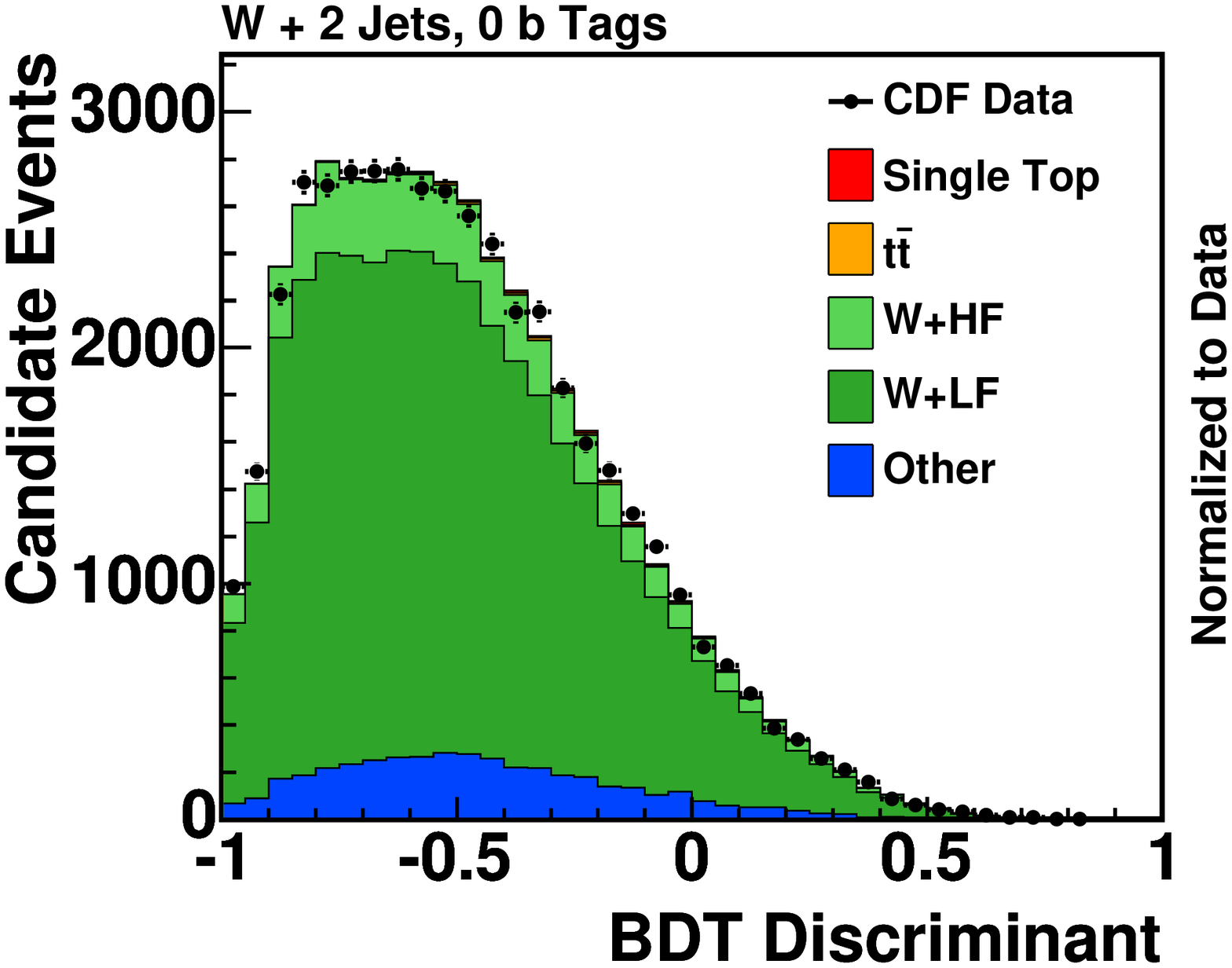}
\label{fig:allBDT_0tag}}
\end{center}
\caption{\label{fig:allBDT} Comparison of the data with the sum of the predictions 
of the BDT output for the sum of all selected data samples (left) and the BDT output for two-jet
one-$b$-tag events applied to the untagged two-jet control sample (right), where the dominant
contributing process is $W$+light-flavored jets.
The data are indicated by points with error bars, and 
the predictions are shown stacked, with the stacking order following that of the legend.
}
\end{figure*}

\begin{figure}
\begin{center}
\includegraphics[width=0.8\columnwidth]{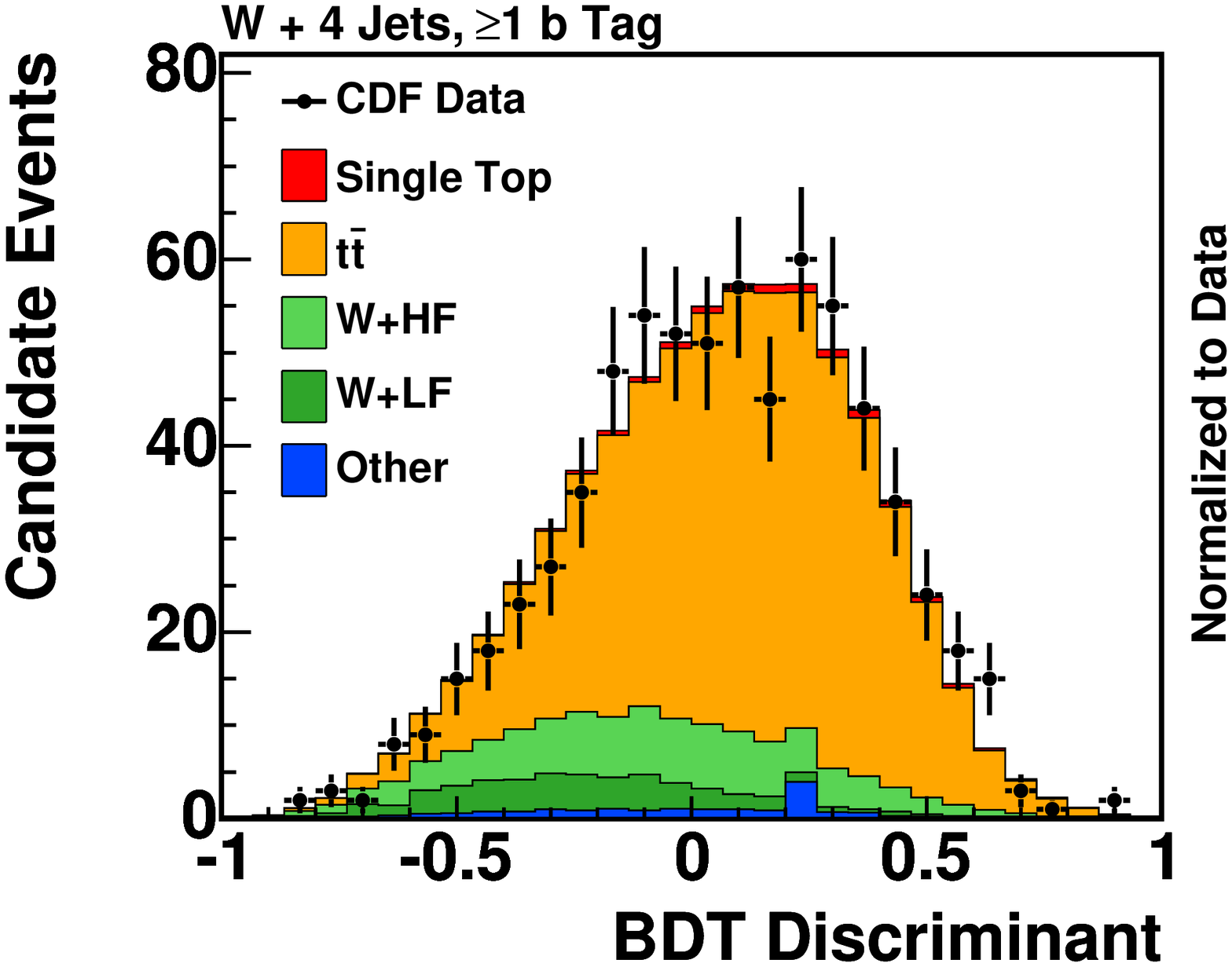}
\end{center}
\caption{\label{fig:BDT4jet} The BDT output for four-jet events containing
one or more $b$~tags.  The dominant source of background is $t{\bar{t}}$ events.
The data are indicated with points and the stacked histograms show the prediction,
scaled to the total data rate, with the stacking order following that of the legend.}
\end{figure}

We validate the modeling of the backgrounds in each boosted tree by checking it in the
sample of events with no $b$~tags, separately for events with two and three jets.
For variables depending on $b$-tagging information like $M_{\ell \nu b}$ and
$Q\times\eta$, the leading jet is chosen as the ``$b$-tagged'' jet, and for the $b_{\mathrm{NN}}$
variable the output value is randomly taken from a $W$+LF template. 
An example is shown in Fig.~\ref{fig:allBDT}\subref{fig:allBDT_0tag}, which shows the two-jet, one $b$-tag
BDT tested with the two-jet, zero $b$-tag sample.  The dominant source of background tested
in Fig.~\ref{fig:allBDT}\subref{fig:allBDT_0tag} is $W$+LF, and the {\sc alpgen} Monte Carlo
predicts the BDT output very well.  We further test the four-jet sample with one or more $b$-tags,
shown in Fig.~\ref{fig:BDT4jet}, taking the leading two jets to test the two-jet, one $b$-tag BDT.
The dominant background in this test is $t{\bar{t}}$, and the good modeling of the distribution of the
output of the BDT by {\sc pythia} raises our confidence that this background, too, is modeled well in the
data samples.

\section{\label{sec:Systematics} Systematic Uncertainties}

The search for single top quark production and the measurement of the
cross section require substantial input from theoretical models, Monte
Carlo simulations, and extrapolations from control samples in data.
We assign systematic uncertainties to our predictions and
include the effects of these uncertainties on the measured cross
sections as well as the significance of the signal.

We consider three categories of systematic uncertainty:
uncertainty in the predicted rates of the signal and background
processes, uncertainty in the shapes of the distributions of the
discriminant variables, and
uncertainty arising from the limited number of Monte Carlo events used
to predict the signal and background expectations in each bin
of each discriminant distribution.  Sources of uncertainty may affect multiple
signal and background components.  The effects of systematic
uncertainty from the same source are considered to be fully
correlated.  For example, the integrated luminosity estimate affects
the predictions of the Monte-Carlo based background processes and the
signal, so the uncertainty on the integrated luminosity
affects all of these processes in a correlated way.  The effects of different
sources of systematic uncertainty are considered to be uncorrelated.

The effects of all systematic uncertainties are included in the
hypothesis tests and
cross section measurements performed by each analysis, as described in
Section~\ref{sec:Interpretation}.  Detailed descriptions of the
sources of uncertainty and their estimation are given below.

\subsection{Rate Uncertainties}

Rate uncertainties affect the expected contributions of the signal and
background samples.  Some sources have asymmetric uncertainties.
All rate uncertainties are assigned
truncated Gaussian priors, where the truncation prevents predictions
from being negative for any source of signal or background.  The
sources of rate uncertainties in this analysis are described below, and their
impacts on the signal and background predictions are summarized in Table~\ref{tab:sys}.

\begin{itemize}

\item {\bf Integrated Luminosity:}  A symmetric uncertainty of $\pm$6\%
is applied to all Monte-Carlo based predictions.  This uncertainty
includes the uncertainty in the \ppbar\ inelastic cross section as
well as the uncertainty in the acceptance of CDF's luminosity
monitor~\cite{CLC}.  The requirement that the primary vertex position
in $z$ is within $\pm$60 cm of the origin causes a small acceptance
uncertainty that is included as well.

\item {\bf Theoretical Cross Sections:}  
Our MC-based background processes are scaled to theoretical predictions at NLO
(or better).  We apply the associated theoretical
uncertainties.  We separate out the effects of the top quark mass from
the other sources of uncertainty affecting the theoretical
predictions.  Not every theoretical cross section uncertainty is used
in each result; details are given in
Section~\ref{sec:Interpretation}.

\item {\bf Monte Carlo Generator:} Different Monte Carlo generators for
the signal result in different acceptances.  The deviations are small
but are still included as a rate uncertainty on the signal
expectation as described in Section~\ref{sec:SignalModel}.

\item {\bf Acceptance and Efficiency Scale Factors:}  The predicted
rates of the Monte Carlo background processes and of the signals are affected by
trigger efficiency, mismodeling of the lepton identification
probability, and the $b$-tagging efficiency.  Known differences
between the data and the simulation are corrected for by scaling the
prediction, and uncertainties on these scale factors are collected
together in one source of uncertainty since they affect the
predictions in the same way.

\item {\bf Heavy Flavor Fraction in $W$+jets:}  The prediction of the \Wbb, \Wcc, and \Wc\ fractions in the
$W+2$ jets and $W+3$ jets samples are extrapolated from the $W+1$
jet sample as described in Section~\ref{sec:Background}.
  It is found that {\sc alpgen} underpredicts the \Wbb\ and \Wcc\
fractions in the $W+1$ jet sample by a factor of $1.4\pm 0.4$. We
assume that the \Wbb\ and \Wcc\ predictions are correlated.  The
uncertainty on this scale factor comes from the spread in the measured
heavy-flavor fractions using different variables to fit the data,
and in the difference between the \Wbb\ and \Wcc\ scale
factors.  The \Wc\ prediction from {\sc alpgen} is compared with CDF's
measurement~\cite{Wcharm:2007dm} and is found not to require scaling, but
a separate, uncorrelated uncertainty is assigned to the \Wc\
prediction, with the same relative magnitude as the \Wbb+\Wcc\ uncertainty.

\item {\bf Mistag Estimate:}  The method for estimating the yield
of events with incorrectly $b$-tagged events is described in
Section~\ref{sec:mistag}.  The largest source of systematic
uncertainty in this estimate comes from extrapolating from the
negative tag rate in the data to positive tags by estimating the
asymmetry between positive light-flavor tags and negative light-flavor
tags.  Other sources of uncertainty come from differences in the
negative tag rates of different data samples used to construct the
mistag matrix.  

\item {\bf Non-$W$ Multijet Estimate:}  The Non-$W$ rate prediction
varies when the $\EtMiss$\ distribution is constructed with a different
number of bins or if different models are used for the Non-$W$
templates.  The $\EtMiss$\ fits also suffer from small data samples,
particularly in the double-tagged samples.  A relative uncertainty of
$\pm 40$\% is assesed on all Non-$W$ rate predictions.

\item {\bf Initial State Radiation (ISR):}  The model used for ISR is 
{\sc pythia}'s ``backwards evolution'' method~\cite{Sjostrand:2000wi}.  This
uncertainty is evaluated by generating new Monte Carlo samples for
\ttbar\ and single top quark signals with \LambdaQCD\ doubled or divided in
half, to generate samples with more ISR and less ISR, respectively.
Simultaneously, the initial transverse momentum scale is multiplied by
four or divided by four, and the hard scattering scale of the shower
is multiplied by four or divided by four, for more ISR and less ISR,
respectively.  
These variations are chosen by comparing Drell-Yan Monte Carlo and
data samples.  The \pt\ distributions of dileptons are compared
as a function of the dilepton invariant mass, and the ISR more/less prescriptions
generously bracket the available data~\cite{Abulencia:2005aj}.
Since the ISR prediction must be extrapolated from the $Z$ mass scale to the
higher-$Q^2$ scales of \ttbar\ and single top quark events, the variation chosen
is much more than is needed to bracket the $p_{\rm T}^Z$ data.

\item {\bf Final State Radiation (FSR):}  {\sc pythia}'s model of gluon
radiation from partons emitted from 
the hard-scattering interaction has been tuned with high precision to
LEP data~\cite{Sjostrand:2000wi}.  Nonetheless, uncertainty remains in
the radiation from beam remnants, and parameters analogous to those
adjusted for ISR are
adjusted in {\sc pythia} for the final-state showering, except for the
hard-scattering scale parameter.  The effects of variations in ISR and
FSR are treated as 100\% correlated with each other.
ISR and FSR rate uncertainties are not
evaluated for the $W$+jets Monte Carlo samples because the rates
are scaled to data-driven estimates with associated uncertainties, and
the kinematic shapes of all predictions have factorization and 
renormalization scale uncertainties applied, as
discussed below.

\item {\bf Jet Energy Scale (JES):}  The calibration of the calorimeter
response to jets is a multi-step process,
and each step involves an uncertainty which is propagated to the final
jet-energy scale~\cite{Bhatti:2005ai}.  Raw measurements of the 
jet energies are corrected according
to test beam calibrations, detector non-uniformity, multiple interactions, and energy
that is not assigned to the jet because it lies outside of the jet
cone.  The uncertainties in the jet energy scale are incorporated by
processing all events in all Monte Carlo samples with the jet energy
scale varied upwards and again downwards.  The kinematic properties
of each event are affected, and some events are
re-categorized as having a different number of jets as jets change
their \ET\, inducing correlated rate and shape
uncertainties.  An example of the shape uncertainty to the NN analysis's discriminant
is shown in Fig.~\ref{fig:jessys}.

\begin{figure}
\begin{center}
\includegraphics[width=0.8\columnwidth]{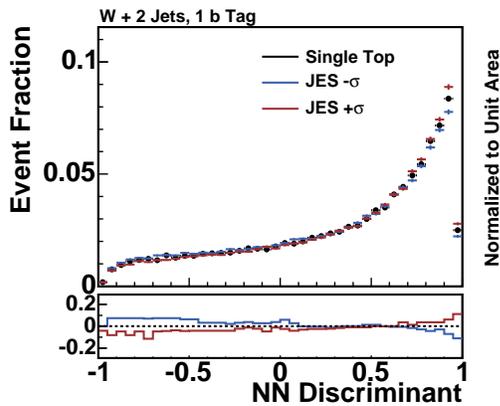}
\end{center}
\caption{\label{fig:jessys}An example of systematically shifted
shape templates.  This figure shows the jet energy scale shifted
histograms for the single top quark signal in two-jet one-$b$-tag events for the
NN discriminant. The plot below shows the relative difference between the central shape and the two
alternate shapes.}
\end{figure}

\item {\bf Parton Distribution Functions (PDF):}  The PDFs
used in this analysis are the CTEQ5L set of leading-order PDFs~\cite{Lai:1999wy}.
To evaluate the systematic uncertainties on the rates due to
uncertainties in these PDFs, we add in quadrature the differences
between the predictions of the following pairs of PDFs:
\begin{itemize}
\item CTEQ5L and MRST72~\cite{Martin:1998sq}, PDF sets computed by
different groups.  MRST72 is also a leading-order PDF set.
\item MRST72 and MRST75, which differ in their value of $\alpha_s$.
The former uses 0.1125; the latter uses 0.1175.
\item CTEQ6L and CTEQ6L1, of which the former has a 1-loop $\alpha_s$
correction, and the latter has a 2-loop $\alpha_s$ correction.
\item The 20 signed eigenvectors of CTEQ6M, each compared with the
default CTEQ5L PDFs.
\end{itemize}
The PDF uncertainty induces a correlated rate and shape uncertainty
in the applicable templates.

\end{itemize} 

\subsection{Shape-Only Uncertainties}

Many of the sources of rate uncertainty listed above also induce
distortions in the shapes of the templates for the signals and
background processes used to model the data.  These include ISR, FSR, JES, and
PDF uncertainties.  Here we list the
sources of shape uncertainties which do not have associated rate
uncertainties. 

Shape uncertainty templates are all smoothed with a median smoothing
algorithm.  This procedure takes the ratio of the systematically
shifted histograms to the central histograms and replaces the contents of
each bin with the median of the ratios of a five-bin window 
around the bin.  The first two bins
and the last two bins are left unaffected by this procedure.  The five-bin
window was chosen as the minimum size that provides adequate smoothing, as judged
from many shape variation ratio histograms.  The
smoothed ratio histograms are then multiplied by the central histograms
to obtain the new varied template histograms.  This procedure reduces
the impact of limited Monte Carlo statistics in the bins of the
central and varied templates.

\begin{itemize}

\item {\bf Jet Flavor Separator Modeling:}  
The distribution of $b_{\mathrm{NN}}$ for light-flavor jets 
is found to require a small correction, as described in
Section~\ref{sec:btagger}.  The full difference between the
uncorrected light-flavor Monte Carlo prediction and the data-derived
corrected distribution is taken as a one-sided systematic uncertainty.
Since a pure sample of charm jets is not available in the data, a
systematic uncertainty is also assessed on the shape of the charm prediction,
taking the difference between the distribution predicted by the Monte Carlo
simulation and the Monte Carlo
distribution altered by the light-flavor correction function.
These shifts in the distributions of $b_{\rm NN}$ for these samples
are propagated through to the
predictions of the shapes of the corresponding discriminant output histograms.

\item {\bf Mistag Model:} To cover uncertainty in modeling the shape of
the analysis discriminant output histograms for 
mistagged events, the untagged data, weighted by the mistag
matrix weights, are used to make an alternate shape
template for the mistags.  The untagged data largely consist of
$W+$light flavored jets, but there is a contamination from \Wbb,
\Wcc, \ttbar, and even single top quark signal events, making the 
estimate of the systematic uncertainty conservative.

\item {\bf Factorization and Renormalization Scale:}  Because
{\sc alpgen} performs fixed-order calculations to create $W$+jets diagrams, it
requires factorization and renormalization scales as inputs.  Both of these
scales are set for each event in our {\sc alpgen} samples to 
\begin{equation}
\sqrt{M_{W}^{2} + \sum_{\rm partons} m_{\rm T}^2},
\end{equation}
where $m_{\rm T}^2 = m^2 + p_{\rm T}^2/c^2$ is the transverse mass of the generated parton.  For light
partons, $u,d,s,g$, the mass $m$ is approximately zero; $m_b$ is set to 4.7~GeV/$c^2$ and $m_c$ is set
to 1.5~GeV/$c^2$.
The sum is over all final-state partons excluding the $W$ boson decay products.
In addition, {\sc alpgen} evaluates $\alpha_s$ separately at each $gqq$ and $ggg$ vertex, and the scale at which
this is done is set to the transverse momentum of the vertex.
The three scales are halved and doubled together in order to produce templates that cover the scale uncertainty.
Although {\sc alpgen}'s $W$+heavy-flavor cross section predictions are strongly dependent on
the input scales, we do not assign additional rate uncertainties on the
$W$+heavy flavor yields because we do not use {\sc alpgen} to 
predict rates; the yields are calibrated using the data.  We do not consider the
calibrations of these yields to constrain the values of the scales for purposes
of estimating the shape uncertainty; we prefer to take
the customary variation described above.

\item {\bf Non-{\it W} Flavor Composition:}  The distribution of $b_{\mathrm{NN}}$
is used to fit the flavor
fractions in the low-$\EtMiss$\ control samples in order to estimate the
central predictions of the flavor composition of $b$-tagged jets in non-$W$ events,
as described in Section~\ref{sec:btagger}.  The limited statistical
precision of these fits and the necessity of extrapolating to the
higher-$\EtMiss$\ signal region motivates an uncertainty on the flavor
composition.  The central predictions for the flavor composition are 45\% $b$ jets,
40\% $c$ jets, and 15\% light-flavored jets.  The
``worst-case'' variation of the flavor composition is 60\% $b$ jets,
30\% $c$ jets, and 10\% light-flavor jets, which we use to set our uncertainty.
The predictions of the yields are unchanged by this uncertainty, but the distribution
of $b_{\mathrm{NN}}$ is varied in a correlated way for each analysis, and propagated
to the predictions of the discriminant output histograms.

\item {\bf Jet {\boldmath $\eta$} Distribution:}  
Checks of the untagged $W+2$ jet control region show that the rate of appearance of
jets at high $|\eta |$ in the data is underestimated by the prediction (Fig.~\ref{fig:mismodeled}~(a)).  
Inaccurate  modeling of the distribution of this variable has a potentially
significant impact on the analysis because of use of the sensitive
variable $Q\times\eta$, which is highly
discriminating for events with jets at large $|\eta|$.  Three 
explanations for the discrepancies between data and MC are possible---beam halo overlapping
with real $W+$jets events, miscalibration of the jet energy scale in
the forward calorimeters, and {\sc alpgen} mismodeling.  We cannot
distinguish between these possibilities with the data, and thus choose
to reweight all Monte Carlo samples by a weighting factor based on the
ratio of the data and Monte Carlo in the untagged sideband, to make alternate
shape templates for the discriminants 
for all Monte Carlo samples.  No corresponding rate uncertainty is applied.

\item {\bf Jet {\boldmath $\Delta R$} Distribution:}  Similarly, the distribution
of $\Delta R(j_1,j_2)=\sqrt{(\Delta\eta)^2+(\Delta\phi)^2}$,
a measure of the angular separation between two jets, is found to be
mismodeled in the untagged control sample (Fig.~\ref{fig:mismodeled}~(b)).  
Modeling this distribution correctly is important because of the use of the
input variable $M_{jj}$, which is highly correlated with $\Delta R(j_1,j_2)$ in our discriminants.
The mismodeling of $\Delta R(j_1,j_2)$ is
believed to be due to the gluon splitting fraction in {\sc alpgen}, but
since this conclusion is not fully supported, we take as a
systematic uncertainty the difference in predictions of all Monte
Carlo based templates after reweighting them using the ratio of the
untagged data to the prediction.

\begin{figure*}
\begin{center}
\subfigure[]{
\includegraphics[width=0.8\columnwidth]{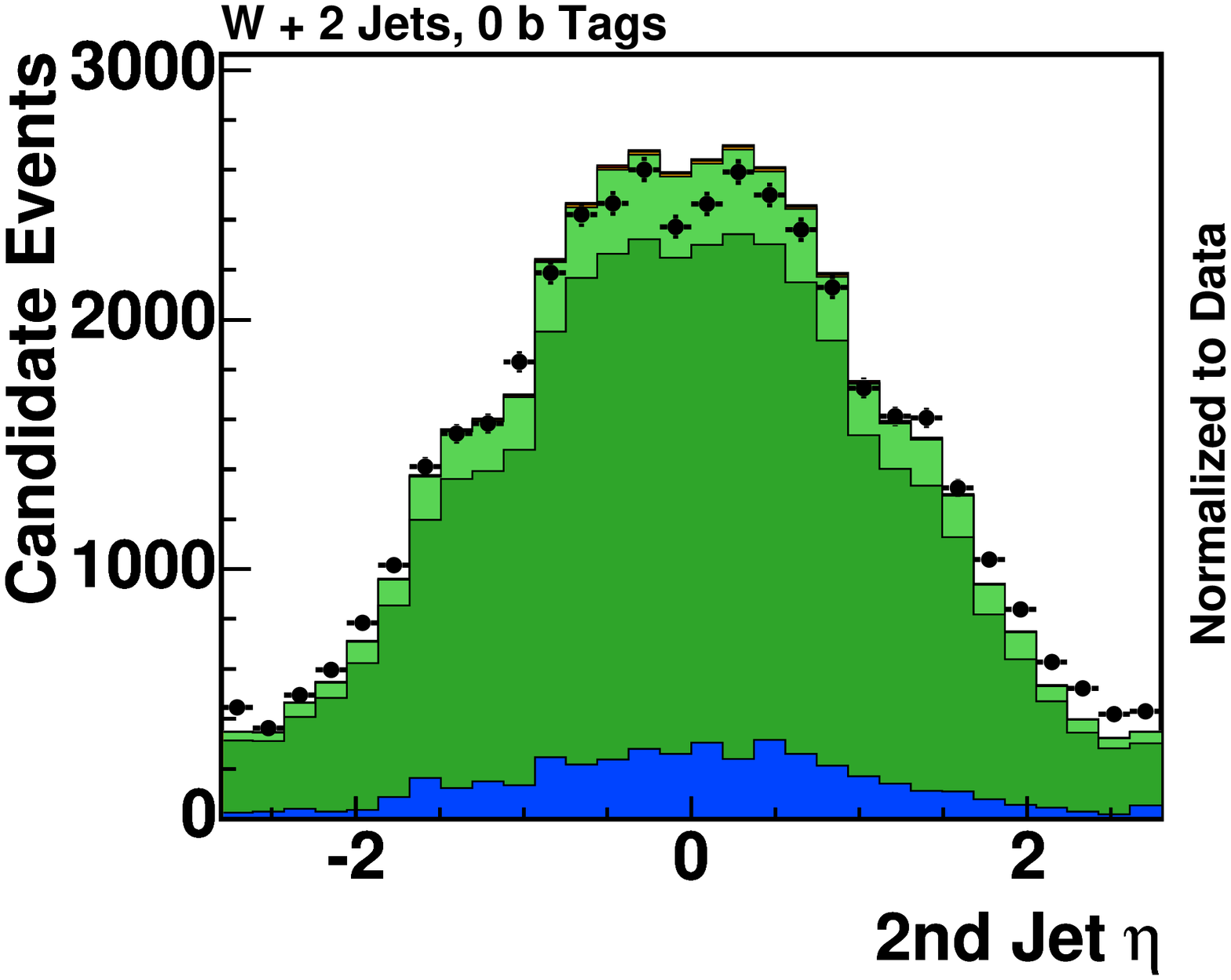}
\label{fig:j2eta}}
\subfigure[]{
\includegraphics[width=0.8\columnwidth]{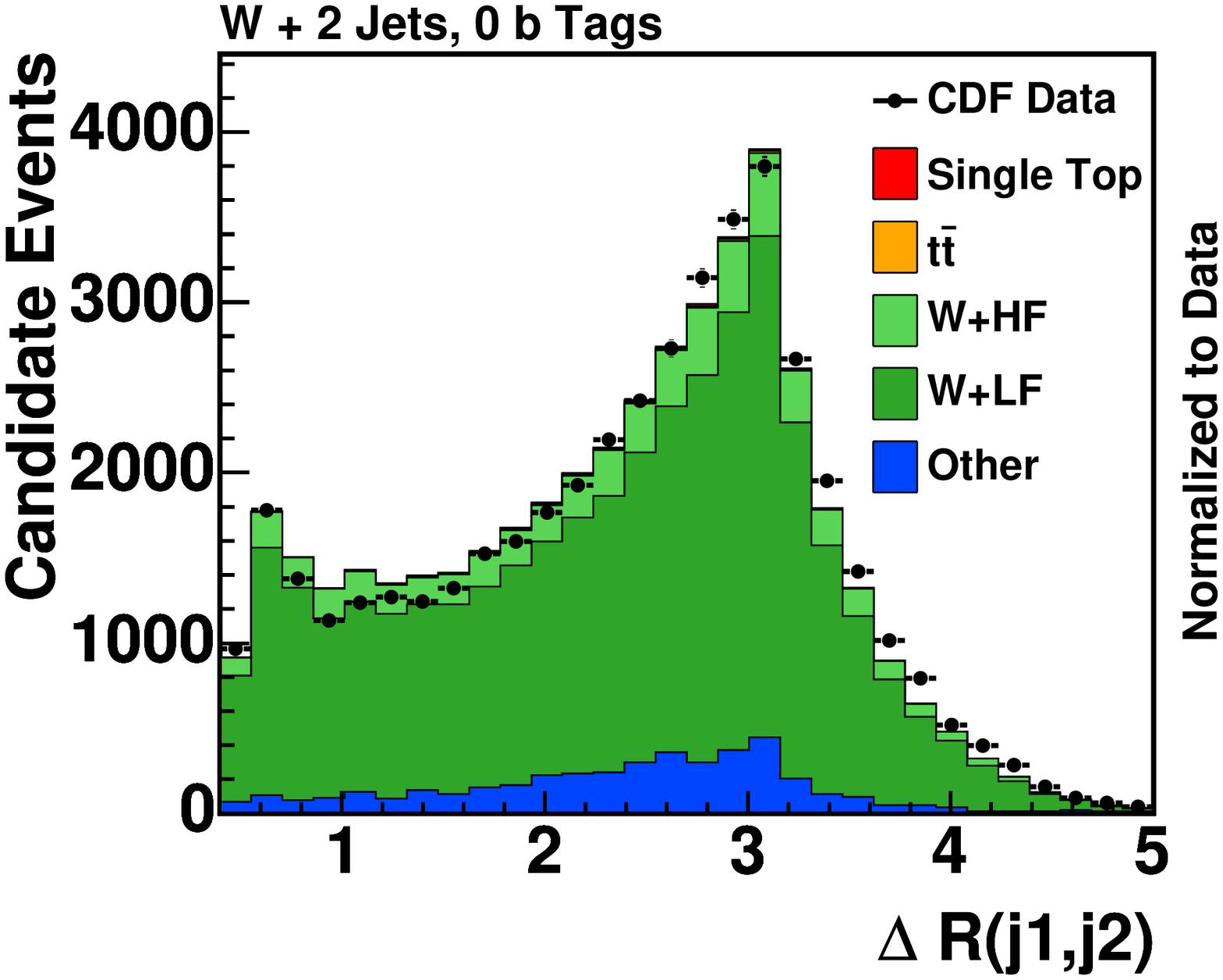}
\label{fig:deltaR}}
\end{center}
\caption{\label{fig:mismodeled}Graphs showing the poor modeling of the
second jet pseudorapidity and the distance between the two jets in the
$\eta$-$\phi$ plane. These are accounted for with systematic
uncertainties on the shapes of the $W$+jets predictions.
The data are indicated by points with error bars, and 
the predictions are shown stacked, with the stacking order following that of the legend.
}
\end{figure*}

\end{itemize}

\begin{table*}[tbh]
\caption{\label{tab:sys}Sources of systematic uncertainty considered
in this analysis.  Some uncertainties are listed as ranges, as the impacts
of the uncertain parameters depend on the numbers of jets and $b$~tags, and
which signal or background component is predicted.
Sources listed below the double line are used only
in the calculation of the $p$-value.}
\begin{center}
\begin{tabular}{lccc}\hline\hline
Source of Uncertainty & Rate & Shape & Processes affected\\
\hline
Jet energy scale & 0--16\% & \Checkmark & all \\
Initial state radiation & 0--11\% & \Checkmark & single top,
$t\bar{t}$ \\
Final state radiation & 0--15\% & \Checkmark & single top,
$t\bar{t}$ \\
Parton distribution functions & 2--3\% & \Checkmark & single top,
$t\bar{t}$ \\
Acceptance and efficiency scale factors & 0--9\% & & single top,
$t\bar{t}$, diboson, $Z/\gamma^*$+jets \\
Luminosity & 6\% & & single top, $t\bar{t}$, diboson, $Z/\gamma^*$+jets \\
Jet flavor separator & & \Checkmark & all \\
Mistag model & & \Checkmark & $W$+light \\
Non-$W$ model & & \Checkmark & Non-$W$\\
Factorization and renormalization scale & & \Checkmark & $Wb\bar{b}$ \\
Jet $\eta$ distribution & & \Checkmark & all \\
Jet $\Delta R$ distribution & & \Checkmark & all \\
Non-$W$ normalization & 40\% & & Non-$W$ \\
$Wb\bar{b}$ and $Wc\bar{c}$ normalization & 30\% & & $Wb\bar{b}$,
$Wc\bar{c}$ \\
$Wc$ normalization & 30\% & & $Wc$ \\
Mistag normalization & 17--29\% & & $W$+light \\
$t\bar{t}$ normalization & 12\% & & $t\bar{t}$ \\
Monte Carlo generator & 1--5\% & & single top \\
\hline\hline
Single top normalization & 12\% & & single top \\
Top mass & 2--12\% & \Checkmark & single top, $t\bar{t}$ \\\hline\hline
\end{tabular}
\end{center}
\end{table*}

\section{\label{sec:Interpretation} Interpretation}

The analyses presented in this paper have two goals:
to evaluate the significance of the excess of events compared with the background prediction,
and to make a precise measurement of the cross section.
These goals have much in common: better separation of signal events from
background events and the reduction of uncertainties help improve both
the cross section measurements and the expected significance if a
signal is truly present.
But there are
also differences.  For example, the 
systematic uncertainty on the signal acceptance
affects the precision of the cross section measurement, but it has almost no
effect on the observed significance level, and only a minor effect on
the predicted significance level; Section~\ref{sec:significancecalc} discusses
this point in more detail.  More importantly, a precision cross
section measurement relies most on increasing acceptance and
understanding the background in a larger sample.  The significance of
an excess, however, can be much larger if one bin in an analysis has a
very low expected background yield and has data in it that are incompatible with that
background, even though that bin may not contribute much information
to the cross section measurement.

The contents of the low signal-to-background bins are important
for the proper interpretation of the high signal-to-background bins.
They serve as signal-depleted control samples which can be used to
help constrain the background predictions.  Not all bins are fully
depleted in signal, and the signal-to-background ratio varies from
very small to about 2:1 in some analyses.  Simultaneous use of all
bins' contents, comparing the observations to the predictions, is
needed to optimally measure the cross section and to 
compute the significance.  
Systematic uncertainties on the predicted rates and
shapes of each component of the background and the two signals ($s$-channel and
$t$-channel), and
also bin-by-bin systematic uncertainties, affect the extrapolation of
the background fits to the signal regions.

These considerations are addressed below, and the procedures for
measuring the cross section and the significance of the excess are
performed separately.  The handling of the systematic
uncertainties is Bayesian, in that priors are assigned for the values
of the uncertain nuisance parameters, the impacts of the nuisance
parameters on the predictions are evaluated, and integrals are
performed as described below over the values of the nuisance
parameters.

\subsection{Likelihood Function}

The likelihood function we use in the extraction of the
cross section and in the determination of the significance is the
product of Poisson probabilities for each bin in each histogram of 
the discriminant output variable of
each channel.  Here, the channels are the non-overlapping data samples
defined by the number of jets, the number of $b$ tags, and whether the charged lepton candidate
is a triggered electron or muon, or whether it was an extended muon coverage candidate
event.  We do not simply add the distributions of the discriminants in these very different
samples because doing so would collect bins with a higher signal purity with those of lower
signal purity, diluting our sensitivity.
The Poisson probabilities are functions of
the number of observed data events in each bin $d_i$ and the predictions in
each bin $\mu_i$, where $i$ ranges from 1 to $n_{\rm{bins}}$.  The
likelihood function is given by
\begin{equation}
L = \prod_{i=1}^{n_{\rm{bins}}} \frac{\mu_i^{d_i}e^{-\mu_i}}{d_i!}.
\end{equation}
The prediction in each bin is a sum over signal and background contributions:
\begin{equation}
\mu_i = \sum_{k=1}^{n_{\rm bkg}} b_{ik} + \sum_{k=1}^{n_{\rm sig}} s_{ik} 
\end{equation}
where $b_{ik}$ is the background prediction in bin $i$ for background
source $k$; $n_{\rm{bkg}}$ is the total number of background
contributions.  The signal is the sum of the $s$-channel and $t$-channel
contributions; $n_{\rm{sig}}=2$ is the number of signal sources, and
the $s_{ik}$ are their predicted yields in each bin.
The predictions $b_{ik}$ and $s_{ik}$ depend on $n_{\rm nuis}$
uncertain nuisance parameters $\theta_m$, where $m = 1 ... n_{\rm nuis}$, one for each
independent source of systematic uncertainty.
These nuisance parameters
are given Gaussian priors centered on zero with unit width, and their impacts
on the signal and background predictions are described in the steps below.

In the discussion below, the procedure for applying systematic shifts to the signal and
background predictions is given step by step, for each kind of systematic uncertainty.
Shape uncertainties are applied first, then bin-by-bin uncertainties, and
finally rate uncertainties.  The
bin-by-bin uncertainties arise from limited Monte Carlo (or data from a
control sample) statistics and are taken to be independent of each other
and all other sources of systematic uncertainty.
The steps are labeled $b^0$ for the central, unvaried
background prediction in each bin, and $b^4$ for the prediction with all systematic uncertainties
applied.

The contribution to a bin's prediction from a given source of shape uncertainty
is modified by linearly
interpolating and extrapolating the difference between the central
prediction $b_{ik}^0$ and the prediction in a histogram corresponding
to a $+1\sigma$ variation $\kappa_{b,ik}^{m+}$ if $\theta_m>0$, and
performing a similar operation using a $-1\sigma$ varied histogram if
$\theta_m<0$:
\begin{equation}
b_{ik}^1 = b_{ik}^0 + \sum_{m=1}^{n_{\rm nuis}}\left\{ 
\begin{array}{r@{\quad:\quad}l}
 (\kappa_{b,ik}^{m+}-b_{ik}^0)\theta_m & \theta_m \ge 0 \\
 (b_{ik}^0-\kappa_{b,ik}^{m-})\theta_m & \theta_m < 0 
\end{array}
  \right. .
\end{equation}
The parameter list is shared between the signal and background predictions because
some sources of systematic uncertainty affect both in a correlated way.
The application of shape uncertainties is not allowed to produce a
negative prediction in any bin for any source of background or signal:
\begin{equation}
b_{ik}^2 = {\rm max}(0,b_{ik}^1).
\end{equation}

Each template histogram, including the systematically varied
histograms, has a statistical
uncertainty in each bin.  These bin-by-bin uncertainties are linearly
interpolated in each bin in the same way as the predicted
values.  This procedure works well when the shape-variation templates
share all or most of the same events, but it overestimates the bin-by-bin
uncertainties when the alternate shape templates are filled with
independent samples.  If the bin-by-bin uncertainty on
$b_{ik}^0$ is $\delta_{b,ik}^0$, and the bin-by-bin uncertainty on
$b_{ik}^{m\pm}$ is $\delta_{b,ik}^{m\pm}$, then
\begin{equation}
\delta_{b,ik}^1 = \delta_{b,ik}^0 + \sum_{m=1}^{n_{\rm nuis}}\left\{ 
\begin{array}{r@{\quad:\quad}l}
 (\delta_{b,ik}^{m+}-\delta_{b,ik}^0)\theta_m & \theta_m \ge 0 \\
 (\delta_{b,ik}^0-\delta_{b,ik}^{m-})\theta_m & \theta_m < 0 
\end{array}
  \right. .
\end{equation}
 Each bin of each background has a
nuisance parameter $\eta_{b,ik}$ associated with it.
\begin{equation}
b_{ik}^3 = b_{ik}^2 + \delta_{b,ik}^1\eta_{b,ik},
\end{equation}
where $\eta_{b,ik}$ is drawn from a Gaussian centered on zero with unit width when integrating over it.
If $b_{ik}^3 < 0$, then $\eta_{b,ik}$ is re-drawn from that Gaussian.

Finally, rate uncertainties are applied multiplicatively.  If the
fractional uncertainty on $b_{ik}^0$ due to nuisance parameter $m$ is
$\rho_{b,ik}^{m+}$ for a $+1\sigma$ variation and it is
$\rho_{b,ik}^{m-}$ for a negative variation, then a quadratic function
is determined to make a smooth application of the nuisance parameter
to the predicted value:
\begin{widetext}
\begin{equation}
b_{ik} = b^4_{ik} = b_{ik}^3\prod_{m=1}^{n_{\rm nuis}}\left(1+\frac{\rho_{b,ik}^{m+}+\rho_{b,ik}^{m-}}{2}\theta_m^2
+  \frac{\rho_{b,ik}^{m+}-\rho_{b,ik}^{m-}}{2}\theta_m \right).
\end{equation}
\end{widetext}
The rate uncertainties are applied multiplicatively because most of
them affect the rates by scale factors, such as the luminosity and
acceptance uncertainties, and they are applied last because they affect
the distorted shapes in the same way as the undistorted shapes.  Multiple shape
uncertainties are treated additively because most of them correspond
to events migrating from one bin to another.

The signal predictions are based on their Standard
Model rates.  These are scaled to test other values of the single top quark
production cross sections:
\begin{equation}
s_{ik} = s_{ik}^4\beta_{k},
\end{equation}
where $\beta_s$ scales the $s$-channel signal and $\beta_t$ scales the $t$-channel signal, and
the ``4'' superscript indicates that the same chain of application of nuisance parameters is applied
to the signal prediction as is applied to the background.

The likelihood is a function of the observed data ${\bf D} =
\{d_i\}$, the signal scale factors
$\mbox{\boldmath$\beta$}=\{\beta_s,\beta_t\}$, the nuisance parameters
$\mbox{\boldmath$\theta$} = \{\theta_m\}$ and $\mbox{\boldmath$\eta$}
= \{\eta_{s,ik},\eta_{b,ik}\}$, the central values of the
signal and background predictions $\mbox{\boldmath$s$} = \{s^0_{ik}\}$
and $\mbox{\boldmath$b$} = \{b^0_{ik}\}$, and the rate, shape, and
bin-by-bin uncertainties $\mbox{\boldmath$\rho$} =
\{\rho^{m\pm}_{b,ik},\rho^{m\pm}_{s,ik}\}$, $\mbox{\boldmath$\kappa$}
= \{\kappa^{m\pm}_{b,ik},\kappa^{m\pm}_{s,ik}\}$,
$\mbox{\boldmath$\delta$} =
\{\delta^0_{b,ik},\delta^{m\pm}_{b,ik},\delta^0_{s,ik},\delta^{m\pm}_{s,ik}\}$:
\begin{equation}
L = L({\bf D}|\mbox{\boldmath $\beta,\theta,\eta,s,b,\rho,\kappa,\delta$}).
\end{equation}

\subsection{Cross Section Measurement}

Because the signal template shapes and the
\ttbar\ background template rates and shapes are functions of $m_t$, we quote the single top quark cross section assuming a top quark mass of
$m_t=175$ \gevcc\  and also evaluate $\partial\sigma_{s+t}/\partial m_t$.  We therefore do not include the uncertainty on the top quark
mass when measuring the cross section.

\subsubsection{Measurement of $\sigma_{s+t}$}

We measure the total cross section of single top quark production
$\sigma_{s+t}$, assuming the SM ratio between $s$-channel and
$t$-channel production: $\beta_s=\beta_t\equiv\beta$.  We use a
Bayesian marginalization technique~\cite{pdgstats} to incorporate the
effects of systematic uncertainty:
\begin{widetext}
\begin{equation}
L^\prime(\beta) = \int L({\bf D}|\beta,\mbox{\boldmath $\theta,\eta,s,b,\rho,\kappa,\delta$})
\pi(\mbox{\boldmath $\theta$})\pi(\mbox{\boldmath $\eta$})
d\mbox{\boldmath $\theta$}d\mbox{\boldmath $\eta$},
\end{equation}
\end{widetext}
where the $\pi$ functions are the Bayesian priors assigned to each
nuisance parameter.  The priors are unit Gaussian functions centered on zero which
are truncated whenever the value of a nuisance parameter would result in a non-physical
prediction.
The measured cross section
corresponds to the maximum of $L^\prime$, which occurs at
$\beta^{\rm{max}}$:
\begin{equation}
\sigma_{s+t}^{\rm{meas}} = \sigma_{s+t}^{\rm{SM}}\beta^{\rm{max}}.
\end{equation}
The uncertainty corresponds to the shortest
interval $\left[\beta_{\rm{low}},\beta_{\rm{high}}\right]$
containing 68\% of the integral of the posterior, assuming a
uniform positive prior in $\beta$ $\pi(\beta)=1$:
\begin{equation}
0.68 = \frac{\int_{\beta_{\rm{low}}}^{\beta_{\rm{high}}}
L^\prime(\beta)\pi(\beta) d\beta} {\int_0^\infty
L^\prime(\beta)\pi(\beta) d\beta}.
\end{equation}
This prescription has the property that the numerical value of the posterior on the
low end of the interval is equal to that on the high end of the interval.

Following the example of other top quark properties analyses, the
single top quark cross section is measured assuming a top quark mass of
175~\gevcc.  This measurement is repeated with separate Monte Carlo
samples and background estimates generated with masses of 
170~\gevcc\ and 180~\gevcc, and the result is used to find $d\sigma_{s+t}/dm_{t}$.

\subsubsection{Extraction of Bounds on $|V_{tb}|$}

The parameter 
\begin{equation}
\beta =  \frac{\sigma_{s+t}^{\rm{meas}}}{\sigma_{s+t}^{\rm{SM}}}
\end{equation}
is identified in the Standard Model as $|V_{tb}|^2$, under the
assumption that $|V_{td}|^2+|V_{ts}|^2 \ll |V_{tb}|^2$, and that new
physics contributions affect only $|V_{tb}|$.  The theoretical uncertainty on
$\sigma_{s+t}^{\rm{SM}}$ must be introduced for this calculation.  The
95\% confidence lower limit on $|V_{tb}|$ is calculated by requiring $0 \le
|V_{tb}| \le 1$ and finding the point at which 95\% of the likelihood
curve lies to the right of the point.  This calculation uses a prior
which is flat in $|V_{tb}|^2$.

\subsection{Check for Bias}

As a cross-check of the cross-section measurement method, simulated
pseudoexperiments were generated, randomly fluctuating the systematically uncertain nuisance
parameters, propagating their impacts on the predictions of each signal and background
source in each bin of each histogram, and drawing random Poisson pseudodata in those bins from the fluctuated means.
 Samples of pseudoexperiments were generated assuming different signal cross sections,
and the cross section posterior was formed for each one in the same way as
it is for the data.  We take the value of the cross section that maximizes the
posterior as the best fit value, and calculate the total uncertainty on it in the
same way as for the data.  The resulting pull distribution is a unit Gaussian, provided
that the input cross section for the pseudoexperiments is sufficiently far away from zero.

\begin{figure}
\begin{center}
\vskip 0.5cm
\includegraphics[width=0.9\columnwidth]{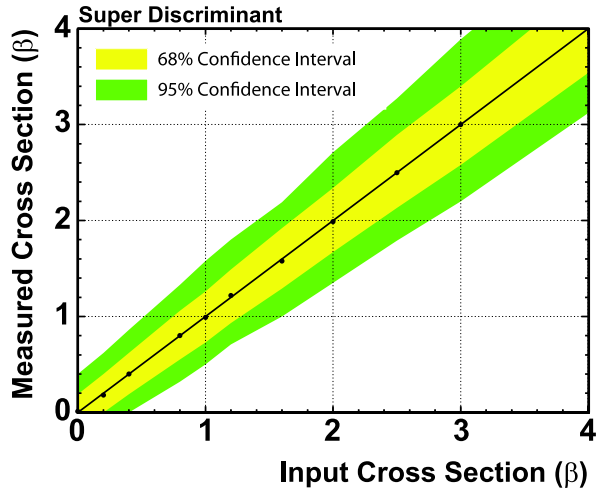}
\end{center}
\caption{\label{fig:biaschecks} Check of the bias of the cross-section measurement method
using pseudoexperiments, for the super discriminant
combination described in Section~\protect{\ref{sec:Combination}}.  The points indicate the median
fit cross section, and the bands show the 68\% and 95\% quantiles of the distribution of the fitted
cross section, as functions of the input cross section.  A line is drawn showing
equal input and fitted cross sections; it is not a fit to the points.}
\end{figure}

Because the prior for the cross section does not allow negative values,
the procedure described here cannot produce a negative cross section
measurement.  For an input cross section of zero, half of the pseudoexperiments
will have measured
cross sections that are exactly zero, and the other half form a
distribution of positive cross sections.  We therefore compare the
median measured cross section with the input cross section of the pseudoexperiments
because the average measured cross section is biased.  Distributions
of 68\% and 95\% of extracted cross sections centered on the median
are shown as a function of the input cross section in
Fig.~\ref{fig:biaschecks}, demonstrating that the measurement technique does not
introduce bias for any value of the cross section used as input to the
pseudoexperiments.  These checks were performed for each
analysis; Figure~\ref{fig:biaschecks} shows the results for the super discriminant
combination, which is described in Section~\ref{sec:Combination}. 
Some nuisance parameters have asymmetric
priors, and the inclusion of their corresponding systematic
uncertainties will shift the fitted cross section.  This is not a
bias which must be corrected but rather it is a consequence of our
belief that the values of the uncertain parameters are not
centered on their central values.

\subsection{Significance Calculation}
\label{sec:significancecalc}

The other goal of the search is to establish observation of single top quark
production.  The
significance is summarized by a $p$-value, the probability of observing an
outcome of an experiment at least as signal-like as the one
observed, assuming that a signal is absent.  We follow the
convention that a $p$-value less than $1.35\times 10^{-3}$
constitutes evidence for a signal, and that a $p$-value less
than $2.87\times 10^{-7}$ constitutes a discovery.  These are the
one-sided integrals of the tails of a unit Gaussian distribution
beyond $+3\sigma$ and $+5\sigma$, respectively.

We rank experimental outcomes on a one-dimensional scale using the
likelihood ratio~\cite{Neyman:1933}
\begin{widetext}
\begin{equation}
-2\ln Q = -2\ln\frac{L({\bf D}|\beta,\mbox{\boldmath
${\hat{\theta}}$}_{\rm SM},\mbox{\boldmath{$\hat{\eta}$}}_{\rm SM},{\mbox{\boldmath $s$}}={\mbox{\boldmath $s$}}_{\rm{SM}},
\mbox{\boldmath $b,\rho,\kappa,\delta$})}
{L({\bf D}|\beta,\mbox{\boldmath ${\hat{\theta}_{0}},{\hat{\eta}_{0}},s$}=\mbox{\boldmath  $0, b,\rho,\kappa,\delta$})},
\end{equation}
\end{widetext}
where $\mbox{\boldmath ${\hat{\theta}}$}_{\rm SM}$ and $\mbox{\boldmath${\hat{\eta}}$}_{\rm SM}$ 
are the best-fit values of the nuisance parameters
which maximize $L$ given the data ${\bf{D}}$, assuming the single top quark
signal is present at its SM rate, and $\mbox{\boldmath
${\hat{\theta}_{0}}$}$ and $\mbox{\boldmath ${\hat{\eta}_{0}}$}$
are the best-fit values of the nuisance parameters which maximize $L$
assuming that no single top quark signal is present.  These fits are
employed not  to incorporate
systematic uncertainties, 
but to optimize the sensitivity.  Fits to other nuisance
parameters do not appreciably improve the sensitivity of
the search and are not performed.  Therefore, only the most important
nuisance parameters are fit for: the
heavy-flavor fraction in $W+$jets events and the mistag rate.

The desired $p$-value is then
\begin{equation}
p = p(-2\ln Q\le -2\ln Q_{\rm{obs}}|{\bf s}={\bf 0}),
\end{equation}
since signal-like outcomes have smaller values of $-2\ln Q$ than
background-like outcomes.  Systematic uncertainties are included not
in the definition of $-2\ln Q$, which is a known function of the
observed data and is not uncertain, but rather in the expected
distributions of $-2\ln Q$ assuming $\mbox{\boldmath $s$} = \mbox{\boldmath $0$}$ or 
$\mbox{\boldmath $s$} = \mbox{\boldmath $s$}_{\rm SM}$, 
since our expectation is what is uncertain.
These uncertainties are included in a Bayesian fashion by averaging
the distributions of $-2\ln Q$ over variations of the nuisance
parameters, weighted by their priors.  In practice, this is done by
filling histograms of $-2\ln Q$ with the results of simulated
pseudoexperiments, each one of which is drawn from predicted
distributions after varying the nuisance parameters according to their prior
distributions.  The fit to the main nuisance parameters insulates
$-2\ln Q$ from the fluctuations in the values of the
nuisance parameters and optimizes our sensitivity in the
presence of uncertainty.

The measured cross section and the $p$-value depend
on the observed data.  We gauge the
performance of our techniques not based on the single random outcome
observed in the data but rather by the sensitivity -- the 
distribution of outcomes expected if a signal is present.  The
sensitivity of the cross section measurement is given by the median
expected total uncertainty on the cross section, and the sensitivity
of the significance calculation is given by the median expected
significance.  The distributions from which these sensitivities are
computed are Monte Carlo pseudoexperiments with all nuisance
parameters fluctuated according to their priors.  Optimizations 
of the analyses were based on the median expected
$p$-values, without reference to the observed data.  Indeed, the data events passing the event selection
requirements were hidden during the analysis optimization.

In the computation of the observed and expected $p$-values, we include all sources of
systematic uncertainty in the pseudoexperiments, including the theoretical
uncertainty in the signal cross sections and the top quark mass.  Because the observed $p$-value
is the probability of an upward fluctuation of the background prediction to the observed data, with the
outcomes ordered as signal-like based on $-2\ln Q$, the observed $p$-value
depends only weakly on the predicted signal model, and in particular, almost 
not at all on the predicted signal rate.  Hence the inclusion of the signal rate systematic uncertainty
in the observed $p$-value has practically no impact, and the shape uncertainties in the signal model also
have little impact (the background shape uncertainties are quite important though).
On the other hand, the expected $p$-value and the cross
section measurement depend on the signal model and its uncertainties; a large signal is expected
to be easier to discover than a small signal, for example.

\section{\label{sec:Combination} Combination}

The four analyses presented in Section~\ref{sec:Multivariate} each seek to
establish the existence of single top quark production and to measure
the production cross section, each using the same set of selected events.
Furthermore, the same models of the signal and background expectations are
shared by all four analyses.
We therefore expect the results to have a high degree of statistical and
systematic correlation.  Nonetheless, the techniques used to separate the signal
from the background are different and are not guaranteed to be fully optimal
for observation or cross section measurement purposes; the figures of merit
optimized in the construction of each of the discriminants are not directly related
to either of our goals, but instead are synthetic functions designed to be easy to
use during the training, such as the Gini function~\cite{gini} used by the BDT analysis, and
a sum of classification errors squared used by the neural network analysis.

The discriminants all perform well in separating the expected signal from the
expected background, and in fact their values are highly correlated, event to event, as is expected,
since they key on much of the same input information, but in different ways.  The coefficients
of linear correlation between the four discriminants vary between 0.55 and 0.8, depending
on the pair of discriminants chosen and the data or Monte Carlo sample used to
evaluate the correlation.  Since any invertible function of a discriminant variable
has the same separating power as the variable itself, and since the coefficients of linear
correlation between pairs of variables change if the variables are transformed, these
coefficients are not particularly useful except to verify that indeed the results are
highly, but possibly not fully, correlated.  

As a more relevant indication of how correlated the analyses are,
pseudoexperiments are performed with fully simulated Monte Carlo
events analyzed by each of the analyses, and the correlations between the best-fit
cross section values are computed.  The coefficients of linear correlation of the
output fit results are given in Table~\ref{tab:corrxs}.

\begin{table}[ht]
\begin{center}
\caption[]{Correlation coefficients between pairs of cross section
  measurements evaluated on Monte Carlo pseudoexperiments.}
\label{tab:corrxs}
\begin{tabular}{lcccc} \hline\hline
   & LF     & ME    & NN    & BDT  \\ 
LF & 1.0    & 0.646 & 0.672 & 0.635 \\ 
ME & ---    & 1.0   & 0.718 & 0.694 \\ 
NN & ---    & ---   & 1.0   & 0.850 \\ 
BDT& ---    & ---   & ---   & 1.0 \\\hline\hline
\end{tabular}
\end{center}
\end{table}

The four discriminants, LF, ME, NN, BDT make use of different observable quantities as inputs.
In particular, the LF, NN, and BDT discriminants use variables that make assignments
of observable particles to hypothetical partons from single top quark production, while the ME
method integrates over possible interpretations.   Furthermore, since the correlations
between pairs of the four discriminants are different for the different physics processes,
we expect this information also to be useful in separating the signal from the background processes.
In order to extract a cross section and a significance, we need to interpret each event
once, and not four times, in order for Poisson statistics to apply.  We therefore choose
to combine the analyses by forming a super discriminant, which is a scalar function of the four
input discriminants, and which can be evaluated for each event in the data and each event in
the simulation samples.  The functional form we choose is a neural network, similar to that
used in the 2.2 fb$^{-1}$ single top quark combination at CDF~\cite{Aaltonen:2008sy} as well as the recent $H \to
WW$ search at CDF~\cite{Aaltonen:2008ec}.  The distributions of the super discriminant are used to
compute a cross section and a significance in the same way as is done for the component analyses.

In order to train, evaluate, and make predictions which can be compared with the observations for the
super discriminant, a common set of events must be analyzed in the ME, NN, LF, and BDT frameworks.
The discriminant values are collected from the separate analysis teams
for each data event and for each event simulated in Monte Carlo.
Missing events or extra events in one or more
analyses are investigated and are restored or omitted as discrepancies are found and understood.
The $W+$jets predictions in particular involve weighting Monte Carlo events by mistag probabilities
and by generator luminosity weights, and these event weights are also unified across four
analysis teams.  The procedure of making a super discriminant combination provides a strong level
of cross checks between analysis teams.  It has identified many kinds of simple mistakes
and has required us to correct them before proceeding.  All of these crosschecks were performed
at the stage in which event data were exchanged and before the training
of the fnal discriminant, preserving the blindness of the result.

We further take the opportunity during the combination procedure to optimize our final
discriminant for the goal that we set, that is, to maximize the probability of observing single top quark
production.   A typical approach to neural network training uses a gradient descent
method, such as back-propagation, to minimize the classification error,
defined by $\sum{(o_i-t_i)^2}$, where $o_i$ is the output of the
neural network and $t_i$ is the desired output, usually zero for
background and one for signal.  Although back-propagation is a powerful
and fast technique for training neural networks, it is not necessarily
true that minimizing the classification error will
provide the greatest sensitivity in a search.  The best choice is to use the
median expected $p$-value for discovery of single top quark production as the figure of merit to optimize,
but it cannot be computed quickly.  Once a candidate network is proposed, the Monte Carlo
samples must be run through it, the distributions made, and many millions of pseudoexperiments
run in order to evaluate its discovery potential.  Even if a more lightweight figure of merit
can be computed from the predicted distributions of the signals and background processes, the step
of reading through all of the Monte Carlo samples limits the number of candidate neural networks that
can be practically considered. 

We therefore use the novel neural network training method of Neuro-Evolution,
which uses genetic algorithms instead of back-propagation, to optimize our
networks.  This technique allows us to compute an arbitrary figure of merit for a particular
network configuration which depends on all of the training events and not just
one at a time.  The software package we use
here is Neuro-Evolution of Augmenting Topologies
({\sc neat})~\cite{neat}.  {\sc neat} has the ability to optimize both the inter-node
weights and the network topology, adding and rearranging nodes as 
needed to improve the performance. 

We train the {\sc neat} networks using half of the events in each Monte Carlo sample,
reserving the other half for use in predicting the outcomes in an unbiased way,
and to check for overtraining.  All background processes are included in the training
except non-$W$ because the non-$W$ sample suffers from extremely low
statistics.  The output values are stored
in histograms which are used for the figure of merit calculation. 
We use two figures of merit which are closely related to the median expected
$p$-value, but which can be calculated much more quickly:
\begin{description}
   \item[``$o$-value'']  This figure of merit (so named because it is
   closely related to the expected $p$-value) is obtained from an
   ensemble of pseudoexperiments by taking the difference in the
   median of the test statistic $-2\ln Q$ for the background-only and signal
   plus background hypotheses, divided by the quadrature sum of the
   widths of those distributions:
   \begin{equation}
      o = \frac{-2\ln Q_B^{\rm{med}} +2\ln Q_{S+B}^{\rm{med}}}{\sqrt{(\Delta2\ln Q_B)^2+(\Delta 2\ln Q_{S+B})^2}}.
   \end{equation}
  Figure~\ref{fig:results}(c) shows the distributions of $-2\ln Q$ separately for $S+B$ and $B$-only
  pseudoexperiments for the final network chosen.
  Typically, 2500 pseudoexperiments give a precision of roughly 1-2\%
   and require
   one to two minutes to calculate.  
   This is still too slow to be used directly in the evolution, but it
   is used at the end to select the best network from a sample of
   high-performing networks identified during the evolution.  This
   figure
   of merit includes all rate and shape systematic uncertainties. 
   \item[Analytic Figure of Merit]  As a faster alternative to the
   figure of merit defined above, we calculate the quadrature sum of
   expected signal divided by the square root of the expected
   background ($s/\sqrt{b}$) in each bin of each histogram.
   To account for the effects of finite Monte Carlo
   statistics, this figure of merit is calculated repeatedly,
   each time letting the value of the expected signal and
   background processes fluctuate according to a Gaussian
   distribution with a width corresponding to the Monte Carlo statistical error
   on each bin.  The median of these trials is
   quoted as the figure of merit.   This figure of merit does not include rate
   and shape systematic uncertainties.
\end{description}

The network training procedure also incorporates an 
optimization of the binning of the histograms of the network output.  In general,
the sensitivity is increased by separating events into bins of different purity;
combining the contents of bins of different purity degrades our ability 
to test for the existence of the signal and to measure the cross section.
Competing against our desire for fine
gradations of purity is our need to have solid predictions of the signal and background
yields in each bin with reliable uncertainties -- binning the output histogram 
too finely can result in an overestimate
of the sensitivity due to downward fluctuations in the Monte Carlo background predictions.
Care is taken here, as described below, to allow the automatic binning optimization to maximize
our sensitivity without overestimating it.

The procedure, applied to each channel separately, is to first use a fixed binning of 100 bins
in the neural network output from zero to one.  The network output may not necessarily fill
all 100 bins; different choices of network parameters, which are optimized by the training,
will fill different subsets of these bins.
To avoid problems with Monte Carlo statistics at the extreme ends of the distributions, bins
at the high end of the histogram are grouped together, and similarly at the low end,
sacrificing a bit of separation of signal from background for more robust predictions.  At each step, the horizontal
axis is relabeled so that the histogram is defined between zero (lowest signal purity) and one
(highest purity).
The bins are grouped first so that there are no bins with a total background prediction of zero.
Next, we require that the histograms have a monotonically decreasing purity as the output variable
decreases from one towards zero.  If a bin shows an anomalously high purity, its contents are collected with those
of all bins with higher network outputs to form a new end bin.  Finally, we require that on the high-purity
side of the histogram, the background prediction does not drop off too quickly.
We expect $\ln{\int_{x}^{1}{B}}\propto\ln{\int_{x}^{1}{S}}$ for all $x$ in the highest purity region of the histogram.
If the background decreases at a faster rate, we group the bins on the high 
end together until this condition is met. After this procedure, we achieve a signal-to-background ratio exceeding 
5:1 in the highest-discriminant output bins in the two-jet, one $b$-tag sample.

\begin{figure*}
\begin{center}
\subfigure[]{
\includegraphics[width=0.65\columnwidth]{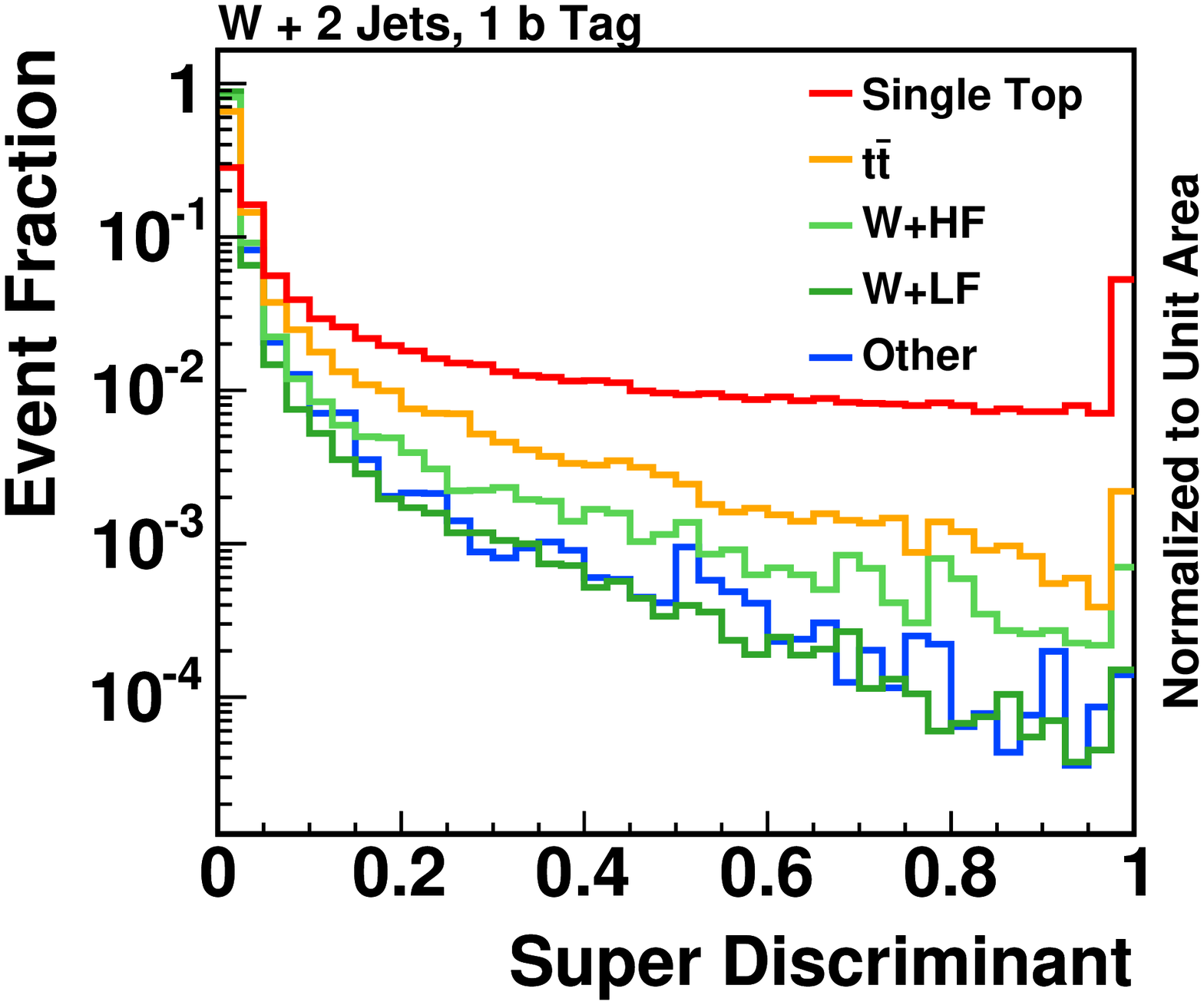}
\label{fig:COMBO2j1t_shape}}
\subfigure[]{
\includegraphics[width=0.65\columnwidth]{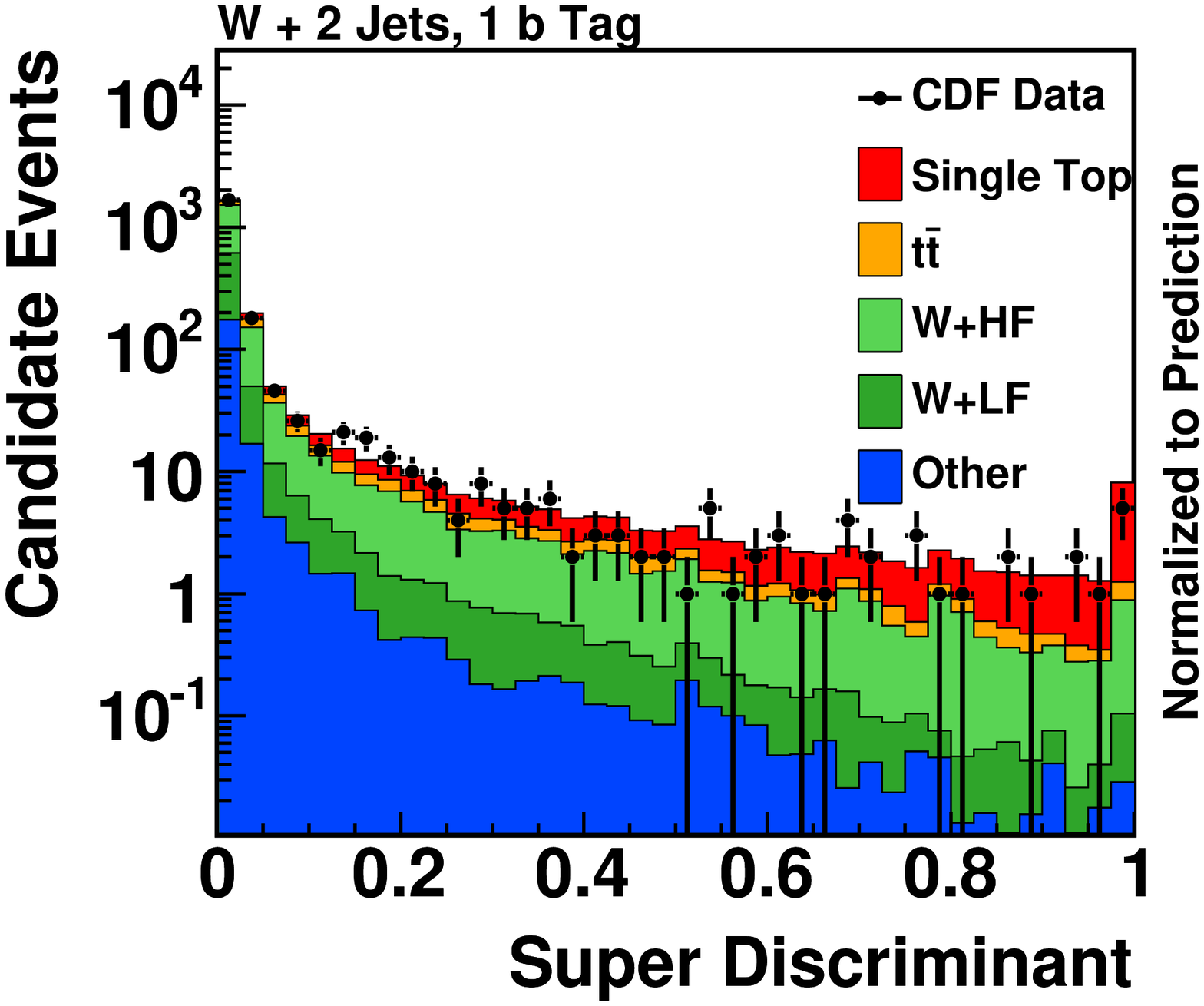}
\label{fig:COMBO2j1t}} \\
\subfigure[]{
\includegraphics[width=0.65\columnwidth]{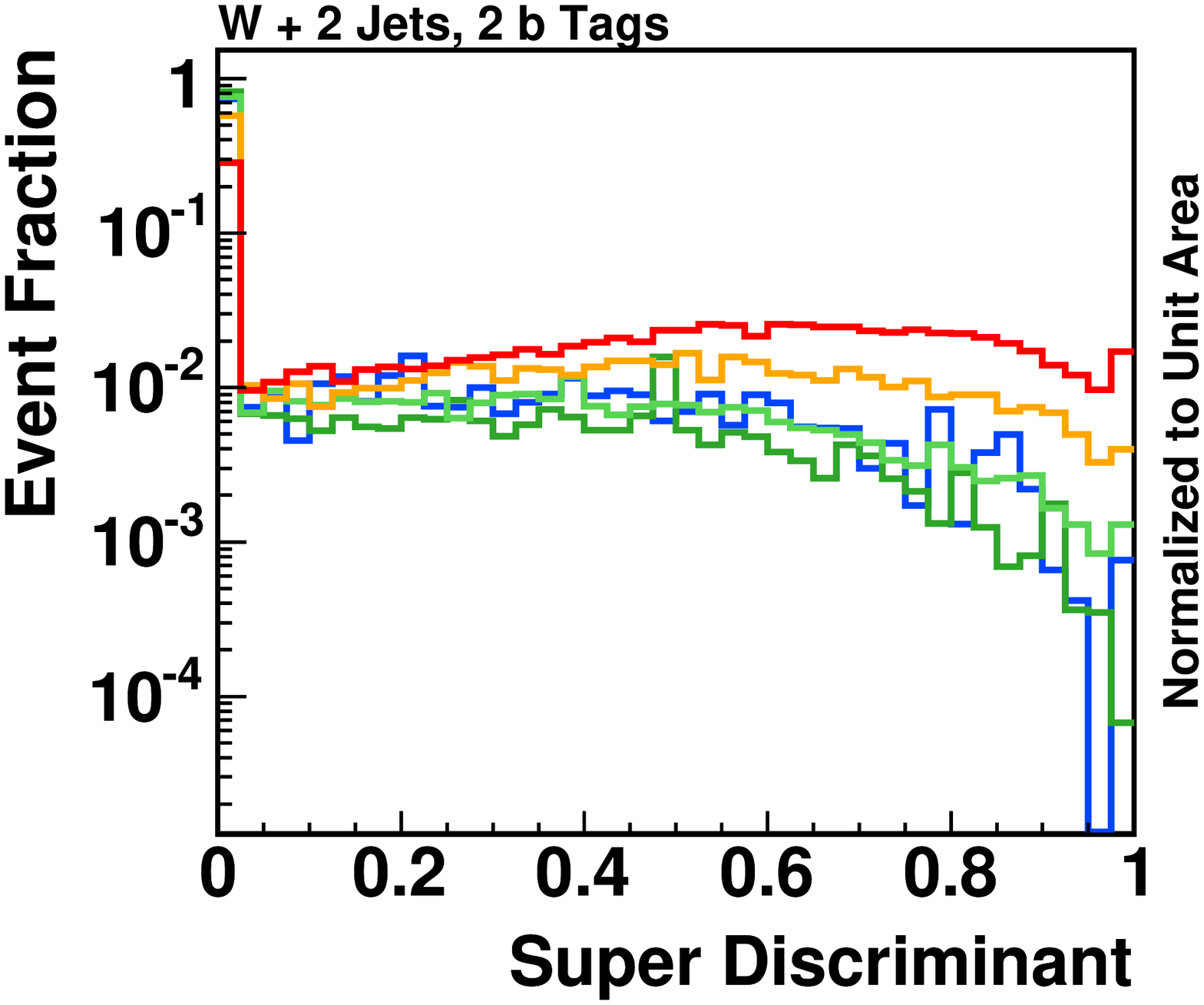}
\label{fig:COMBO2j2t_shape}}
\subfigure[]{
\includegraphics[width=0.65\columnwidth]{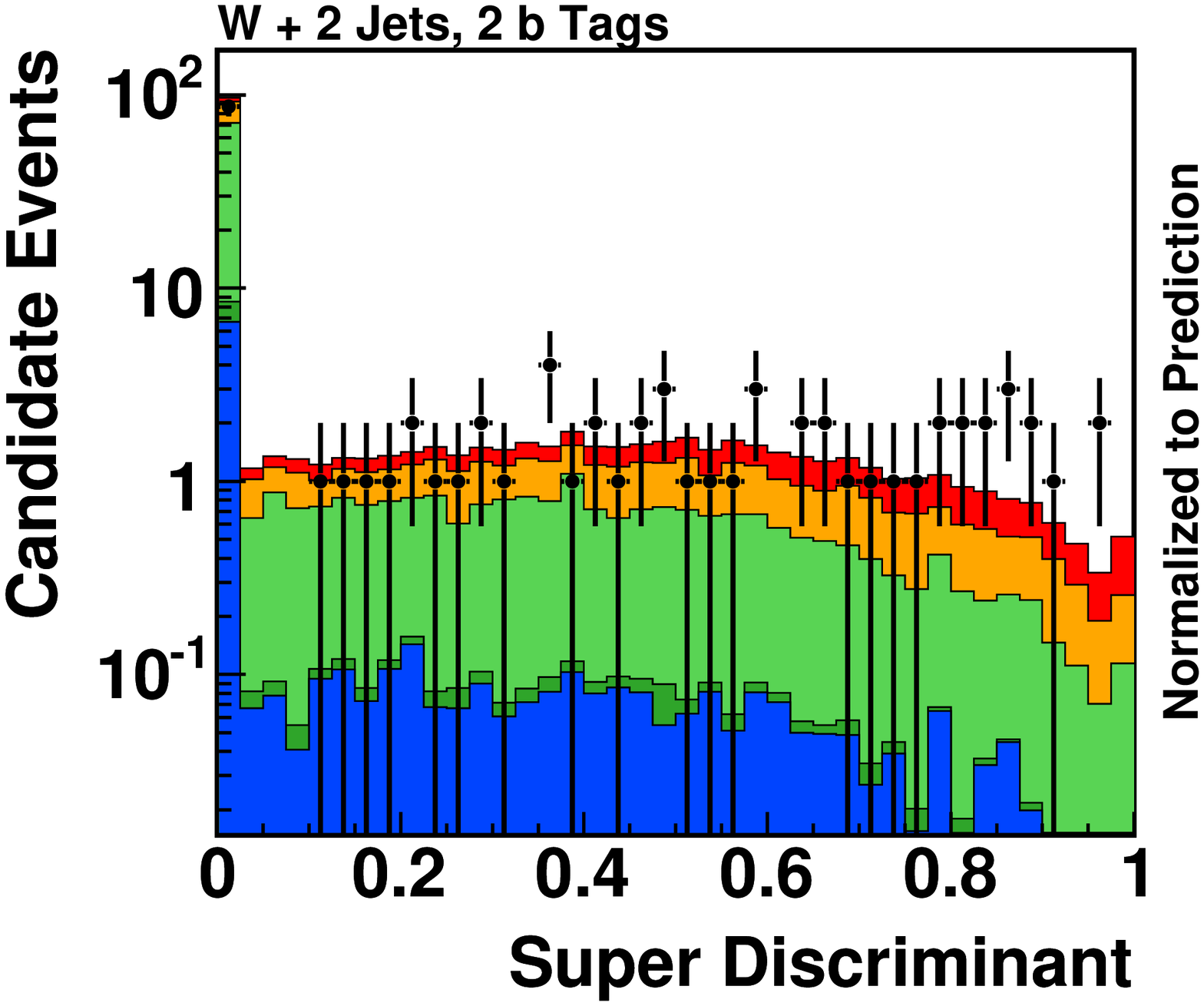}
\label{fig:COMBO2j2t}} \\
\subfigure[]{
\includegraphics[width=0.65\columnwidth]{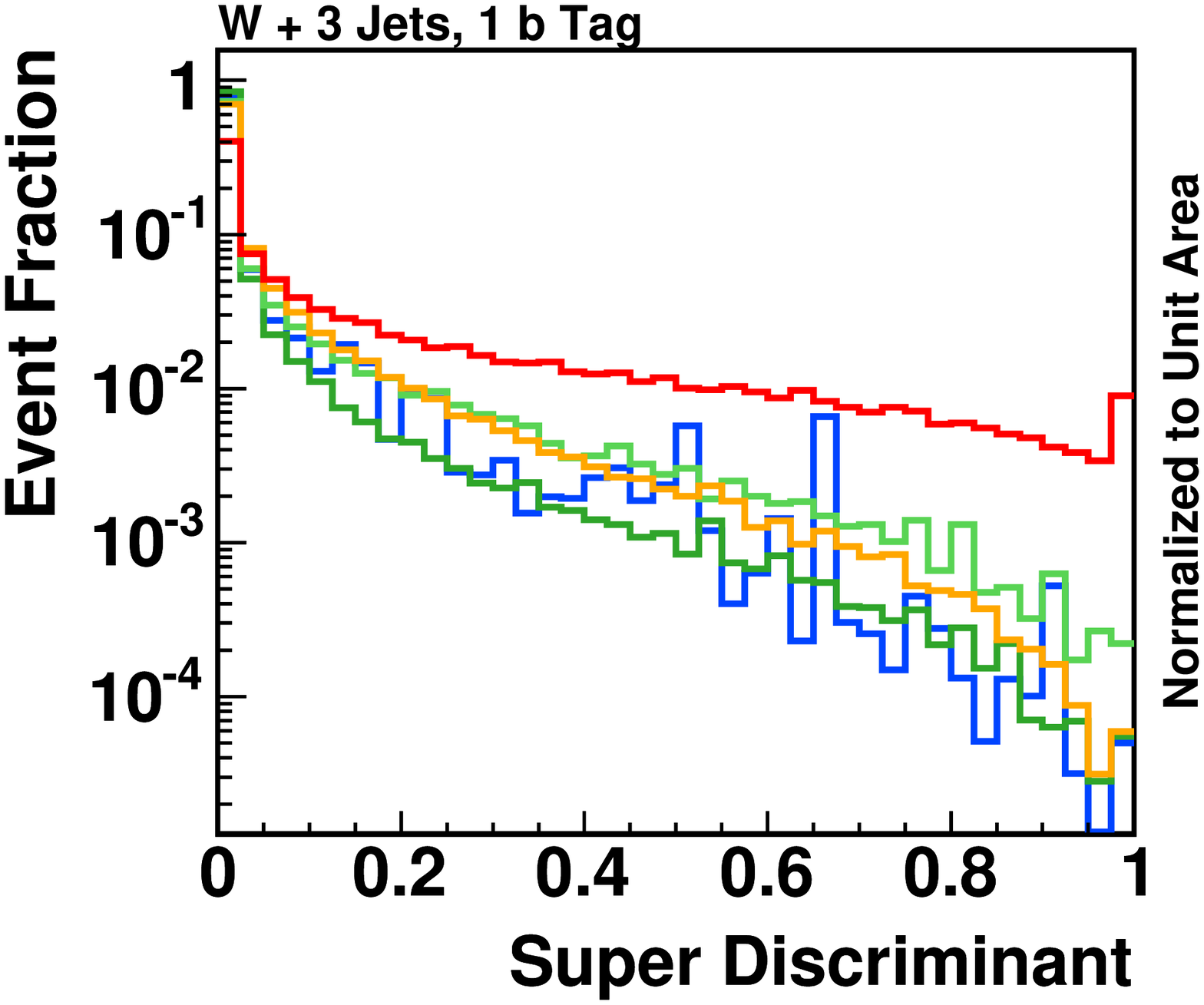}
\label{fig:COMBO3j1t_shape}}
\subfigure[]{
\includegraphics[width=0.65\columnwidth]{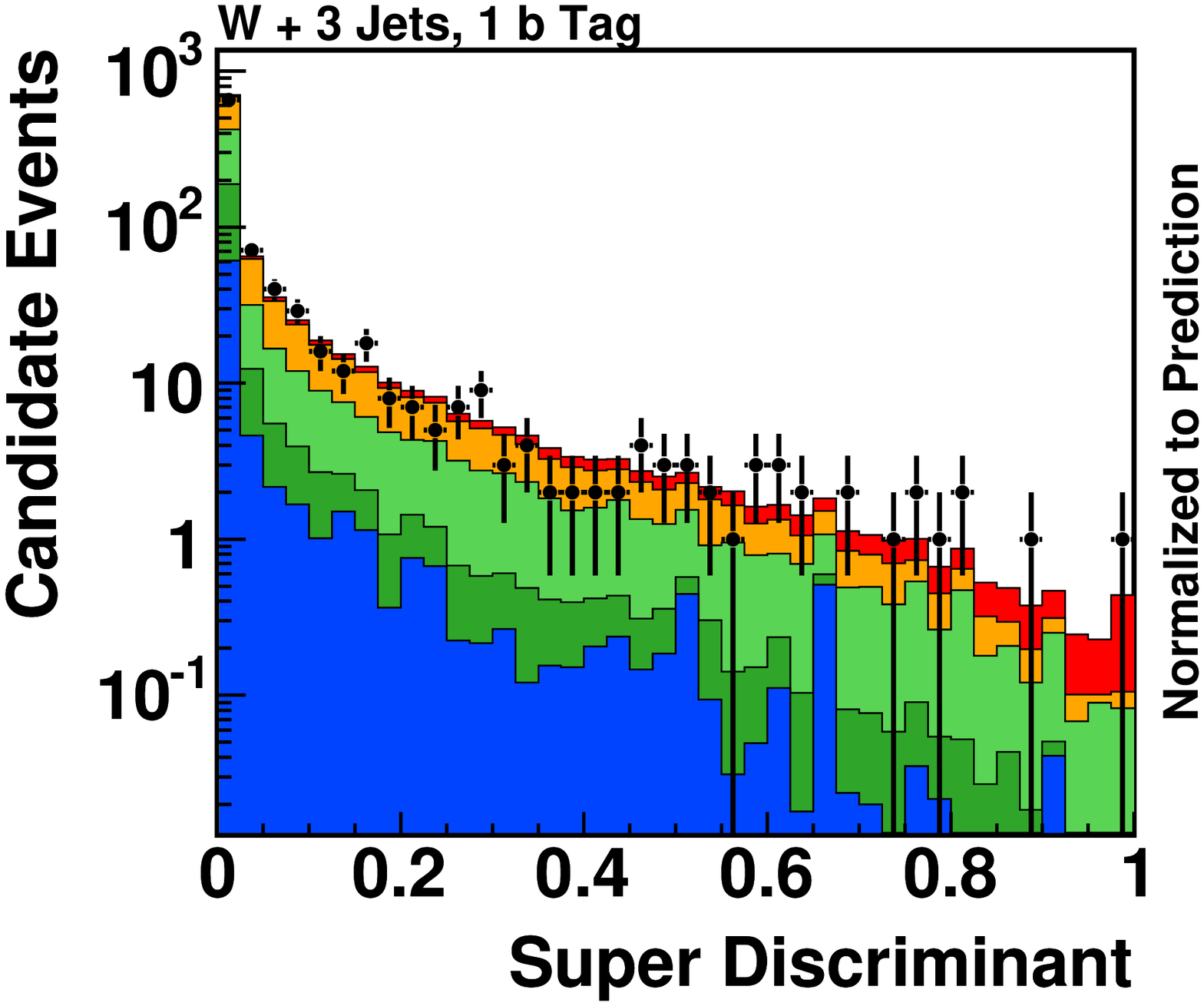}
\label{fig:COMBO3j1t}} \\
\subfigure[]{
\includegraphics[width=0.65\columnwidth]{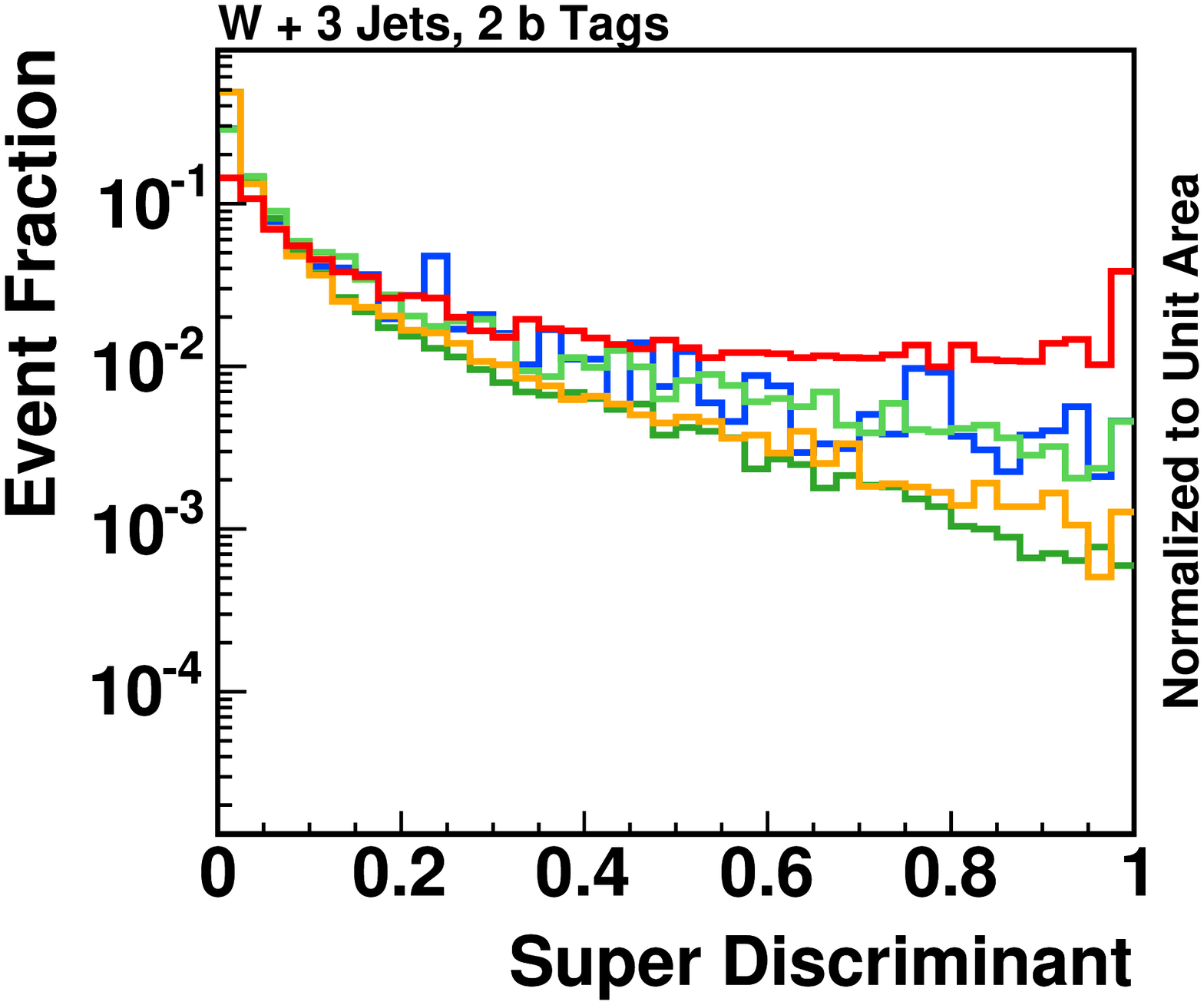}
\label{fig:COMBO3j2t_shape}}
\subfigure[]{
\includegraphics[width=0.65\columnwidth]{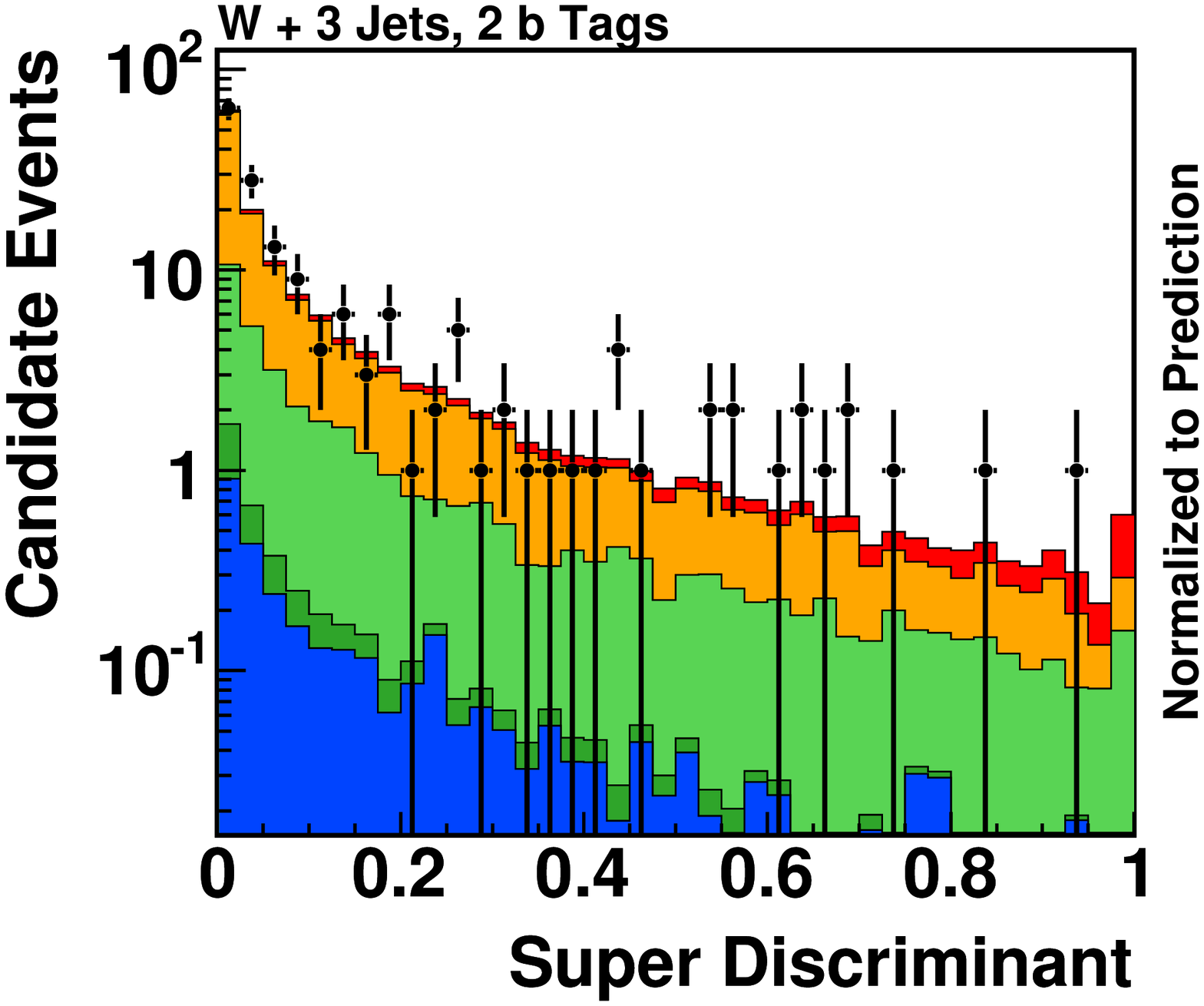}
\label{fig:COMBO3j2t}}
\end{center}
\caption{\label{fig:COMBO} Normalized templates (left) and plots comparing the predicted distributions
with data (right) of the final combined neural network output for each selected data sample.  
These distributions are more sensitive than any single analysis.
The data are indicated by points with error bars, and 
the predictions are shown stacked, with the stacking order following that of the legend.
}
\end{figure*}

The resulting templates and distributions are shown for all four
selected data samples in Fig.~\ref{fig:COMBO}.  In the comparisons of the predictions to the
data, the  predictions are normalized to our signal and
background models, which are described in Sections~\ref{sec:Background}
and~\ref{sec:SignalModel}, respectively.  Each distribution is more
sensitive than any single analysis.  

\section{\label{sec:1D} One-Dimensional Fit Results}

We use the methods described in Section~\ref{sec:Interpretation} to
extract the single top cross section, the significance of the excess
over the background prediction, and the sensitivity, defined to be the
median expected significance, separately for each component analysis
described in Section~\ref{sec:Multivariate}, and for the super
discriminant combined analysis (SD), which is described in
Section~\ref{sec:Combination}.  The results are listed in
Table~\ref{tab:results}.  The cross section measurements of the
individual analyses are quite similar, which is not surprising due to
the overlap in the selected data samples.  The measurements are only
partially correlated, though, as shown in Table~\ref{tab:corrxs},
indicating that the separate analyses extract highly correlated but
not entirely identical information from each event.

Because the super discriminant has access to the most information on
each event, and because it is optimized for the expected sensitivity,
it is the most powerful single analysis.  It is followed by the Neural
Network (NN) and Boosted Decision Tree (BDT) analyses, and the Matrix
Element (ME) analysis.  The Likelihood Function (LF) analysis result
in the table is shown only for the $t$-channel optimized likelihood
functions, although the $s$-channel signals were included in the
templates.

A separate result, a measurement just of the $s$-channel signal cross
section, is extracted from just the two-jet, two-$b$-tag LF analysis,
assuming the $t$-channel signal cross section is at its SM value.  The
result thus obtained is $\sigma_s^{\rm{LF}} = 1.5^{+0.9}_{-0.8}$~pb,
with an observed significance of $2.0\,\sigma$ and an expected
significance of $1.1\,\sigma$.

The super discriminant analysis, like the component analyses, fits
separately the distributions of events in eight non-overlapping
categories, defined by whether the events have two or three jets
passing the selection requirements, one or two $b$-tags, and whether
the charged lepton was a triggered $e$ or $\mu$ candidate (TLC), as opposed to
a non-triggered extended muon coverage lepton candidate (EMC).  A
separate cross section fit is done for each of these categories, and
the results are shown in Table~\ref{tab:xsbyfinalstate}.  The dominant
components of the uncertainties are statistical, driven by the
small data sample sizes in the most pure bins of our discriminant distributions.
The cross sections
extracted for each final state are consistent with each other within their
uncertainties.

The results described above are obtained from the $\ell+\EtMiss+$jets selection.  An
entirely separate analysis conducted by CDF is the search for single top
quark events in the $\EtMiss$ plus two- and three-jets sample~\cite{MET_jets} (MJ), which uses
a data sample corresponding to 2.1~fb$^{-1}$ of data.
The events selected by the MJ analysis do not overlap with those described
in this paper because the MJ analysis imposes a charged lepton veto and an isolated
high-$p_{\rm T}$ track veto.   The MJ analysis
separates its candidate events into three subsamples based on the
$b$-tagging requirements~\cite{MET_jets}, and the results are
summarized in Table~\ref{tab:xsbyfinalstate}.

\begin{figure*}
\begin{center}
\subfigure[]{
\includegraphics[width=0.8\columnwidth]{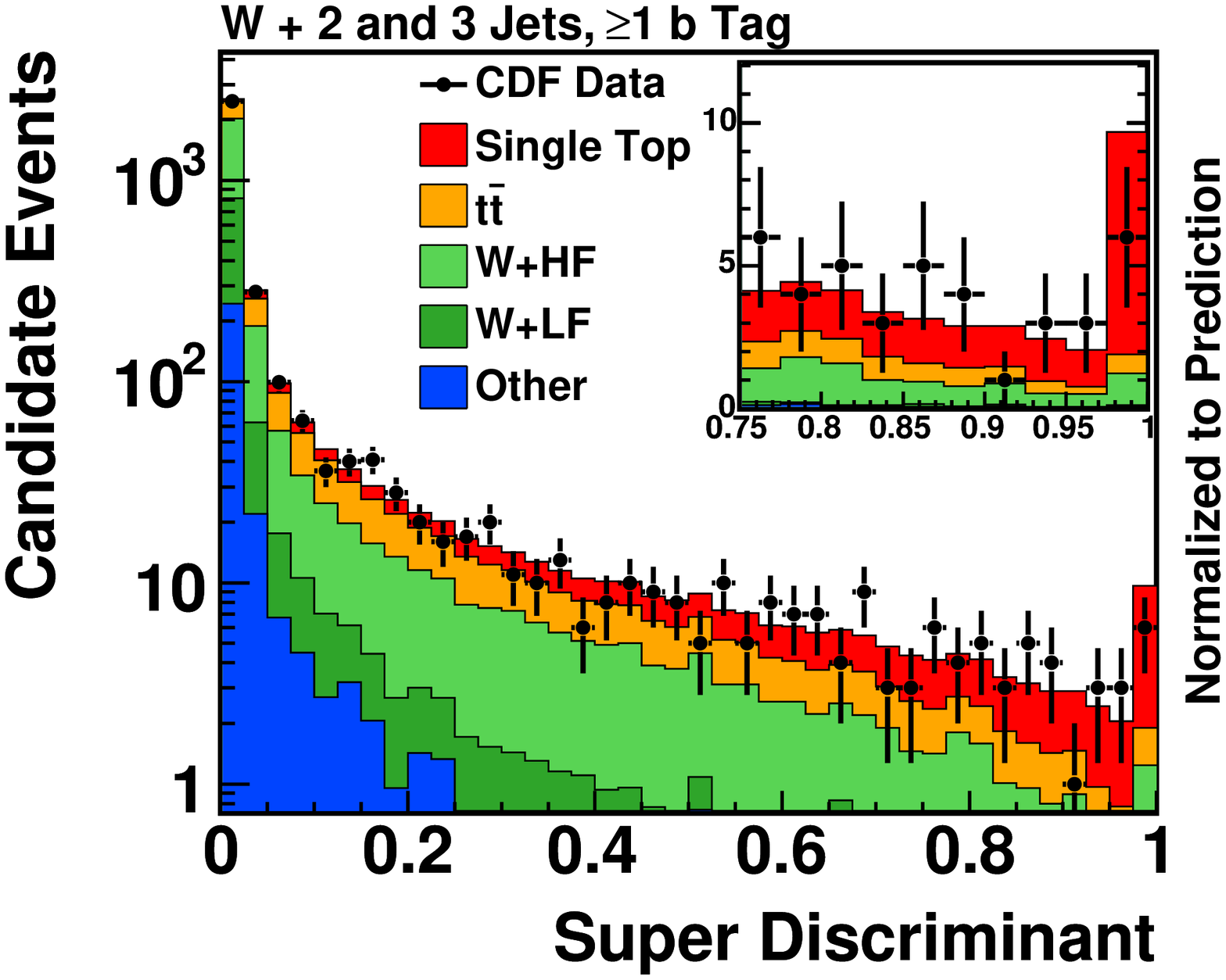}
\label{fig:SDandMJ1}}
\subfigure[]{
\includegraphics[width=0.8\columnwidth]{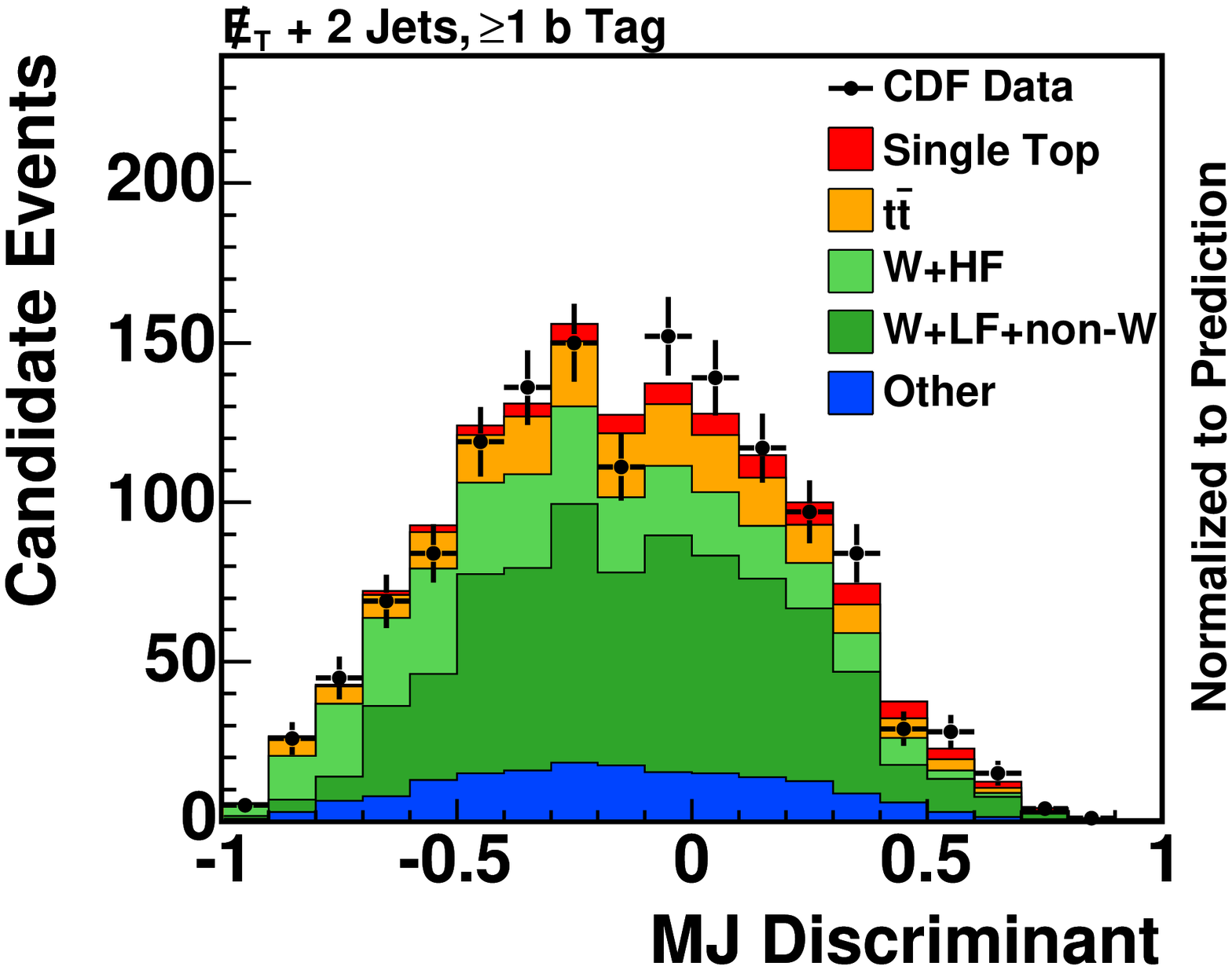}
\label{fig:SDandMJ2}} \\
\end{center}
\caption{\label{fig:SDandMJ} Comparison of the predicted distributions
with data summed over all selected data samples of the super discriminant (left) and the MJ discriminant (right).
Points with error bars indicate the observed data, while the stacked, shaded histograms show
the predictions, including a standard model single top signal.  In each panel, 
the order of the stacked components follows that of the legend.}
\end{figure*}

\begin{figure*}
\begin{center}
\subfigure[]{
\includegraphics[width=0.8\columnwidth]{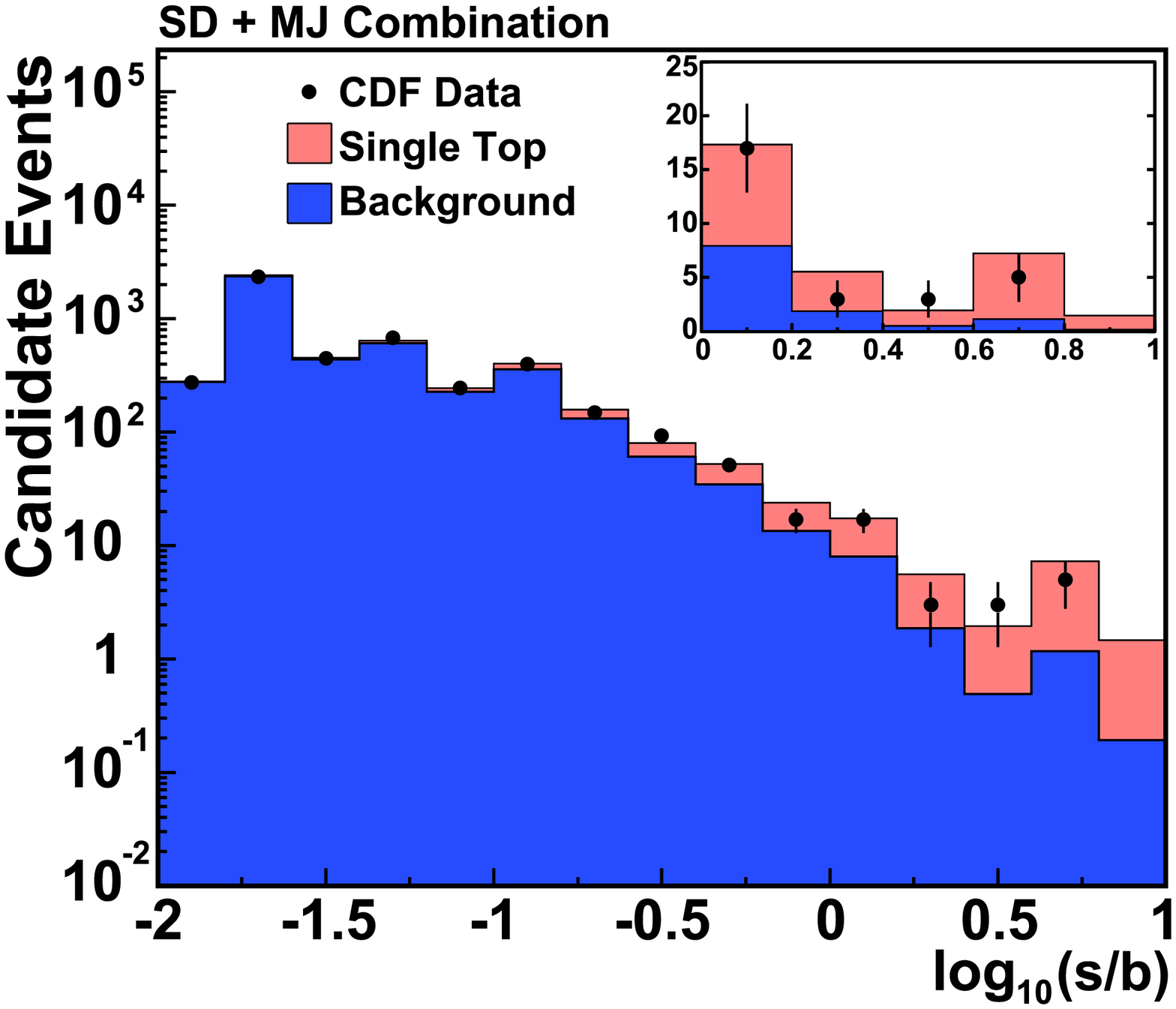}}
\subfigure[]{
\includegraphics[width=0.8\columnwidth]{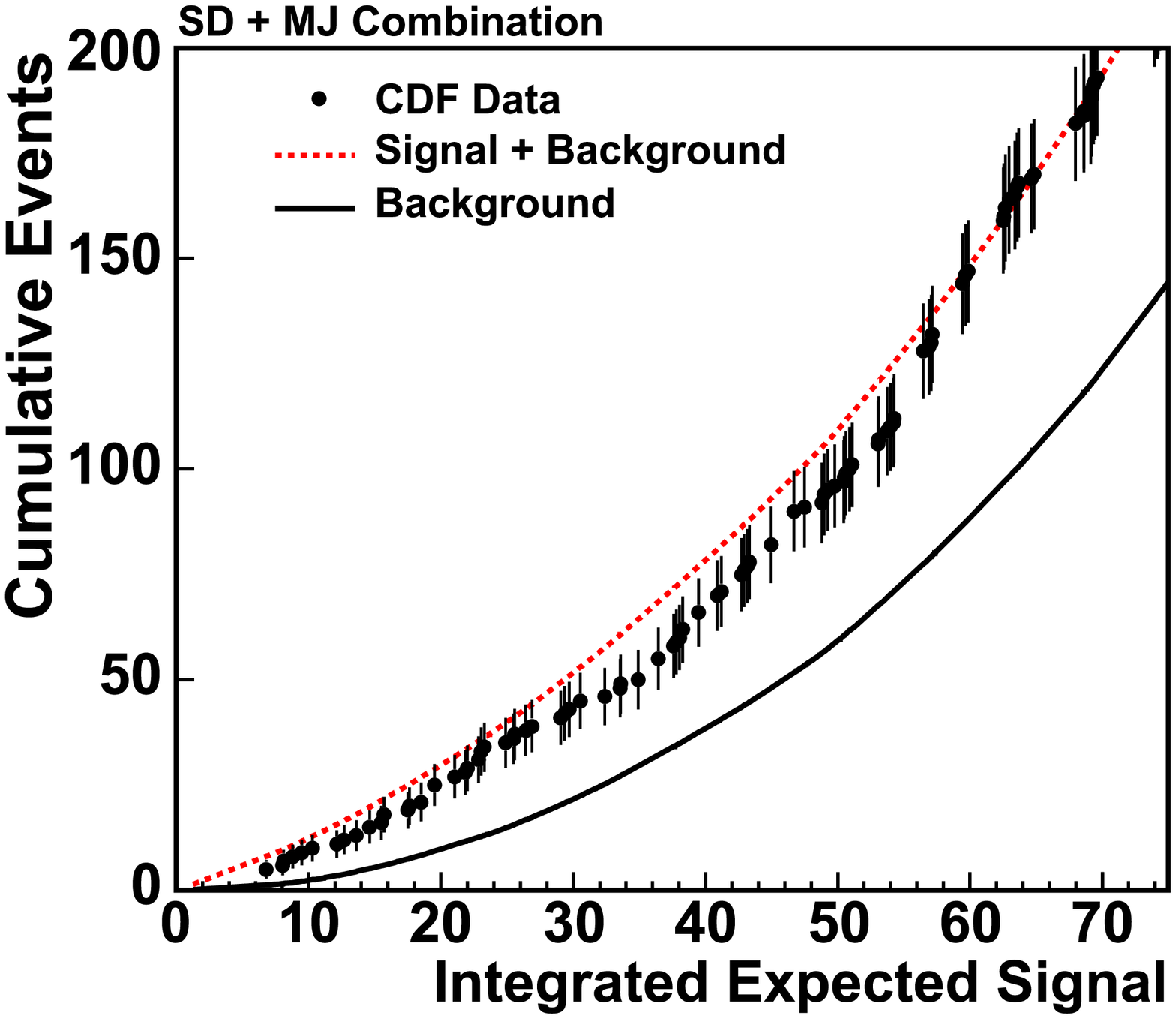}}
\end{center}
\caption{\label{fig:logsb} Distributions of data and predictions for the SD and MJ analyses,
where bins of similar $s/b$ have been collected together (left).  The points with error
bars indicate the observed data, while the stacked, shaded histograms show the predictions,
including a standard model single top signal.  These distributions are integrated starting on
the high-$s/b$ side, and the resulting cumulative event counts are shown on the right, separately
for the observed data, for the background-only prediction, and the signal-plus-background prediction.
}
\end{figure*}

The distributions of the super discriminant in the $\ell+\EtMiss+$jets sample and
the MJ neural network discriminant in the $\EtMiss+$jets sample
are shown in Fig.~\ref{fig:SDandMJ}, summed
over the event categories, even though the cross section fits are performed 
and the significances are calculated
separating the categories.  The sums over event categories add the contents of
bins of histograms with different $s/b$ together and thus do not show the
full separation power of the analyses.  Another way to show the combined data
set is to collect bins with similar $s/b$ in all of the channels of the SD and MJ discriminant
histograms and graph the resulting distribution as a function of $\log_{10}(s/b)$, which
is shown in Fig.~\ref{fig:logsb}(a).  This distribution isolates, at the high $s/b$ side, the
events that contribute the most to the cross section measurement and the significance.
Figure~\ref{fig:logsb}(b) shows the integral of this distribution, separately for the background
prediction, the signal plus background prediction, and the data.  The distributions are integrated
from the highest $s/b$ side downwards, accumulating events and predictions in the highest $s/b$ bins.
The data points are updated on the plot as bins with data entries in them are added to the integral,
and thus are highly correlated from point to point.  A clear excess of data is seen over the background
prediction, not only in the most pure bins, but also as the $s/b$ requirement is loosened, and the excess
is consistent with the standard model single top prediction.

\begin{figure*}
\begin{center}
\subfigure[]{
\includegraphics[width=0.8\columnwidth]{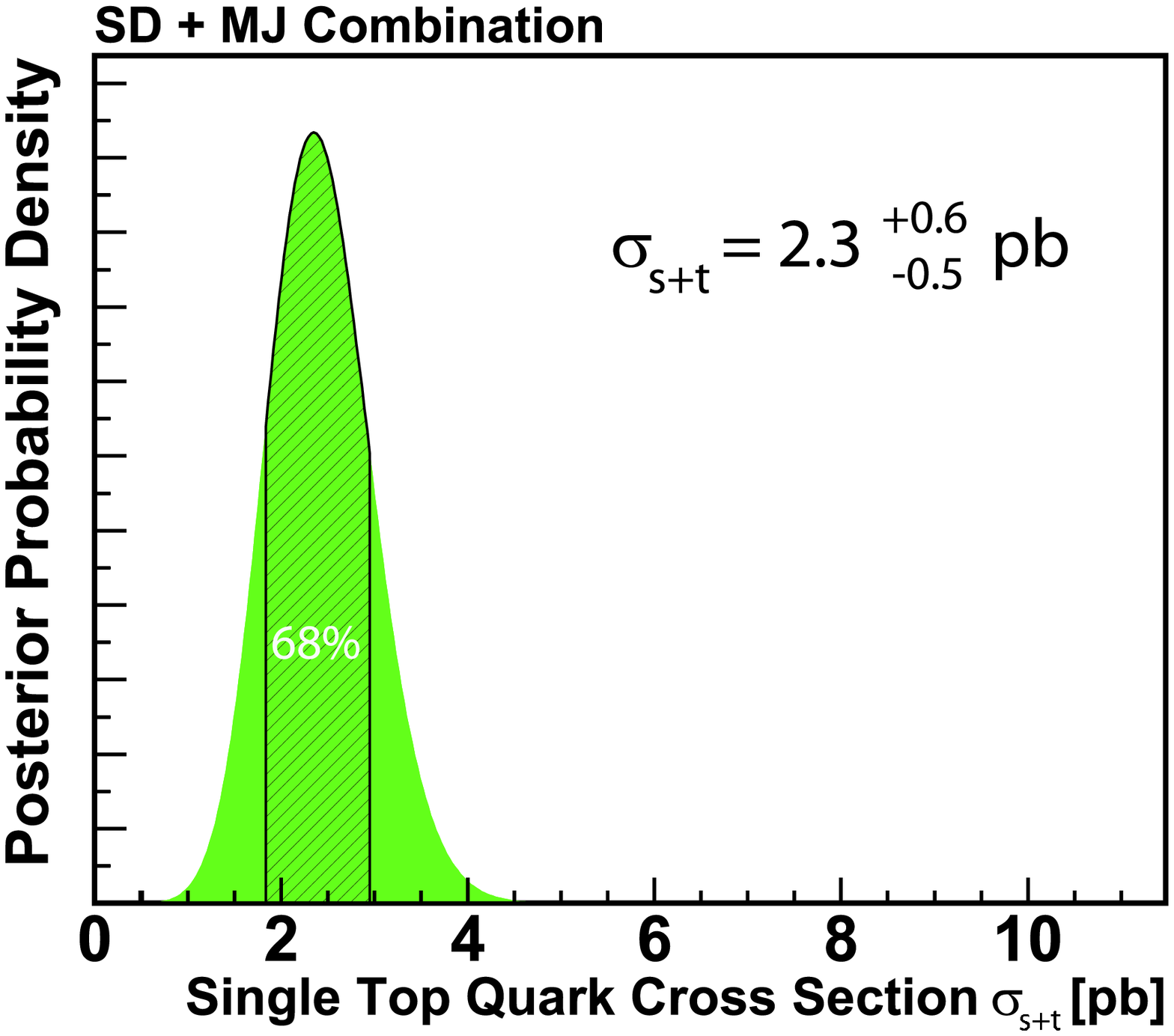}
\label{fig:crossection}}
\subfigure[]{
\includegraphics[width=0.8\columnwidth]{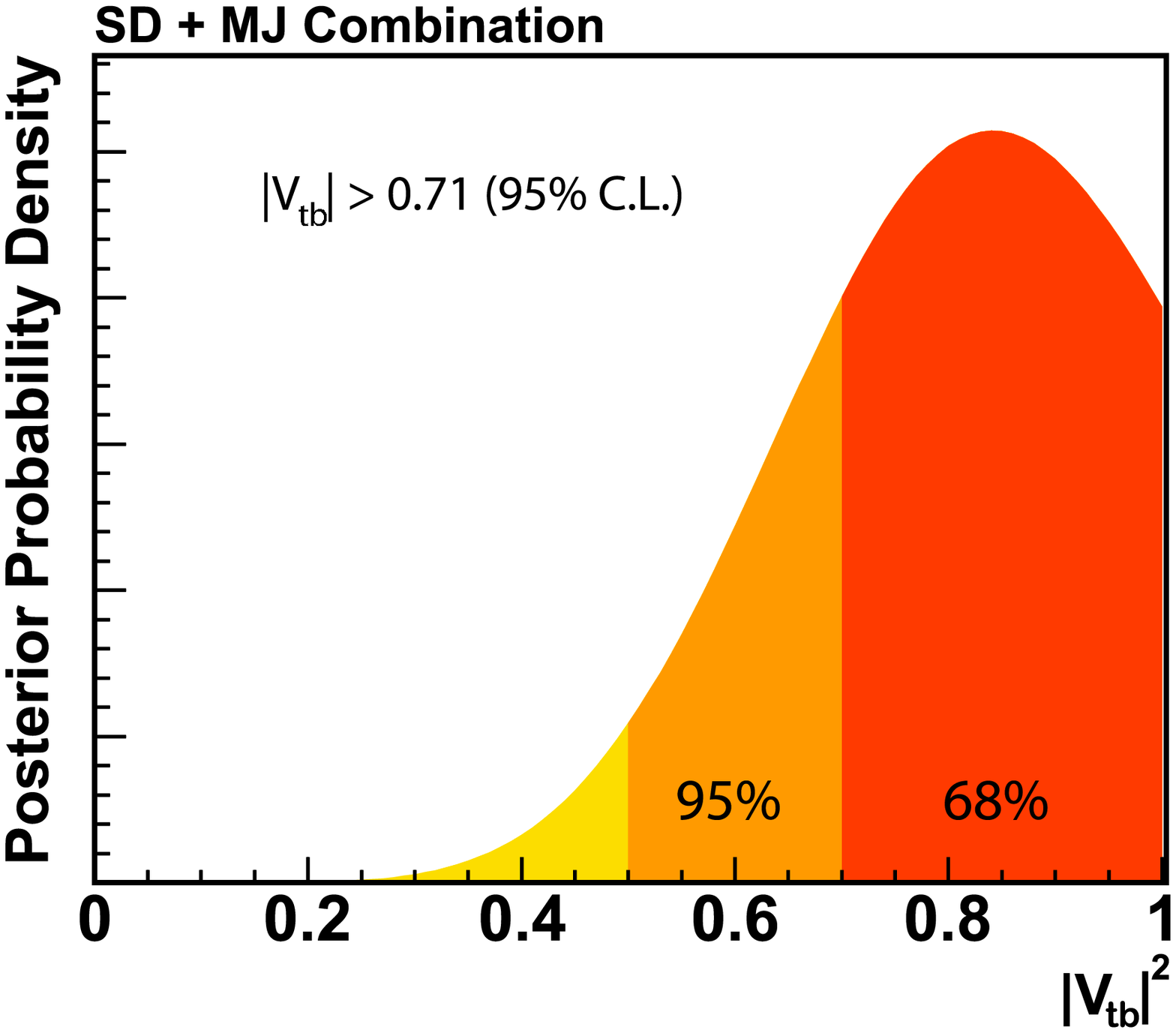} 
\label{fig:vtb}} \\
\subfigure[]{
\includegraphics[width=0.8\columnwidth]{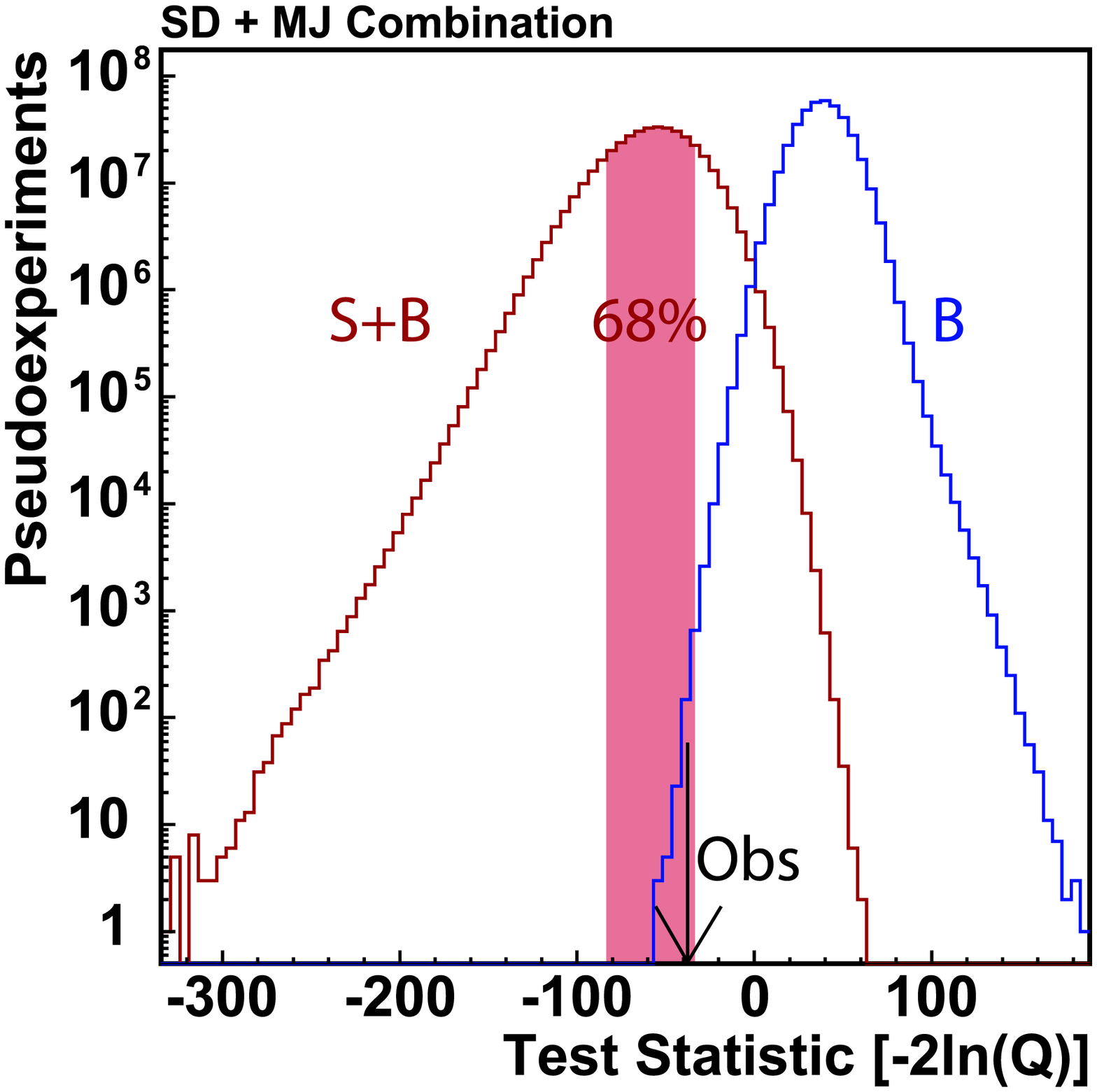}
\label{fig:pvalue}}
\end{center}
\caption{\label{fig:results} The posterior curve of the cross
section measurement calculated with the super discriminant histograms
as inputs (a), the posterior curve for the $|V_{tb}|$ calculation (b), and the distributions
of $-2\ln Q$ in simulated $S+B$ and $B$-only pseudoexperiments, assuming a Standard Model
single top quark signal (c).  The value of $-2\ln Q$ observed in the data is indicated with an arrow.}
\end{figure*}

Because the $\ell+\EtMiss+$jets sample and the $\EtMiss+$jets sample have 
no overlapping events, they can be combined
as separate channels using the same likelihood technique
described in Section~\ref{sec:Interpretation}.  
The joint posterior
distribution including all eleven independent categories simultaneously is shown in Figure~\ref{fig:results}(a).
From this distribution,
we obtain a single top quark cross section measurement of
$\sigma_{s+t} = 2.3 ^ {+0.6}_{-0.5}$ pb, assuming a top quark mass of
175 \gevcc.  The dependence of the measured cross section
on the assumed top quark mass is
$\partial\sigma_{s+t}/\partial m_t=+0.02$~pb/(\gevcc).  Table~\ref{tab:xsbyfinalstate} shows the results
of fitting for $\sigma_s$ and $\sigma_t$ in the separate jet, $b$-tag,
and lepton categories.  The dominant source of uncertainty is the statistical component
from the data sample size.
Our best-fit single top quark cross section is approximately one standard deviation
below the Standard Model prediction of~\cite{Harris:2002md,Sullivan:2004ie}.
The prediction of~\cite{Kidonakis:2006bu} is somewhat higher, but it is also consistent with
our measurement.

To extract $|V_{tb}|$ from the combined measurement, we take advantage of the fact that
the production cross section $\sigma_{s+t}$ is directly proportional to $|V_{tb}|^2$.  We use
the relation
\begin{equation}
|V_{tb}|^2_{\rm{measured}} = \sigma_{s+t}^{\rm{measured}}|V_{tb}|^2_{\rm{SM}}/\sigma_{s+t}^{\rm{SM}},
\label{eq:vtb}
\end{equation}
where $|V_{tb}|^2_{\rm{SM}}\approx 1$ and $\sigma_{s+t}^{\rm{SM}}=2.86\pm0.36$~\cite{Harris:2002md,Sullivan:2004ie}.  
Equation~\ref{eq:vtb} further assumes that $|V_{tb}|^2\gg |V_{ts}|^2+|V_{td}|^2$, because we are assuming
that the top quark decays to $Wb$ 100\% of the time, and because we assume that the production cross section
scales with $|V_{tb}|^2$, while the other CKM matrix elements may contribute as well if they were not very small.
We drop the ``measured'' subscripts and superscripts elsewhere.
Figure~\ref{fig:results}(b) shows the joint posterior distribution of all of our independent channels as a function
of $|V_{tb}|^2$ (which includes the theoretical uncertainty on the predicted production rate, which is not
part of the cross section posterior), from which
we obtain $|V_{tb}|=0.91 \pm 0.11$(stat.+syst.)$\pm 0.07$(theory) and a 95\% confidence level lower
limit of $|V_{tb}| > 0.71$.  

We compute the $p$-value for the significance of this result as described in Section~\ref{sec:significancecalc}.
The distributions of $-2\ln Q$ from which the $p$-value is obtained, are shown in Fig.~\ref{fig:results}(c).
We obtain a $p$-value of $3.1 \times 10^{-7}$ which corresponds to a 4.985 standard deviation excess of data above
the background prediction.  We quote this to two significant digits as a 5.0 standard deviation excess.
The median expected $p$-value is in excess of 5.9 standard deviations; the precision of this estimate is limited by
the number of pseudoexperiments which were fit.  The fact that the observed significance is approximately one sigma
below its SM expectation is not surprising given that our cross section measurement is also approximately one sigma
below its expectation, although this relation is not strictly guaranteed.

\begin{table*}[tbh]
\caption{A summary of the analyses covered in this paper, with their
measured cross sections, observed significances, and sensitivities,
defined to be their median expected $p$-values, converted into Gaussian
standard deviations. The analyses are combined into a super discriminant
(SD), which is combined with the orthogonal $\EtMiss$+jets sample (MJ) to
make the final CDF combination.  
}
\begin{center}
\begin{tabular}{lccc}\hline\hline
Analysis & Cross Section & Significance & Sensitivity \\
         & [pb]          & [$\sigma$]   & [$\sigma$]  \\
\hline
LF & $1.6^{+0.8}_{-0.7}$ & 2.4 & 4.0 \\
ME & $2.5^{+0.7}_{-0.6}$ & 4.3 & 4.9 \\
NN & $1.8^{+0.6}_{-0.6}$ & 3.5 & 5.2 \\
BDT & $2.1^{+0.7}_{-0.6}$ & 3.5 & 5.2 \\
\hline
SD & $2.1^{+0.6}_{-0.5}$ & 4.8 & $>5.9$ \\
MJ & $4.9^{+2.5}_{-2.2}$ & 2.1 & 1.4 \\
\hline
SD + MJ Combination & $2.3^{+0.6}_{-0.5}$ & 5.0 & $>5.9$ \\
\hline\hline
\end{tabular}
\label{tab:results}
\end{center}
\end{table*}

\begin{table}[tbh]
\caption{A summary of the measured values of the single
top production cross section $\sigma_s+\sigma_t$ using the super discriminant
analysis, separately for
each of the non-overlapping final state categories, based on the number of jets,
the number of $b$~tags, and the lepton category.  Also
listed are the MJ cross section fit results by $b$-tagging category.}
\begin{center}
\begin{tabular}{lc}\hline\hline
Category & Cross Section [pb] \\ \hline 
SD 2-Jet, 1-Tag, TLC & $1.7^{+0.7}_{-0.6}$ \\
SD 2-Jet, 2-Tag, TLC & $4.1^{+2.3}_{-1.9}$ \\
SD 3-Jet, 1-Tag, TLC & $2.4^{+2.1}_{-1.7}$ \\
SD 3-Jet, 2-Tag, TLC & $6.3^{+4.9}_{-4.2}$ \\
SD 2-Jet, 1-Tag, EMC & $2.3^{+1.4}_{-1.1}$ \\
SD 2-Jet, 2-Tag, EMC & $9.8^{+5.7}_{-4.4}$ \\
SD 3-Jet, 1-Tag, EMC & $7.2^{+5.5}_{-4.6}$ \\
SD 3-Jet, 2-Tag, EMC & $0.0^{+8.8}_{-0.0}$ \\ \hline
SD                   & $2.1^{+0.6}_{-0.5}$ \\ \hline
MJ 2-Tag             & $5.9^{+4.2}_{-3.7}$ \\
MJ 1-Tag +{\sc jetprob} & $2.7^{+4.6}_{-2.7}$ \\
MJ 1-Tag             & $4.3^{+2.6}_{-2.3}$ \\ \hline
MJ                   & $4.9^{+2.5}_{-2.2}$ \\ \hline
SD + MJ Combination           & $2.3^{+0.6}_{-0.5}$ \\ \hline\hline
\end{tabular}
\label{tab:xsbyfinalstate}
\end{center}
\end{table}

Recently, the cross section measurement shown here has been combined with that measured
by D0~\cite{Abazov:2009ii}.  The same technique for extracting the cross section
in combination as for each individual measurement is used~\cite{cdfd0combo}, and the
best-fit cross section is $\sigma_{s+t}=2.76^{+0.58}_{-0.47}$~pb, assuming $m_t=170$~GeV/$c^2$.

\section{\label{sec:2D} Two-Dimensional Fit Results}

The extraction of the combined signal cross section $\sigma_{s+t}$
proceeds by constructing a one-dimensional Bayesian posterior with a
uniform prior in the cross section to be measured.  An extension of
this is to form the posterior in the two-dimensional plane, $\sigma_s$
vs. $\sigma_t$, and to extract the $s$-channel and the $t$-channel
cross sections separately.  We assume a uniform prior in the
$\sigma_s$ vs. $\sigma_t$ plane, and integrate over the nuisance
parameters in the same way as we did for the one-dimensional cross
section extraction.  The input histograms for this extraction are the
distributions of the super discriminant for the $W+$jets analyses, and
the MJ discriminant histograms are also included, exactly as is done
for the one-dimensional cross section fit.

\begin{figure}
\begin{center}
\includegraphics[width=0.9\columnwidth]{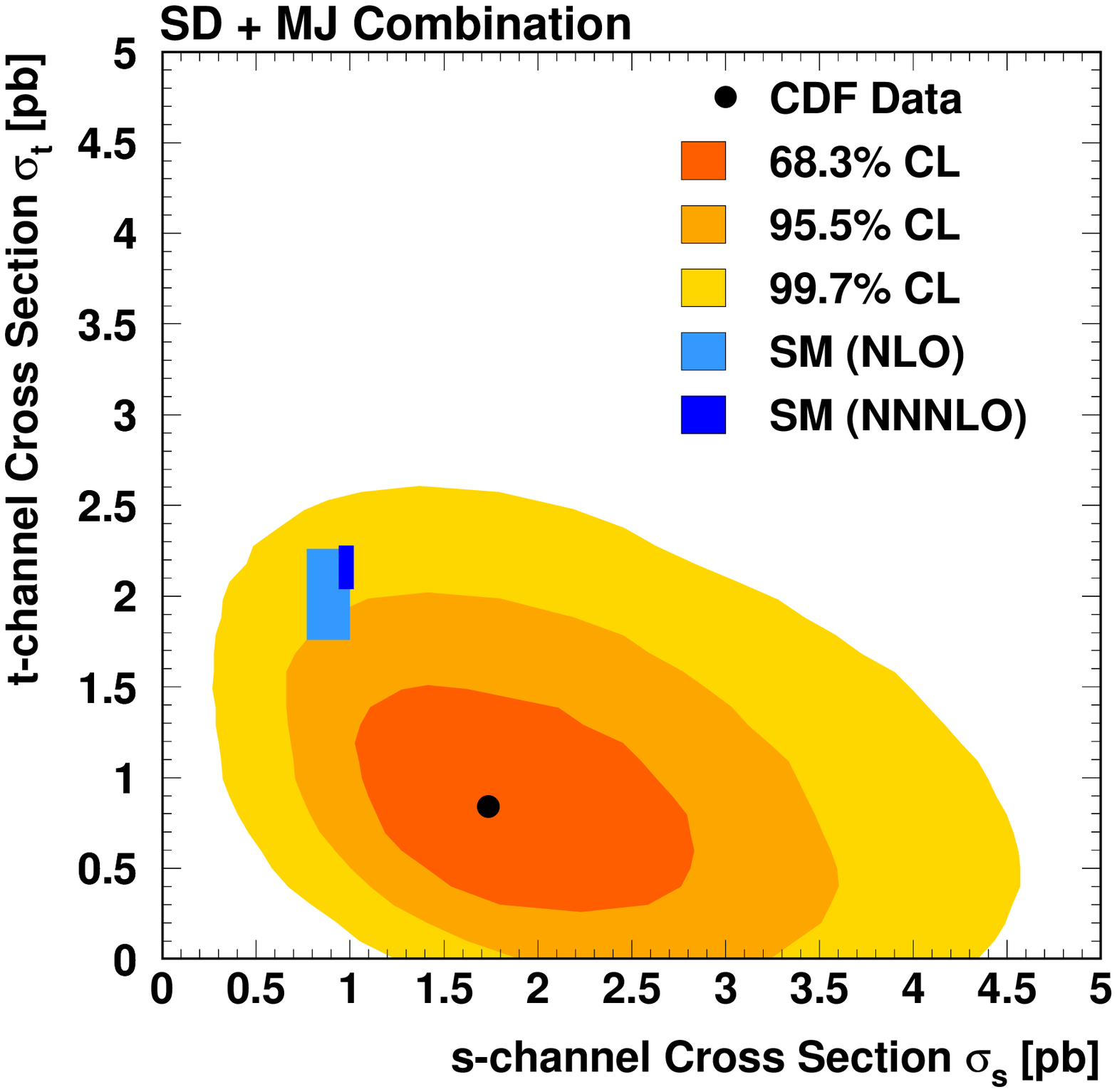}

\end{center}
\caption{\label{fig:2dxs}The results of the two-dimensional fit for 
$\sigma_s$ and $\sigma_t$.  The black point
shows the best fit value, and the 68.3\%, 95.5\%, and 99.7\% credibility
regions are shown as shaded areas.  The SM predictions are also
indicated with their theoretical uncertainties.  The SM predictions
shown are those of~\protect{\cite{Harris:2002md,Sullivan:2004ie}}
(NLO) and~\protect{\cite{Kidonakis:2006bu}} (NNNLO).}
\end{figure}

The best-fit cross section is the one for which the posterior is
maximized, and corresponds to $\sigma_{s} = 1.8 ^ {+0.7}_{-0.5}$~pb
and $\sigma_{t} = 0.8 ^ {+0.4}_{-0.4}$~pb.  The uncertainties on the
measurements of $\sigma_s$ and $\sigma_t$ are correlated with each
other because $s$-channel and $t$-channel signals both populate the
signal-like bins of each of our discriminant variables.  Regions of
68.3\%, 95.5\%, and 99.7\% credibility are derived from the
distribution of the posterior by evaluating the smallest region in
area that contains 68.3\%, 95.5\% or 99.7\% of the integral of the
posterior.  Each region has the property that the numerical values of
the posterior along the boundary of the region are equal to each
other.  The best-fit values, the credibility regions, and the SM
predictions of $\sigma_s$ and $\sigma_t$ are shown in
Fig.~\ref{fig:2dxs}.  We compare these with the NLO SM predictions
of $\sigma_t=1.98\pm 0.25$~pb and $\sigma_s=0.88\pm 0.11$~pb~\cite{Harris:2002md,Sullivan:2004ie},
and also with the NNNLO predictions of $\sigma_t=2.16\pm 0.12$~pb and 
$\sigma_s=0.98\pm 0.04$~pb~\cite{Kidonakis:2006bu}.

The coverage of the technique is checked by generating 1500
pseudo-datasets randomly drawn from systematically-varied predictions
assuming that a single top signal is present as predicted by the SM,
and performing the two-dimensional extraction of $\sigma_s$ and
$\sigma_t$ for each one in the same way as is done for the data.  No
bias is seen in the median fit $\sigma_s$ and $\sigma_t$ values.  Each
pseudo-dataset has a corresponding set of regions at 68.3\%, 95.5\%,
and 99.7\% credibility.  The fractions of the pseudo-datasets' fit
bands that contain the input prediction for $\sigma_s$ and $\sigma_t$
is consistent with the credibility levels at which the bands are
quoted.

The two-dimensional fit result is not in good agreement with the SM
prediction; the difference is at approximately the two standard
deviation level of significance.  The differences between the measured
values of the $s$- and $t$-channel cross sections and their SM
predictions are driven by the deficit of events observed in the high-discriminant
output regions of the two-jet, one-$b$-tag channels relative to the SM
signal-plus background prediction as shown in
Fig.~\ref{fig:COMBO}~(b), and the excess of events observed in the two-jet,
two-b-tag distributions, as shown in Fig.~\ref{fig:COMBO}~(d).  The
measured total cross sections in these jet and $b$-tagging categories,
listed in Table~\ref{tab:xsbyfinalstate}, show the effects of these
discrepancies with respect to the SM predictions.

The newer calculation of the $t$-channel kinematic
distributions~\cite{Campbell:2009ss,Campbell:2009gj} predicts a larger fraction of
$t$-channel signal events with a visible recoiling $b$ jet, which is
normally not reconstructed because it is beyond the forward acceptance
of the detector or because the jet $E_{\rm T}$ is too small.  This
calculation has almost the same overall cross section prediction for
$\sigma_t$ as the one we use elsewhere in this
paper~\cite{Harris:2002md}, but it reduces the two-jet, one $b$-tag
prediction for the $t$-channel signal and raises the two-jet two-$b$-tag
and 3-jet predictions.  After fully simulating and reconstructing the
signal events, the effects on the predicted yields are small; the
3-jet channels' contribution to our measurement sensitivity is also
small.  The change to the 1D and 2D fit results is not noticeable when
using the model of~\cite{Campbell:2009ss,Campbell:2009gj} compared to our central
prediction within the rounding precision of the results we quote.

The $t$-channel process is sensitive to the $b$ quark PDF of the proton,
while the $s$-channel process is not.  The low measured value of $\sigma_t$
reported here is not in good agreement with the SM predictions.  The D0 collaboration
has recently measured $\sigma_t=3.14^{+0.94}_{-0.80}$~pb using a data sample
corresponding to 2.3~fb$^{-1}$ of integrated luminosity~\cite{Abazov:2009pa}, which is
larger than the standard model prediction.  Taken together, there is insufficient
evidence to exclude a standard model explanation of the results.

\section{\label{sec:Summary} Summary}

The observation of single top quark production poses many difficult experimental
challenges.  CDF performs this analysis in proton-antiproton collisions
at 1.96~TeV in events with a leptonically decaying $W$ boson and jets.
The low signal-to-background ratio in the data samples passing our
selection requirements necessitates precise modeling of
the signal and background
kinematic distributions with matrix-element-based Monte Carlo generators using
full parton showering and detailed detector simulation, and also requires the
normalization of the dominant background rates
to measured rates in sideband data samples.  The small signals and large,
uncertain background processes also require us to take maximum advantage of
the expected kinematic and flavor differences between the signals and the
background processes.  We develop novel, powerful
techniques for combining information from several observable
quantities computed for each event.  We purify
a subsample of single top quark events with a predicted signal to background ratio
exceeding 5:1 from a sample starting with a signal to background ratio of 1:16
after $b$-tagging.  

Our final discriminant variables are functions of many 
kinematic and $b$-tagging variables.  Incorrect modeling of one or more variables,
or even of the correlations between variables, can bias the results.
We therefore evaluate an exhaustive list of systematic uncertainties
which affect the predicted signal and background components'
rates and kinematic distributions, including both theoretical uncertainties
and uncertainties which arise from discrepancies observed between 
the data and the simulations in control regions.  The correlations
between the systematic uncertainties on the
rate and shape predictions of the signal and background processes in several
data samples are taken into account in all of the
results and in computing the expected sensitivities presented in this paper.
 We also consider Monte Carlo statistical
uncertainties in each bin of each template histogram in each channel
independently.  We constrain the major background rates {\it in situ} in the
selected event samples to further reduce the uncertainties in their values
and to improve the sensitivity of our results.  

Our analyses were
optimized based on predictions and were blinded to the data during
their development.  The analyses were cross-checked using the data in control samples before looking
at the data in the signal regions.  We perform many checks of our
methods -- we compare the observed and predicted distributions of the discriminant
input and output variables in independent control samples, and we also train
discriminants that enrich samples of each background as if it
were signal.  The vast majority of our cross checks show that the predictions model
the data very well, and those that show discrepancies contribute to our systematic
uncertainties.

The four analyses in the $\ell+\EtMiss+$jets sample described in this paper 
are combined with a statistically independent analysis in the $\EtMiss+$jets sample~\cite{MET_jets}
to maximize the total sensitivity.
We report an observation of electroweak single top quark production with a
$p$-value of $3.1 \times 10^{-7}$, which corresponds to a significance of
5.0 standard deviations.  The measured value of the combined $s$- and $t$-channel cross section
is $\sigma_{s+t} = 2.3 ^ {+0.6}_{-0.5}$~pb assuming the top quark mass is
175~\gevcc, and also assuming the SM value of $\sigma_s/\sigma_t$.
The dependence of the measured cross section on the assumed top quark mass
is $\partial\sigma_{s+t}/\partial m_t=+0.02$~pb$\,c^2$/GeV.  We extract a value of $|V_{tb}|=0.91 \pm
0.11$(stat.+syst.)$\pm 0.07$(theory) and a 95\% confidence level lower
limit of $|V_{tb}| > 0.71$, using the prediction of~\cite{Harris:2002md,Sullivan:2004ie} for the
SM cross section, and also assuming that $|V_{tb}|^2\gg |V_{ts}|^2+|V_{td}|^2$.  With a two-dimensional
fit for $\sigma_s$ and $\sigma_t$, using the same combination of analyses as the one-dimensional fit,
we obtain $\sigma_{s} = 1.8 ^ {+0.7}_{-0.5}$~pb and $\sigma_{t} = 0.8 ^ {+0.4}_{-0.4}$~pb.

\begin{center}
{\bf Acknowledgments}
\end{center}

We thank the Fermilab staff and the technical staffs of the
participating institutions for their vital contributions. This work
was supported by the U.S. Department of Energy and National Science
Foundation; the Italian Istituto Nazionale di Fisica Nucleare; the
Ministry of Education, Culture, Sports, Science and Technology of
Japan; the Natural Sciences and Engineering Research Council of
Canada; the National Science Council of the Republic of China; the
Swiss National Science Foundation; the A.P. Sloan Foundation; the
Bundesministerium f\"ur Bildung und Forschung, Germany; the World
Class University Program, the National Research Foundation of Korea;
the Science and Technology Facilities Council and the Royal Society,
UK; the Institut National de Physique Nucleaire et Physique des
Particules/CNRS; the Russian Foundation for Basic Research; the
Ministerio de Ciencia e Innovaci\'{o}n, and Programa
Consolider-Ingenio 2010, Spain; the Slovak R\&D Agency; and the
Academy of Finland.

\bibliographystyle{apsrev}

\begin{thebibliography}{102}
\expandafter\ifx\csname natexlab\endcsname\relax\def\natexlab#1{#1}\fi
\expandafter\ifx\csname bibnamefont\endcsname\relax
  \def\bibnamefont#1{#1}\fi
\expandafter\ifx\csname bibfnamefont\endcsname\relax
  \def\bibfnamefont#1{#1}\fi
\expandafter\ifx\csname citenamefont\endcsname\relax
  \def\citenamefont#1{#1}\fi
\expandafter\ifx\csname url\endcsname\relax
  \def\url#1{\texttt{#1}}\fi
\expandafter\ifx\csname urlprefix\endcsname\relax\def\urlprefix{URL }\fi
\providecommand{\bibinfo}[2]{#2}
\providecommand{\eprint}[2][]{\url{#2}}

\bibitem[{\citenamefont{{Tevatron Electroweak Working Group, arXiv:1007.3178
  [hep-ex]}}(2010)}]{topmass10}
\bibinfo{author}{\bibnamefont{{Tevatron Electroweak Working Group,
  arXiv:1007.3178 [hep-ex]}}} (\bibinfo{year}{2010}).

\bibitem[{\citenamefont{Abe {\it et~al}.}(1995)}]{Abe:1995hr}
\bibinfo{author}{\bibfnamefont{F.}~\bibnamefont{Abe}} \bibnamefont{{\it et~al}.}
  (\bibinfo{collaboration}{CDF Collaboration}), \bibinfo{journal}{Phys. Rev.
  Lett.} \textbf{\bibinfo{volume}{74}}, \bibinfo{pages}{2626}
  (\bibinfo{year}{1995}).

\bibitem[{\citenamefont{Abachi {\it et~al}.}(1995)}]{Abachi:1995iq}
\bibinfo{author}{\bibfnamefont{S.}~\bibnamefont{Abachi}} \bibnamefont{{\it et~al}.}
  (\bibinfo{collaboration}{D0 Collaboration}), \bibinfo{journal}{Phys. Rev.
  Lett.} \textbf{\bibinfo{volume}{74}}, \bibinfo{pages}{2632}
  (\bibinfo{year}{1995}).

\bibitem[{\citenamefont{Abulencia
  {\it et~al}.}(2006{\natexlab{a}})}]{Abulencia:2006in}
\bibinfo{author}{\bibfnamefont{A.}~\bibnamefont{Abulencia}}
  \bibnamefont{{\it et~al}.} (\bibinfo{collaboration}{CDF Collaboration}),
  \bibinfo{journal}{Phys. Rev. Lett.} \textbf{\bibinfo{volume}{97}},
  \bibinfo{pages}{082004} (\bibinfo{year}{2006}{\natexlab{a}}).

\bibitem[{\citenamefont{Acosta {\it et~al}.}(2005{\natexlab{a}})}]{Acosta:2005hr}
\bibinfo{author}{\bibfnamefont{D.~E.} \bibnamefont{Acosta}}
  \bibnamefont{{\it et~al}.} (\bibinfo{collaboration}{CDF Collaboration}),
  \bibinfo{journal}{Phys. Rev. Lett.} \textbf{\bibinfo{volume}{95}},
  \bibinfo{pages}{102002} (\bibinfo{year}{2005}{\natexlab{a}}).

\bibitem[{\citenamefont{Abulencia
  {\it et~al}.}(2006{\natexlab{b}})}]{Abulencia:2005xf}
\bibinfo{author}{\bibfnamefont{A.}~\bibnamefont{Abulencia}}
  \bibnamefont{{\it et~al}.} (\bibinfo{collaboration}{CDF Collaboration}),
  \bibinfo{journal}{Phys. Rev. D} \textbf{\bibinfo{volume}{73}},
  \bibinfo{pages}{111103} (\bibinfo{year}{2006}{\natexlab{b}}).

\bibitem[{\citenamefont{Amsler {\it et~al}.}(2008{\natexlab{a}})}]{Amsler:2008zz}
\bibinfo{author}{\bibfnamefont{C.}~\bibnamefont{Amsler}} \bibnamefont{{\it et~al}.}
  (\bibinfo{collaboration}{Particle Data Group}), \bibinfo{journal}{Phys. Lett.
  B} \textbf{\bibinfo{volume}{667}}, \bibinfo{pages}{1}
  (\bibinfo{year}{2008}{\natexlab{a}}).

\bibitem[{\citenamefont{Willenbrock and Dicus}(1986)}]{Willenbrock:1986cr}
\bibinfo{author}{\bibfnamefont{S.~S.~D.} \bibnamefont{Willenbrock}}
  \bibnamefont{and} \bibinfo{author}{\bibfnamefont{D.~A.} \bibnamefont{Dicus}},
  \bibinfo{journal}{Phys. Rev. D} \textbf{\bibinfo{volume}{34}},
  \bibinfo{pages}{155} (\bibinfo{year}{1986}).

\bibitem[{\citenamefont{Harris {\it et~al}.}(2002)\citenamefont{Harris, Laenen, Phaf,
  Sullivan, and Weinzierl}}]{Harris:2002md}
\bibinfo{author}{\bibfnamefont{B.~W.} \bibnamefont{Harris}},
  \bibinfo{author}{\bibfnamefont{E.}~\bibnamefont{Laenen}},
  \bibinfo{author}{\bibfnamefont{L.}~\bibnamefont{Phaf}},
  \bibinfo{author}{\bibfnamefont{Z.}~\bibnamefont{Sullivan}}, \bibnamefont{and}
  \bibinfo{author}{\bibfnamefont{S.}~\bibnamefont{Weinzierl}},
  \bibinfo{journal}{Phys. Rev. D} \textbf{\bibinfo{volume}{66}},
  \bibinfo{pages}{054024} (\bibinfo{year}{2002}).

\bibitem[{\citenamefont{Sullivan}(2004)}]{Sullivan:2004ie}
\bibinfo{author}{\bibfnamefont{Z.}~\bibnamefont{Sullivan}},
  \bibinfo{journal}{Phys. Rev. D} \textbf{\bibinfo{volume}{70}},
  \bibinfo{pages}{114012} (\bibinfo{year}{2004}).

\bibitem[{\citenamefont{Kidonakis}(2006)}]{Kidonakis:2006bu}
\bibinfo{author}{\bibfnamefont{N.}~\bibnamefont{Kidonakis}},
  \bibinfo{journal}{Phys. Rev. D} \textbf{\bibinfo{volume}{74}},
  \bibinfo{pages}{114012} (\bibinfo{year}{2006}).

\bibitem[{\citenamefont{{N. Kidonakis, arXiv:0705.2431
  [hep-ph]}}(2007)}]{Kidonakis:2007wg}
\bibinfo{author}{\bibnamefont{{N. Kidonakis, arXiv:0705.2431 [hep-ph]}}}
  (\bibinfo{year}{2007}).

\bibitem[{\citenamefont{{N. Kidonakis, arXiv:0909.0037
  [hep-ph]}}(2009)}]{Kidonakis:2009mx}
\bibinfo{author}{\bibnamefont{{N. Kidonakis, arXiv:0909.0037 [hep-ph]}}}
  (\bibinfo{year}{2009}).

\bibitem[{\citenamefont{N. Kidonakis}(2010)}]{Kidonakis:2010tc}
\bibinfo{author}{\bibfnamefont{N.~Kidonakis}},
  \bibinfo{journal}{Phys. Rev. D} \textbf{\bibinfo{volume}{81}},
  \bibinfo{pages}{054028} (\bibinfo{year}{2010}).


\bibitem[{\citenamefont{{LEP Electroweak Working Group}}(2006)}]{Z-Pole}
\bibinfo{author}{\bibnamefont{{LEP Electroweak Working Group}}}
  (\bibinfo{collaboration}{ALEPH, DELPHI, L3, OPAL, and SLD Collaborations}),
  \bibinfo{journal}{Phys. Rept.} \textbf{\bibinfo{volume}{427}},
  \bibinfo{pages}{257} (\bibinfo{year}{2006}).

\bibitem[{\citenamefont{Alwall {\it et~al}.}(2007)}]{Alwall:2006bx}
\bibinfo{author}{\bibfnamefont{J.}~\bibnamefont{Alwall}} \bibnamefont{{\it et~al}.},
  \bibinfo{journal}{Eur. Phys. J. C} \textbf{\bibinfo{volume}{49}},
  \bibinfo{pages}{791} (\bibinfo{year}{2007}).

\bibitem[{\citenamefont{Tait and Yuan}(2000)}]{Tait:2000sh}
\bibinfo{author}{\bibfnamefont{T.~M.~P.} \bibnamefont{Tait}} \bibnamefont{and}
  \bibinfo{author}{\bibfnamefont{C.~P.} \bibnamefont{Yuan}},
  \bibinfo{journal}{Phys. Rev. D} \textbf{\bibinfo{volume}{63}},
  \bibinfo{pages}{014018} (\bibinfo{year}{2000}).

\bibitem[{\citenamefont{{K. A. Assamagan, {\it et al}.,
  hep-ph/0406152}}(2004)}]{Assamagan:2004mu}
\bibinfo{author}{\bibnamefont{{K. A. Assamagan, {\it et al}., hep-ph/0406152}}}
  (\bibinfo{year}{2004}).

\bibitem[{\citenamefont{Kidonakis and Vogt}(2003)}]{Kidonakis:2003qe}
\bibinfo{author}{\bibfnamefont{N.}~\bibnamefont{Kidonakis}} \bibnamefont{and}
  \bibinfo{author}{\bibfnamefont{R.}~\bibnamefont{Vogt}},
  \bibinfo{journal}{Phys. Rev. D} \textbf{\bibinfo{volume}{68}},
  \bibinfo{pages}{114014} (\bibinfo{year}{2003}).

\bibitem[{\citenamefont{Mahlon and Parke}(1997)}]{Mahlon:1996pn}
\bibinfo{author}{\bibfnamefont{G.}~\bibnamefont{Mahlon}} \bibnamefont{and}
  \bibinfo{author}{\bibfnamefont{S.~J.} \bibnamefont{Parke}},
  \bibinfo{journal}{Phys. Rev. D} \textbf{\bibinfo{volume}{55}},
  \bibinfo{pages}{7249} (\bibinfo{year}{1997}).

\bibitem[{\citenamefont{Stelzer {\it et~al}.}(1998)\citenamefont{Stelzer, Sullivan,
  and Willenbrock}}]{Stelzer:1998ni}
\bibinfo{author}{\bibfnamefont{T.}~\bibnamefont{Stelzer}},
  \bibinfo{author}{\bibfnamefont{Z.}~\bibnamefont{Sullivan}}, \bibnamefont{and}
  \bibinfo{author}{\bibfnamefont{S.}~\bibnamefont{Willenbrock}},
  \bibinfo{journal}{Phys. Rev. D} \textbf{\bibinfo{volume}{58}},
  \bibinfo{pages}{094021} (\bibinfo{year}{1998}).

\bibitem[{\citenamefont{Abazov {\it et~al}.}(2007)}]{Abazov:2006gd}
\bibinfo{author}{\bibfnamefont{V.~M.} \bibnamefont{Abazov}}
  \bibnamefont{{\it et~al}.} (\bibinfo{collaboration}{D0 Collaboration}),
  \bibinfo{journal}{Phys. Rev. Lett.} \textbf{\bibinfo{volume}{98}},
  \bibinfo{pages}{181802} (\bibinfo{year}{2007}).

\bibitem[{\citenamefont{Abazov {\it et~al}.}(2008)}]{Abazov:2008kt}
\bibinfo{author}{\bibfnamefont{V.~M.} \bibnamefont{Abazov}}
  \bibnamefont{{\it et~al}.} (\bibinfo{collaboration}{D0 Collaboration}),
  \bibinfo{journal}{Phys. Rev. D} \textbf{\bibinfo{volume}{78}},
  \bibinfo{pages}{012005} (\bibinfo{year}{2008}).

\bibitem[{\citenamefont{Abazov {\it et~al}.}(2009)}]{Abazov:2009ii}
\bibinfo{author}{\bibfnamefont{V.~M.} \bibnamefont{Abazov}}
  \bibnamefont{{\it et~al}.} (\bibinfo{collaboration}{D0 Collaboration}),
  \bibinfo{journal}{Phys. Rev. Lett.} \textbf{\bibinfo{volume}{103}},
  \bibinfo{pages}{092001} (\bibinfo{year}{2009}).


\bibitem[{\citenamefont{{V.~M.~Abazov {\it et al.} (D0 Collaboration)}}(2009)}]{Abazov:2009nu}
\bibinfo{author}{\bibfnamefont{V.~M.} \bibnamefont{Abazov}}
  \bibnamefont{{\it et~al}.} (\bibinfo{collaboration}{D0 Collaboration}),
  \bibinfo{journal}{Phys. Lett. B} \textbf{\bibinfo{volume}{690}},
  \bibinfo{pages}{5} (\bibinfo{year}{2010}).

\bibitem[{\citenamefont{Aaltonen {\it et~al}.}(2008{\natexlab{a}})}]{Aaltonen:2008sy}
\bibinfo{author}{\bibfnamefont{T.}~\bibnamefont{Aaltonen}} \bibnamefont{{\it et~al}.}
  (\bibinfo{collaboration}{CDF Collaboration}), \bibinfo{journal}{Phys. Rev.
  Lett.} \textbf{\bibinfo{volume}{101}}, \bibinfo{pages}{252001}
  (\bibinfo{year}{2008}{\natexlab{a}}).

\bibitem[{\citenamefont{Aaltonen {\it et~al}.}(2009{\natexlab{a}})}]{Aaltonen:2009jj}
\bibinfo{author}{\bibfnamefont{T.}~\bibnamefont{Aaltonen}} \bibnamefont{{\it et~al}.}
  (\bibinfo{collaboration}{CDF Collaboration}), \bibinfo{journal}{Phys. Rev.
  Lett.} \textbf{\bibinfo{volume}{103}}, \bibinfo{pages}{092002}
  (\bibinfo{year}{2009}{\natexlab{a}}).

\bibitem[{\citenamefont{Aaltonen {\it et~al}.}(2010)}]{MET_jets}
\bibinfo{author}{\bibfnamefont{T.}~\bibnamefont{Aaltonen}} \bibnamefont{{\it et~al}.}
(\bibinfo{collaboration}{CDF Collaboration}),
  \bibinfo{journal}{Phys. Rev. D} \textbf{\bibinfo{volume}{81}},
  \bibinfo{pages}{072003} (\bibinfo{year}{2010}).


\bibitem[{\citenamefont{Abulencia {\it et~al}.}(2007)}]{Abulencia:2005ix}
\bibinfo{author}{\bibfnamefont{A.}~\bibnamefont{Abulencia}}
  \bibnamefont{{\it et~al}.} (\bibinfo{collaboration}{CDF Collaboration}),
  \bibinfo{journal}{J. Phys. G} \textbf{\bibinfo{volume}{34}},
  \bibinfo{pages}{2457} (\bibinfo{year}{2007}).

\bibitem[{\citenamefont{Acosta {\it et~al}.}(2005{\natexlab{b}})}]{Acosta:2004yw}
\bibinfo{author}{\bibfnamefont{D.~E.} \bibnamefont{Acosta}}
  \bibnamefont{{\it et~al}.} (\bibinfo{collaboration}{CDF Collaboration}),
  \bibinfo{journal}{Phys. Rev. D} \textbf{\bibinfo{volume}{71}},
  \bibinfo{pages}{032001} (\bibinfo{year}{2005}{\natexlab{b}}).

\bibitem[{\citenamefont{Acosta {\it et~al}.}(2005{\natexlab{c}})}]{Acosta:2004hw}
\bibinfo{author}{\bibfnamefont{D.~E.} \bibnamefont{Acosta}}
  \bibnamefont{{\it et~al}.} (\bibinfo{collaboration}{CDF Collaboration}),
  \bibinfo{journal}{Phys. Rev. D} \textbf{\bibinfo{volume}{71}},
  \bibinfo{pages}{052003} (\bibinfo{year}{2005}{\natexlab{c}}).

\bibitem[{\citenamefont{Hill}(2004)}]{Hill:2004qb}
\bibinfo{author}{\bibfnamefont{C.~S.} \bibnamefont{Hill}}
  (\bibinfo{collaboration}{CDF Collaboration}), \bibinfo{journal}{Nucl.
  Instrum. Methods A} \textbf{\bibinfo{volume}{530}}, \bibinfo{pages}{1}
  (\bibinfo{year}{2004}).

\bibitem[{\citenamefont{Sill}(2000)}]{Sill:2000zz}
\bibinfo{author}{\bibfnamefont{A.}~\bibnamefont{Sill}}
  (\bibinfo{collaboration}{CDF Collaboration}), \bibinfo{journal}{Nucl.
  Instrum. Methods A} \textbf{\bibinfo{volume}{447}}, \bibinfo{pages}{1}
  (\bibinfo{year}{2000}).

\bibitem[{\citenamefont{Affolder {\it et~al}.}(2000)}]{Affolder:2000tj}
\bibinfo{author}{\bibfnamefont{A.~A.} \bibnamefont{Affolder}}
  \bibnamefont{{\it et~al}.} (\bibinfo{collaboration}{CDF Collaboration}),
  \bibinfo{journal}{Nucl. Instrum. Methods A} \textbf{\bibinfo{volume}{453}},
  \bibinfo{pages}{84} (\bibinfo{year}{2000}).

\bibitem[{\citenamefont{Affolder {\it et~al}.}(2004)}]{Affolder:2003ep}
\bibinfo{author}{\bibfnamefont{A.~A.} \bibnamefont{Affolder}}
  \bibnamefont{{\it et~al}.} (\bibinfo{collaboration}{CDF Collaboration}),
  \bibinfo{journal}{Nucl. Instrum. Methods A} \textbf{\bibinfo{volume}{526}},
  \bibinfo{pages}{249} (\bibinfo{year}{2004}).

\bibitem[{\citenamefont{Abe {\it et~al}.}(1992)}]{CDF:inclJet}
\bibinfo{author}{\bibfnamefont{F.}~\bibnamefont{Abe}} \bibnamefont{{\it et~al}.}
  (\bibinfo{collaboration}{CDF Collaboration}), \bibinfo{journal}{Phys. Rev. Lett.} 
  \textbf{\bibinfo{volume}{68}}, \bibinfo{pages}{1104}
  (\bibinfo{year}{1992}).

\bibitem[{\citenamefont{Balka {\it et~al}.}(1988)}]{Balka:1987ty}
\bibinfo{author}{\bibfnamefont{L.}~\bibnamefont{Balka}} \bibnamefont{{\it et~al}.}
  (\bibinfo{collaboration}{CDF Collaboration}), \bibinfo{journal}{Nucl.
  Instrum. Methods A} \textbf{\bibinfo{volume}{267}}, \bibinfo{pages}{272}
  (\bibinfo{year}{1988}).

\bibitem[{\citenamefont{Albrow {\it et~al}.}(2002)}]{Albrow:2001jw}
\bibinfo{author}{\bibfnamefont{M.~G.} \bibnamefont{Albrow}}
  \bibnamefont{{\it et~al}.} (\bibinfo{collaboration}{CDF Collaboration}),
  \bibinfo{journal}{Nucl. Instrum. Methods A} \textbf{\bibinfo{volume}{480}},
  \bibinfo{pages}{524} (\bibinfo{year}{2002}).

\bibitem[{\citenamefont{Bertolucci {\it et~al}.}(1988)}]{Bertolucci:1987zn}
\bibinfo{author}{\bibfnamefont{S.}~\bibnamefont{Bertolucci}}
  \bibnamefont{{\it et~al}.} (\bibinfo{collaboration}{CDF Collaboration}),
  \bibinfo{journal}{Nucl. Instrum. Methods A} \textbf{\bibinfo{volume}{267}},
  \bibinfo{pages}{301} (\bibinfo{year}{1988}).

\bibitem[{\citenamefont{Apollinari {\it et~al}.}(1998)\citenamefont{Apollinari,
  Goulianos, Melese, and Lindgren}}]{Apollinari:1998bg}
\bibinfo{author}{\bibfnamefont{G.}~\bibnamefont{Apollinari}},
  \bibinfo{author}{\bibfnamefont{K.}~\bibnamefont{Goulianos}},
  \bibinfo{author}{\bibfnamefont{P.}~\bibnamefont{Melese}}, \bibnamefont{and}
  \bibinfo{author}{\bibfnamefont{M.}~\bibnamefont{Lindgren}},
  \bibinfo{journal}{Nucl. Instrum. Methods A} \textbf{\bibinfo{volume}{412}},
  \bibinfo{pages}{515} (\bibinfo{year}{1998}).

\bibitem[{\citenamefont{Ascoli {\it et~al}.}(1988)}]{Ascoli:1987av}
\bibinfo{author}{\bibfnamefont{G.}~\bibnamefont{Ascoli}} \bibnamefont{{\it et~al}.},
  \bibinfo{journal}{Nucl. Instrum. Methods A} \textbf{\bibinfo{volume}{268}},
  \bibinfo{pages}{33} (\bibinfo{year}{1988}).

\bibitem[{\citenamefont{Blair {\it et~al}.}()}]{Blair:1996kx}
\bibinfo{author}{\bibfnamefont{R.}~\bibnamefont{Blair}} \bibnamefont{{\it et~al}.}
  (\bibinfo{collaboration}{CDF Collaboration}),
  \bibinfo{note}{{F}ERMILAB-PUB-96-390-E (1996)}.

\bibitem[{\citenamefont{Artikov {\it et~al}.}(2005)}]{Artikov:2004ew}
\bibinfo{author}{\bibfnamefont{A.}~\bibnamefont{Artikov}} \bibnamefont{{\it et~al}.},
  \bibinfo{journal}{Nucl. Instrum. Methods A} \textbf{\bibinfo{volume}{538}},
  \bibinfo{pages}{358} (\bibinfo{year}{2005}).

\bibitem[{\citenamefont{Acosta {\it et~al}.}(2002)}]{CLC}
\bibinfo{author}{\bibfnamefont{D.}~\bibnamefont{Acosta}} \bibnamefont{{\it et~al}.},
  \bibinfo{journal}{Nucl. Instrum. Methods A} \textbf{\bibinfo{volume}{494}},
  \bibinfo{pages}{57} (\bibinfo{year}{2002}).

\bibitem[{\citenamefont{Klimenko {\it et~al}.}()\citenamefont{Klimenko, Konigsberg,
  and Liss}}]{inelppbarxs}
\bibinfo{author}{\bibfnamefont{S.}~\bibnamefont{Klimenko}},
  \bibinfo{author}{\bibfnamefont{J.}~\bibnamefont{Konigsberg}},
  \bibnamefont{and} \bibinfo{author}{\bibfnamefont{T.~M.} \bibnamefont{Liss}},
 \bibinfo{note}{{F}ERMILAB-FN-0741 (2003)}.

\bibitem[{\citenamefont{Thomson {\it et~al}.}(2002)}]{Thomson:2002xp}
\bibinfo{author}{\bibfnamefont{E.~J.} \bibnamefont{Thomson}}
  \bibnamefont{{\it et~al}.}, \bibinfo{journal}{IEEE Trans. Nucl. Sci.}
  \textbf{\bibinfo{volume}{49}}, \bibinfo{pages}{1063} (\bibinfo{year}{2002}).

\bibitem[{\citenamefont{Downing {\it et~al}.}(2007)}]{Downing:2006xb}
\bibinfo{author}{\bibfnamefont{R.}~\bibnamefont{Downing}} \bibnamefont{{\it et~al}.},
  \bibinfo{journal}{Nucl. Instrum. Methods A} \textbf{\bibinfo{volume}{570}},
  \bibinfo{pages}{36} (\bibinfo{year}{2007}).

\bibitem[{\citenamefont{Gomez-Ceballos {\it et~al}.}(2004)}]{GomezCeballos:2004jk}
\bibinfo{author}{\bibfnamefont{G.}~\bibnamefont{Gomez-Ceballos}}
  \bibnamefont{{\it et~al}.}, \bibinfo{journal}{Nucl. Instrum. Methods A}
  \textbf{\bibinfo{volume}{518}}, \bibinfo{pages}{522} (\bibinfo{year}{2004}).

\bibitem[{\citenamefont{Bhatti {\it et~al}.}(2006)}]{Bhatti:2005ai}
\bibinfo{author}{\bibfnamefont{A.}~\bibnamefont{Bhatti}} \bibnamefont{{\it et~al}.},
  \bibinfo{journal}{Nucl. Instrum. Methods A} \textbf{\bibinfo{volume}{566}},
  \bibinfo{pages}{375} (\bibinfo{year}{2006}).

\bibitem[{\citenamefont{Maltoni and Stelzer}(2003)}]{Maltoni:2002qb}
\bibinfo{author}{\bibfnamefont{F.}~\bibnamefont{Maltoni}} \bibnamefont{and}
  \bibinfo{author}{\bibfnamefont{T.}~\bibnamefont{Stelzer}},
  \bibinfo{journal}{J. High Energy Phys.} \textbf{\bibinfo{volume}{02}},
  \bibinfo{pages}{027} (\bibinfo{year}{2003}).

\bibitem[{\citenamefont{Lai {\it et~al}.}(2000)}]{Lai:1999wy}
\bibinfo{author}{\bibfnamefont{H.~L.} \bibnamefont{Lai}} \bibnamefont{{\it et~al}.}
  (\bibinfo{collaboration}{CTEQ Collaboration}), \bibinfo{journal}{Eur. Phys.
  J. C} \textbf{\bibinfo{volume}{12}}, \bibinfo{pages}{375}
  (\bibinfo{year}{2000}).

\bibitem[{\citenamefont{Sullivan}(2005)}]{Sullivan:2005ar}
\bibinfo{author}{\bibfnamefont{Z.}~\bibnamefont{Sullivan}},
  \bibinfo{journal}{Phys. Rev. D} \textbf{\bibinfo{volume}{72}},
  \bibinfo{pages}{094034} (\bibinfo{year}{2005}).

\bibitem[{\citenamefont{Sj{\"o}strand {\it et~al}.}(2001)}]{Sjostrand:2000wi}
\bibinfo{author}{\bibfnamefont{T.}~\bibnamefont{Sj{\"o}strand}}
  \bibnamefont{{\it et~al}.}, \bibinfo{journal}{Comput. Phys. Commun.}
  \textbf{\bibinfo{volume}{135}}, \bibinfo{pages}{238} (\bibinfo{year}{2001}).

\bibitem[{\citenamefont{Sj{\"o}strand {\it et~al}.}(2006)\citenamefont{Sj{\"o}strand,
  Mrenna, and Skands}}]{Sjostrand:2006za}
\bibinfo{author}{\bibfnamefont{T.}~\bibnamefont{Sj{\"o}strand}},
  \bibinfo{author}{\bibfnamefont{S.}~\bibnamefont{Mrenna}}, \bibnamefont{and}
  \bibinfo{author}{\bibfnamefont{P.}~\bibnamefont{Skands}},
  \bibinfo{journal}{J. High Energy Phys.} \textbf{\bibinfo{volume}{05}},
  \bibinfo{pages}{026} (\bibinfo{year}{2006}).

\bibitem[{\citenamefont{Boos {\it et~al}.}(2006)\citenamefont{Boos, Bunichev, Dudko,
  Savrin, and Sherstnev}}]{Boos:2006af}
\bibinfo{author}{\bibfnamefont{E.~E.} \bibnamefont{Boos}},
  \bibinfo{author}{\bibfnamefont{V.~E.} \bibnamefont{Bunichev}},
  \bibinfo{author}{\bibfnamefont{L.~V.} \bibnamefont{Dudko}},
  \bibinfo{author}{\bibfnamefont{V.~I.} \bibnamefont{Savrin}},
  \bibnamefont{and} \bibinfo{author}{\bibfnamefont{A.~V.}
  \bibnamefont{Sherstnev}}, \bibinfo{journal}{Phys. Atom. Nucl.}
  \textbf{\bibinfo{volume}{69}}, \bibinfo{pages}{1317} (\bibinfo{year}{2006}).

\bibitem[{\citenamefont{Campbell
  {\it et~al}.}(2009{\natexlab{b}})\citenamefont{Campbell, Frederix, Maltoni, and
  Tramontano}}]{Campbell:2009ss}
\bibinfo{author}{\bibfnamefont{J.~M.} \bibnamefont{Campbell}},
  \bibinfo{author}{\bibfnamefont{R.}~\bibnamefont{Frederix}},
  \bibinfo{author}{\bibfnamefont{F.}~\bibnamefont{Maltoni}}, \bibnamefont{and}
  \bibinfo{author}{\bibfnamefont{F.}~\bibnamefont{Tramontano}},
  \bibinfo{journal}{Phys. Rev. Lett.} \textbf{\bibinfo{volume}{102}},
  \bibinfo{pages}{182003} (\bibinfo{year}{2009}{\natexlab{b}}).

\bibitem[{\citenamefont{Campbell
  {\it et~al}.}(2009{\natexlab{a}})\citenamefont{Campbell, Frederix, Maltoni, and
  Tramontano}}]{Campbell:2009gj}
\bibinfo{author}{\bibfnamefont{J.~M.} \bibnamefont{Campbell}},
  \bibinfo{author}{\bibfnamefont{R.}~\bibnamefont{Frederix}},
  \bibinfo{author}{\bibfnamefont{F.}~\bibnamefont{Maltoni}}, \bibnamefont{and}
  \bibinfo{author}{\bibfnamefont{F.}~\bibnamefont{Tramontano}},
  \bibinfo{journal}{J. High Energy Phys.} \textbf{\bibinfo{volume}{10}},
  \bibinfo{pages}{042} (\bibinfo{year}{2009}{\natexlab{a}}).

\bibitem[{\citenamefont{Dokshitzer}(1977)}]{Dokshitzer:1977sg}
\bibinfo{author}{\bibfnamefont{Y.~L.} \bibnamefont{Dokshitzer}},
  \bibinfo{journal}{Sov. Phys. JETP} \textbf{\bibinfo{volume}{46}},
  \bibinfo{pages}{641} (\bibinfo{year}{1977}).

\bibitem[{\citenamefont{Gribov and Lipatov}(1972)}]{Gribov:1972ri}
\bibinfo{author}{\bibfnamefont{V.~N.} \bibnamefont{Gribov}} \bibnamefont{and}
  \bibinfo{author}{\bibfnamefont{L.~N.} \bibnamefont{Lipatov}},
  \bibinfo{journal}{Sov. J. Nucl. Phys.} \textbf{\bibinfo{volume}{15}},
  \bibinfo{pages}{438} (\bibinfo{year}{1972}).

\bibitem[{\citenamefont{Altarelli and Parisi}(1977)}]{Altarelli:1977zs}
\bibinfo{author}{\bibfnamefont{G.}~\bibnamefont{Altarelli}} \bibnamefont{and}
  \bibinfo{author}{\bibfnamefont{G.}~\bibnamefont{Parisi}},
  \bibinfo{journal}{Nucl. Phys. B} \textbf{\bibinfo{volume}{126}},
  \bibinfo{pages}{298} (\bibinfo{year}{1977}).

\bibitem[{\citenamefont{Lueck}({\natexlab{a}})}]{Lueck:2006hz}
\bibinfo{author}{\bibfnamefont{J.}~\bibnamefont{Lueck}},
  \bibinfo{note}{diplom Thesis, University of Karlsruhe, 2006,
  FERMILAB-MASTERS-2006-01}.

\bibitem[{\citenamefont{Lueck}({\natexlab{b}})}]{Lueck:2009zz}
\bibinfo{author}{\bibfnamefont{J.}~\bibnamefont{Lueck}},
  \bibinfo{note}{Ph.D. Thesis, University of Karlsruhe, 2009,
  FERMILAB-THESIS-2009-33}.

\bibitem[{\citenamefont{Abbiendi {\it et~al}.}(1999)}]{Abbiendi:1998eh}
\bibinfo{author}{\bibfnamefont{G.}~\bibnamefont{Abbiendi}} \bibnamefont{{\it et~al}.}
  (\bibinfo{collaboration}{OPAL Collaboration}), \bibinfo{journal}{Eur. Phys.
  J. C} \textbf{\bibinfo{volume}{8}}, \bibinfo{pages}{217}
  (\bibinfo{year}{1999}).

\bibitem[{\citenamefont{Donini {\it et~al}.}(2008)}]{Donini:2008nt}
\bibinfo{author}{\bibfnamefont{J.}~\bibnamefont{Donini}} \bibnamefont{{\it et~al}.},
  \bibinfo{journal}{Nucl. Instrum. Methods A} \textbf{\bibinfo{volume}{596}},
  \bibinfo{pages}{354} (\bibinfo{year}{2008}).

\bibitem[{\citenamefont{Jadach {\it et~al}.}(1993)\citenamefont{Jadach, Was, Decker,
  and Kuhn}}]{Jadach:1993hs}
\bibinfo{author}{\bibfnamefont{S.}~\bibnamefont{Jadach}},
  \bibinfo{author}{\bibfnamefont{Z.}~\bibnamefont{Was}},
  \bibinfo{author}{\bibfnamefont{R.}~\bibnamefont{Decker}}, \bibnamefont{and}
  \bibinfo{author}{\bibfnamefont{J.~H.} \bibnamefont{Kuhn}},
  \bibinfo{journal}{Comput. Phys. Commun.} \textbf{\bibinfo{volume}{76}},
  \bibinfo{pages}{361} (\bibinfo{year}{1993}).

\bibitem[{\citenamefont{Bonciani {\it et~al}.}(1998)\citenamefont{Bonciani, Catani,
  Mangano, and Nason}}]{Bonciani:1998vc}
\bibinfo{author}{\bibfnamefont{R.}~\bibnamefont{Bonciani}},
  \bibinfo{author}{\bibfnamefont{S.}~\bibnamefont{Catani}},
  \bibinfo{author}{\bibfnamefont{M.~L.} \bibnamefont{Mangano}},
  \bibnamefont{and} \bibinfo{author}{\bibfnamefont{P.}~\bibnamefont{Nason}},
  \bibinfo{journal}{Nucl. Phys. B} \textbf{\bibinfo{volume}{529}},
  \bibinfo{pages}{424} (\bibinfo{year}{1998}).

\bibitem[{\citenamefont{Cacciari {\it et~al}.}(2004)\citenamefont{Cacciari, Frixione,
  Mangano, Nason, and Ridolfi}}]{Cacciari:2003fi}
\bibinfo{author}{\bibfnamefont{M.}~\bibnamefont{Cacciari}},
  \bibinfo{author}{\bibfnamefont{S.}~\bibnamefont{Frixione}},
  \bibinfo{author}{\bibfnamefont{M.~L.} \bibnamefont{Mangano}},
  \bibinfo{author}{\bibfnamefont{P.}~\bibnamefont{Nason}}, \bibnamefont{and}
  \bibinfo{author}{\bibfnamefont{G.}~\bibnamefont{Ridolfi}},
  \bibinfo{journal}{J. High Energy Phys.} \textbf{\bibinfo{volume}{04}},
  \bibinfo{pages}{068} (\bibinfo{year}{2004}).

\bibitem[{\citenamefont{Berger and Contopanagos}(1997)}]{Berger:1997ed}
\bibinfo{author}{\bibfnamefont{E.~L.} \bibnamefont{Berger}} \bibnamefont{and}
  \bibinfo{author}{\bibfnamefont{H.}~\bibnamefont{Contopanagos}}
  (\bibinfo{year}{1997}), \eprint{arXiv:hep-ph/9706356}.

\bibitem[{\citenamefont{Campbell and Ellis}(1999)}]{PhysRevD.60.113006}
\bibinfo{author}{\bibfnamefont{J.~M.} \bibnamefont{Campbell}} \bibnamefont{and}
  \bibinfo{author}{\bibfnamefont{R.~K.} \bibnamefont{Ellis}},
  \bibinfo{journal}{Phys. Rev. D} \textbf{\bibinfo{volume}{60}},
  \bibinfo{pages}{113006} (\bibinfo{year}{1999}).

\bibitem[{\citenamefont{Mangano {\it et~al}.}(2003)\citenamefont{Mangano, Moretti,
  Piccinini, Pittau, and Polosa}}]{Mangano:2002ea}
\bibinfo{author}{\bibfnamefont{M.~L.} \bibnamefont{Mangano}},
  \bibinfo{author}{\bibfnamefont{M.}~\bibnamefont{Moretti}},
  \bibinfo{author}{\bibfnamefont{F.}~\bibnamefont{Piccinini}},
  \bibinfo{author}{\bibfnamefont{R.}~\bibnamefont{Pittau}}, \bibnamefont{and}
  \bibinfo{author}{\bibfnamefont{A.~D.} \bibnamefont{Polosa}},
  \bibinfo{journal}{J. High Energy Phys.} \textbf{\bibinfo{volume}{07}},
  \bibinfo{pages}{001} (\bibinfo{year}{2003}).

\bibitem[{\citenamefont{Aaltonen {\it et~al}.}(2008{\natexlab{b}})}]{Aaltonen:2007cp}
\bibinfo{author}{\bibfnamefont{T.}~\bibnamefont{Aaltonen}} \bibnamefont{{\it et~al}.}
  (\bibinfo{collaboration}{CDF Collaboration}), \bibinfo{journal}{Phys. Rev.
  Lett.} \textbf{\bibinfo{volume}{100}}, \bibinfo{pages}{102001}
  (\bibinfo{year}{2008}{\natexlab{b}}).

\bibitem[{\citenamefont{Bern {\it et~al}.}(1997)\citenamefont{Bern, Dixon, Kosower,
  and Weinzierl}}]{Bern:1996ka}
\bibinfo{author}{\bibfnamefont{Z.}~\bibnamefont{Bern}},
  \bibinfo{author}{\bibfnamefont{L.~J.} \bibnamefont{Dixon}},
  \bibinfo{author}{\bibfnamefont{D.~A.} \bibnamefont{Kosower}},
  \bibnamefont{and}
  \bibinfo{author}{\bibfnamefont{S.}~\bibnamefont{Weinzierl}},
  \bibinfo{journal}{Nucl. Phys. B} \textbf{\bibinfo{volume}{489}},
  \bibinfo{pages}{3} (\bibinfo{year}{1997}).

\bibitem[{\citenamefont{Bern {\it et~al}.}(1998)\citenamefont{Bern, Dixon, and
  Kosower}}]{Bern:1997sc}
\bibinfo{author}{\bibfnamefont{Z.}~\bibnamefont{Bern}},
  \bibinfo{author}{\bibfnamefont{L.~J.} \bibnamefont{Dixon}}, \bibnamefont{and}
  \bibinfo{author}{\bibfnamefont{D.~A.} \bibnamefont{Kosower}},
  \bibinfo{journal}{Nucl. Phys. B} \textbf{\bibinfo{volume}{513}},
  \bibinfo{pages}{3} (\bibinfo{year}{1998}).

\bibitem[{\citenamefont{Giele {\it et~al}.}(1996)\citenamefont{Giele, Keller, and
  Laenen}}]{Giele:1996aa}
\bibinfo{author}{\bibfnamefont{W.~T.} \bibnamefont{Giele}},
  \bibinfo{author}{\bibfnamefont{S.}~\bibnamefont{Keller}}, \bibnamefont{and}
  \bibinfo{author}{\bibfnamefont{E.}~\bibnamefont{Laenen}},
  \bibinfo{journal}{Nucl. Phys. Proc. Suppl.} \textbf{\bibinfo{volume}{51C}},
  \bibinfo{pages}{255} (\bibinfo{year}{1996}).

\bibitem[{\citenamefont{Ellis and Veseli}(1999)}]{Ellis:1998fv}
\bibinfo{author}{\bibfnamefont{R.~K.} \bibnamefont{Ellis}} \bibnamefont{and}
  \bibinfo{author}{\bibfnamefont{S.}~\bibnamefont{Veseli}},
  \bibinfo{journal}{Phys. Rev. D} \textbf{\bibinfo{volume}{60}},
  \bibinfo{pages}{011501} (\bibinfo{year}{1999}).

\bibitem[{\citenamefont{Febres~Cordero
  {\it et~al}.}(2006)\citenamefont{Febres~Cordero, Reina, and
  Wackeroth}}]{FebresCordero:2006sj}
\bibinfo{author}{\bibfnamefont{F.}~\bibnamefont{Febres~Cordero}},
  \bibinfo{author}{\bibfnamefont{L.}~\bibnamefont{Reina}}, \bibnamefont{and}
  \bibinfo{author}{\bibfnamefont{D.}~\bibnamefont{Wackeroth}},
  \bibinfo{journal}{Phys. Rev. D} \textbf{\bibinfo{volume}{74}},
  \bibinfo{pages}{034007} (\bibinfo{year}{2006}).

\bibitem[{\citenamefont{Campbell {\it et~al}.}(2007)\citenamefont{Campbell, Ellis,
  Maltoni, and Willenbrock}}]{Campbell:2006cu}
\bibinfo{author}{\bibfnamefont{J.~M.} \bibnamefont{Campbell}},
  \bibinfo{author}{\bibfnamefont{R.~K.} \bibnamefont{Ellis}},
  \bibinfo{author}{\bibfnamefont{F.}~\bibnamefont{Maltoni}}, \bibnamefont{and}
  \bibinfo{author}{\bibfnamefont{S.}~\bibnamefont{Willenbrock}},
  \bibinfo{journal}{Phys. Rev. D} \textbf{\bibinfo{volume}{75}},
  \bibinfo{pages}{054015} (\bibinfo{year}{2007}).

\bibitem[{\citenamefont{Campbell {\it et~al}.}(2009{\natexlab{b}})}]{Campbell:2008hh}
\bibinfo{author}{\bibfnamefont{J.~M.} \bibnamefont{Campbell}}
  \bibnamefont{{\it et~al}.}, \bibinfo{journal}{Phys. Rev. D}
  \textbf{\bibinfo{volume}{79}}, \bibinfo{pages}{034023}
  (\bibinfo{year}{2009}{\natexlab{b}}).

\bibitem[{\citenamefont{Cordero {\it et~al}.}(2009)\citenamefont{Cordero, Reina, and
  Wackeroth}}]{Cordero:2009kv}
\bibinfo{author}{\bibfnamefont{F.~F.} \bibnamefont{Cordero}},
  \bibinfo{author}{\bibfnamefont{L.}~\bibnamefont{Reina}}, \bibnamefont{and}
  \bibinfo{author}{\bibfnamefont{D.}~\bibnamefont{Wackeroth}},
  \bibinfo{journal}{Phys. Rev. D} \textbf{\bibinfo{volume}{80}},
  \bibinfo{pages}{034015} (\bibinfo{year}{2009}).

\bibitem[{\citenamefont{Aaltonen {\it et~al}.}(2008{\natexlab{c}})}]{Aaltonen:2007ip}
\bibinfo{author}{\bibfnamefont{T.}~\bibnamefont{Aaltonen}} \bibnamefont{{\it et~al}.}
  (\bibinfo{collaboration}{CDF Collaboration}), \bibinfo{journal}{Phys. Rev. D}
  \textbf{\bibinfo{volume}{77}}, \bibinfo{pages}{011108}
  (\bibinfo{year}{2008}{\natexlab{c}}).

\bibitem[{\citenamefont{Aaltonen {\it et~al}.}(2008{\natexlab{d}})}]{Wcharm:2007dm}
\bibinfo{author}{\bibfnamefont{T.}~\bibnamefont{Aaltonen}} \bibnamefont{{\it et~al}.}
  (\bibinfo{collaboration}{CDF Collaboration}), \bibinfo{journal}{Phys. Rev.
  Lett.} \textbf{\bibinfo{volume}{100}}, \bibinfo{pages}{091803}
  (\bibinfo{year}{2008}{\natexlab{d}}).

\bibitem[{\citenamefont{Richter}()}]{Richter:2007zzc}
\bibinfo{author}{\bibfnamefont{S.}~\bibnamefont{Richter}},
  \bibinfo{note}{Ph.D. Thesis, University of Karlsruhe, 2007,
  FERMILAB-THESIS-2007-35}.

\bibitem[{\citenamefont{Feindt and Kerzel}(2006)}]{Feindt:2006pm}
\bibinfo{author}{\bibfnamefont{M.}~\bibnamefont{Feindt}} \bibnamefont{and}
  \bibinfo{author}{\bibfnamefont{U.}~\bibnamefont{Kerzel}},
  \bibinfo{journal}{Nucl. Instrum. Methods A} \textbf{\bibinfo{volume}{559}},
  \bibinfo{pages}{190} (\bibinfo{year}{2006}).

\bibitem[{\citenamefont{Yuan}(1990)}]{Yuan:1989tc}
\bibinfo{author}{\bibfnamefont{C.~P.} \bibnamefont{Yuan}},
  \bibinfo{journal}{Phys. Rev. D} \textbf{\bibinfo{volume}{41}},
  \bibinfo{pages}{42} (\bibinfo{year}{1990}).

\bibitem[{\citenamefont{Ackerstaff {\it et~al}.}(1998)}]{Ackerstaff:1997cz}
\bibinfo{author}{\bibfnamefont{K.}~\bibnamefont{Ackerstaff}}
  \bibnamefont{{\it et~al}.} (\bibinfo{collaboration}{OPAL Collaboration}),
  \bibinfo{journal}{Eur. Phys. J. C} \textbf{\bibinfo{volume}{1}},
  \bibinfo{pages}{425} (\bibinfo{year}{1998}).

\bibitem[{\citenamefont{Budd}()}]{Budd:2008zz}
\bibinfo{author}{\bibfnamefont{S.~R.} \bibnamefont{Budd}},
  \bibinfo{note}{Ph.D. Thesis, University of Illinois, 2008,
  FERMILAB-THESIS-2008-41}.

\bibitem[{\citenamefont{Nakamura}()}]{Nakamura:2009zzd}
\bibinfo{author}{\bibfnamefont{K.}~\bibnamefont{Nakamura}},
  \bibinfo{note}{Ph.D. Thesis, University of Tsukuba, 2009,
  FERMILAB-THESIS-2009-13}.

\bibitem[{\citenamefont{Dong}()}]{Dong:2008zzc}
\bibinfo{author}{\bibfnamefont{P.~J.} \bibnamefont{Dong}},
  \bibinfo{note}{Ph.D. Thesis, University of California at Los Angeles, 2008,
  FERMILAB-THESIS-2008-12}.

\bibitem[{\citenamefont{Murayama and Hagiwara}(1992)}]{Murayama:1992}
\bibinfo{author}{\bibfnamefont{I.}~\bibnamefont{Murayama},
  \bibfnamefont{H.~Watanabe}} \bibnamefont{and}
  \bibinfo{author}{\bibfnamefont{K.}~\bibnamefont{Hagiwara}},
  \bibinfo{type}{Tech. Rep.} \bibinfo{number}{91-11},
  \bibinfo{institution}{KEK} (\bibinfo{year}{1992}).

\bibitem[{\citenamefont{Genz and Malik}(1980)}]{Genz:1980}
\bibinfo{author}{\bibfnamefont{A.}~\bibnamefont{Genz}} \bibnamefont{and}
  \bibinfo{author}{\bibfnamefont{A.}~\bibnamefont{Malik}}, \bibinfo{journal}{J.
  Comput. Appl. Math.} \textbf{\bibinfo{volume}{6}}, \bibinfo{pages}{295}
  (\bibinfo{year}{1980}), \bibinfo{note}{implemented as CERNLIB algorithm D120,
  documented in http://wwwasdoc.web.cern.ch/ \\
  wwwasdoc/shortwrupsdir/d151/top.html.}

\bibitem[{\citenamefont{Hahn}(2005)}]{Hahn:2004fe}
\bibinfo{author}{\bibfnamefont{T.}~\bibnamefont{Hahn}},
  \bibinfo{journal}{Comput. Phys. Commun.} \textbf{\bibinfo{volume}{168}},
  \bibinfo{pages}{78} (\bibinfo{year}{2005}).

\bibitem[{\citenamefont{Neyman and Pearson}(1933)}]{Neyman:1933}
\bibinfo{author}{\bibfnamefont{J.}~\bibnamefont{Neyman}} \bibnamefont{and}
  \bibinfo{author}{\bibfnamefont{E.}~\bibnamefont{Pearson}},
  \bibinfo{journal}{Phil. Trans. of the Royal Soc. of London A}
  \textbf{\bibinfo{volume}{31}}, \bibinfo{pages}{289} (\bibinfo{year}{1933}).

\bibitem[{\citenamefont{Papaikonomou}()}]{Papaikonomou:2009zz}
\bibinfo{author}{\bibfnamefont{A.}~\bibnamefont{Papaikonomou}},
  \bibinfo{note}{Ph.D. Thesis, University of Karlsruhe, 2009,
  FERMILAB-THESIS-2009-21}.

\bibitem[{\citenamefont{Hocker {\it et~al}.}(2007)}]{Hocker:2007ht}
\bibinfo{author}{\bibfnamefont{A.}~\bibnamefont{Hocker}} \bibnamefont{{\it et~al}.},
  \bibinfo{journal}{PoS} \textbf{\bibinfo{volume}{ACAT}}, \bibinfo{pages}{040}
  (\bibinfo{year}{2007}).

\bibitem[{\citenamefont{Casal~Larana}()}]{CasalLarana:2010zz}
\bibinfo{author}{\bibfnamefont{B.}~\bibnamefont{Casal~Larana}},
  \bibinfo{note}{Ph.D. Thesis, University of Cantabria, 2010,
  FERMILAB-THESIS-2010-04}.

\bibitem[{\citenamefont{Gini}()}]{gini}
\bibinfo{author}{\bibfnamefont{C.}~\bibnamefont{Gini}},
  \emph{\bibinfo{title}{{Variabilit\`{a} e Mutabilit\`{a}}}},
  \bibinfo{howpublished}{{(1912), reprinted in {\it Memorie di Metodologica
  Statistica}, edited by E.~Pizetti and T.~Savemini, Rome: Libreria Eredi
  Virgilio Veschi, 1955}}.

\bibitem[{\citenamefont{Freund and Schapire}(1997)}]{Freund:1997}
\bibinfo{author}{\bibfnamefont{Y.}~\bibnamefont{Freund}} \bibnamefont{and}
  \bibinfo{author}{\bibfnamefont{R.}~\bibnamefont{Schapire}},
  \bibinfo{journal}{Computer and System Science} \textbf{\bibinfo{volume}{55}},
  \bibinfo{pages}{119} (\bibinfo{year}{1997}).

\bibitem[{\citenamefont{Breiman {\it et~al}.}(1984)\citenamefont{Breiman, Friedman,
  Olshen, and Stone}}]{Breiman:1984}
\bibinfo{author}{\bibfnamefont{L.}~\bibnamefont{Breiman}},
  \bibinfo{author}{\bibfnamefont{J.}~\bibnamefont{Friedman}},
  \bibinfo{author}{\bibfnamefont{R.}~\bibnamefont{Olshen}}, \bibnamefont{and}
  \bibinfo{author}{\bibfnamefont{C.}~\bibnamefont{Stone}},
  \emph{\bibinfo{title}{Classification and Regression Trees}}
  (\bibinfo{publisher}{Wadsworth}, \bibinfo{year}{1984}).

\bibitem[{\citenamefont{Abulencia
  {\it et~al}.}(2006{\natexlab{c}})}]{Abulencia:2005aj}
\bibinfo{author}{\bibfnamefont{A.}~\bibnamefont{Abulencia}}
  \bibnamefont{{\it et~al}.} (\bibinfo{collaboration}{CDF Collaboration}),
  \bibinfo{journal}{Phys. Rev. D} \textbf{\bibinfo{volume}{73}},
  \bibinfo{pages}{032003} (\bibinfo{year}{2006}{\natexlab{c}}).

\bibitem[{\citenamefont{Martin {\it et~al}.}(1998)\citenamefont{Martin, Roberts,
  Stirling, and Thorne}}]{Martin:1998sq}
\bibinfo{author}{\bibfnamefont{A.~D.} \bibnamefont{Martin}},
  \bibinfo{author}{\bibfnamefont{R.~G.} \bibnamefont{Roberts}},
  \bibinfo{author}{\bibfnamefont{W.~J.} \bibnamefont{Stirling}},
  \bibnamefont{and} \bibinfo{author}{\bibfnamefont{R.~S.}
  \bibnamefont{Thorne}}, \bibinfo{journal}{Eur. Phys. J. C}
  \textbf{\bibinfo{volume}{4}}, \bibinfo{pages}{463} (\bibinfo{year}{1998}).

\bibitem[{\citenamefont{Amsler {\it et~al}.}(2008{\natexlab{b}})}]{pdgstats}
\bibinfo{author}{\bibfnamefont{C.}~\bibnamefont{Amsler}} \bibnamefont{{\it et~al}.}
  (\bibinfo{collaboration}{Particle Data Group}), \bibinfo{journal}{Phys. Lett.
  B} \textbf{\bibinfo{volume}{667}}, \bibinfo{pages}{324}
  (\bibinfo{year}{2008}{\natexlab{b}}).

\bibitem[{\citenamefont{Aaltonen {\it et~al}.}(2009{\natexlab{b}})}]{Aaltonen:2008ec}
\bibinfo{author}{\bibfnamefont{T.}~\bibnamefont{Aaltonen}} \bibnamefont{{\it et~al}.}
  (\bibinfo{collaboration}{CDF Collaboration}), \bibinfo{journal}{Phys. Rev.
  Lett.} \textbf{\bibinfo{volume}{102}}, \bibinfo{pages}{021802}
  (\bibinfo{year}{2009}{\natexlab{b}}).

\bibitem[{\citenamefont{Stanley and Miikkulainen}(2002)}]{neat}
\bibinfo{author}{\bibfnamefont{K.}~\bibnamefont{Stanley}} \bibnamefont{and}
  \bibinfo{author}{\bibfnamefont{R.}~\bibnamefont{Miikkulainen}},
  \bibinfo{journal}{Evolutionary Compuation} \textbf{\bibinfo{volume}{10}}, \bibinfo{pages}{99}
  (\bibinfo{year}{2002}).

\bibitem[{\citenamefont{{Tevatron Electroweak Working Group,
  arXiv:0908.2171}}(2009)}]{cdfd0combo}
\bibinfo{author}{\bibnamefont{{Tevatron Electroweak Working Group,
  arXiv:0908.2171}}} (\bibinfo{year}{2009}).

\bibitem[{\citenamefont{Abazov {\it et~al}.}(2010)}]{Abazov:2009pa}
\bibinfo{author}{\bibfnamefont{V.~M.} \bibnamefont{Abazov}}
  \bibnamefont{{\it et~al}.} (\bibinfo{collaboration}{D0 Collaboration}),
  \bibinfo{journal}{Phys. Lett. B} \textbf{\bibinfo{volume}{682}},
  \bibinfo{pages}{363} (\bibinfo{year}{2010}).

\end{thebibliography}

\end{document}